# SPECTROSCOPIC OBSERVATIONS OF THE GALEX NEARBY YOUNG STAR SURVEY SAMPLE. I. NEARBY MOVING GROUP CANDIDATES

Navya Nagananda[1], Laura Vican[2], Ben Zuckerman[2], David Rodriguez[3], Alexander Binks[5], and Joel Kastner[1,4]
[1]Chester F. Carlson Center for Imaging Science, Rochester Institute of Technology, Rochester, NY 14623, USA
[2]Department of Physics and Astronomy, University of California, Los Angeles, CA 90095, USA
[3]Space Telescope Science Institute, Baltimore, MD, USA
[4]School of Physics & Astronomy and Laboratory for Multiwavelength Astrophysics, Rochester Institute of Technology, Rochester, NY 14623, USA and
[5]Institut für Astronomie und Astrophysik, Eberhard-Karls Universität Tübingen, Sand 1, 72076 Tübingen, Germany

## ABSTRACT

The GALEX Nearby Young Star Search (GALNYSS) yielded the identification of more than 2000 late-type stars that, based on their ultraviolet and infrared colors and pre-Gaia proper motions, are potentially of age $< 200$ Myr and lie within $\sim 120$ pc of Earth. We present the results of a campaign of medium- and high-resolution optical spectroscopy of 471 GALNYSS stars aimed at confirming their youth and their potential membership in nearby young stellar moving groups. We present radial velocity (RV), Li absorption, and H-$\alpha$ emission measurements for these spectroscopically observed GALNYSS stars, and assess their Li absorption and optical emission-line properties and infrared excesses. Our RV measurements are combined with literature and Gaia DR3 RV measurements and Gaia DR3 astrometry and photometry to obtain the spatiokinematics and color-magnitude positions of GALNYSS stars. We use these results to assess membership in the TW Hya, Tuc-Hor, Carina, Columba, and Argus Associations and the $\beta$PMG and AB Dor moving groups. We have identified 132 stars as candidate members of these seven groups; roughly half of these candidates are newly identified on the basis of data presented here. At least one-third of the 132 candidates are spectroscopic and/or photometric binaries and/or have comoving (visual) binary companions in Gaia DR3. The contingent of young, low-mass stars in the solar vicinity we identify here should provide excellent subjects for future direct imaging exoplanet surveys and studies of the early evolution of low-mass stars and their planetary progeny.

## 1. INTRODUCTION

Low-mass stars of late K- and M-type have enjoyed a burst of attention as a result of the releases of Gaia Space Astrometry mission data (Gaia Collaboration et al. 2018a, 2020, 2023) and the realization that many and perhaps most such stars are hosts to planets of mass similar to that of Earth (e.g., Muirhead et al. 2015). Young *low-mass* stars in the solar vicinity are particularly attractive targets for direct imaging exoplanet surveys, due to the favorable combination of relatively low contrast and large angular separation (**for a given physical separation**) between star and planet. As with solar-type stars, the main challenge is to identify those low-mass stars that are young enough that any planets around them would still be self-luminous and, hence, bright enough to detect via direct (thermal infrared) imaging.

Methods used to determine ages for solar-type stars, such as those from Asteroseismology and/or upper-main sequence fitting, are not applicable for low-mass stars, however. Most low-mass stars are magnetically active so, while intense chromospheric (UV or Balmer line) emission or coronal (X-ray) emission can be used to isolate candidate young, low-mass stars from the field, the mere presence of such chromospheric and/or coronal emission sources is not enough to establish youth since, e.g., older tidally-locked binary systems are often highly active (e.g., Stelzer et al. 2013). Fully convective early/mid-M stars have burned their full lithium content by an age of $\sim 20$ Myr (Jeffries 2014; Lyubimkov 2016; Randich & Magrini 2021). Thus, except in the very youngest stars, depletion of photospheric lithium cannot be used as an age indicator.

One particularly effective means to establish stellar youth, and ascertain the age of a star, is to identify that star as a member of a nearby young moving group (e.g., Zuckerman & Song 2004; Torres et al. 2008; Gagné 2024, hereafter NYMG). These are loose collections of stars that share common Galactic space motions (kinematics) and, hence likely formed together (i.e., are coeval). Since the initial mass functions of both field stars and young Galactic clusters are heavily weighted towards low mass (e.g., Lada 2006), NYMGs should be heavily populated with such stars. However, the known membership of NYMGs remains demonstrably deficient in low-mass stars (Bowler et al. 2019, and references therein). To address this deficiency, which was already well-established more than a decade ago (Torres et al. 2008), Rodriguez et al. (2011) developed a method for identifying candidate low-mass NYMG members on the basis of

Corresponding author: nn3264@rit.edu

both (pre-Gaia) Galactic space motions ($UVW$ velocities) and young-star near-UV (NUV) characteristics, with the latter obtained by the Galaxy Evolution Explorer (GALEX; Martin et al. 2005). This method takes advantage of the observation that young stars stand out in NUV–IR color-color plots of late-type stars, due to excess NUV emission from active chromospheres (e.g., Rodriguez et al. 2011, their Fig. 2). The distinctive NUV excesses characteristic of youthful low-mass stars was also noted by other, parallel, investigations (e.g. Shkolnik et al. 2011, Guinan & Engle 2009). Subsequently, Rodriguez et al. (2013) exploited this methodology and the broad sky coverage of GALEX and the Two-Micron All-Sky Survey (2MASS; Cutri et al. 2003) to generate a list of over 2,000 candidate young, nearby, low-mass stars (hereafter, the GALEX Nearby Young Star Survey [GALNYSS]).

Here we aim to confirm or refute the youth and possible NYMG membership of a substantial sample of GALNYSS stars. First, we used six instruments at four telescopes in both the Northern and Southern hemispheres to obtain high- and medium- resolution spectroscopy for a subsample of more than 450 GALNYSS stars. These data allow us to access spectral diagnostics of youth, such as measurements of lithium abundance and assessments of levels of magnetic activity (e.g., H$\alpha$ and Ca II H&K emission), and to measure radial velocities. The results from this spectroscopic survey presented here are a subset of those presented and analyzed in Vican (2016). Second, we leverage the data releases from the Gaia Space Astrometry Mission (Data Releases 2 and 3 and Early Data Release 3, hereafter DR2, DR3, and EDR3; Gaia Collaboration et al. 2018a, 2020, 2023), which have greatly accelerated the pace of identification and study of NYMGs and their members (see, e.g., Gagné & Faherty 2018; Lee & Song 2019; Schneider et al. 2019; Kerr et al. 2021; Luhman 2023). As the Vican (2016) thesis work was executed pre-Gaia, we searched the Gaia DR3 catalog (Gaia Collaboration et al. 2023) at the positions of these spectroscopically observed GALNYSS stars, so as to determine their distances, kinematics, and color-magnitude diagram positions.

In the present paper, we present the results of our spectroscopic campaign targeting GALNYSS stars, focusing on detections and measurements of spectral features that are diagnostic of youth (i.e., photospheric Li and emission lines indicative of accretion and/or magnetic activity) and measurements of stellar radial velocities. We then use the combination of our spectroscopic data and Gaia data to evaluate GALNYSS star membership in seven well-studied nearby young stellar groups: the TW Hya, Tuc-Hor, Carina, Columba, and Argus Associations and the $\beta$PMG and AB Dor moving groups. In a subsequent paper (Binks et al., in prep.; hereafter Paper II) we evaluate the likelihood of membership of the remaining GALNYSS sample stars in additional NYMGs and nearby star-forming regions (e.g., Sco-Cen, Upper Sco, Lower Centaurus Crux), and we investigate the natures of Li-rich GALNYSS stars that are not clearly associated with known NYMGs.

## 2. SAMPLE AND DATA

### 2.1. *Sample: GALNYSS Catalog Stars*

The construction of the GALNYSS star catalog is described in detail in Rodriguez et al. (2013). In brief, the stars were assembled by cross-matching GALEX UV sources against 2MASS and WISE IR sources, using a 3$''$ cross-match radius. The resulting matches were then subject to a set of down-selection criteria that ensure uniqueness of UV–IR source matches. Stars were initially selected based on their UV fluxes relative to their IR fluxes, wherein 9.5$\leq$NUV-W1<12.5 with NUV the GALEX near-UV magnitude ($\sim$0.23$\mu$m) and W1 the magnitude in WISE Band 1 (3.4 $\mu$m). A cut was also made at J-W2 $\geq$ 0.8 to select late-type stars (late-K through early-L) where J is the 2MASS magnitude at 1.25 $\mu$m and W2 is the WISE Band 2 magnitude at 4.6 $\mu$m. The resulting GALNYSS catalog includes over 2,000 stars.

We selected stars for spectroscopic observation based on their apparent magnitudes, proper motions (as high proper motion stars are likely to be closer to Earth and of more interest for follow-up), and potential for moving group membership based on their position in the sky and (pre-Gaia) proper motions.

### 2.2. *Spectroscopic Observations*

TABLE 1 Summary of Observations

| Observatory | Telescope | Instrument | Resolution | $\lambda$ range [Å] | # Nights | # Stars |
|---|---|---|---|---|---|---|
| Northern Hemisphere Observations | | | | | | |
| Keck Observatory | Keck 10m | HIRES | 45,000 | 4300-9000 | 5 | 65 |
| Lick Observatory | Shane 3m | Hamilton | 45,000 | 3500-9500 | 12 | 40 |
| Lick Observatory | Shane 3m | Kast Dual Spectrometer (blue) | 1,500 | 3500-5500 | 12 | 153 |
| Lick Observatory | Shane 3m | Kast Dual Spectrometer (red) | 2,500 | 5500-7200 | 12 | 153 |
| Southern Hemisphere Observations | | | | | | |
| European Southern Observatory (ESO) | MPG 2.2m | FEROS | 48,000 | 3500-9200 | 15 | 118 |
| Las Campanas Observatory | du Pont 2.5m | Echelle | 45,000 | 5300-9100 | 3 | 37 |
| Las Campanas Observatory | du Pont 2.5m | B&C Spectrograph (600)[a] | 1,200 | 5500-8700 | 1 | 21 |
| Las Campanas Observatory | du Pont 2.5m | B&C Spectrograph (832)[a] | 2,400 | 6000-8700 | 5 | 118 |

NOTES:
a) The B&C spectrograph at Las Campanas Observatory was used with both the 600 grooves/mm and 832 grooves/mm gratings.

Spectroscopic observations of GALNYSS stars were carried out on four telescopes (Keck 10-m, Shane 3-m, MPG 2.2m, and the duPont 2.5m) over a total of 53 nights. Our general protocol was to observe stars with medium-resolution spectrographs (R∼2,000) to look for evidence of Li line absorption (indicative of stellar youth), and then to obtain high resolution spectra (R∼40,000) for those stars that showed evidence of Li in their medium-resolution spectra or that had a high probability of being members of NYMGs on the basis of (pre-Gaia) proper motions, or both. A summary of observations is presented in Table 1. In all, we obtained spectroscopic observations of 471 individual stars; the instrumentation used as well as dates of observation for each star are listed in Table 8. For a half-dozen stars (§3.1.2), followup spectra were obtained on June 8, 2016 (UT) with the ESI spectrometer on the Keck II telescope at Mauna Kea Observatory (Sheinis et al. 2002). With the 0.3″ slit, the spectral resolution of Keck/ESI was ∼14,000.

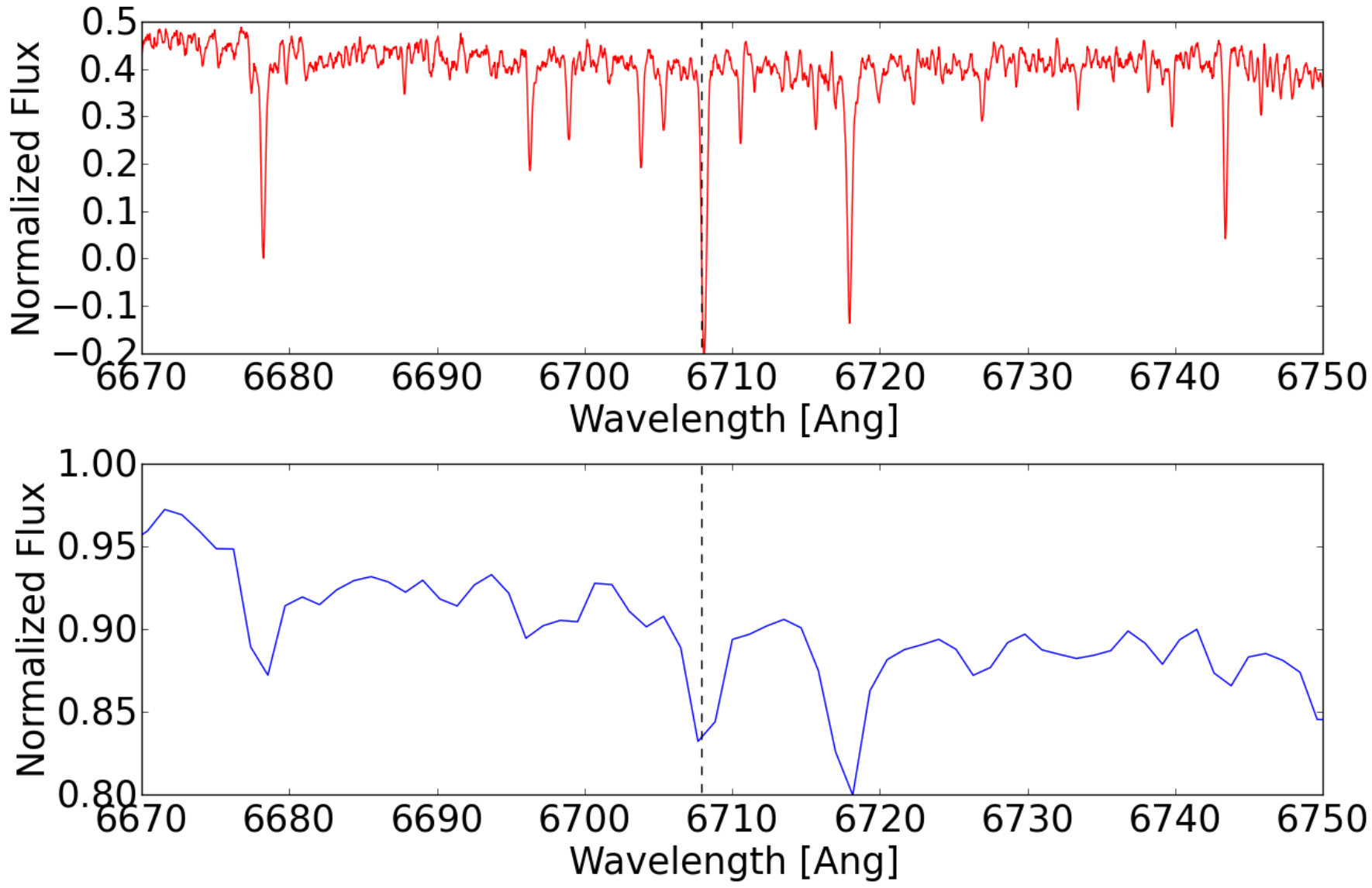

FIG. 1.—: Spectra of J0441-1947 (spectral type M1) as obtained with HIRES (high resolution; top panel) and Kast (medium resolution; bottom panel). The 6708 Å Li absorption line, detected in this star, is indicated with a vertical dashed line. Abscissa wavelengths are in air.

Shane and duPont spectra were reduced using standard IRAF packages. FEROS and HIRES spectra were reduced through an automated pipeline. Figure 1 shows example spectra for a Li-detected sample star (J0441-1947, spectral type M1) as obtained with medium- and high-resolution spectrograph (Lick/Kast and Keck/HIRES, respectively).

### 2.3. *Gaia Counterparts to GALNYSS Stars Observed Spectroscopically*

The subset of (471) GALNYSS stars observed spectroscopically were cross-matched with the GAIA DR3 catalog (Gaia Collaboration et al. 2023) so as to extract astrometric and photometric data. To obtain Gaia DR3 counterparts, the WISE designations (i.e., WISE coordinates) of the 471 target GALNYSS stars were used to search the GAIA DR3 archives for possible matches within a 5″ radius. We found that all of the spectroscopically observed GALNYSS stars have Gaia DR3 counterparts. In Fig. 2, we present a Gaia DR3 color-magnitude (absolute $G$ vs. $G - R_P$) diagram (CMD) for the full sample of 471 stars. The CMD is overlaid with (1) the Gaia DR3 CMD positions of IC 2391 members (age 50 Myr; Gaia Collaboration et al. 2018b) and (2) a field star sample consisting of all stars in Gaia DR2 (Gaia Collaboration et al. 2018a) with high-quality (i.e., S/N ≥ 10) parallaxes $\pi > 40$ mas (distances <25 pc). The latter (field star) sample serves to mark the CMD locus of main-sequence stars (e.g., Zuckerman et al. 2019, their Fig. 2). Fig. 2 immediately confirms that the spectroscopically observed GALNYSS sample is dominated by young late-type stars, i.e., stars with $G - R_P > 0.8$ whose CMD positions place them well above the main sequence. For purposes of this paper — that is, the identification and investigation of young stars near the Sun — we have flagged and rejected from further consideration all spectroscopically observed GALNYSS stars with EDR3 parallaxes $\pi < 5$ mas (distances >200 pc). Fig. 2 demonstrates that many of these stars are cool subgiants or giants with absolute $G$ magnitudes <5. After removing stars with Gaia DR3 parallaxes $\pi < 5$ mas, we are left with 431 spectroscopically observed GALNYSS stars that were subject to further analysis. The NYMG candidates considered here are restricted to distances of $\lesssim$100 pc; some more distant candidates will be discussed in Paper II.

## 3. ANALYSIS

In this section we describe the analysis of the spectroscopic observations of GALNYSS stars with Gaia DR3 parallaxes ≥ 5 mas, as well as analysis of the Gaia astrometric and photometric data for these stars. A summary of results of the spectral line equivalent width (EW) analysis for all stars observed spectroscopically, described in § 3.1, can be found in Table 9 in the Appendix. Photometry for the spectroscopically observed stars, compiled from catalogs available in Vizier, is listed in Table 10 in the Appendix. Table 11 in the Appendix presents a compilation of GALNYSS sample star radial velocities (RVs) as obtained from literature sources and Gaia DR3 (see §3.3). Our RV measurements and the final adopted RV values are listed in Table 12 in the Appendix. Gaia photometric and astrometric data are presented in Table 13 in the Appendix for the subset of stars we identify as candidate members of the seven NYMGs of interest (Table 2; see § 4.1). Gaia-derived Galactic positions ($XYZ$) and kinematics ($UVW$) for candidate members of the seven NYMGs of interest are presented in Table 3. The spectroscopic properties of these candidate NYMG members as well as references to previous NYMG membership associations, if any, are presented in Table 4.

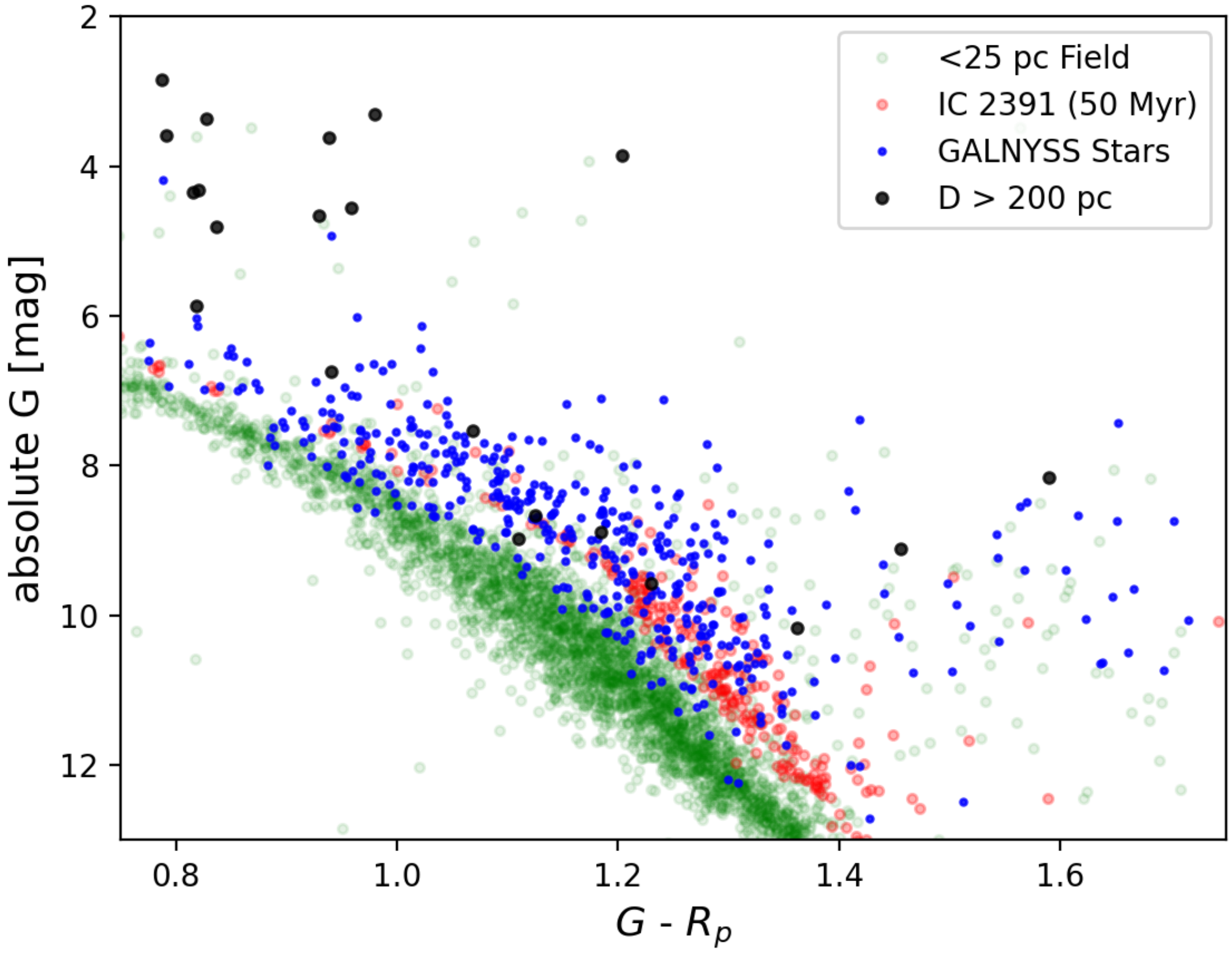

FIG. 2.—: Gaia DR3 absolute $G$ vs. $G - R_P$ color-magnitude diagram for all spectroscopically observed GALNYSS sample stars (blue symbols), overlaid with the Gaia DR3 color-magnitude positions of IC 2391 members and a nearby (distance <25 pc) field star sample (orange and green symbols, respectively; see text). The $G - R_P$ colors of stars in the lower right corner of the diagram are likely affected by anomalies (see § 4.2). Spectroscopically observed GALNYSS stars with distances > 200 pc (or plx < 5 mas) are flagged as larger filled black circles; these stars have been removed from further consideration in this paper.

### 3.1. *Spectroscopic Properties; Signatures of Youth*

#### 3.1.1. *Lithium*

The presence of the 6708Å Li absorption line is a well-established signature of youth for some late K and M stars of age $\lesssim$50 Myr, and the EW of this line serves as a stringent age diagnostic for such stars (e.g., Binks & Jeffries 2014; Kraus et al. 2014). To ascertain which stars in our spectroscopically observed sample display such Li-based evidence of youth, we measured the EW of the 6708Å Li line (in-air wavelength, without attempting to de-blend the doublet), with two bands on either side of the feature (6700–6705Å and 6710–6714Å) defining the surrounding continuum; 78 stars have detectable Li. The ensemble of EW measurements is listed in Table 9. Of the 78 Li-detected stars, we have identified 26 that are probable or possible members of the seven NYMGs of interest here; most of these stars are candidate members of the TWA, $\beta$PMG, and Tuc-Hor (see § 4.1). Results for Li-detected candidate members of the $\beta$PMG are discussed in § 4.1.2. The Li-detected stars in our spectroscopically observed sample that are not candidate members of the seven NYMGs of interest will be discussed in Paper II. Most of these stars are candidate members of the LCC, UCL, or Upper Sco Associations, although a significant number cannot be kinematically identified with any known moving group and hence may belong to the “kinematically hot” nearby young star population studied by Binks et al. (2020).

In Fig. 3, we present a Gaia DR3 CMD in which GALNYSS stars with detectable Li absorption lines are indicated. This Figure confirms that the GALNYSS stars with detectable Li lie well above the main sequence (field star locus). The majority of these Li-detected stars have ages $\lesssim$50 Myr, based on the comparison with the CMD positions of IC

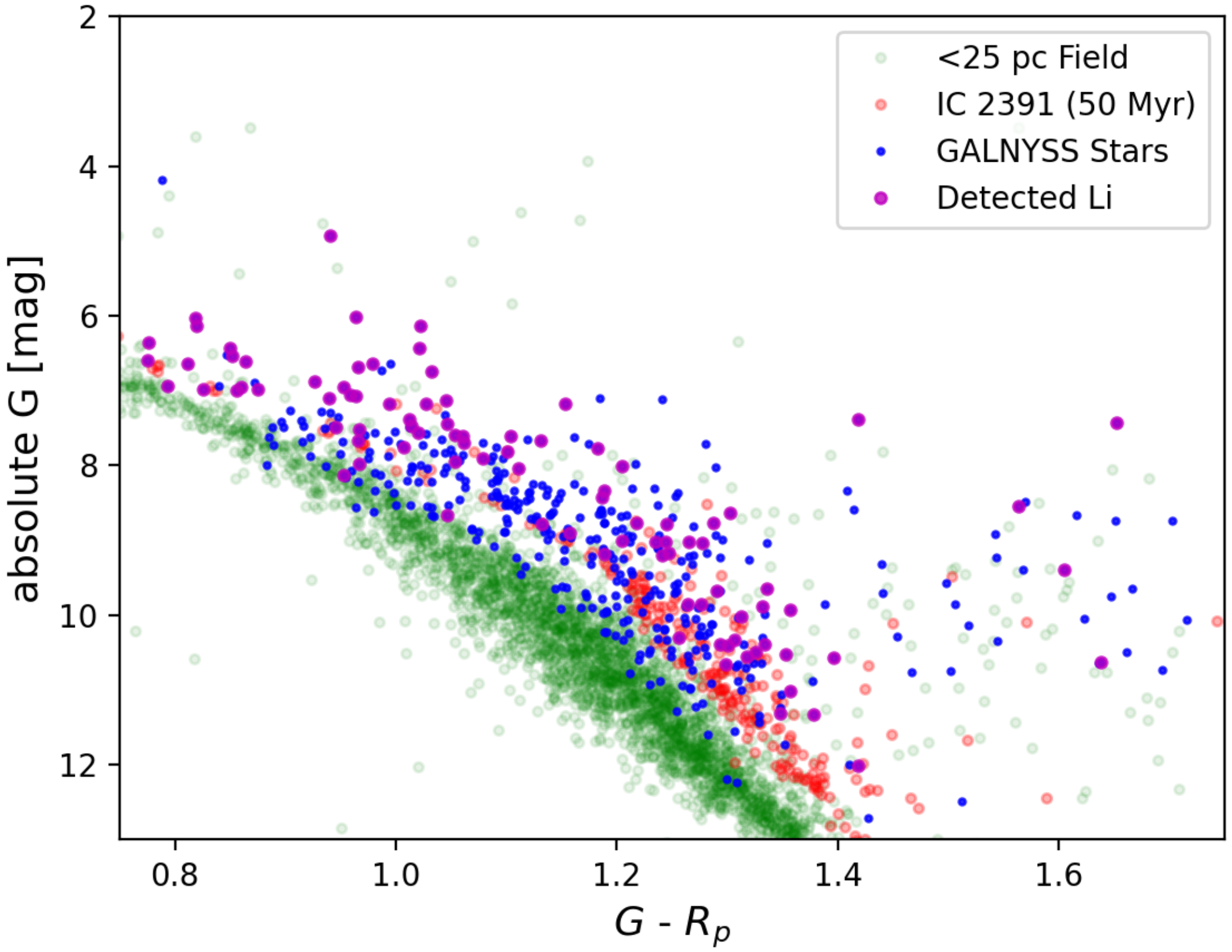

FIG. 3.—: Gaia DR3 color-magnitude diagram for spectroscopically observed GALNYSS stars with parallaxes $\pi > 5$ mas (blue symbols), highlighting the stars for which the 6707Å Li absorption line was detected (magenta symbols). As in Fig. 2, IC 2391 members and a nearby (distance <25 pc) field star sample are also indicated (orange and green symbols, respectively). The $G - R_P$ colors of stars in the lower right corner of the diagram are likely affected by anomalies (see § 4.2).

2391 stars.

#### 3.1.2. *Emission Features*

*Hα emission:* For those stars observed at high spectral resolution ($R \sim 45000$), we measured the EW of Hα (Table 9) by fitting a Voigt line profile to the emission feature. We measured the EW of each individual line several times, varying the location of the continuum. The EWs presented in Table 9 represent the average EW resulting from this process, while the errors represent the standard deviation of all measurements. While many of the observed stars show double-peaked Hα emission features, we measured EW in the same way that we would for a single-peaked emission feature. We only deblended this feature if the spectrum indicates the presence of a spectroscopic binary (see Section 4.3).

Figure 4 compares the Hα EW to the 10% width of the Hα line for those stars observed at high spectral resolution that show Hα in emission. The 10% linewidth is a widely used diagnostic of ongoing mass accretion from a circumstellar disk. White & Basri (2003) used a 10% width of $> 270$ km s$^{-1}$ as the threshold to establish the presence of accretion. We, instead, set this threshold at 200 km s$^{-1}$, since we are aware of at least two stars — i.e., the components of the LDS 5606 binary system, listed in this work as J0448+1439AB — that show signs of accretion, but have Hα 10% widths <270 km s$^{-1}$ (see Zuckerman et al. 2014 and Rodriguez et al. 2014 for details). Nine of our stars show 10% widths $\gtrsim$200 km s$^{-1}$, including J0448+1439AB. Among these stars, three (J0448+1439, J1250-4231, and J2051-1548) also show evidence of metal line emission (see Figure 4 caption).

*Other emission lines:* In addition to Hα, some accreting pre-main sequence stars also display emission in He I and in lines of various metals, such as Fe II, Na I, and Ca II (White & Basri 2003; Zuckerman & Song 2004) Such metal-line and/or He I emission features were detected in spectra of a dozen of our stars for which Hα and other Balmer lines are seen in emission: J0100+0250, J0448+1439AB, J0524-1601, J1250-4231, J1542+5936, J1915-2847 (ESI data), J1953-0707 (ESI data), J2008-2545 (ESI data), J2051-1548 (ESI data), J2110-2710 (ESI data), J2137-6036, and J2302-1215. Excerpts of spectra for six of these stars, covering lines of He I, Na I, and Fe II, are presented in Figure 5.

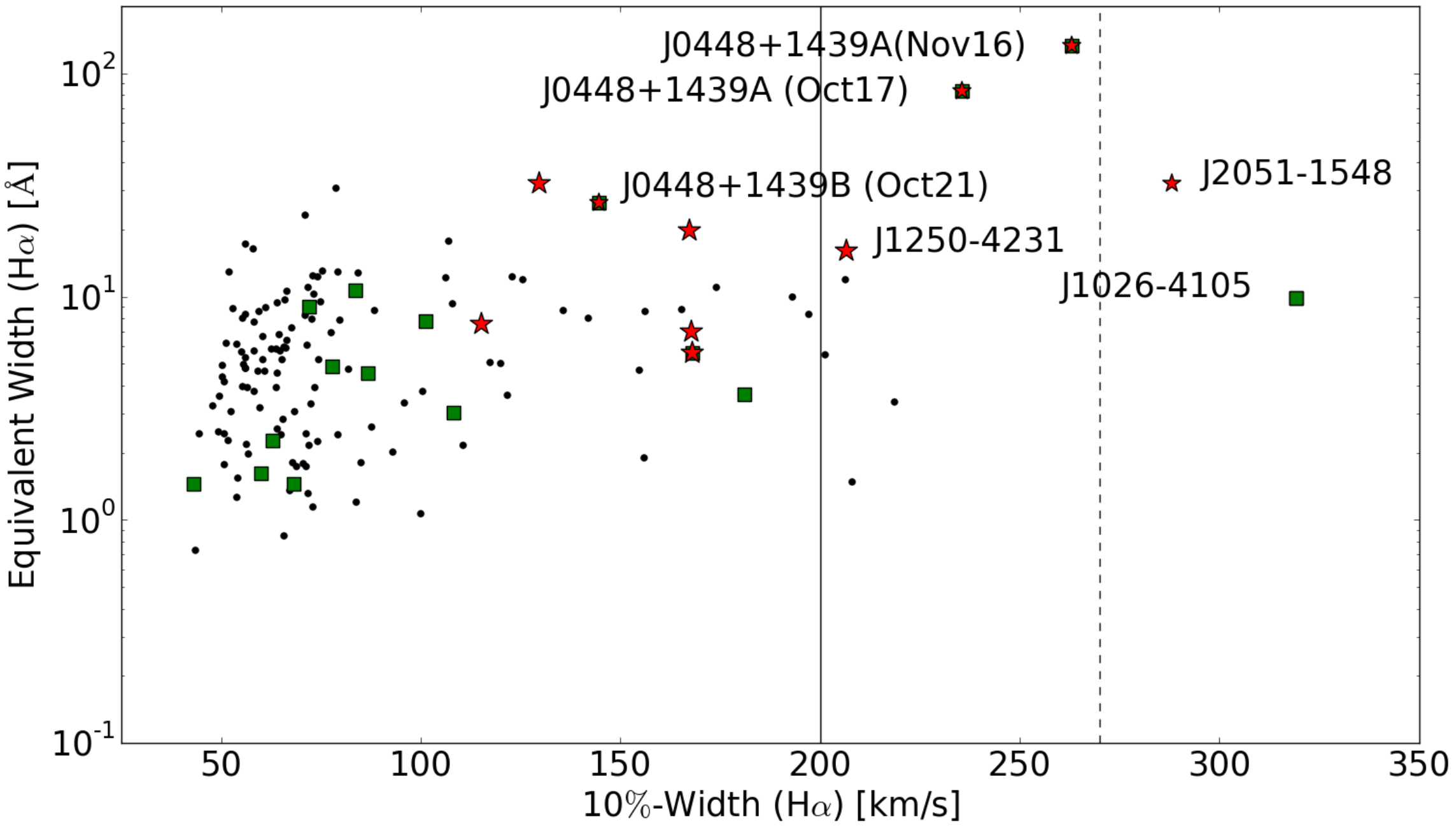

FIG. 4.—: Hα emission EWs and 10% widths for stars observed with high-resolution spectrographs (black dots). Green squares represent spectroscopic binary systems. Red stars represent stars that show metal emission features. Stars discussed in the text are labeled, as are the 10% width accretion thresholds (vertical solid and dashed lines at 10% widths of 200 and 270 km $s^{-1}$, respectively).

As described in §3.1.3, two of the aforementioned emission-line stars, J0448+1439AB and J2051-1548, also show IR excesses indicative of circumstellar dust. The emission-line spectrum of J0448+1439AB was presented and analyzed in detail in Zuckerman et al. 2014). Like J0448+1439AB, H$\alpha$–$\epsilon$ are all detected strongly in emission in J2051-1548 (ESI has essentially no throughput for shorter wavelength Balmer lines). The Hα EW was about 20 Å, about half that measured three years earlier with the duPont/B&C spectrometer. Some helium lines (air wavelengths 4026.2, 4471.5, 5875.7, and 6678.15 Å) are also seen in emission. As noted below, the Ca II H and K lines are also in emission, as is the [O I] electric quadrupole line at 6300.3 Å. The upper limit on Li obtained from the ESI spectrum of J2051-1548 is 300 mÅ, a factor ∼2 smaller than obtained from the duPont spectrum. Perhaps most intriguing, our RV measurement from the ESI spectrum, $-10.5\pm3$ km $s^{-1}$, yields $UVW$ velocities of $-6.9, -21.5, -6.7$ which, given the star's sky plane coordinates, are not compatible with membership in any known NYMG. We further consider this and other particularly interesting "homeless" GALNYSS stars in a followup paper (Paper II; Binks et al., in prep.).

Curiously, the star with the broadest Hα profile (J1026-4105) does not show any evidence of metal emission lines in its spectrum. This star does show other Balmer emission lines (Hα - H$\epsilon$), all of which exhibit significant broadening (200–300 km $s^{-1}$). However the lack of He I and Na I emission indicates that the star is not actively accreting. We also see no emission in the Ca IR triplet at ∼8500 Å. We conclude that the large Hα emission width of J1026-4105 may be due to a combination of unresolved binarity and rapid rotation, rather than accretion. This star, like J2051-1548, has $UVW$ velocities incompatible with membership in a known NYMG. We further consider these and other "homeless" spectroscopically observed GALNYSS stars in a followup paper (Binks et al., in prep.).

### 3.1.3. *IR Excesses*

The 2MASS/WISE infrared colors for our sample can be found in Table 10. Following Schneider et al. (2012), we identified stars that are IR excess candidates as those with a WISE $W1 - W4$ color >1.0. Of the 471 stars we observed, 333 were detected in W4, but only 25 have W1-W4 >1.0 (see Figure 6). To rule out possible contamination, we examined the WISE data products of those 25 stars. Only two systems, J0448+1439AB and J2051-1548, appeared to have clean WISE images in W4 (22$\mu$m). These stars are the only two among our sample of spectroscopically observed stars with real evidence for the presence of warm dust in circumstellar disks. The apparent IR excesses at the other 23 stars are likely the result of confusion with nearby background sources.

The spectral energy distributions (SEDs) of the A and B components of J0448+1439 (= LDS 5606; see § 4.2.3) are presented in Zuckerman et al. (2014). The SED of J2051-1548 — which, as noted earlier, does not clearly belong to any known NYMG — is shown in Figure 7. It is one of our latest-type stars, and (unlike J0448+1439) does not appear to have a common proper motion companion (within 3′). Fitting a simple blackbody to the observed excess, we find a dust temperature of 238 K and fractional IR luminosity of $\tau = 0.047$. These dust properties are similar to those of the secondary in the J0448+1439 system, for which $T_{dust} \sim 220$ K and $\tau \sim 0.06$ (Zuckerman et al. 2014). As noted above, the numerous strong emission features in the spectra of this star provide strong evidence for ongoing accretion

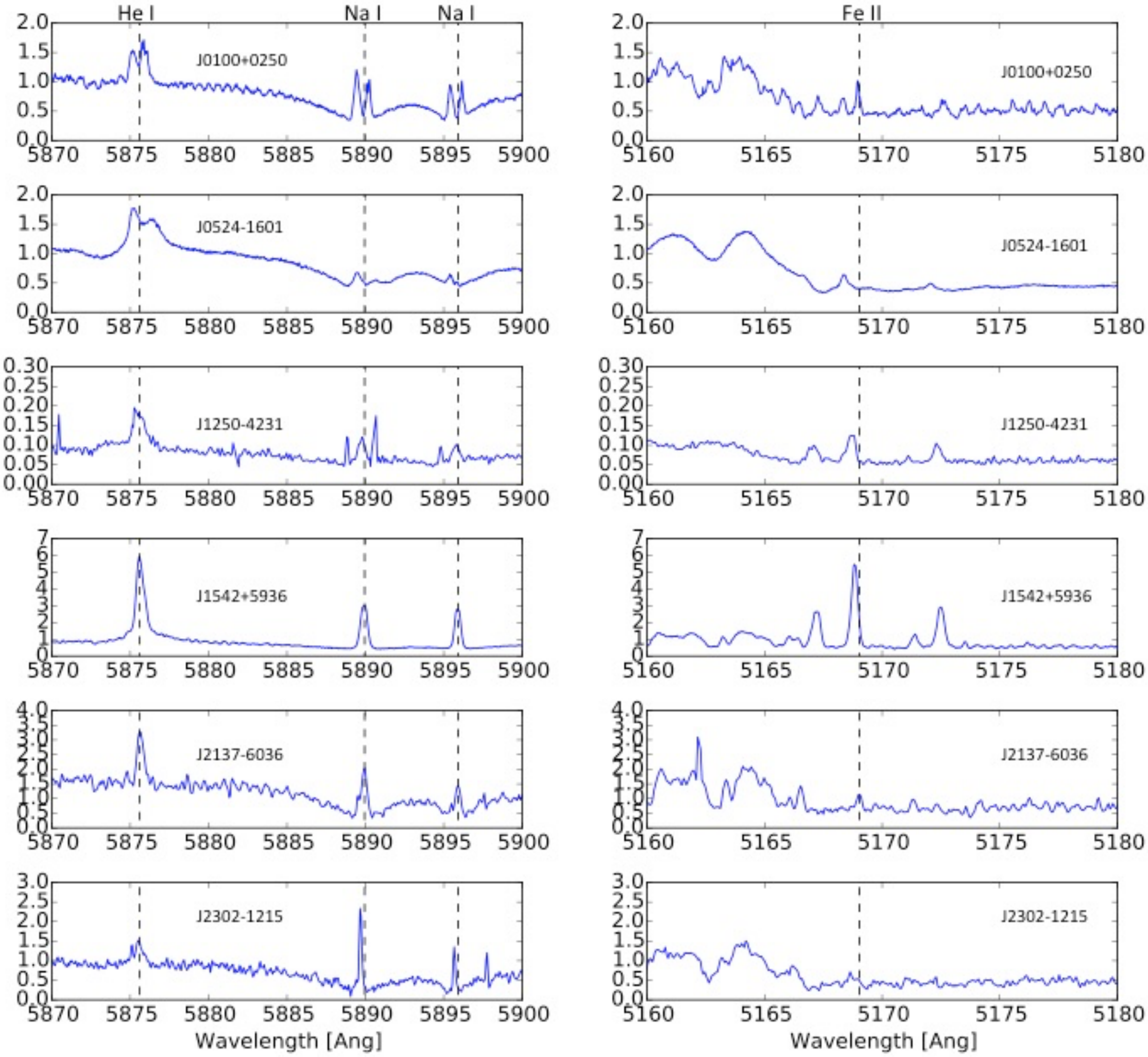

FIG. 5.—: Spectra of six stars displaying emission lines of Fe II, He I, and Na I. J0100+0250 is a spectroscopic binary; J0524-1601, which displays rapid rotation, is a possible spectroscopic binary.

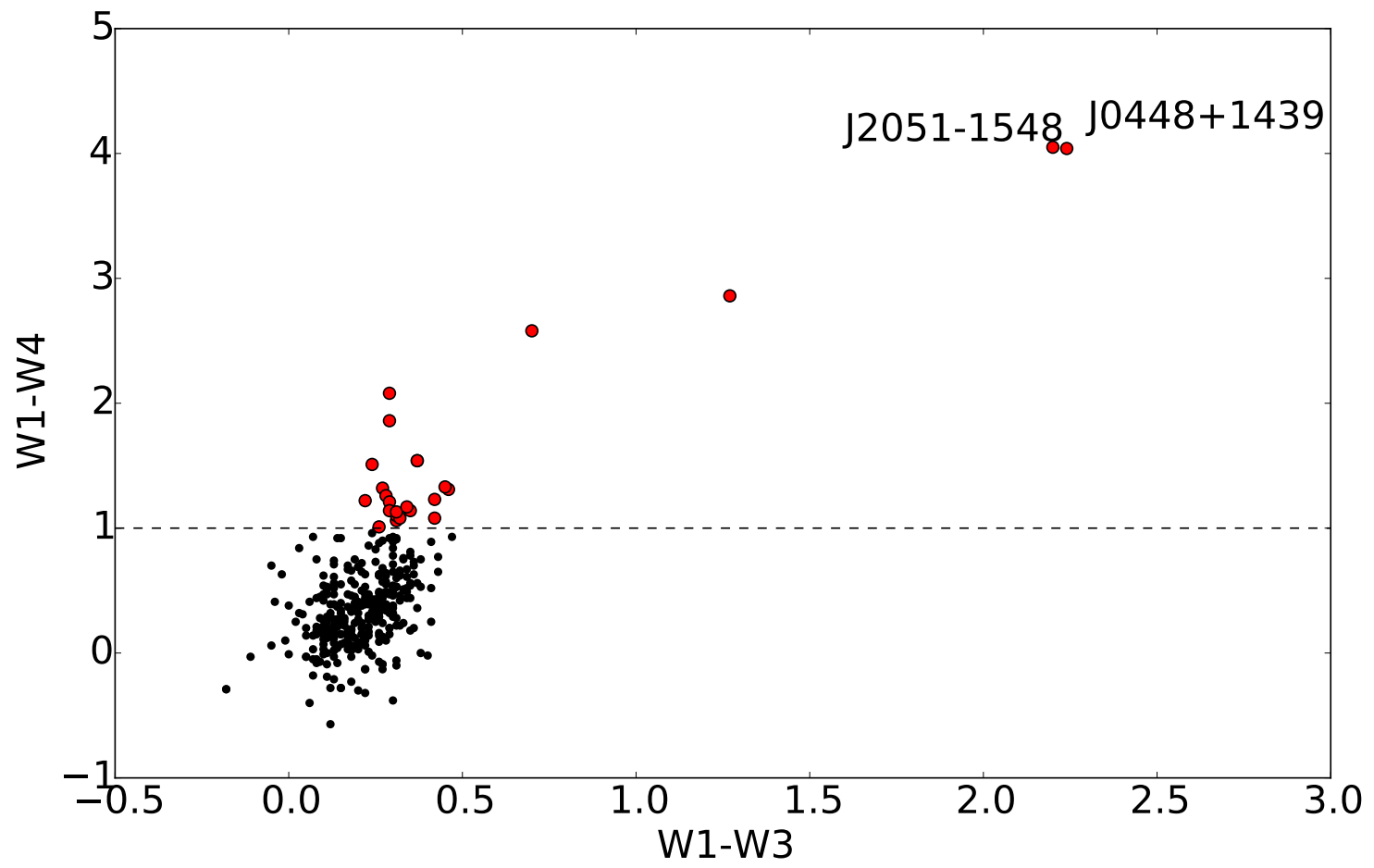

FIG. 6.—: WISE colors of stars in our spectrocopically observed sample. Stars with W1-W4>1.0 are indicated with red points. The two stars found to have *bona fide* IR excesses, J0448+1439 and J2051-1548, are labeled.

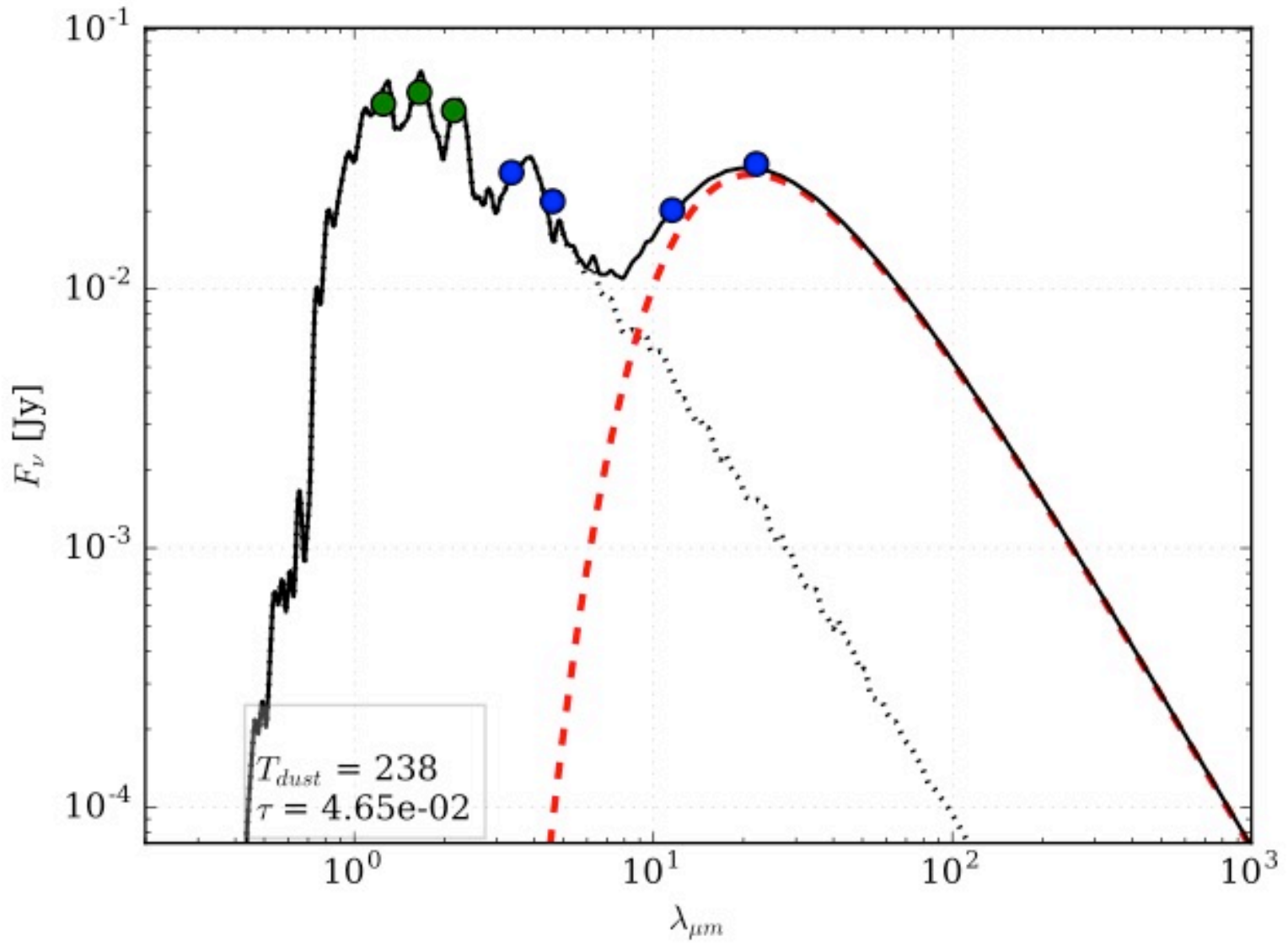

Fig. 7.—: Spectral energy distribution (SED) of J2051-1548. The black dotted curve is a photospheric model for a 3700 K star from Hauschildt et al. (1999). The red dashed curve is a simple blackbody with T=238 K. The black solid curve represents the combined contribution of the stellar photosphere and dust blackbody. $\tau$ is the fractional IR luminosity ($L_{IR}/L_{bol}$).

from this dusty disk.

Our finding that only two of 333 stars have a measurable mid-IR excess is consistent with previous studies that place the debris disk incidence among field M stars at near 0% (e.g. Binks 2016). On the other hand, M dwarfs of age similar to and younger than that of the $\beta$PMG ($\sim$24 Myr) appear to have a higher incidence of excess IR emission ($\sim$10%; Flaherty et al. 2019 and refs. therein). In addition, a remarkable $\sim$27% of M-type members located in the vicinity of the tidal radius of the $\sim$40 Myr-old $\chi^1$ For cluster display mid-IR excesses (Zuckerman et al. 2019).

### 3.2. *Spectral Indices*

Table 4 lists spectral types for stars that we associate with one of the Table 2 NYMGs (see §4.1). These types are photometric and are based on the four quantities given in Table 4 of Zuckerman et al. (2019): absolute $G$ and $K_s$ magnitudes ($M_G, M_{K_s}$), $G - K_s$, and $K_s - W2$. Argus, Tuc-Hor, Columba and Carina are all of similar age ($\sim$40 Myr), so we use these Zuckerman et al. (2019) Table 4 entries, appropriate for stars of this age, directly in estimating spectral types for these four groups. For younger (e.g., BPMG) and older (AB Dor) moving groups we checked various papers for spectral types, based on spectra, of previously accepted members of these groups or clusters of similar age (Steele & Jameson 1995; Pecaut & Mamajek 2013; Binks & Jeffries 2016a), and calculated the equivalent photometric quantities ($M_G, M_{K_s}$, $G - K_s$, and $K_s - W2$) for these reference stars. As expected, $M_G$ and $M_K$ increase (become fainter with age), but there is little change in $G - K_s$ and $K_s - W2$ with age. The increase in $M_G$ with age can be seen by comparison of the locations of the single star main sequence in the color-magnitude diagrams presented in §4.1.

We also assessed spectral types based on TiO indices, for stars where these could be measured. These spectral types were measured from the $\mathbf{TiO_5}$ molecular band features using equation 1 in Gizis (1997), which provides an accuracy of approximately half a sub-class. Fig. 8 compares spectral types derived from TiO with those determined from photometry for stars where the former is measured. It is apparent that the agreement between the two methods is generally good, with the TiO-based spectral types typically $\sim$0.5 earlier than those determined photometrically. Therefore, we regard the listed spectral types as generally accurate to 0.5 subclasses.

### 3.3. *Radial Velocities*

#### 3.3.1. *Literature RVs*

The stellar radial velocities (RVs) used to calculate $UVW$ velocities — and, hence, to assess the likelihood of membership of a given GALNYSS star in a given NYMG — have been obtained via the combination of our own RV measurements and RVs available in the literature and Gaia DR3. We searched for previous literature RV measurements for each target using the Vizier online database[1], and a semi-automated routine developed by coauthor A. Binks[2]. Since there are many different labels given to RV measurements in the myriad catalogues containing such measurements, we acquisitioned any catalogue that contains both the Vizier keyword "Stars", and a column name that matched a list of widely-used RV monikers (e.g.,"RV","RV","HRV","Vrad"). The routine then outputs a table containing (1) the target name, (2) the Vizier catalogue name, (3) the angular distance from the target (in arcseconds), (4) the RV and (5) the RV error. There are many Vizier catalogues that contain archival RV data, i.e., the measurements are themselves derived from a literature search rather than being the "provenance", and this can lead to multiple-counting of the same measurement and a biased final RV measurement. Therefore we set a constraint that if two or more measurements of the same star are within 0.02 km/s, then the (inverse square) error-weighted mean and mean error bar are taken as a single measurement. For literature measurements without error bars we assign these a value of twice the median error bar from all entries that have a quoted uncertainty. Approximately 5% of low-mass stars selected without any kinematic bias are expected to reside in multiple systems whose RV amplitudes vary by more than a few km s$^{-1}$ (Binks et al. 2015). Therefore, we assign a multiplicity test for each target by measuring the RV difference between all pairs and flagging multiple systems as those satisfying the condition $\Delta RV > \epsilon \times (\sigma_{\rm RV_1} + \sigma_{\rm RV_2})$ for any pair (we select $\epsilon = 3$). The resulting literature-derived RVs as well as Gaia RV measurements, along with their associated errors, are compiled in Table 11.

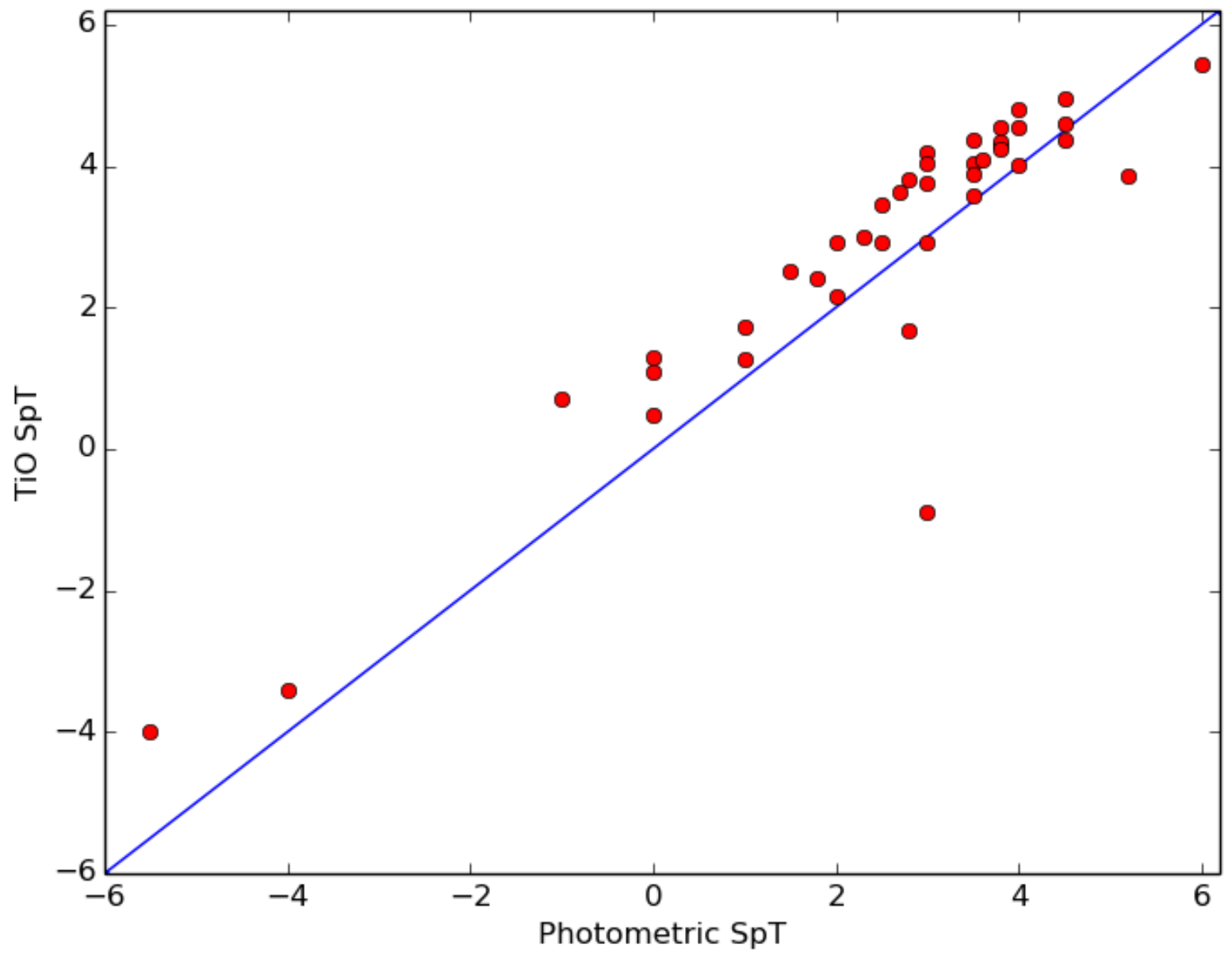

Fig. 8.—: Comparison between spectral types determined photometrically vs. by TiO indices. The diagonal line indicates perfect agreement between the two methods. Here -2 represents K8.

### 3.3.2. *Newly measured RVs*

For those stars observed with high-resolution spectrographs, we have measured heliocentric radial velocities using absorption features near 6430Å. We used RV standards from Nidever et al. (2002), observed each night, to calibrate and validate our RV measurements. Barycentric corrections were calculated using the IRAF task `rvcorrect`. Errors were calculated from the standard deviation in the measurement of radial velocity across several orders of the spectrum. Typical errors for our radial velocity measurements are ~1.5 km s$^{-1}$ (~3 km s$^{-1}$ for ESI spectra).

### 3.3.3. *Candidate NYMG members: merged and adopted RVs*

The resulting RVs measured for the subset of stars that are candidate members of the seven NYMGs of interest (§ 4.1) are listed, along with their associated errors, in Table 12. Table 12 also lists the adopted RV for purposes of $UVW$ calculations. For stars with more than one RV measurement available — from the literature, Gaia DR3, and/or our own measurements — the adopted RV is obtained as the weighted mean of the available RVs and their associated errors. In some cases, noted in the Table, we find that either the literature or our spectroscopically measured RVs are discrepant with Gaia DR3 RVs. These stars may be RV variables due to the presence of lower-mass binary companions; followup RV observations are warranted.

## 3.4. *Astrometric Data from Gaia DR3*

[1] `https://vizier.cds.unistra.fr/`

[2] Code available at `https://github.com/alexbinks/RV_selection`

Gaia DR3 photometric and astrometric data are listed in Table 13 for the subset of stars that we found to be candidate members of one of the seven NYMGs of interest (§ 4.1). The parallaxes, positions, and proper motions obtained from the Gaia EDR3 catalog were used as the basis to obtain Galactic positions ($XYZ$) and space motions ($UVW$), where the latter also required measurements of radial velocities (see § 3.3). The resulting $XYZ$ and $UVW$ are compiled in Table 3 for stars that are candidate members of the seven NYMGs of interest (§ 4.1). Uncertainties in $UVW$ are listed in Table 3; relative uncertainties in $XYZ$ tend to be negligible (thanks to Gaia's high astrometric precision) and are hence not listed.

## 4. RESULTS AND DISCUSSION

### 4.1. *Moving Group Membership*

TABLE 2 Moving Groups of Interest

| Moving Group | Age<br>Myr | UVW<br>km/s | $\sigma$(UVW)<br>km/s |
|---|---|---|---|
| TW Hydrae | 10±3 | (-11.6, -17.9, -5.6) | (1.8, 1.8, 1.6) |
| $\beta$ Pictoris | 24±3 | (-10.9, -16.0, -9.0) | (2.2, 1.2, 1.0) |
| Tuc Hor | 45±4 | (-9.8, -20.9, -1.0) | (0.9, 0.8, 0.7) |
| Columba | $42^{+6}_{-4}$ | (-12.1, -21.3, -5.7) | (1.0, 1.3, 0.8) |
| Argus | 45±5 | (-22.5, -14.6, -5.0) | (1.2, 2.1, 1.6) |
| Carina | $45^{+11}_{-7}$ | (-10.7, -21.9, -5.5) | (0.7, 1.0, 1.0) |
| AB Dor | $149^{+51}_{-19}$ | (-7.2, -27.6, -14.2) | (1.4, 1.0, 1.8) |

NOTES:
Mean UVW, one-sigma ranges in UVW ($\sigma$(UVW)), and ages for NYMGs, other than Argus, as listed in Gagné et al. (2018, and references therein). Data for Argus from Zuckerman (2019).

The moving groups we considered for possible GALNYSS star membership assignments are listed in Table 2, which also lists their ages and $UVW$ domains as compiled by Gagné et al. (2018). The NYMG ages listed in Table 2 have largely held up in the more recent literature, albeit with some refinements and debate. In particular, the ∼10 Myr age of the TW Hya Association is supported by available Gaia data (Luhman 2023), while recent estimates of the age of the $\beta$PMG range from ∼11 Myr to ∼33 Myr, with a strong dependence on methodology and models (Lee & Song 2024; Lee et al. 2024).

To make NYMG membership assignments, we calculated Galactic positions ($XYZ$) and space motions ($UVW$) for stars in our sample, where the latter were obtained by adopting the astrometric data listed in Table 13 and the radial velocities listed in Table 12 (column "Adopted"). We entered the resulting $XYZ$ and $UVW$ into the BANYAN $\Sigma$ tool (Gagné et al. 2018), and examined the membership probabilities. If a star with a measurable radial velocity was found to have a probability $\geq$ 50% of belonging to a Table 2 moving group according to publicly available version of BANYAN[3], we considered the star a candidate member. As the BANYAN NYMG membership models are frequently updated, we have also considered membership probabilities as calculated by an as-yet unpublished set of BANYAN models (J. Gagne, private communication).

In addition to these purely BANYAN-based membership assignments, we have identified a few dozen additional candidate Table 2 NYMG members among the spectroscopically observed GALNYSS sample, despite these stars having low BANYAN probabilities of membership. The NYMG membership assignments in these cases have been made on the basis of a reasonable kinematic ($UVW$) match to a given Table 2 NYMG and/or a previous (literature) assignment of a star to a Table 2 NYMG. In a few cases (for the younger Table 2 groups), a star's relatively large Li EW (>200 mÅ) supports in its membership assignment. Conversely, a number of stars with ≥50% membership probabilities in one of the Table 2 NYMGs were found to have Gaia color-magnitude diagram positions and/or measured Li EWs that are inconsistent with membership. We describe many of these individual cases in §§4.2.

TABLE 3 Galactic Positions and Space Motions of Candidate Members of NYMGs

| WISE Desig | BANYAN<br>prob. | $X$<br>pc | $Y$<br>pc | $Z$<br>pc | $U$<br>km/s | $V$<br>km/s | $W$<br>km/s | $\sigma_U$<br>km/s | $\sigma_V$<br>km/s | $\sigma_W$<br>km/s |
|---|---|---|---|---|---|---|---|---|---|---|
| AB Dor | | | | | | | | | | |
| J003903.51+133016.0 | 0.72 | -16.72 | 31.03 | -40.92 | -7.87 | -29.05 | -8.78 | 0.5 | 0.92 | 1.21 |
| J012118.22-543425.1 | 0.23 | 8.95 | -20.39 | -41.91 | -4.08 | -29.81 | -14.0 | 0.15 | 0.15 | 0.18 |
| J041807.76+030826.0 | 0.37 | -72.48 | -12.8 | -45.53 | -6.63 | -28.2 | -11.35 | 1.43 | 0.32 | 0.91 |
| J054709.88-525626.1 | 0.09 | -12.12 | -72.64 | -43.93 | -10.2 | -28.04 | -12.31 | 0.42 | 2.54 | 1.54 |
| J072911.26-821214.3[a] | 0.85 | 19.19 | -42.31 | -22.25 | -4.17 | -25.36 | -15.1 | 0.19 | 0.41 | 0.22 |
| J101543.44+660442.3 | 0.32 | -21.89 | 16.04 | 26.42 | -3.21 | -27.47 | -12.5 | 0.83 | 0.64 | 0.99 |
| J112547.46-441027.4 | 0.98 | 13.11 | -42.56 | 12.81 | -5.61 | -29.2 | -13.51 | 0.61 | 1.84 | 0.58 |
| J121511.25-025457.1[b] | 0.0 | 13.9 | -50.82 | 86.69 | -6.02 | -27.47 | -11.98 | 0.54 | 1.91 | 3.25 |
| J133509.40+503917.5 | 0.11 | -5.27 | 17.94 | 40.18 | -8.04 | -22.94 | -8.46 | 0.46 | 1.43 | 3.17 |
| J155947.24+440359.6 | 0.03 | 9.53 | 26.02 | 31.67 | -9.23 | -24.57 | -8.79 | 0.46 | 1.24 | 1.51 |

[3] https://www.exoplanetes.umontreal.ca/banyan/banyansigma.php

| | | | | | | | | | | |
|---|---|---|---|---|---|---|---|---|---|---|
| J194834.58-760546.9 | 0.47 | 38.67 | -34.24 | -29.46 | -2.97 | -26.11 | -14.12 | 0.26 | 0.23 | 0.2 |
| J210338.46+075330.3[b] | 0.0 | 49.68 | 76.11 | -41.84 | -6.06 | -27.47 | -13.41 | 0.52 | 0.77 | 0.45 |
| J230327.73-211146.2[b] | 0.0 | 28.58 | 24.5 | -79.72 | -6.79 | -25.62 | -6.99 | 1.95 | 1.67 | 5.43 |
| Argus | | | | | | | | | | |
| J030824.14+234554.2[c,d] | 0.15 | -58.89 | 22.45 | -35.39 | -20.32 | -14.53 | -9.52 | 1.1 | 0.59 | 0.74 |
| J103952.70-353402.5[c] | 0.88 | 6.71 | -78.56 | 28.76 | -21.8 | -18.72 | -3.15 | 0.1 | 0.75 | 0.28 |
| J135913.33-292634.2[c] | 0.0 | 49.3 | -41.1 | 38.79 | -18.21 | -20.63 | -2.77 | 0.54 | 0.48 | 0.37 |
| J191534.83-083019.9[c] | 0.25 | 58.43 | 31.12 | -10.73 | -24.34 | -14.55 | 0.26 | 2.09 | 1.11 | 0.39 |
| J195227.23-773529.4[c] | 0.98 | 24.62 | -23.17 | -19.29 | -21.38 | -9.29 | -5.23 | 1.39 | 1.31 | 1.09 |
| J195331.72-070700.5[a,c] | 0.51 | 25.91 | 17.2 | -9.53 | -19.47 | -15.59 | -1.56 | 1.51 | 1.0 | 0.56 |
| J213835.44-505111.0 | 1.0 | 30.15 | -7.35 | -32.96 | -24.64 | -12.43 | -3.15 | 2.4 | 0.6 | 2.62 |
| $\beta$ Pic | | | | | | | | | | |
| J001723.69-664512.4 | 1.0 | 14.61 | -18.6 | -28.22 | -10.52 | -16.06 | -8.57 | 0.12 | 0.15 | 0.23 |
| J004826.70-184720.7[b] | 0.03 | -3.52 | 6.61 | -50.93 | -9.4 | -18.34 | -8.7 | 0.06 | 0.1 | 0.79 |
| J010711.99-193536.4[b,d] | 0.01 | -7.09 | 4.2 | -56.08 | -9.24 | -16.93 | -8.99 | 0.17 | 0.14 | 0.59 |
| J015350.81-145950.6 | 1.0 | -10.94 | 0.73 | -32.01 | -11.74 | -16.11 | -7.45 | 0.2 | 0.02 | 0.57 |
| J022240.88+305515.4 | 0.99 | -32.51 | 22.72 | -21.09 | -12.46 | -15.93 | -7.95 | 0.94 | 0.66 | 0.61 |
| J025154.17+222728.9 | 1.0 | -20.86 | 9.17 | -14.52 | -12.58 | -15.63 | -8.91 | 0.15 | 0.07 | 0.11 |
| J034236.95+221230.2 | 1.0 | -34.69 | 7.91 | -17.06 | -11.85 | -16.35 | -8.01 | 1.23 | 0.28 | 0.61 |
| J035100.83+141339.2 | 0.47 | -34.56 | 2.98 | -19.98 | -8.51 | -18.82 | -4.12 | 6.04 | 0.52 | 3.49 |
| J035733.95+244510.2[c] | 0.97 | -62.6 | 13.31 | -25.1 | -13.76 | -16.08 | -7.35 | 0.64 | 0.14 | 0.26 |
| J044356.87+372302.7[c] | 0.0 | -68.82 | 17.65 | -6.89 | -10.75 | -18.9 | -8.34 | 0.29 | 0.08 | 0.04 |
| J044721.05+280852.5[a,c] | 0.65 | -60.44 | 7.16 | -11.73 | -21.81 | -14.87 | -8.21 | 5.07 | 0.6 | 0.98 |
| J044800.86+143957.7[b] | 0.01 | -79.28 | -6.1 | -27.43 | -10.24 | -19.34 | -8.13 | 0.57 | 0.06 | 0.2 |
| J044802.59+143951.1[b] | 0.02 | -79.09 | -6.09 | -27.36 | -11.76 | -19.62 | -8.56 | 0.38 | 0.05 | 0.14 |
| J050827.31-210144.3[b,d] | 0.18 | -30.55 | -27.53 | -25.33 | -12.94 | -18.91 | -6.17 | 0.19 | 0.17 | 0.16 |
| J052419.14-160115.5[a] | 0.9 | -21.96 | -17.31 | -13.82 | -10.62 | -14.87 | -5.34 | 0.57 | 0.46 | 0.38 |
| J052944.69-323914.1 | 1.0 | -14.11 | -21.42 | -15.1 | -11.99 | -16.51 | -9.14 | 0.38 | 0.58 | 0.41 |
| J053747.56-424030.8 | 1.0 | -9.48 | -23.97 | -15.51 | -10.61 | -14.23 | -7.42 | 1.04 | 2.63 | 1.7 |
| J061313.30-274205.6[d] | 1.0 | -18.38 | -25.85 | -11.57 | -12.32 | -16.25 | -9.77 | 0.11 | 0.15 | 0.08 |
| J081738.97-824328.8 | 1.0 | 10.42 | -21.84 | -10.91 | -10.06 | -18.58 | -10.12 | 0.51 | 1.07 | 0.53 |
| J180554.92-570431.3 | 0.99 | 49.68 | -21.36 | -16.22 | -9.79 | -15.05 | -7.73 | 0.35 | 0.15 | 0.12 |
| J180929.71-543054.2 | 1.0 | 35.03 | -13.18 | -10.8 | -10.43 | -15.15 | -8.14 | 0.72 | 0.27 | 0.22 |
| J184206.97-555426.2 | 1.0 | 45.09 | -16.61 | -18.39 | -7.9 | -15.71 | -8.55 | 0.18 | 0.07 | 0.07 |
| J192434.97-344240.0 | 1.0 | 47.88 | 3.21 | -18.82 | -7.09 | -15.91 | -9.8 | 0.37 | 0.04 | 0.15 |
| J193411.46-300925.3 | 0.99 | 62.07 | 10.02 | -25.18 | -6.52 | -14.85 | -9.35 | 1.37 | 0.23 | 0.56 |
| J194816.54-272032.3 | 0.99 | 57.6 | 13.48 | -26.08 | -7.41 | -14.95 | -9.22 | 0.71 | 0.17 | 0.32 |
| J195602.95-320719.3[d] | 1.0 | 45.13 | 6.92 | -23.14 | -6.74 | -15.9 | -9.72 | 0.24 | 0.66 | 0.35 |
| J200137.19-331314.5 | 1.0 | 52.29 | 7.25 | -28.44 | -7.17 | -15.75 | -9.39 | 0.17 | 0.03 | 0.1 |
| J200556.44-321659.7 | 1.0 | 42.82 | 6.93 | -23.98 | -9.9 | -15.33 | -8.25 | 0.6 | 0.1 | 0.34 |
| J200837.87-254526.2 | 0.99 | 47.61 | 14.08 | -25.94 | -7.61 | -14.75 | -9.67 | 1.36 | 0.4 | 0.74 |
| J200853.72-351949.3 | 1.0 | 38.45 | 4.03 | -22.59 | -8.9 | -15.73 | -10.62 | 0.17 | 0.05 | 0.1 |
| J201000.06-280141.6 | 1.0 | 39.52 | 9.95 | -22.2 | -9.98 | -15.44 | -9.63 | 0.69 | 0.26 | 0.4 |
| J203337.63-255652.8[d] | 1.0 | 34.78 | 11.47 | -23.78 | -10.27 | -15.51 | -8.54 | 0.4 | 0.13 | 0.27 |
| J211004.67-192031.2[d] | 1.0 | 22.8 | 12.76 | -21.12 | -9.31 | -16.1 | -10.71 | 1.04 | 0.64 | 0.98 |
| J211005.41-191958.4 | 1.0 | 21.96 | 12.3 | -20.34 | -9.51 | -15.01 | -10.27 | 0.2 | 0.11 | 0.19 |
| J211031.49-271058.1 | 1.0 | 28.62 | 10.1 | -26.65 | -9.33 | -15.06 | -9.21 | 0.14 | 0.05 | 0.13 |
| J211031.49-271058.1 | 1.0 | 28.62 | 10.1 | -26.65 | -10.74 | -15.56 | -7.89 | 1.63 | 0.58 | 1.52 |
| J212007.84-164548.2 | 0.99 | 30.72 | 20.3 | -31.2 | -8.72 | -14.54 | -10.16 | 0.38 | 0.25 | 0.39 |
| J212128.89-665507.1 | 1.0 | 20.41 | -13.46 | -19.93 | -12.27 | -15.85 | -9.14 | 0.45 | 0.36 | 0.44 |
| J213740.24+013713.2[c] | 0.35 | 16.15 | 24.48 | -20.73 | -7.69 | -9.24 | -12.22 | 0.9 | 1.36 | 1.16 |
| J220850.39+114412.7 | 0.99 | 9.45 | 29.27 | -21.25 | -10.11 | -15.38 | -9.32 | 0.33 | 1.02 | 0.74 |
| J224500.20-331527.2 | 1.0 | 9.51 | 2.11 | -18.42 | -10.51 | -15.85 | -9.51 | 0.05 | 0.01 | 0.09 |
| Carina | | | | | | | | | | |
| J040827.01-784446.7 | 1.0 | 18.4 | -44.09 | -32.34 | -10.55 | -22.5 | -5.12 | 0.16 | 0.38 | 0.28 |
| J070657.72-535345.9 | 1.0 | -4.34 | -43.93 | -15.59 | -10.76 | -20.52 | -6.14 | 0.07 | 0.66 | 0.23 |
| J075233.22-643630.5 | 0.25 | 11.7 | -93.55 | -30.89 | -10.82 | -18.86 | -3.18 | 0.47 | 3.77 | 1.25 |
| J080636.05-744424.6 | 1.0 | 19.11 | -60.95 | -25.0 | -10.66 | -22.19 | -4.8 | 0.28 | 0.89 | 0.36 |
| Columba | | | | | | | | | | |
| J004210.98-425254.8 | 0.95 | 9.12 | -11.21 | -50.84 | -10.92 | -21.59 | -5.93 | 0.07 | 0.09 | 0.38 |
| J010629.32-122518.4[b] | 0.0 | -12.77 | 11.85 | -64.47 | -12.46 | -20.75 | -5.28 | 0.1 | 0.09 | 0.48 |
| J024852.67-340424.9[d] | 1.0 | -10.37 | -15.13 | -37.44 | -11.76 | -21.18 | -4.44 | 0.1 | 0.15 | 0.36 |
| J032047.66-504133.0 | 1.0 | -3.2 | -26.05 | -35.13 | -11.53 | -21.47 | -5.37 | 0.03 | 0.24 | 0.32 |
| J033640.91+032918.3 | 0.72 | -20.57 | -0.69 | -17.12 | -14.23 | -22.17 | -9.84 | 1.31 | 0.05 | 1.09 |
| J034115.60-225307.8 | 0.99 | -35.87 | -26.31 | -55.86 | -13.06 | -21.5 | -4.37 | 0.6 | 0.44 | 0.94 |
| J034116.16-225244.0 | 0.95 | -35.98 | -26.38 | -56.02 | -12.02 | -21.01 | -3.99 | 1.26 | 0.92 | 1.96 |
| J035345.92-425018.0[e] | 0.35 | -21.67 | -54.32 | -69.86 | -12.09 | -22.21 | -5.46 | 0.05 | 0.12 | 0.15 |
| J040649.38-450936.3[d] | 1.0 | -16.4 | -48.3 | -55.51 | -13.57 | -22.75 | -7.03 | 0.18 | 0.51 | 0.59 |
| J040711.50-291834.3[e] | 0.50 | -34.66 | -38.22 | -55.6 | -11.46 | -20.58 | -5.17 | 0.58 | 0.66 | 0.92 |
| J042736.03-231658.8 | 1.0 | -22.81 | -19.85 | -26.58 | -11.72 | -20.03 | -5.28 | 0.79 | 0.69 | 0.92 |
| J045114.41-601830.5[e] | 0.22 | -0.1 | -78.97 | -61.89 | -11.76 | -22.15 | -4.46 | 0.01 | 0.63 | 0.49 |
| J051026.38-325307.4 | 1.0 | -37.69 | -55.4 | -46.05 | -13.08 | -21.42 | -4.31 | 0.19 | 0.27 | 0.23 |
| J051403.20-251703.8 | 1.0 | -36.92 | -39.93 | -33.67 | -13.42 | -21.95 | -5.8 | 0.81 | 0.87 | 0.74 |
| J051650.66+022713.0[d,e] | 0.47 | -51.3 | -18.03 | -19.47 | -6.83 | -18.12 | -4.08 | 4.09 | 1.44 | 1.55 |
| J054433.76-200515.5[d,e] | 0.0 | -74.19 | -72.66 | -44.93 | -12.48 | -20.19 | -2.9 | 1.18 | 1.16 | 0.72 |
| J055008.59+051153.2[a] | 0.54 | -58.63 | -22.66 | -12.36 | -5.74 | -16.9 | -3.97 | 3.66 | 1.42 | 0.77 |
| J072821.16+334511.6[d] | 0.14 | -42.14 | -3.74 | 16.95 | -8.18 | -22.5 | -6.49 | 3.7 | 0.33 | 1.49 |
| J073138.47+455716.5[a] | 0.93 | -49.9 | 6.62 | 24.37 | -15.65 | -21.26 | -4.01 | 3.39 | 0.71 | 1.66 |
| Tuc Hor | | | | | | | | | | |
| J001527.62-641455.2 | 1.0 | 18.43 | -22.49 | -37.8 | -9.3 | -20.86 | -0.45 | 0.15 | 0.19 | 0.32 |
| J001536.79-294601.2 | 1.0 | 5.12 | 1.12 | -35.88 | -9.43 | -20.9 | -2.8 | 0.1 | 0.03 | 0.69 |
| J003057.97-655006.4[d] | 0.43 | 16.83 | -22.94 | -35.34 | -7.16 | -23.47 | -4.86 | 0.38 | 0.51 | 0.78 |
| J004528.25-513734.4[c,d] | 0.96 | 9.84 | -13.97 | -37.45 | -6.82 | -25.03 | -11.36 | 0.17 | 0.24 | 0.64 |
| J012245.24-631845.0 | 1.0 | 12.27 | -24.05 | -36.43 | -9.49 | -20.76 | -0.57 | 0.43 | 0.85 | 1.29 |
| J012332.89-411311.4 | 1.0 | 1.84 | -10.57 | -38.52 | -9.77 | -21.06 | -1.85 | 0.08 | 0.45 | 1.64 |
| J012532.11-664602.6[c] | 0.99 | 13.82 | -26.31 | -35.44 | -6.04 | -27.41 | -9.54 | 1.37 | 2.62 | 3.52 |
| J015057.01-584403.4 | 1.0 | 8.03 | -23.48 | -37.84 | -9.38 | -21.53 | -1.54 | 0.12 | 0.36 | 0.59 |
| J020012.84-084052.4 | 0.98 | -14.95 | 3.3 | -33.45 | -10.48 | -20.79 | -2.32 | 0.2 | 0.05 | 0.45 |
| J021258.28-585118.3 | 1.0 | 6.84 | -26.43 | -39.37 | -10.01 | -20.02 | 0.63 | 0.13 | 0.5 | 0.74 |
| J022424.69-703321.2 | 1.0 | 11.9 | -29.02 | -30.93 | -9.55 | -20.89 | -0.88 | 0.11 | 0.26 | 0.28 |
| J024552.65+052923.8[a] | 0.91 | -39.56 | 8.76 | -43.81 | -12.48 | -20.49 | -1.81 | 0.73 | 0.16 | 0.81 |
| J024746.49-580427.4 | 1.0 | 3.58 | -26.63 | -35.62 | -9.86 | -21.25 | -1.27 | 0.05 | 0.36 | 0.48 |
| J030251.62-191150.0[d] | 1.0 | -19.61 | -9.25 | -35.9 | -10.87 | -21.29 | -2.95 | 0.56 | 0.27 | 1.03 |
| J031650.45-350937.9[d] | 1.0 | -13.61 | -20.62 | -39.63 | -11.41 | -21.73 | -3.75 | 0.5 | 0.76 | 1.44 |
| J035829.67-432517.2[b,d] | 0.01 | -21.97 | -57.13 | -70.74 | -9.25 | -16.51 | 1.26 | 0.94 | 2.44 | 3.03 |
| J040539.68-401410.5[d] | 0.8 | -14.71 | -30.13 | -37.35 | -12.61 | -20.82 | -1.36 | 0.42 | 0.82 | 1.0 |
| J041255.78-141859.2 | 0.99 | -37.98 | -20.37 | -38.14 | -12.18 | -21.3 | -1.85 | 0.13 | 0.07 | 0.13 |
| J041336.14-441332.4 | 1.0 | -12.48 | -33.69 | -37.64 | -10.57 | -20.9 | -0.46 | 0.34 | 0.91 | 1.01 |
| J042139.19-723355.7 | 1.0 | 11.62 | -41.39 | -32.01 | -9.56 | -21.11 | -1.12 | 0.07 | 0.23 | 0.18 |
| J043657.44-161306.7[b] | 0.0 | -26.32 | -17.34 | -23.67 | -10.08 | -20.24 | 0.05 | 0.33 | 0.22 | 0.3 |
| J044700.46-513440.4 | 1.0 | -8.6 | -43.92 | -37.57 | -10.71 | -22.31 | -1.96 | 0.06 | 0.3 | 0.26 |
| J050610.44-582828.5[a] | 0.99 | -2.05 | -42.9 | -31.8 | -10.12 | -22.0 | -2.1 | 0.01 | 0.16 | 0.12 |
| J053925.08-424521.0 | 0.07 | -15.49 | -39.46 | -25.24 | -10.59 | -21.33 | -2.58 | 0.06 | 0.16 | 0.1 |
| J192250.70-631058.6[a,d] | 0.98 | 48.93 | -24.73 | -28.54 | -8.84 | -21.01 | -1.29 | 1.21 | 0.75 | 0.7 |
| J194309.89-601657.8 | 1.0 | 44.9 | -19.33 | -27.83 | -7.74 | -20.7 | -0.94 | 0.8 | 0.34 | 0.49 |
| J200409.19-672511.7[b,d] | 0.02 | 38.19 | -23.53 | -27.86 | -5.8 | -22.86 | -0.01 | 2.9 | 1.8 | 2.11 |
| J210722.53-705613.4 | 1.0 | 30.89 | -23.62 | -28.67 | -8.1 | -21.3 | -1.29 | 0.38 | 0.29 | 0.36 |
| J211635.34-600513.4 | 1.0 | 31.66 | -14.64 | -30.48 | -8.06 | -21.1 | -1.27 | 1.03 | 0.47 | 0.99 |
| J213708.89-603606.4 | 1.0 | 29.21 | -14.67 | -30.85 | -8.48 | -20.86 | -1.03 | 0.13 | 0.07 | 0.14 |
| J214905.04-641304.8[d] | 1.0 | 27.26 | -17.0 | -29.91 | -8.09 | -21.02 | -1.16 | 1.99 | 1.24 | 2.18 |
| J220216.29-421034.0[a] | 0.99 | 27.01 | -0.92 | -35.3 | -8.3 | -20.91 | -2.79 | 0.18 | 0.02 | 0.24 |
| J220254.57-644045.0 | 1.0 | 26.18 | -17.37 | -30.33 | -9.66 | -20.27 | 0.39 | 0.9 | 0.6 | 1.04 |
| J224448.45-665003.9[d] | 1.0 | 23.44 | -19.56 | -31.42 | -11.35 | -18.84 | 2.28 | 0.91 | 0.76 | 1.22 |
| J224634.82-735351.0 | 1.0 | 26.38 | -27.49 | -32.52 | -8.05 | -21.83 | -1.79 | 0.37 | 0.38 | 0.45 |
| J230209.10-121522.0 | 0.97 | 12.18 | 19.27 | -40.12 | -9.45 | -20.78 | -0.2 | 0.4 | 0.63 | 1.3 |
| J232857.75-680234.5 | 1.0 | 21.66 | -22.34 | -33.66 | -8.92 | -21.0 | -1.07 | 1.18 | 1.22 | 1.84 |
| J232917.64-675000.6 | 1.0 | 21.61 | -22.2 | -33.76 | -8.4 | -21.52 | -1.95 | 0.61 | 0.63 | 0.96 |
| J234243.45-622457.1 | 1.0 | 18.91 | -18.24 | -34.78 | -9.32 | -20.79 | -0.56 | 2.0 | 1.92 | 3.67 |
| J234326.88-344658.5[a,c] | 0.57 | 10.87 | 0.19 | -36.85 | -6.73 | -19.58 | -10.95 | 0.88 | 0.06 | 2.97 |
| TWA | | | | | | | | | | |
| J111128.13-265502.9 | 1.0 | 5.09 | -42.05 | 25.3 | -13.44 | -17.86 | -5.52 | 0.27 | 2.13 | 1.28 |
| J112105.43-384516.6 | 1.0 | 15.04 | -59.26 | 23.24 | -12.38 | -18.5 | -6.51 | 0.23 | 0.91 | 0.36 |
| J112651.28-382455.5 | 1.0 | 16.73 | -61.39 | 25.11 | -12.07 | -19.43 | -6.23 | 0.25 | 0.9 | 0.37 |
| J115927.82-451019.3[d] | 0.99 | 27.48 | -63.5 | 20.81 | -10.58 | -19.88 | -6.42 | 0.77 | 1.76 | 0.58 |
| J120237.94-332840.4 | 0.99 | 20.09 | -51.3 | 29.66 | -12.2 | -17.46 | -6.11 | 0.64 | 1.64 | 0.95 |
| J123005.17-440236.1 | 1.0 | 35.78 | -64.85 | 25.01 | -9.94 | -20.42 | -5.91 | 0.92 | 1.66 | 0.64 |
| J125049.12-423123.6 | 0.98 | 49.49 | -76.77 | 33.88 | -7.68 | -21.06 | -4.88 | 1.02 | 1.58 | 0.7 |

NOTES:

a) Stars for which new membership probabilities from an as-yet unpublished version of BANYAN (J. Gagne, private comm.) differ by more than 30% from those listed.

b) Stars for which we judge $XYZ$, $UVW$, and Gaia CMD position to be consistent with membership, despite low BANYAN membership probabilities.

c) Low-confidence candidate member and/or literature membership conflict; see § 4.2.

d) Candidate photometric binary.

e) Stars that have a $> 50\%$ probability of belonging to the NYMG based on the unpublished version of BANYAN.

TABLE 4 Spectroscopic Properties of Candidate Members of NYMGs

| WISE Desig | SpT | EW(Hα) [Å] | σ [Å] | EW(Li) [Å] | σ [Å] | Instrument | Prev. ID[a] |
|---|---|---|---|---|---|---|---|
| AB Dor | | | | | | | |
| J003903.51+133016.0 | M5.2 | -9.0 | 0.1 | <0.07 | | HIRES | S12 |
| J012118.22-543425.1 | K7 | 0.7 | 0.0 | <0.03 | | FEROS | V16 |
| J041807.76+030826.0 | K8 | | | | | FEROS | |
| J054709.88-525626.1 | M3.5 | -1.9 | 0.2 | <0.2 | | FEROS | V16 |
| J072911.26-821214.3 | M0 | -1.9 | 0.1 | <0.05 | | FEROS | V16 |
| J101543.44+660442.3 | M0 | -3.9 | 0.4 | <0.15 | | Kast | V16 |
| J112547.46-441027.4 | M4.5 | -10.7 | 0.2 | <0.15 | | FEROS | M14, V16 |
| J121511.25-025457.1 | M1.5 | 0.0 | | <0.02 | | Kast | |
| J133509.40+503917.5 | M1 | -4.5 | 0.4 | <0.18 | | Kast | S12, V16 |
| J155947.24+440359.6 | K8 | -2.9 | 0.3 | <0.18 | | Kast | B16, V16 |
| J194834.58-760546.9 | M0.5 | -1.2 | 0.3 | <0.15 | | B&C-832 | |
| J210338.46+075330.3 | K5.5 | -1.1 | 0.1 | 0.24 | 0.01 | Hamilton | |
| J230327.73-211146.2 | M3.7 | -5.6 | 0.4 | <0.3 | | B&C-832 | |
| Argus | | | | | | | |
| J030824.14+234554.2 | K9 | -2.0 | 0.3 | <0.21 | | Kast | |
| J103952.70-353402.5 | M2 | -2.3 | 0.2 | <0.12 | | B&C-832 | |
| J135913.33-292634.2 | K8 | 0.3 | 0.0 | <0.07 | | echelle | |
| J191534.83-083019.9 | K7 | -1.3 | 0.1 | 0.12 | 0.01 | Hamilton | |
| J195227.23-773529.4 | M3.5 | -3.0 | 1.4 | <0.3 | | FEROS | V16 |
| J195331.72-070700.5 | M4 | -4.6 | 0.4 | <0.33 | | B&C-832 | |
| J213835.44-505111.0 | M4 | -7.1 | 0.4 | <0.15 | | B&C-832 | |
| $\beta$ Pic | | | | | | | |
| J001723.69-664512.4 | M3 | -5.3 | 0.3 | <0.07 | | echelle | M14 |
| J004826.70-184720.7 | M5 | -8.4 | 0.3 | 0.59 | 0.02 | Kast/HIRES | S17 |
| J010711.99-193536.4 | M5 | -2.4 | 0.1 | 0.35 | 0.0 | HIRES | S17 |
| J015350.81-145950.6 | M5 | -7.7 | 0.4 | <0.21 | | Kast | M14 |
| J022240.88+305515.4 | M3.5 | -5 | 0.3 | <0.24 | | Kast | S12, V16 |
| J025154.17+222728.9 | M3.7 | -5.8 | 0.1 | <0.07 | | Hamilton | V16, G18 |
| J034236.95+221230.2 | M6.5 | -12.9 | 0.3 | 0.68 | 0.01 | HIRES | V16 |
| J035100.83+141339.2 | M5.5 | -10 | 0.5 | <0.36 | | Kast | |
| J035733.95+244510.2 | K9 | -1.7 | 0.2 | <0.03 | | Kast | G12, V16 |
| J044356.87+372302.7 | M2.5 | -6.9 | 0.3 | 0.21 | 0.01 | Hamilton/HIRES | S12, M14, V16, B19 |
| J044721.05+280852.5 | M3.5 | -9.6 | 0.3 | <0.12 | | Kast | V16 |
| J044800.86+143957.7 | M5 | DBL | | 0.45 | | HIRES | R14, V16 |
| J044802.59+143951.1 | M5.5 | -23.2 | | 0.63 | | HIRES | R14, V16 |
| J050827.31-210144.3 | M5.6 | -17.8 | | 0.49 | | HIRES | M14, S17 |
| J052419.14-160115.5 | M5.5 | -19.8 | 0.5 | 0.15 | 0.03 | HIRES | V16, S17 |
| J052944.69-323914.1 | M4 | -6.1 | 0.2 | <0.09 | | HIRES | V16, S19 |
| J053747.56-424030.8 | M5.5 | -12.8 | 0.3 | 0.84 | 0.02 | HIRES | V16, G18 |
| J061313.30-274205.6 | M3.5 | -4.9 | 0.4 | <0.07 | | FEROS | M14, V16 |
| J081738.97-824328.8 | M4 | -8 | 0.2 | <0.03 | | echelle | V16, S17 |
| J180554.92-570431.3 | M3.3 | -6.7 | 0.3 | <0.24 | | B&C-832 | S17 |
| J180929.71-543054.2 | M5 | -8.9 | 0.5 | 0.24 | 0.08 | B&C-832 | V16, S17 |
| J184206.97-555426.2 | M3.5 | -7 | 1.1 | <0.14 | | FEROS | V16, S17 |
| J192434.97-344240.0 | M4 | -12.2 | 0.6 | <0.14 | | B&C-832 | V16, S17 |
| J193411.46-300925.3 | M5.2 | -13.1 | 1.7 | 0.65 | 0.14 | FEROS | V16, G18 |
| J194816.54-272032.3 | M1.2 | -4.3 | 0.3 | <0.09 | | Kast | V16, G18 |
| J195602.95-320719.3 | M3.5 | -6.1 | 0.3 | <0.33 | | B&C-832 | M14, V16 |
| J200137.19-331314.5 | M0 | -3.8 | 0.4 | 0.13 | 0.02 | B&C-832 | V16, S17 |
| J200556.44-321659.7 | M0 | -2.5 | 0.2 | <0.09 | | B&C-832 | M13, V16 |
| J200837.87-254526.2 | M4.5 | -11.4 | 0.4 | 0.35 | 0.09 | B&C-832 | S17 |
| J200853.72-351949.3 | M5.5 | -6.9 | 0.3 | <0.36 | | B&C-832 | G18 |
| J201000.06-280141.6 | M5.5 | -13.7 | 0.6 | <0.08 | | B&C-832 | M14, V16 |
| J203337.63-255652.8 | M5 | -12.3 | 0.3 | 0.52 | 0.02 | B&C-832 | M14, V16 |
| J211004.67-192031.2 | M4.5 | -8.7 | 0.2 | <0.2 | | FEROS | V16, S17 |
| J211005.41-191958.4 | M3 | -4.4 | 0.4 | <0.07 | | FEROS | V16, S17 |
| J211031.49-271058.1 | M5 | -30.5 | 2.9 | 0.48 | 0.01 | B&C-832 | V16, S17 |
| J211031.49-271058.1 | M5 | -11.9 | 1.1 | 0.86 | 0.01 | HIRES | V16, S17 |
| J212007.84-164548.2 | M3.5 | -6.6 | 0.5 | <0.2 | | B&C-832 | V16, S17 |
| J212128.89-665507.1 | K7 | | | <0.06 | | B&C-832 | M14, V16 |
| J213740.24+013713.2 | M5 | -12.1 | 0.9 | <0.08 | | HIRES | V16, S17 |
| J220850.39+114412.7 | M4 | -6.8 | 0.0 | <0.05 | | HIRES | V16, S17 |
| J224500.20-331527.2 | M4.5 | -8.1 | 0.2 | <0.15 | | B&C-832 | S03, V16 |
| Carina | | | | | | | |
| J040827.01-784446.7 | K9 | -2.2 | 0.1 | 0.08 | 0.01 | FEROS | M14, V16 |
| J070657.72-535345.9 | K8 | -2.4 | 0.5 | <0.27 | | B&C-600 | S19 |
| J075233.22-643630.5 | K8 | -2.0 | 0.2 | 0.26 | 0.03 | B&C-600 | M14 |
| J080636.05-744424.6 | M1.5 | -2.0 | 0.3 | <0.06 | | FEROS | V16, G18 |
| Columba | | | | | | | |
| J004210.98-425254.8 | M2 | -2.5 | 0.1 | <0.07 | | echelle | |
| J010629.32-122518.4 | M2.5 | -9.7 | 0.8 | <0.11 | | FEROS | V16 |
| J024852.67-340424.9 | M2.5 | -9.1 | 0.1 | <0.03 | | HIRES | |
| J032047.66-504133.0 | M1.5 | -0.8 | 0.2 | <0.03 | | echelle | G18 |
| J033640.91+032918.3 | M3.8 | -12.2 | 0.9 | <0.08 | | FEROS | |
| J034115.60-225307.8 | K9 | -2.8 | 0.2 | <0.08 | | FEROS | V16, G18 |
| J034116.16-225244.0 | K9 | -2.8 | 0.3 | <0.09 | | B&C-832 | V16 |
| J035345.92-425018.0[c] | M1.5 | -2.4 | 0.1 | <0.1 | | FEROS | V16 |
| J040649.38-450936.3 | M2.5 | -5.7 | 0.3 | <0.1 | | FEROS | V16 |
| J040711.50-291834.3 | K7 | -2.6 | 0.1 | 0.4 | 0.05 | HIRES | M14B, V16 |
| J042736.03-231658.8 | M4.2 | -8.9 | | 0.39 | | HIRES | G15, V16 |
| J045114.41-601830.5 | K8 | -1.2 | 0.2 | <0.04 | | FEROS | V16 |
| J051026.38-325307.4 | M3 | | | | | HIRES | G18 |
| J051403.20-251703.8 | M2.5 | -4.9 | 0.2 | <0.06 | | FEROS | V16, G18 |
| J051650.66+022713.0 | M3.7 | -11.2 | 0.5 | <0.24 | | B&C-832 | |
| J054433.76-200515.5 | K7 | -2.2 | 0.1 | <0.04 | | FEROS | |
| J055008.59+051153.2 | M1 | -1.7 | 0.4 | <0.39 | | Kast | V16 |
| J072821.16+334511.6 | M2.2 | -7.0 | 0.5 | <0.08 | | Hamilton | |
| J073138.47+455716.5 | M3 | -7.3 | 0.4 | <0.27 | | Kast | |
| Tuc Hor | | | | | | | |
| J001527.62-641455.2 | M2 | -3.3 | 0.2 | <0.08 | | FEROS | K14, V16 |
| J001536.79-294601.2 | M4 | -4.5 | 0.5 | <0.04 | | echelle | S19 |
| J003057.97-655006.4 | M3.8 | -3.1 | 0.1 | <0.03 | | echelle | |
| J004528.25-513734.4 | M1 | -0.9 | 0.1 | <0.03 | | FEROS | L19 |
| J012245.24-631845.0 | M3 | -12.9 | 0.8 | <0.08 | | FEROS | K14 |
| J012332.89-411311.4 | M4 | -5.9 | 0.5 | <0.27 | | B&C-832 | K14, V16 |
| J012532.11-664602.6 | M3.8 | -7.8 | 0.4 | <0.21 | | B&C-832 | K14, V16 |
| J015057.01-584403.4 | M2.8 | -8.8 | 0.3 | <0.06 | | FEROS | K14, V16 |
| J020012.84-084052.4 | M2 | -4.0 | 0.1 | <0.08 | | Hamilton | K14 |
| J021258.28-585118.3 | M1.8 | -3.2 | 0.2 | <0.04 | | echelle | K14 |
| J022424.69-703321.2 | M4 | -3.1 | 0.1 | <0.03 | | echelle | K14 |
| J024552.65+052923.8 | M1.5 | -3.0 | 0.3 | <0.21 | | Kast | |
| J024746.49-580427.4 | M2 | -3.1 | 0.2 | <0.05 | | FEROS | K14, V16 |
| J030251.62-191150.0 | M4.2 | -7.3 | 0.6 | <0.05 | | HIRES | G15, V16 |
| J031650.45-350937.9 | M3 | -8.6 | 0.2 | <0.12 | | FEROS | |
| J035829.67-432517.2 | M1 | -8.5 | 0.3 | <0.12 | | B&C-832 | |
| J040539.68-401410.5 | M0.4 | -8.6 | 0.3 | <0.08 | | FEROS | V16, G18 |
| J041255.78-141859.2 | M0.7 | -2.0 | 0.1 | <0.08 | | FEROS | V16 |
| J041336.14-441332.4 | M3.6 | -2.4 | 0.2 | <0.03 | | echelle | K14 |
| J042139.19-723355.7 | M2.2 | -4.4 | 0.2 | <0.07 | | FEROS | K14, V16 |
| J043657.44-161306.7 | M3.3 | -8.2 | 0.3 | <0.1 | | FEROS | K14 |
| J044700.46-513440.4 | M2 | -2.9 | 0.2 | <0.05 | | FEROS | K14, V16 |
| J050610.44-582828.5 | M1.5 | -4.6 | 0.5 | <0.07 | | FEROS | V16, G18 |
| J053925.08-424521.0 | M1.7 | -2.3 | 0.4 | <0.05 | | FEROS | K14 |
| J192250.70-631058.6 | M2.5 | -6.0 | 0.4 | <0.1 | | FEROS | M14, V16 |
| J194309.89-601657.8 | M3.6 | -9.2 | 0.4 | <0.21 | | B&C-832 | V16, G18 |
| J200409.19-672511.7 | M1 | -5.2 | 0.2 | <0.12 | | B&C-832 | V16 |
| J210722.53-705613.4 | M2.8 | -3.9 | 0.2 | 0.14 | 0.03 | FEROS | R17 |
| J211635.34-600513.4 | M3.5 | -6.4 | 0.2 | <0.22 | | FEROS | K14, V16 |
| J213708.89-603606.4 | M2.8 | -7.5 | 0.4 | <0.11 | | FEROS | K14, V16 |
| J214905.04-641304.8 | M4.3 | -7.7 | 0.5 | <0.24 | | B&C-832 | K14, V16 |
| J220216.29-421034.0 | M0 | -2.0 | 0.1 | <0.06 | | FEROS | K14, V16 |
| J220254.57-644045.0 | M1.8 | -3.3 | 0.3 | <0.06 | | FEROS | K14, V16 |
| J224448.45-665003.9 | M5 | -10.2 | 0.4 | <0.48 | | B&C-832 | V16 |
| J224634.82-735351.0 | M2 | -5.0 | 0.3 | <0.15 | | FEROS | V16, M22 |
| J230209.10-121522.0 | M3.6 | -7.0 | 0.8 | <0.15 | | FEROS | V16 |
| J232857.75-680234.5 | M2.3 | -6.1 | 0.3 | <0.11 | | FEROS | K14, V16 |
| J232917.64-675000.6 | M4 | -7.4 | 0.4 | <0.24 | | B&C-832 | K14, V16 |
| J234243.45-622457.1 | M4.5 | -11.2 | 0.7 | <0.39 | | B&C-832 | K14, V16 |
| J234326.88-344658.5 | M1 | -2.3 | 0.2 | <0.12 | | B&C-832 | V16 |
| TWA | | | | | | | |
| J111128.13-265502.9 | M6 | -17.2 | 0.8 | 0.92 | 0.05 | FEROS | TWA 37; G15 |
| J112105.43-384516.6 | M2 | -3.5 | 0.1 | 0.54 | 0.02 | FEROS | TWA 12; Z01 |
| J112651.28-382455.5 | M3 | -7.5 | 0.4 | 0.57 | 0.03 | FEROS | V16, G18 |
| J115927.82-451019.3 | M4 | -10.6 | 0.2 | 0.72 | 0.03 | FEROS | TWA 45; R11 |
| J120237.94-332840.4 | M4.5 | -9.9 | 0.6 | 0.73 | 0.06 | FEROS | TWA 36; M15 |
| J123005.17-440236.1 | M3.8 | -7.4 | 0.9 | 0.67 | 0.05 | FEROS | R11, V16 |
| J125049.12-423123.6 | M4 | -16.0 | 1.2 | 0.23 | 0.05 | FEROS | V16, G18 |

NOTES:
a) References for previous NYMG identifications: B16 = Binks & Jeffries (2016b); B19 = Bowler et al. (2019, and refs. therein); D16 = Donaldson et al. (2016); G15 = Gagné et al. (2015); G18 = Gagné & Faherty (2018); K14 = Kraus et al. (2014); L11 = Looper (2011); L19 = Lee & Song (2019); L23 = Luhman (2023) (and references therein); M13 = Moór et al. (2013); M14 = Malo et al. (2014a); M14B = Malo et al. (2014b); M15 = Murphy & Lawson (2015); M22 = Moranta et al. (2022); R11 = Rodriguez et al. (2011); R14 = Rodriguez et al. (2014); R17 = Riedel et al. (2017); S12 = Schlieder et al. (2012a); S03 = Song et al. (2003); S99 = Sterzik et al. (1999); S17 = Shkolnik et al. (2017); S19 = Schneider et al. (2019); V16 = Vican (2016); Z01 = Zuckerman et al. (2001).
b) Outlier in $M_G$ vs. $G - R_P$ CMD (see § 4.1.2).

The resulting NYMG membership assignments and BANYAN membership probabilities are compiled in Table 3. In all, we identify 132 candidate members of the Table 2 moving groups. These candidates do not include stars that have BANYAN membership probabilities of ≥ 50% but that we have rejected on the basis of closer examination of their $UVW$ and/or suspicious positions in Gaia-based color-magnitude diagrams. Plots of the $XYZ$ and $UVW$ positions of the candidates with respect to the boundaries of the Table 2 moving groups (as listed in Gagné et al. 2018) are presented in Fig.9. It is apparent that our candidates are clustered within and around the NYMG $XYZ$ and $UVW$ regions, as expected, albeit with considerable scatter in some cases. In particular, AB Dor and Argus candidates tend to be more widely dispersed in $UVW$, suggesting that some of these stars may belong to the field population. Individual cases are discussed in § 4.2.

Of our 132 NYMG candidate stars, 63 have been previously assigned membership in the same NYMG by studies previous either to the present work or to Vican (2016), on which the present work is partly based. These previous (and, in some cases, subsequent) identifications of NYMG candidates, as well as those by Vican (2016), are indicated in Table 4. Table 5 provides a summary of the NYMG membership assignments of the spectroscopically observed GALNYSS sample stars.

Among the 132 NYMG candidates listed in Table 3 and Table 4, 27 stars would appear to qualify as "bona-fide" members of their associated NYMG on the basis of their having detectable Li absorption lines, as opposed to upper limits on Li EWs (see column 5 of Table 4). Most of these Li-rich stars are $\beta$PMG and TWA members. A handful of Li-detected stars that are candidates of other NYMGs appear to have anomalous color-magnitude diagram positions, however (see §4.2).

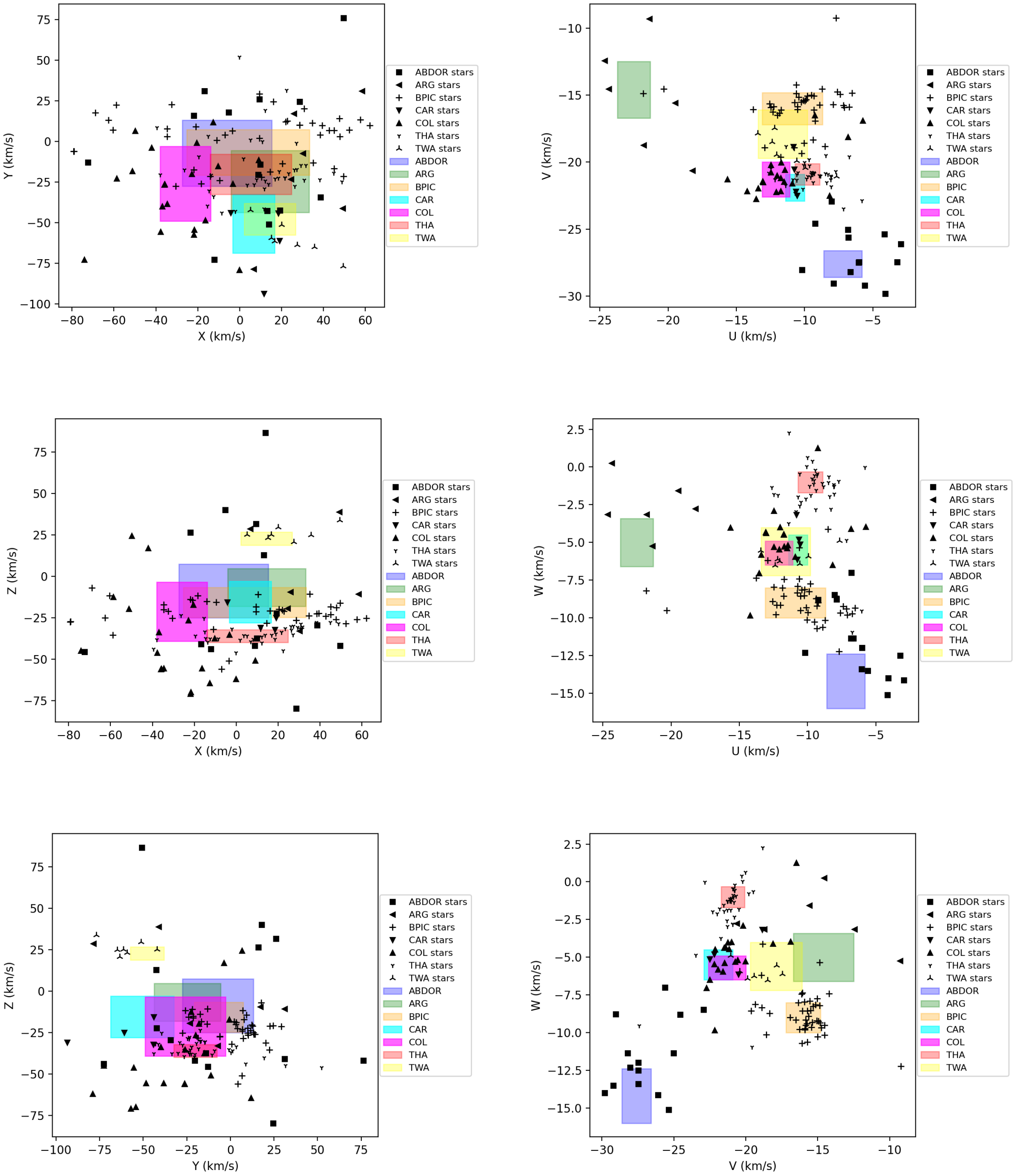

FIG. 9.—: Positions of NYMG candidates listed in Table 3 in $XYZ$ and $UVW$ spaces (left and right columns, respectively), overlaid on the $XYZ$ and $UVW$ boundaries of the Table 2 NYMGs (shaded color regions).

### 4.1.1. *Color-magnitude diagrams*

In Figs. 10-16 we present Gaia absolute $G$ ($= M_G$) vs. $B_P - R_P$ and absolute $G$ vs. $G - R_P$ color-magnitude diagrams (CMDs) constructed for the various moving group candidates which we have identified via the preceding methodology. As in the Gaia CMDs presented earlier in this paper, each CMD in Figs. 10-16 is overlaid with (1) the positions of IC 2391 members (age 50 Myr; Gaia Collaboration et al. 2018b) and (2) the main-sequence star ($D \leq 25$ pc field star) reference sample from Zuckerman et al. (2019, their Fig. 2). It is readily apparent from these CMDs that all of the moving group candidates identified here (Tables 3, 4) lie well above the loci of field stars in both $G$ vs. $B_P - R_P$ and $G$ vs. $G - R_P$ CMDs, confirming their uniformly young ages. Furthermore, as expected, the candidate members of the youngest group, TWA (age ∼10 Myr) — all of which have been previously identified as belonging to the TWA (Vican 2016; Luhman 2023) — also lie well above the $M_G$ vs. $B_P - R_P$ and $M_G$ vs. $G - R_P$ loci of IC 2391 stars. All but a handful of the candidate members of $\beta$PMG (the second youngest group, at age ∼23 Myr) lie above these IC 2391 loci as well (exceptions are discussed in § 4.2). In contrast, the loci of candidate members of the ∼40–50 Myr-old groups Columba, Tuc-Hor, and Carina hew close to those of IC 2391; while about half of the roughly half-dozen candidate stars of AB Dor, the oldest group, lie below the IC 2391 loci but, as noted, sit above the loci of nearby field stars.

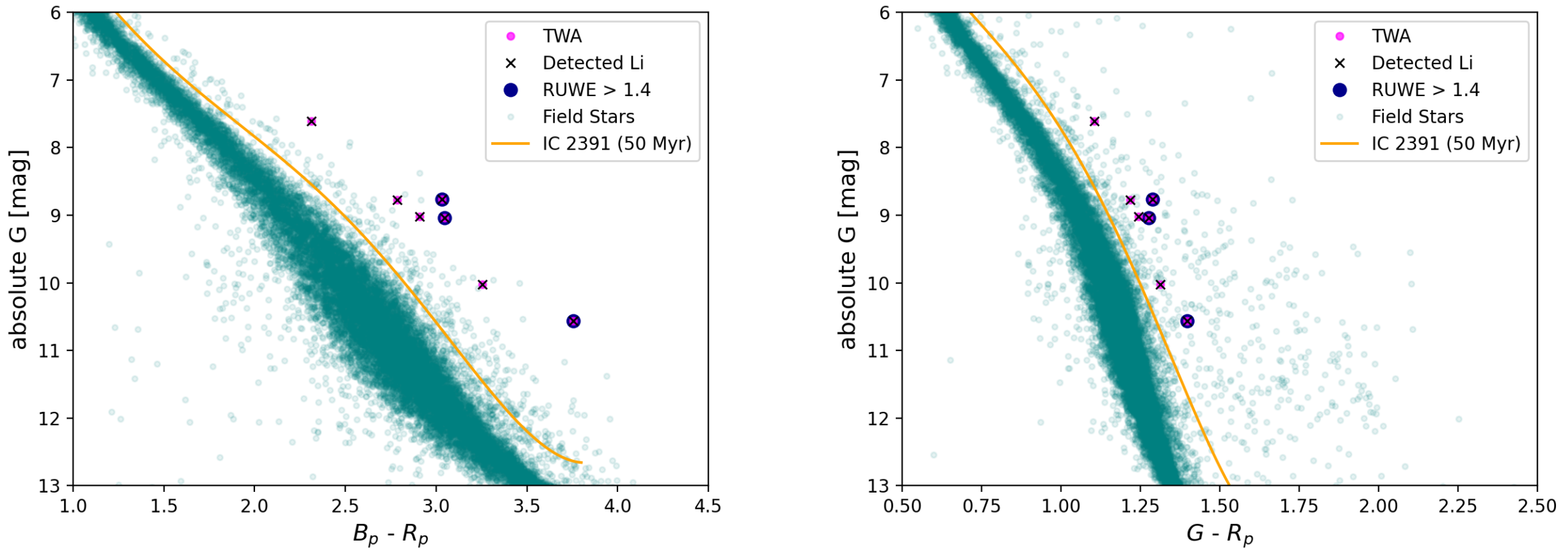

FIG. 10.—: Gaia DR3 CMDs plotting absolute $G$ magnitude vs. $B_P - R_P$ (left) and $G - R_p$ (right) CMDs for TWA members (X's and circles), overlaid with (1) a reference empirical isochrone obtained from fitting Gaia DR3 CMD data for stars of the young cluster IC 2391, age ∼50 Myr; and (2) a nearby ($D < 25$ pc) field star sample (the same as appears in Zuckerman et al. 2019, their Fig. 2) that traces the loci of main sequence stars (small green circles). The black X's indicate stars with Li detected in absorption. Stars with RUWE values > 1.4 in Gaia DR3 data, which is possibly indicative of binarity, are flagged with larger, bright green circles.

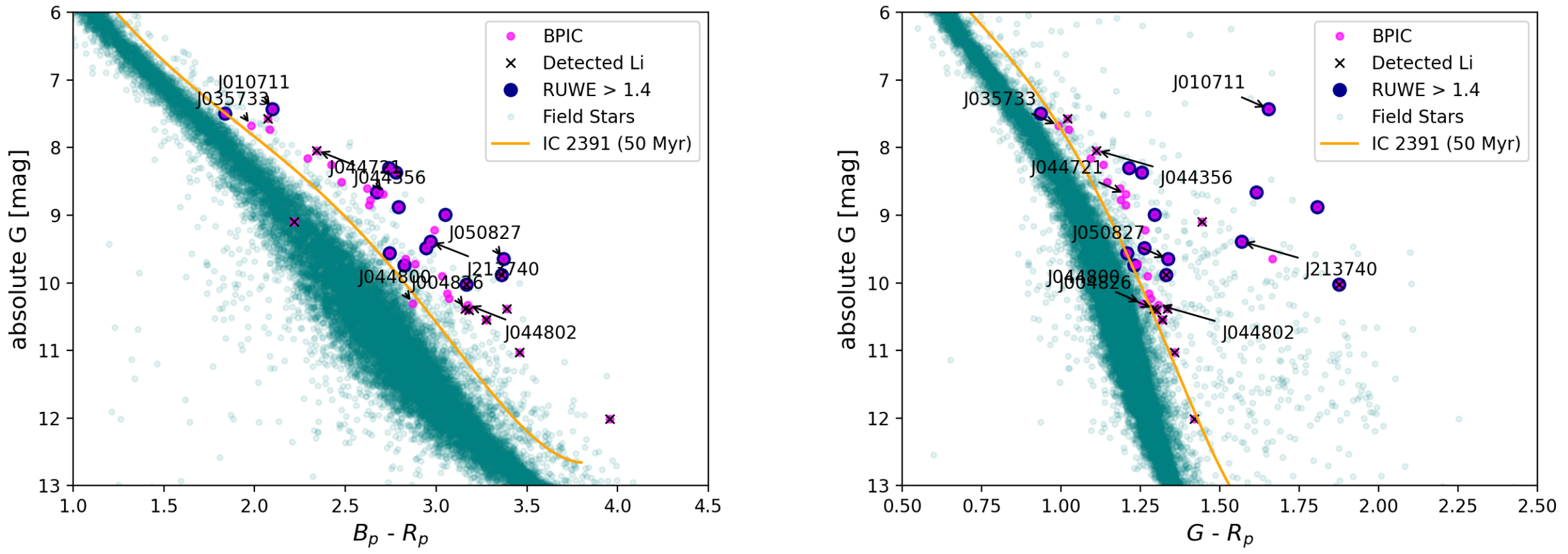

FIG. 11.—: As in Figure 10, for $\beta$PMG members (pink circles). In the right-hand panel of this Figure and Figs. 12, 13, 14, 16, stars with very red $G - R_P$ colors (some with RUWE values > 1.4; green circled points) likely have anomalous Gaia EDR3 photometry. Labeled stars are discussed in § 4.2 and/or are flagged in Table 3.

The CMDs in Figs. 10-16 hence support the proposed membership of the vast majority of candidates listed in Tables 3–4. These CMDs furthermore indicate that some of our candidate NYMG stars are likely photometric binaries, i.e., binary pairs that are unresolved by Gaia and hence lie up to a factor 2 in luminosity (0.75 mag in absolute $G$) above other members of the same NYMG. The fact that some of these same stars are also spectroscopic binaries (§4.3.1), and some also display high values of Gaia RUWE (a data quality metric that is well correlated with unresolved binarity; Belokurov et al. 2020), further supports their probable binary status.

### 4.1.2. *Candidate βPMG Members: Li EWs*

One particularly populous group of interest, the $\beta$PMG, has been the subjects of various investigations of pre-main sequence surface Li depletion as a function of spectral type and age, i.e., to establish the age dependence of the "Li depletion boundary" (LDB; e.g., Binks & Jeffries 2014, 2016b; Galindo-Guil et al. 2022; Lee et al. 2024). In

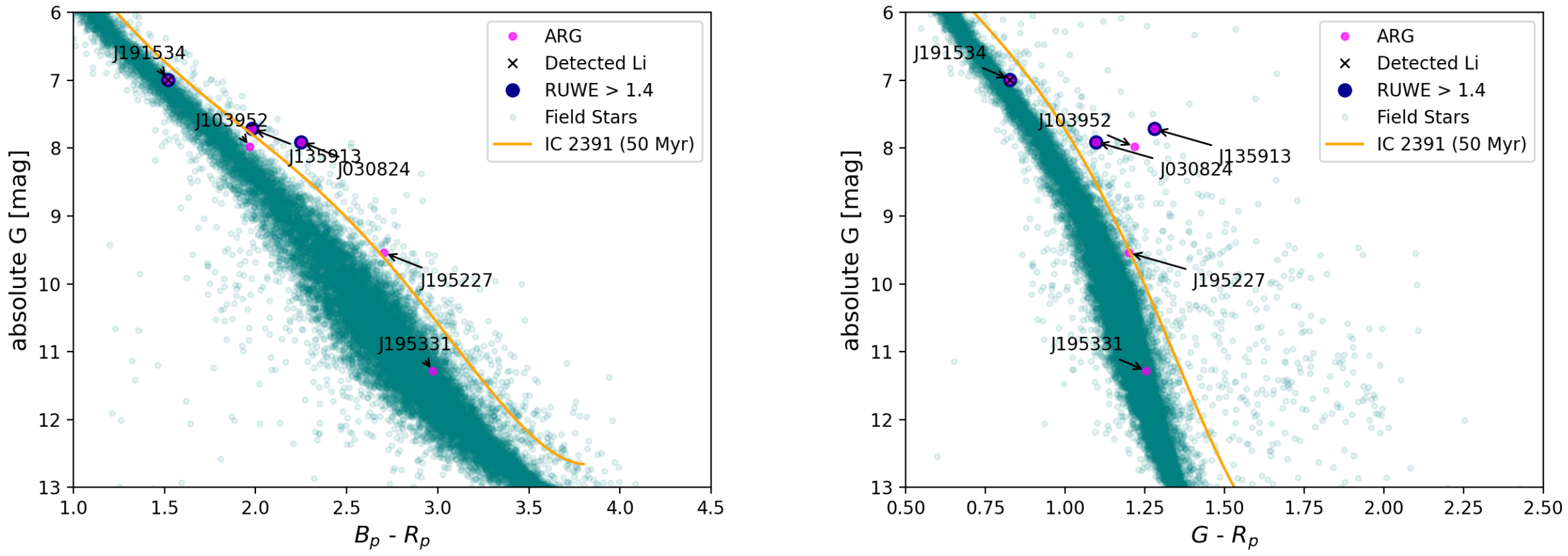

FIG. 12.—: As in Figure 10 for new possible and probable Argus members. Labeled stars are discussed in § 4.2 and/or are flagged in Table 3.

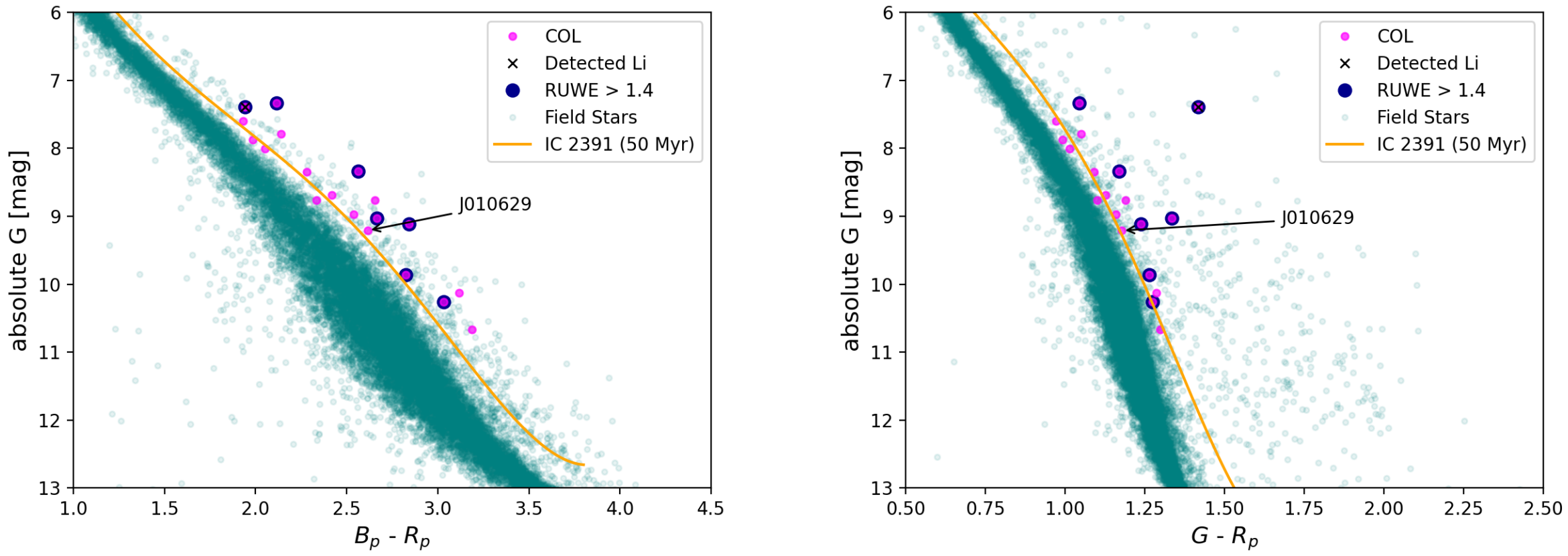

FIG. 13.—: As in Figure 10 for new candidate Columba Association members. Labeled stars are discussed in § 4.2 and/or are flagged in Table 3.

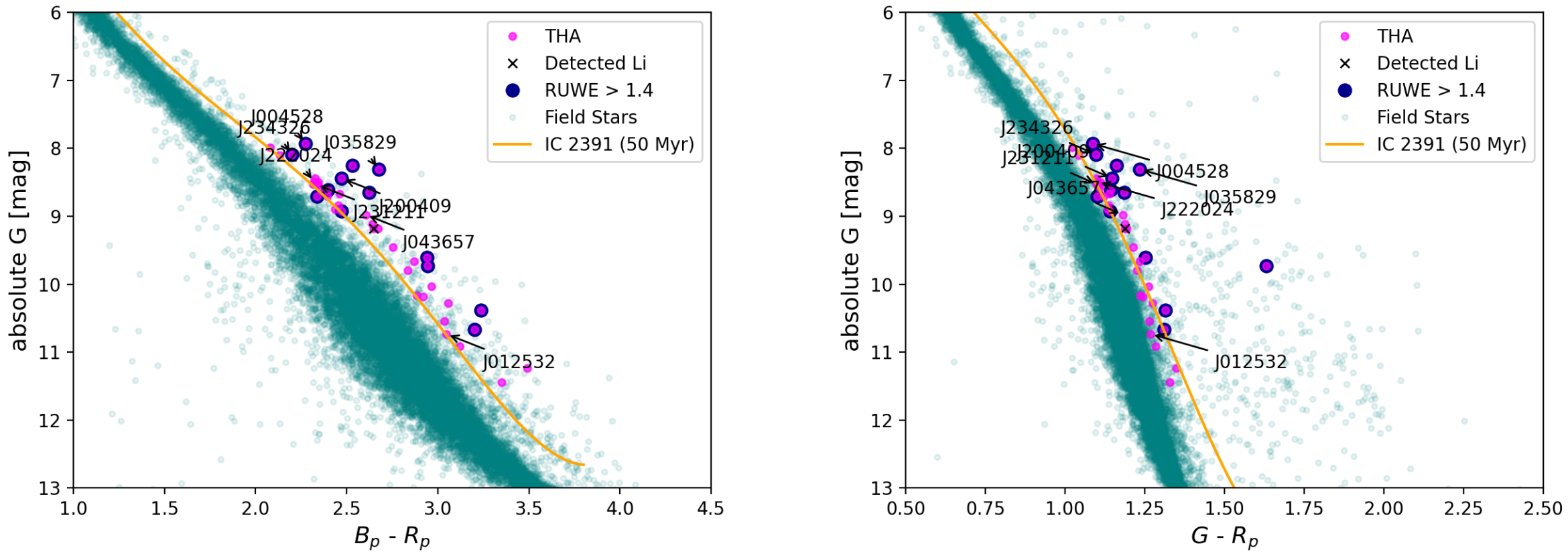

FIG. 14.—: As in Figure 10 for new candidate Tuc-Hor Association members. The labeled stars are discussed in § 4.2 and/or are flagged in Table 3.

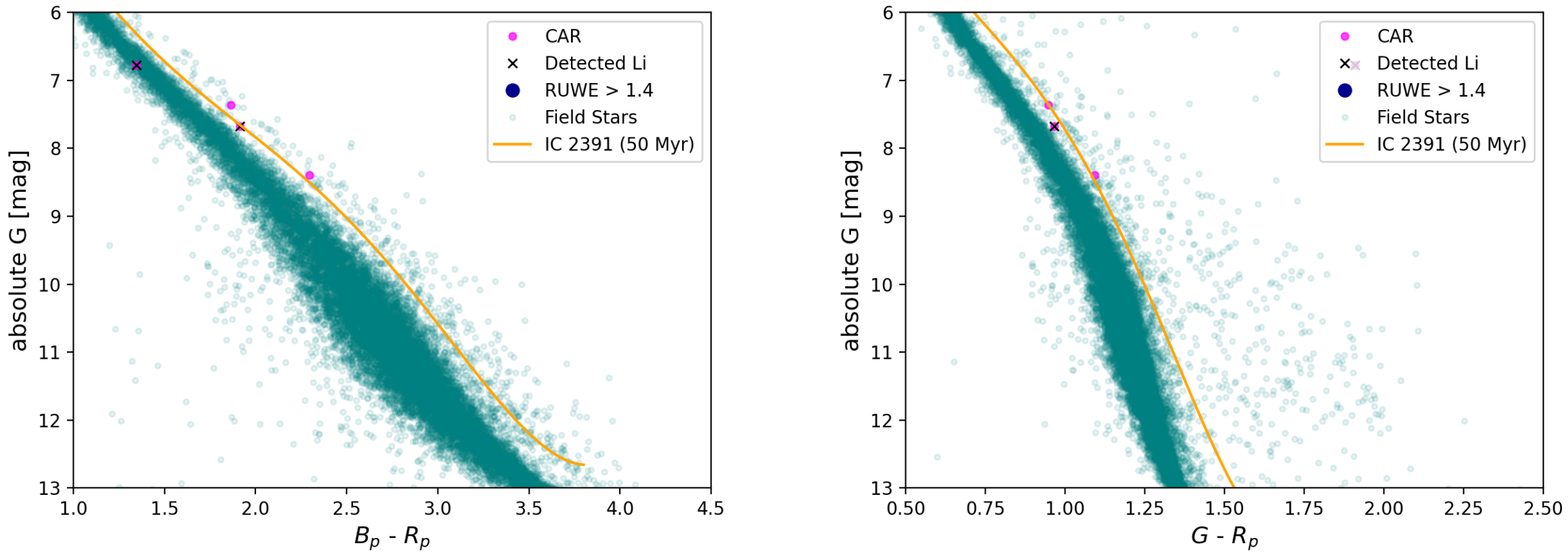

FIG. 15.—: As in Figure 10 for new candidate Carina Association members.

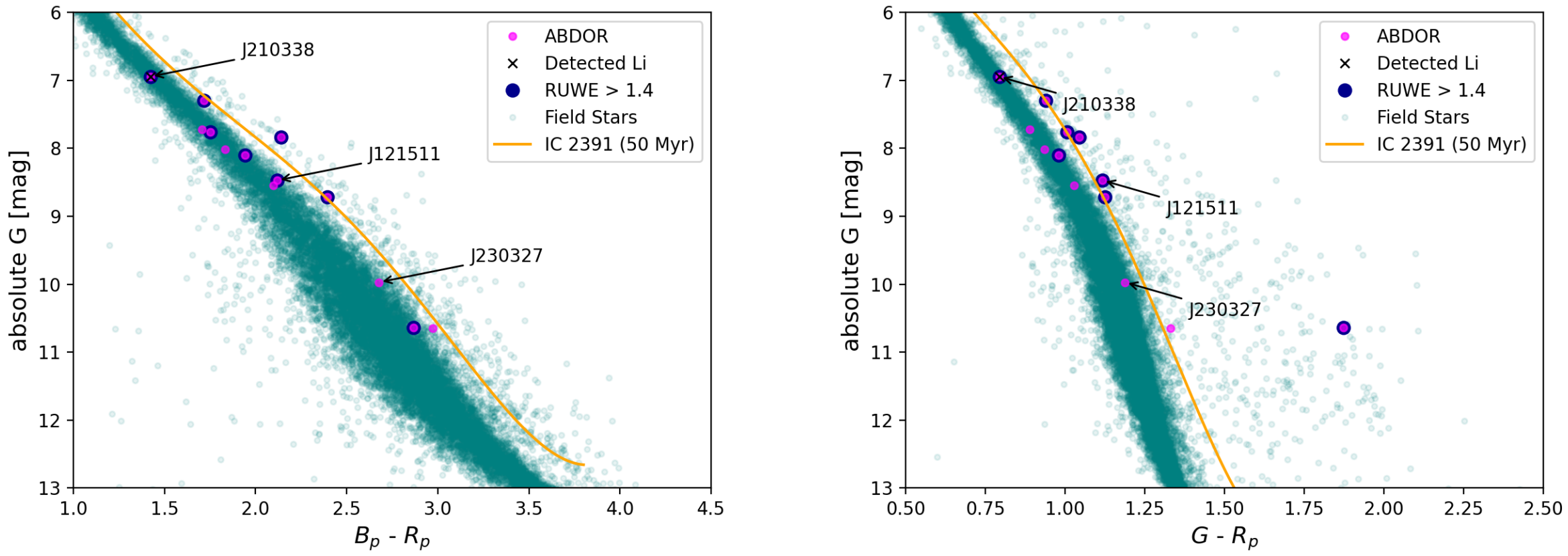

FIG. 16.—: As in Figure 10 for new candidate AB Dor Association members. The labeled stars are discussed in § 4.2.

Fig. 17, we compare our Li EW measurements for candidate members of the $\beta$PMG (Table 4) with those of previously identified members used in these LDB studies. Our 18 new measurements of $\beta$PMG candidates with spectral types ∼M4 and later increases the number of Li EW measurements in this range by a factor ∼2.5. Collectively, these Li EWs appear consistent with previous age determinations based on the LDB method (i.e., ∼15–25 Myr Binks & Jeffries 2014; Galindo-Guil et al. 2022).”

TABLE 5 GALNYSS Star NYMG Candidates: Summary

| NYMG | Total | previously IDed[a] | Low-confidence[b] |
|---|---|---|---|
| AB Dor | 13 | 4 | 7 |
| Argus | 7 | 0 | 3 |
| $\beta$ Pic | 41 | 20 | 9 |
| Carina | 4 | 3 | 1 |
| Columba | 19 | 4 | 2 |
| Tuc-Hor | 41 | 27 | 7 |
| TWA | 7 | 5 | 1 |
| Totals | 132 | 63 | 30 |

NOTES:
a) Number of candidate GALNYSS stars identified as members of the listed NYMG prior to Vican (2016) or this work.
b) Stars with <50% BANYAN membership probabilities and/or whose membership is questionable for other reasons. See §4.2.

### 4.2. *Comments on individual objects and low-confidence NYMG candidates*

In this section, we discuss individual GALNYSS NYMG candidates listed in Table 3 and Table 4, focusing on stars whose NYMG membership is uncertain (as indicated in Table 3, footnotes b or c). These stars are highlighted in the CMDs presented in Figs. 10–16.

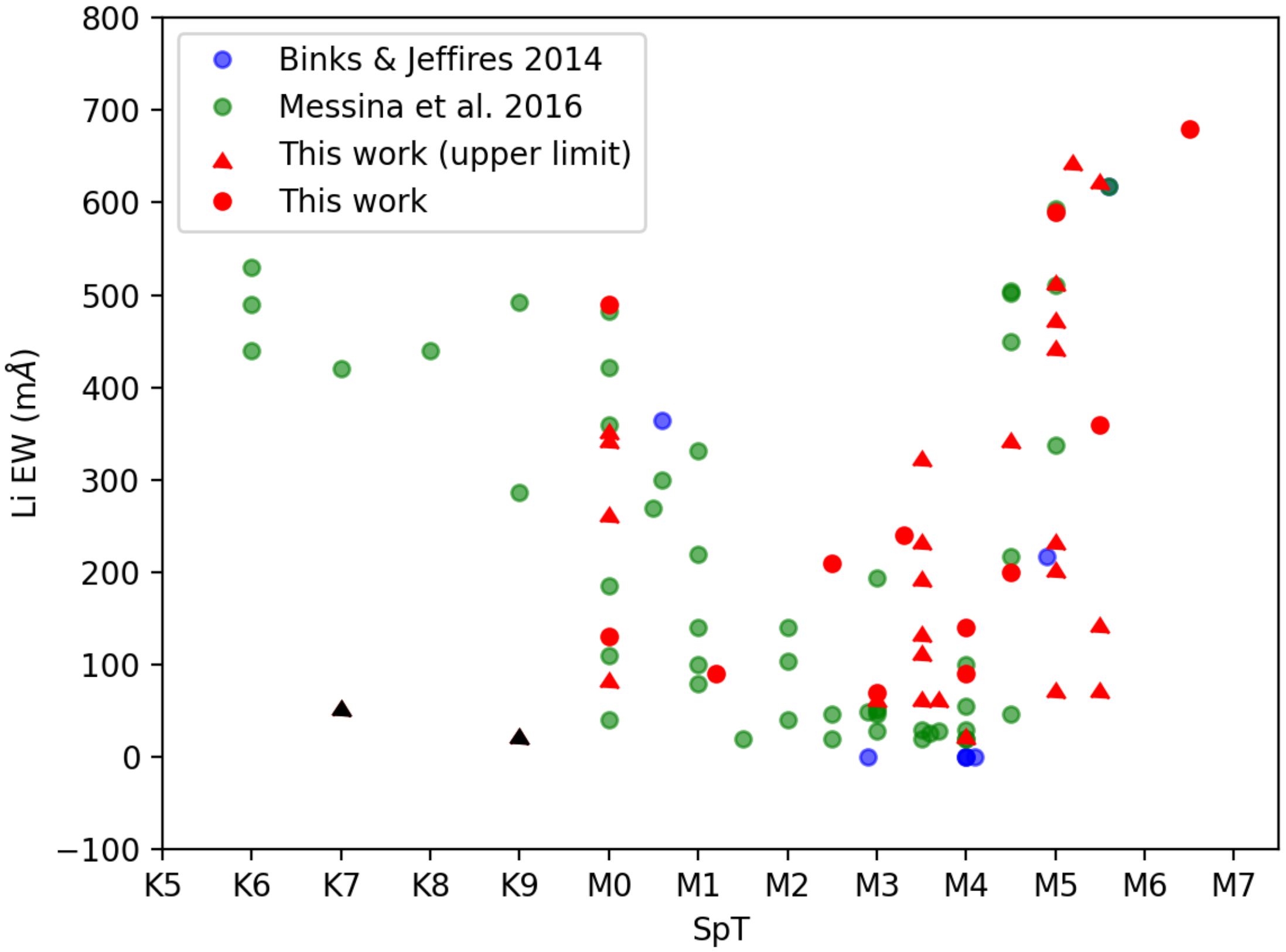

FIG. 17.—: Li λ6708 absorption line equivalent width (EW) measurements for candidate members of βPMG from Table 4, plotted (as red symbols) along with Li EW measurements (as green and blue circles) for previously established members of each group (data from Binks & Jeffries 2014; Messina et al. 2016; Kraus et al. 2014). Red triangles denote Li EW upper limits.

#### 4.2.1. *AB Dor Moving Group*

A majority of the GALNYSS stars we place in the AB Dor Moving Group have BANYAN membership probabilities of less than ∼50%. We have made these AB Dor assignments on the basis of reasonable kinematic ($UVW$) agreement with membership, considering the errors in $UVW$. Cases in point are the stars J121511.25-025457.1, J210338.46+075330.3, and J230327.73-211146.2, all of which have near-zero BANYAN probabilities of membership. In many of these cases we cannot use CMD position as an additional AB Dor membership discriminant; as the oldest NYMG considered here (age ∼150 Myr), only the CMD positions of AB Dor candidate stars with $B_P - R_P > 2.2$ are useful, for purposes of confirming membership (Fig. 16).

We note that three stars we consider here as AB Dor candidates (J054709.88-525626.1, J072911.26-821214.3, J194834.58-760546.9) were previously placed in “greater Scorpius” by Kerr et al. (2021).

The star J210338.46+075330.3 is somewhat enigmatic, and merits followup. This star displays a detectable Li absorption line (EW 240±10 mÅ), indicative of an age <100 Myr — yet has $UVW$ consistent with membership in the ∼150 Myr-old AB Dor Moving Group, and it sits in the midst of the field-star main sequence, in the Gaia DR3 CMD (Fig. 16). Its color-magnitude diagram position is unlikely to be affected by circumstellar dust, as it displays no evidence of IR excess in WISE data.

#### 4.2.2. *Argus Association*

All seven Argus candidates are newly identified as members of this NYMG; however, all but J213835.44-505111.0 are questionable in one way or another and will require confirmation. Two of the three stars with BANYAN membership probabilities >50%, J103952.70-353402.5 and J195331.72-070700.5, are found near the field star MS in the Argus $B_P - R_P$ CMD (Fig. 12). The third, J195227.23-773529.4, has a 98% BANYAN probability of membership, but also has high probability of belonging to the βPMG. The elevated CMD position of J030824.14+234554.2 strongly suggests this is a binary system. The star J191534.83-083019.9 (like AB Dor candidate J210338.46+075330.3) is detected in Li absorption (EW 120±10 mÅ) despite lying very near the field star MS in the $G$ vs. $B_P - R_P$ CMD.

### 4.2.3. *βPMG*

The vast majority of our $\beta$ Pic Moving Group ($\beta$PMG) candidates have previously been identified as such, in the literature (see Table 4). Several $\beta$PMG candidates are problematic and/or interesting. Several stars have BANYAN membership probabilities <50% (Table 3), but in the majority of these cases — e.g., J004826.70-184720.7, J010629.32−122518.4, J050827.31-210144.3, and the wide binary J044800.86+143957.7, J044802.59+143951.1 (= LDS 5606AB; Rodriguez et al. 2014, see below) — the $UVW$ appear compatible with membership and/or the star has previously been assigned to the $\beta$PMG. The $\beta$PMG candidate J053747.56-424030.8 is another example of a Li-rich star (EW 840±20 mÅ) that lies within the field star MS CMD locus. The star J155046.47+305406.9 has $UVW$ velocities that make it equally likely to belong to either $\beta$PMG or AB Dor, but its CMD position is more compatible with the former.

The $\beta$PMG candidacy of the wide binary LDS 5606AB (= J044800.86+143957.7 and J044802.59+143951.1) is deserving of special mention here. This system has recently been considered for membership in the (only slightly younger) 32 Ori stellar association by Luhman (2022) on the basis of the compatibility of the pair's $UVW$ with that of 32 Ori candidates, as well as its distance from Earth, which is larger than previously established $\beta$PMG members. Indeed, the BANYAN $\beta$PMG membership probabilities of the two components of LDS 5606AB are near zero; furthermore, both the $X$ position and $V$ velocity components of the pair are the most negative of the $\beta$PMG candidates listed in Table 3, and are indeed within the respective ranges characteristic of 32 Ori members (Luhman 2022). On the other hand, the LDS 5606AB system is only ∼15 pc more distant than our $\beta$PMG candidate J035733.95+244510.2, which lies in more or less the same direction in the sky, has a high $\beta$PMG membership probability, and is not among the new 32 Ori candidates in Luhman (2022). We hence retain J044800.86+143957.7 and J044802.59+143951.1 as $\beta$PMG candidates, but note that the ($\beta$PMG vs. 32 Ori) membership status of this wide binary, along with J035733.95+244510.2 and other objects found along the boundary of these two groups, should be revisited.

### 4.2.4. *Carina Association*

The Carina candidate J075233.22-643630.5 is another example of a Li-rich star (EW 260±30 mÅ) that lies within the field star MS CMD locus. Although this star has a low (25%) BANYAN probability of Carina membership, its $UVW$ appear compatible, and it was previously placed in the group by Malo et al. (2014a).

### 4.2.5. *Columba Association*

The Columba candidates J010629.32-122518.4, J035345.92-425018.0, J045114.41-601830.5, and J054433.76-200515.5 are questionable due to the well-established overlap with the Tuc-Hor Association in both $UVW$ and age (hence stellar CMD positions) (e.g., Zuckerman et al. 2019). The CMD position of star J051650.66+022713.0 may point toward possible membership in $\beta$PMG. The candidate J181725.08+482202.8 sits rather low, near the field star MS, in the CMD. The stars J055008.59+051153.2 and J221842.70+332113.5 were previously placed in $\beta$PMG (by Bowler et al. 2019; Schlieder et al. 2012b, respectively), but we find their $UVW$ are more compatible with Columba membership.

### 4.2.6. *Tuc-Hor Association*

As is the case for the $\beta$PMG, the vast majority of our Tuc-Hor candidates have been previously identified as such, in the literature (see Table 4). Several of our new candidates (e.g., J035829.67-432517.2) have low BANYAN membership probabilities, but we find that their $UVW$ as well as their CMD positions are consistent with membership. Conversely, as is also noted in Table 3, the stars J004528.25-513734.4, J012532.11-664602.6, and J234326.88-344658.5 have high probabilities of membership, but appear to have anomalous $XYZ$.

### 4.2.7. *TW Hya Association*

All TWA candidates have previously been identified as such in the literature (see, e.g., Vican 2016; Luhman 2023, and references therein). One star, J125049.12-423123.6, was also identified as an LCC candidate by Goldman et al. (2018); given its relatively large distance ∼100 pc), this is the only TWA candidate whose membership status is somewhat uncertain.

## 4.3. *Binary Systems*

### 4.3.1. *Spectroscopic Binaries*

For the 242 stars with high resolution data available, we assessed the evidence for spectroscopic binaries (SBs), i.e., double-line spectra or (in the case of those stars observed more than once; see Table 8) RV variability exceeding ∼25 km s$^{-1}$. We find that 30 stars are members of SBs, the majority of which are previously unidentified. Nine of these SBs are candidate members of the seven NYMGs of interest here; these SB stars are listed in Table 6. Furthermore, as noted in Sec. 3.3, eight stars that are candidate members of the Table 2 NYMGs (J001723.69-664512.4, J004210.98-425254.8, J004528.25-513734.4, J021258.28-585118.3, J032047.66-504133.0, J041749.66+001145.4, J111229.74-461610.1, and J232857.75-680234) have spectroscopically measured RVs that are discrepant with their Gaia-measured RVs, indicative of the presence of lower-mass binary companions.

TABLE 6 Spectroscopic Binaries

| WISE Designation | ΔRV [km s$^{-1}$] | EW(Li)$_A$ [Å] | EW(Li)$_B$ [Å] | NYMG | ref |
|---|---|---|---|---|---|
| J004528.25-513734.4 | 25 | | | THA | |
| J024852.67-340424.9 | | | | COL | |
| J061313.30-274205.6 | | | | BPIC | M14 |
| J072821.16+334511.6 | | <0.08 | | COL | S12 |
| J112547.46-441027.4 | | <0.15 | | ABDOR | J12 |
| J195227.23-773529.4 | 75 | | | ARG | |
| J201000.06-280141.6 | | <0.08 | | BPIC | M14 |
| J211031.49-271058.1B | | | | BPIC | M14 |
| J213740.24+013713.2 | | <0.08 | | BPIC | |

NOTES:

a) Difference in measured RV between two measurements. b) If a known binary system, "ref" is the most recent reference in which the binarity is mentioned. M14=Malo et al. (2014a), S12=Shkolnik et al. (2012), J12=Janson et al. (2012)

#### 4.3.2. *Photometric binaries*

In Table 3 we have flagged 22 NYMG candidates that have elevated positions in their NYMG's Gaia CMD (Figs. 10–16) with respect to the majority of candidates in the group. We regard these stars as candidate photometric binaries, i.e., binary systems that are unresolved by Gaia and so appear overluminous for their colors. Most of these 22 stars also have RUWE values >1.4, supporting their status as likely unresolved binaries (Belokurov et al. 2020). In addition, three candidate photometric binaries, Tuc-Hor candidate J004528.25-513734.4 and Columba candidates J024852.67-340424.9 and J072821.16+334511.6, are among the (nine) NYMG candidates that are SBs (Table 6). One candidate photometric binary, $\beta$PMG candidate J211004.67-192031.2, has a comoving companion that is also among our candidate $\beta$PMG members (see next).

#### 4.3.3. *Visual and Wide (Comoving) Binaries*

We also searched Gaia DR3 for comoving companions to the candidate NYMG members. To compile a list of comoving companions, we queried the DR3 catalog for all Gaia sources within a 500" radius of the position of each star (corresponding to projected separation ∼50 kau at a typical sample star distance of ∼100 pc) and identifying any stars whose parallaxes and proper motions are within a few percent of the star searched. Although each such search typically returns a few thousand stars, most of these field stars have parallaxes far smaller than the (nearby) target star; thus, once reordered by parallax, any comoving companion candidates can be readily and unambiguously identified. Results are presented in Table 7. We find 19 of our NYMG candidates reside in comoving systems, three of which consist of pairs of Table 3 NYMG candidates.

The component stars of young, wide binaries frequently exhibit mid-IR excesses indicative of warm debris disks (Zuckerman 2015; Silverberg et al. 2018; Zuckerman et al. 2019). Hence, we searched the WISE database for evidence of an IR excess around the components of these 31 visual and/or wide binary systems. If a star had W1-W3 or W1-W4$\gtrsim$1, we examined the WISE image itself to make sure the field is clean. We found that two stars have evidence of an IR excess in WISE, and clean fields: J0448+1439 (both A and B components) and J1215-7537 (secondary component). As noted earlier (§ 6, § 4.2), J0448+1439AB (= LDS 5606AB) is a $\beta$PMG or 32 Ori member (Rodriguez et al. 2014; Luhman 2022) and was the subject of a detailed, dedicated study (Zuckerman et al. 2014). The *UVW* of J1215-7537 are incompatible with any known moving group; this star and its companion will be considered further in the followup paper devoted to such cases among the spectroscopically observed GALNYSS stars (Binks et al., in prep.).

TABLE 7  Candidate NYMG members: Comoving Companions[a]

| GALNYSS star (WISE Desig) | companion RA, dec | primary, comp. $G$ (mag) | primary, comp. RV (km s$^{-1}$) | NYMG |
|---|---|---|---|---|
| J012118.22-543425.1 | 01:21:18.21, -54:34:23.6 | 10.65, 14.19 | ..., ... | ABDOR |
| J015350.81-145950.6 | 01:53:50.67, -14:59:49.8 | 11.49, 11.52 | ..., ... | BPMG |
| J034115.60-225307.8[b] | | | | COL |
| J034116.16-225244.0[b] | | | | COL |
| J035345.92-425018.0 | 03:53:46.03, -42:50:16.3 | 13.16, 17.58 | 16.70±4.13, ... | COL |
| J044356.87+372302.7 | 04 43 57.51, +37 23 03.1 | 12.32, 16.16 | ..., ... | BPMG |
| J044800.86+143957.7[b] | | | | BPMG |
| J044802.59+143951.1[b] | | | | BPMG |
| J051026.38-325307.4 | 05:10:26.41, -32:53:09.2 | 13.46, 14.41 | ..., ... | COL |
| J052419.14-160115.5 | 05 24 19.17, -16 01 15.08 | 12.49, 12.77 | ...,... | BPMG |
| J073138.47+455716.5 | 07:31:09.04, +45:56:57.2 | 12.65, 15.25 | ..., ... | COL |
| J103952.70-353402.5 | 10:39:52.78, -35:34:03.5 | 12.88, 12.59 | ..., 14.75±0.73 | ARG |
| J133509.40+503917.5 | 13 35 09.62, +50 39 20.26 | 11.96, 13.25 | -12.82±2.76, ... | ABDOR |
| J184536.02-205910.8 | 18 45 36.12, -20 59 08.44 | 12.27, 13.86 | -30.09±0.14, ... | ARG |
| J191534.83-083019.9 | 19 15 39.02, -08 30 11.62 | 11.12, 13.54 | -27.78±3.27, ... | ARG |
| J210338.46+075330.3 | 21 03 38.39, +07 53 32.4 | 11.91, 15.36 | -21.09±3.49, ... | ABDOR |
| J211004.67-192031.2[b] | | | | BPMG |
| J211005.41-191958.4[b] | | | | BPMG |
| J224500.20-331527.2 | 22:44:57.96, -33:15:01.7 | 11.84, 10.74 | ..., ... | BPMG |

NOTES:

a) Stars in Table 3 with equidistant/comoving companions within 500″ in DR2.

b) Both components of each of these pairs are spectroscopically observed GALNYSS stars that are candidate NYMG members (i.e., both stars are listed in Table 3).

### 4.4. *Signatures of Magnetic Activity*

M dwarfs are known to emit in UV and X-rays due to their magnetic activity. It is important to understand the magnetic behavior of M-dwarfs, given their potential for hosting habitable exoplanets at small orbital semimajor axes. In particular, the extreme UV (EUV) radiation from an M star can lead to photoevaporative mass loss from close-in, otherwise potentially habitable exoplanets (Rugheimer et al. 2015). Since there is no instrument currently available to study the EUV spectral range, we must rely on NUV, FUV, and X-ray flux as well as H$\alpha$ emission to understand the magnetic activity of young, low-mass stars so as to extrapolate to the EUV.

Table 14 lists various magnetic activity indicators for the spectroscopically observed GALNYSS sample. These include optical emission-line diagnostics (relative H$\alpha$ line luminosities, $L_{H\alpha}/L_{bol}$, and Ca II H & K line EWs) and relative NUV and X-ray luminosities ($L_{NUV}/L_{bol}$ and $L_X/L_{bol}$, respectively). Vican (2016) converted H$\alpha$ EWs to $L_{H\alpha}/L_{bol}$ via color-dependent conversion factors (Walkowicz et al. 2004); the conversion factor ($\chi$) adopted for each star is listed in Table 14. To obtain $L_{NUV}/L_{bol}$, Vican (2016) used GALEX and 2MASS fluxes (following Shkolnik & Barman 2014). Although the spread of $L_{NUV}/L_{bol}$ is large, the mean value is roughly $-3.6$, as expected for young M stars (e.g. Rodriguez et al. 2013). For those stars with X-ray data available in the ROSAT All-sky Survey, the values of $L_X/L_{bol}$ listed in Table 14 are also in the range expected for young M stars (roughly $-3.5$ to $-3.0$; e.g., Stelzer et al. 2013).

Fig. 18 displays the mean values of $L_{H\alpha}/L_{bol}$ and $L_{NUV}/L_{bol}$ (and errors on these means) vs. age for the candidate members of NYMGs listed in Table 4. It is evident from Fig. 18 that both luminosity ratios remain more or less constant over the first ~150 Myr of an M-type star's lifetime. These results are consistent with previous assessments of the early evolution of chromospheric UV emission indicators in young M stars, wherein $L_{NUV}/L_{bol}$ remains "saturated" for the first ~300 Myr of low-mass star evolution (e.g., Rodriguez et al. 2013; Shkolnik & Barman 2014). Fig. 18 also appears to hint at a decline in mean M-type star $L_{H\alpha}/L_{bol}$ over the age range ~10–150 Myr, which would be consistent with elevated chromospheric H$\alpha$ luminosity in pre-MS stars relative to MS stars (e.g., Zuckerman & Song 2004). However, this apparent trend rests mostly on the elevated mean $L_{H\alpha}/L_{bol}$ of the TWA (age 10 Myr) given the large uncertainty in mean $L_{H\alpha}/L_{bol}$ for the AB Dor MG (age 150 Myr); furthermore the trend could reflect different spectral type (hence mass) distributions among the different groups, rather than a dependence of $L_{H\alpha}/L_{bol}$ on pre-MS age.

## 5. SUMMARY AND CONCLUSIONS

The GALEX Nearby Young Star Search (GALNYSS) yielded the identification of more than 2000 late-type stars that, based on their ultraviolet and infrared colors and pre-Gaia proper motions, are potentially of age $< 200$ Myr and lie within $\sim 120$ pc of Earth. We used seven instruments at four telescopes in the Northern and Southern hemispheres to obtain high- and medium- resolution spectra of 471 candidate nearby, young stars included in the GALNYSS sample, with the goals of confirming their youth and their potential membership in nearby young stellar moving groups. We detected Li in 79 stars, with ~60 of these stars having equivalent widths (EWs) larger than 200 mÅ. Nine stars have broad H$\alpha$ lines indicative of ongoing accretion, and roughly a dozen stars display metal and/or He I lines in emission. Two stars have infrared excesses that are apparent in WISE photometry and images.

Gaia Data Release 3 (DR3) provides distances to and kinematics of these spectroscopically observed GALNYSS stars. We obtained radial velocities (RVs) from our high-resolution spectra for 232 stars i.e., 49% of the spectroscopically observed sample. A significant fraction of the spectroscopically observed GALNYSS stars, including many for which we have measured RVs, also have Gaia DR3 RVs and/or RVs reported in the literature. The resulting merged set of RVs for these GALNYSS stars, combined with Gaia DR3 proper motion data, yields Galactic space motions ($UVW$).

In this paper, we use the combination of these space motions, Galactic positions ($XYZ$), spectral signatures of youth, and positions in Gaia-based color-magnitude diagrams to test for membership of the GALNYSS stars in seven nearby young moving groups (NYMGs): the TW Hya, Tuc-Hor, and Argus Associations and the $\beta$PMG, Carina, Columba, and AB Dor moving groups. We identify 132 of the spectroscopically observed GALNYSS stars as candidate members of these seven NYMGs; 69 of these candidates are newly identified in this paper or in Vican (2016), on which this paper is partly based. We find 26 of these 132 stars are among those with detectable (in most cases, strong) Li absorption lines; most of these stars are hence "bona-fide" NYMG members. Among our total of 132 candidate members of the aforementioned NYMGs, more than a third reside in binary systems: nine stars are spectroscopic binaries (SBs), eight stars are potential RV variables, 22 are candidate photometric binaries (three of which are also SBs), and 16 stars have common proper motion companions within 500″ in Gaia DR3.

In a followup paper (Binks et al., in preparation; Paper II), we investigate the spectroscopically observed GALNYSS stars whose $UVW$ are clearly inconsistent with membership in the aforementioned seven NYMGs. A subset of these stars appear to be members of somewhat more distant groups, such as UCL, LCC, and the Greater Upper Sco complex. However, a significant fraction do not belong to any previously identified moving group. In Paper II, we consider whether these "rogue" GALNYSS stars, especially those with detectable photospheric Li, might be part of the "hot" population of nearby young stars studied by Binks et al. (2020). We also exploit TESS light curves to investigate the rotational behavior of the spectroscopically observed GALNYSS star sample.

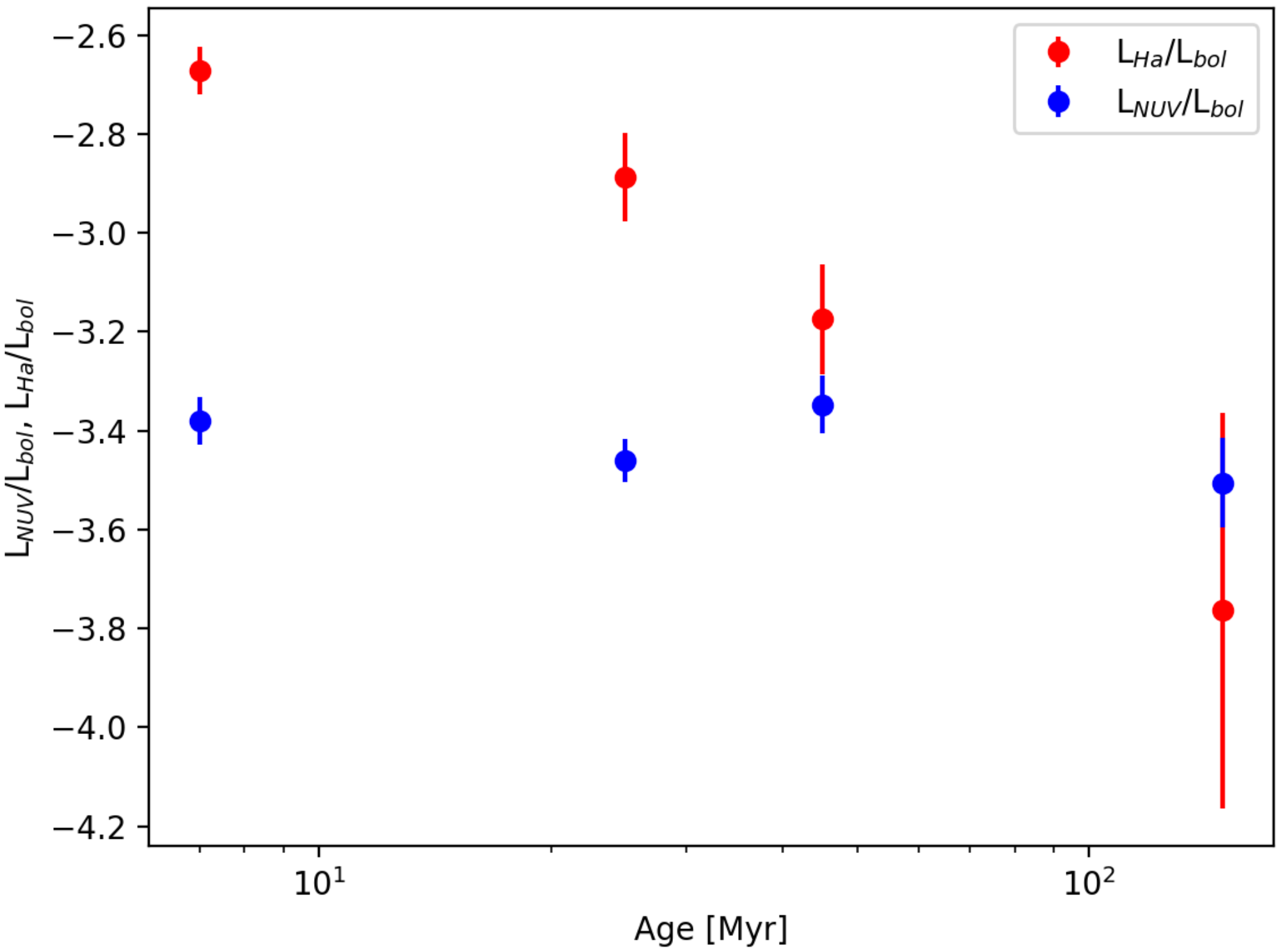

FIG. 18.—: Mean $L_{H\alpha}/L_{bol}$ and $L_{NUV}/L_{bol}$ vs. age for candidate members of the Table 2 NYMGs, i.e., for stars whose spectroscopic properties are listed in Tables 4.

Thanks to their youth and proximity, these spectroscopically observed GALNYSS stars — and, in particular, the newly identified NYMG candidates featured in this paper — offer prime targets for direct-imaging searches for young exoplanets and followup spectroscopic studies of any exoplanets so discovered. The GALNYSS sample stars also provide crucial pre-MS age benchmarks with which various competing theories of planetary (and stellar) evolution can be rigorously tested. Further investigations of the environments and evolutionary tracks of these young, late-type stars in the solar vicinity will hence improve our understanding the early evolution of low-mass stars and their planetary progeny.

## ACKNOWLEDGEMENTS

We thank Dr. Siyi Xu for reducing the ESI spectra and Christine Cheng for assistance with the table of literature radial velocity measurements. This research was supported in part by NASA Astrophysics Data Analysis Program (ADAP) grants NNX09AC96G and 80NSSC22K0625 to RIT, by National Science Foundation Research Experience for Undergraduates Program Grant 2349530 to RIT, by NASA grants to UCLA, and by a National Science Foundation pre-doctoral fellowship to Laura Vican. This research made use of the Montreal Open Clusters and Associations (MOCA) database, operated at the Montréal Planétarium (J. Gagné et al., in preparation).

REFERENCES

Belokurov, V., Penoyre, Z., Oh, S., et al. 2020, MNRAS, 496, 1922, doi: 10.1093/mnras/staa1522

Binks, A. 2016, in IAU Symposium, Vol. 314, IAU Symposium, ed. J. H. Kastner, B. Stelzer, & S. A. Metchev, 159–162, doi: 10.1017/S1743921315006158

Binks, A. S., & Jeffries, R. D. 2014, MNRAS, 438, L11, doi: 10.1093/mnrasl/slt141

—. 2016a, MNRAS, 455, 3345, doi: 10.1093/mnras/stv2431

—. 2016b, MNRAS, 455, 3345, doi: 10.1093/mnras/stv2431

Binks, A. S., Jeffries, R. D., & Maxted, P. F. L. 2015, MNRAS, 452, 173, doi: 10.1093/mnras/stv1309

Binks, A. S., Jeffries, R. D., & Wright, N. J. 2020, MNRAS, 494, 2429, doi: 10.1093/mnras/staa909

Bowler, B. P., Hinkley, S., Ziegler, C., et al. 2019, ApJ, 877, 60, doi: 10.3847/1538-4357/ab1018

Cutri, R. M., Skrutskie, M. F., van Dyk, S., et al. 2003, VizieR Online Data Catalog, 2246

Donaldson, J. K., Weinberger, A. J., Gagné, J., et al. 2016, ApJ, 833, 95, doi: 10.3847/1538-4357/833/1/95

Flaherty, K., Hughes, A. M., Mamajek, E. E., & Murphy, S. J. 2019, ApJ, 872, 92, doi: 10.3847/1538-4357/aaf794

Gagné, J. 2024, arXiv e-prints, arXiv:2405.12860. https://arxiv.org/abs/2405.12860

Gagné, J., & Faherty, J. K. 2018, ApJ, 862, 138, doi: 10.3847/1538-4357/aaca2e

Gagné, J., Lafrenière, D., Doyon, R., Malo, L., & Artigau, É. 2015, ApJ, 798, 73, doi: 10.1088/0004-637X/798/2/73

Gagné, J., Mamajek, E. E., Malo, L., et al. 2018, The Astrophysical Journal, 856, 23

Gaia Collaboration, Brown, A. G. A., Vallenari, A., et al. 2020, arXiv e-prints, arXiv:2012.01533. https://arxiv.org/abs/2012.01533

—. 2018a, A&A, 616, A1, doi: 10.1051/0004-6361/201833051

Gaia Collaboration, Babusiaux, C., van Leeuwen, F., et al. 2018b, A&A, 616, A10, doi: 10.1051/0004-6361/201832843

Gaia Collaboration, Vallenari, A., Brown, A. G. A., et al. 2023, A&A, 674, A1, doi: 10.1051/0004-6361/202243940

Galindo-Guil, F. J., Barrado, D., Bouy, H., et al. 2022, A&A, 664, A70, doi: 10.1051/0004-6361/202141114

Gizis, J. E. 1997, AJ, 113, 806, doi: 10.1086/118302

Goldman, B., Röser, S., Schilbach, E., Moór, A. C., & Henning, T. 2018, ApJ, 868, 32, doi: 10.3847/1538-4357/aae64c

Guinan, E. F., & Engle, S. G. 2009, in IAU Symposium, Vol. 258, The Ages of Stars, ed. E. E. Mamajek, D. R. Soderblom, & R. F. G. Wyse, 395–408, doi: 10.1017/S1743921309032050

Hauschildt, P. H., Allard, F., & Baron, E. 1999, ApJ, 512, 377, doi: 10.1086/306745

Janson, M., Hormuth, F., Bergfors, C., et al. 2012, ApJ, 754, 44, doi: 10.1088/0004-637X/754/1/44

Jeffries, R. D. 2014, in EAS Publications Series, Vol. 65, EAS Publications Series, 289–325, doi: 10.1051/eas/1465008

Kerr, R. M. P., Rizzuto, A. C., Kraus, A. L., & Offner, S. S. R. 2021, ApJ, 917, 23, doi: 10.3847/1538-4357/ac0251

Kraus, A. L., Shkolnik, E. L., Allers, K. N., & Liu, M. C. 2014, AJ, 147, 146, doi: 10.1088/0004-6256/147/6/146

Lada, C. J. 2006, ApJ, 640, L63, doi: 10.1086/503158

Lee, J., & Song, I. 2019, MNRAS, 486, 3434, doi: 10.1093/mnras/stz1044

—. 2024, ApJ, 967, 113, doi: 10.3847/1538-4357/ad3cd9

Lee, R. A., Gaidos, E., van Saders, J., Feiden, G. A., & Gagné, J. 2024, MNRAS, 528, 4760, doi: 10.1093/mnras/stae007

Looper, D. L. 2011, PhD thesis, University of Hawai'i at Manoa

Luhman, K. L. 2022, AJ, 164, 151, doi: 10.3847/1538-3881/ac85e2

—. 2023, AJ, 165, 269, doi: 10.3847/1538-3881/accf19

Lyubimkov, L. S. 2016, Astrophysics, 59, 411, doi: 10.1007/s10511-016-9446-5

Malo, L., Artigau, É., Doyon, R., et al. 2014a, ApJ, 788, 81, doi: 10.1088/0004-637X/788/1/81

Malo, L., Doyon, R., Feiden, G. A., et al. 2014b, ApJ, 792, 37, doi: 10.1088/0004-637X/792/1/37

Martin, D. C., Fanson, J., Schiminovich, D., et al. 2005, The Astrophysical Journal Letters, 619, L1

Messina, S., Lanzafame, A. C., Feiden, G. A., et al. 2016, A&A, 596, A29, doi: 10.1051/0004-6361/201628524

Moór, A., Szabó, G. M., Kiss, L. L., et al. 2013, MNRAS, 435, 1376, doi: 10.1093/mnras/stt1381

Moranta, L., Gagné, J., Couture, D., & Faherty, J. K. 2022, ApJ, 939, 94, doi: 10.3847/1538-4357/ac8c25

Muirhead, P. S., Mann, A. W., Vanderburg, A., et al. 2015, ApJ, 801, 18, doi: 10.1088/0004-637X/801/1/18

Murphy, S. J., & Lawson, W. A. 2015, MNRAS, 447, 1267, doi: 10.1093/mnras/stu2450

Nidever, D. L., Marcy, G. W., Butler, R. P., Fischer, D. A., & Vogt, S. S. 2002, ApJS, 141, 503, doi: 10.1086/340570

Pecaut, M. J., & Mamajek, E. E. 2013, ApJS, 208, 9, doi: 10.1088/0067-0049/208/1/9

Randich, S., & Magrini, L. 2021, Frontiers in Astronomy and Space Sciences, 8, 6, doi: 10.3389/fspas.2021.616201

Riedel, A. R., Alam, M. K., Rice, E. L., Cruz, K. L., & Henry, T. J. 2017, ApJ, 840, 87, doi: 10.3847/1538-4357/840/2/87

Rodriguez, D. R., Bessell, M. S., Zuckerman, B., & Kastner, J. H. 2011, ApJ, 727, 62, doi: 10.1088/0004-637X/727/2/62

Rodriguez, D. R., Zuckerman, B., Faherty, J. K., & Vican, L. 2014, A&A, 567, A20, doi: 10.1051/0004-6361/201423604

Rodriguez, D. R., Zuckerman, B., Kastner, J. H., et al. 2013, ApJ, 774, 101, doi: 10.1088/0004-637X/774/2/101

Rugheimer, S., Kaltenegger, L., Segura, A., Linsky, J., & Mohanty, S. 2015, ApJ, 809, 57, doi: 10.1088/0004-637X/809/1/57

Schlieder, J. E., Lépine, S., & Simon, M. 2012a, AJ, 143, 80, doi: 10.1088/0004-6256/143/4/80

—. 2012b, AJ, 144, 109, doi: 10.1088/0004-6256/144/4/109

Schneider, A., Song, I., Melis, C., Zuckerman, B., & Bessell, M. 2012, ApJ, 757, 163, doi: 10.1088/0004-637X/757/2/163

Schneider, A. C., Shkolnik, E. L., Allers, K. N., et al. 2019, AJ, 157, 234, doi: 10.3847/1538-3881/ab1a26

Sheinis, A. I., Bolte, M., Epps, H. W., et al. 2002, PASP, 114, 851, doi: 10.1086/341706

Shkolnik, E. L., Allers, K. N., Kraus, A. L., Liu, M. C., & Flagg, L. 2017, AJ, 154, 69, doi: 10.3847/1538-3881/aa77fa

Shkolnik, E. L., Anglada-Escudé, G., Liu, M. C., et al. 2012, ApJ, 758, 56, doi: 10.1088/0004-637X/758/1/56

Shkolnik, E. L., & Barman, T. S. 2014, AJ, 148, 64, doi: 10.1088/0004-6256/148/4/64

Shkolnik, E. L., Liu, M. C., Reid, I. N., Dupuy, T., & Weinberger, A. J. 2011, ApJ, 727, 6, doi: 10.1088/0004-637X/727/1/6

Silverberg, S. M., Kuchner, M. J., Wisniewski, J. P., et al. 2018, ApJ, 868, 43, doi: 10.3847/1538-4357/aae3e3

Song, I., Zuckerman, B., & Bessell, M. S. 2003, ApJ, 599, 342, doi: 10.1086/379194

Steele, I. A., & Jameson, R. F. 1995, MNRAS, 272, 630, doi: 10.1093/mnras/272.3.630

Stelzer, B., Marino, A., Micela, G., López-Santiago, J., & Liefke, C. 2013, MNRAS, 431, 2063, doi: 10.1093/mnras/stt225

Sterzik, M. F., Alcalá, J. M., Covino, E., & Petr, M. G. 1999, A&A, 346, L41

Torres, C. A. O., Quast, G. R., Melo, C. H. F., & Sterzik, M. F. 2008, Young Nearby Loose Associations, ed. B. Reipurth, 757

Vican, L. 2016, PhD Thesis,

Walkowicz, L. M., Hawley, S. L., & West, A. A. 2004, PASP, 116, 1105, doi: 10.1086/426792

White, R. J., & Basri, G. 2003, ApJ, 582, 1109, doi: 10.1086/344673

Zuckerman, B. 2015, ApJ, 798, 86, doi: 10.1088/0004-637X/798/2/86

—. 2019, ApJ, 870, 27, doi: 10.3847/1538-4357/aaee66

Zuckerman, B., Klein, B., & Kastner, J. 2019, ApJ, 887, 87, doi: 10.3847/1538-4357/ab45ea

Zuckerman, B., & Song, I. 2004, ARA&A, 42, 685, doi: 10.1146/annurev.astro.42.053102.134111

Zuckerman, B., Song, I., & Webb, R. A. 2001, ApJ, 559, 388, doi: 10.1086/322305

Zuckerman, B., Vican, L., & Rodriguez, D. R. 2014, ApJ, 788, 102, doi: 10.1088/0004-637X/788/2/102

# APPENDIX

# GALNYSS STARS OBSERVED SPECTROSCOPICALLY: TABLES

TABLE 8 Observations: Targets and Instrumentation

| WISE Desig | Binary[a] | Obs Date (UT) | Telescope | Instrument |
|---|---|---|---|---|
| J000453.05-103220.0 | | 31-Oct-12 | Lick 3-m | Kast |
| J000453.05-103220.0 | | 21-Oct-13 | Keck 1 | HIRES |
| J001527.62-641455.2 | | 21-Dec-13 | MPG 2.2m | FEROS |
| J001527.62-641455.2 | | 21-Dec-13 | MPG 2.2m | FEROS |
| J001536.79-294601.2 | | 31-Oct-12 | Lick 3-m | Kast |
| J001536.79-294601.2 | | 12-Sep-14 | du Pont | echelle |
| J001552.28-280749.4 | | 31-Aug-13 | Lick 3-m | Kast |
| J001555.65-613752.2 | | 15-Nov-14 | du Pont | B&C-832 |
| J001709.96+185711.8 | NS | 25-Aug-12 | Lick 3-m | Hamilton |
| J001723.69-664512.4 | | 15-Dec-13 | du Pont | echelle |
| J002101.27-134230.7 | | 31-Oct-12 | Lick 3-m | Kast |
| J003057.97-655006.4 | | 12-Sep-14 | du Pont | echelle |
| J003234.86+072926.4 | | 16-Nov-13 | Keck 1 | HIRES |
| J003903.51+133016.0 | | 16-Oct-14 | Keck 1 | HIRES |
| J004210.98-425254.8 | | 12-Sep-14 | du Pont | echelle |
| J004524.84-775207.5 | | 23-Dec-13 | MPG 2.2m | FEROS |
| J004528.25-513734.4 | AB | 20-Dec-13 | MPG 2.2m | FEROS |
| J004826.70-184720.7 | | 30-Oct-12 | Lick 3-m | Kast |
| J004826.70-184720.7 | | 21-Oct-13 | Keck 1 | HIRES |
| J005633.96-225545.4 | | 31-Aug-13 | Lick 3-m | Kast |
| J010047.97+025029.0 | | 30-Oct-12 | Lick 3-m | Kast |
| J010047.97+025029.0 | | 21-Oct-13 | Keck 1 | HIRES |
| J010126.59+463832.6 | | 25-Aug-12 | Lick 3-m | Hamilton |
| J010126.59+463832.6 | | 30-Oct-12 | Lick 3-m | Kast |
| J010243.86-623534.8 | | 26-Sep-13 | MPG 2.2m | FEROS |
| J010251.05+185653.7 | | 30-Oct-12 | Lick 3-m | Kast |
| J010629.32-122518.4 | | 30-Oct-12 | Lick 3-m | Kast |
| J010629.32-122518.4 | | 23-Sep-13 | MPG 2.2m | FEROS |
| J010711.99-193536.4 | | 16-Nov-13 | Keck 1 | HIRES |
| J011440.20+205712.9 | | 31-Aug-13 | Lick 3-m | Kast |
| J011846.91+125831.4 | | 31-Aug-13 | Lick 3-m | Kast |
| J012118.22-543425.1 | | 18-Dec-13 | MPG 2.2m | FEROS |
| J012245.24-631845.0 | | 26-Sep-13 | MPG 2.2m | FEROS |
| J012332.89-411311.4 | | 14-Nov-14 | du Pont | B&C-832 |
| J012532.11-664602.6 | | 15-Nov-14 | du Pont | B&C-832 |
| J013110.69-760947.7 | | 12-Sep-14 | du Pont | echelle |
| J014156.94-123821.6 | | 1-Sep-13 | Lick 3-m | Kast |
| J014431.99-460432.1 | | 14-Nov-14 | du Pont | B&C-832 |
| J015057.01-584403.4 | | 23-Dec-13 | MPG 2.2m | FEROS |
| J015257.41+083326.3 | | 31-Aug-13 | Lick 3-m | Kast |
| J015257.41+083326.3 | | 21-Oct-13 | Keck 1 | HIRES |
| J015350.81-145950.6 | | 31-Aug-13 | Lick 3-m | Kast |
| J015455.24-295746.0 | | 24-Sep-13 | MPG 2.2m | FEROS |
| J020012.84-084052.4 | | 28-Aug-12 | Lick 3-m | Hamilton |
| J020302.74+221606.8 | | 1-Sep-13 | Lick 3-m | Kast |
| J020305.46-590146.6 | | 15-Nov-14 | du Pont | B&C-832 |
| J020805.55-474633.7 | | 15-Dec-13 | du Pont | echelle |
| J021258.28-585118.3 | | 12-Sep-14 | du Pont | echelle |
| J021330.24-465450.3 | | 23-Sep-13 | MPG 2.2m | FEROS |
| J021935.52-455106.2 | EW | $Oct2014_16$ | Keck 1 | HIRES |
| J022240.88+305515.4 | | 31-Oct-12 | Lick 3-m | Kast |
| J022424.69-703321.2 | | 12-Sep-14 | du Pont | echelle |
| J023005.14+284500.0 | | 31-Aug-13 | Lick 3-m | Kast |
| J023139.36+445638.1 | | 30-Oct-12 | Lick 3-m | Kast |
| J024552.65+052923.8 | | 31-Aug-13 | Lick 3-m | Kast |
| J024746.49-580427.4 | | 18-Dec-13 | MPG 2.2m | FEROS |
| J024852.67-340424.9 | | 16-Nov-13 | Keck 1 | HIRES |
| J025154.17+222728.9 | | 27-Aug-12 | Lick 3-m | Hamilton |
| J025913.40+203452.6 | | 26-Nov-12 | Lick 3-m | Hamilton |
| J030002.98+550652.4 | | 27-Nov-12 | Lick 3-m | Hamilton |
| J030251.62-191150.0 | | 21-Oct-13 | Keck 1 | HIRES |
| J030444.10+220320.8 | | 30-Oct-12 | Lick 3-m | Kast |
| J030444.10+220320.8 | | 17-Oct-13 | Keck 1 | HIRES |
| J030824.14+234554.2 | | 31-Aug-13 | Lick 3-m | Kast |
| J031650.45-350937.9 | | 24-Sep-13 | MPG 2.2m | FEROS |
| J032047.66-504133.0 | | 12-Sep-14 | du Pont | echelle |
| J033235.82+284354.6 | | 31-Oct-12 | Lick 3-m | Kast |
| J033431.66-350103.3 | | 14-Nov-14 | du Pont | B&C-832 |
| J033640.91+032918.3 | | 26-Sep-13 | MPG 2.2m | FEROS |
| J034115.60-225307.8 | | 23-Sep-13 | MPG 2.2m | FEROS |
| J034116.16-225244.0 | | 14-Nov-14 | du Pont | B&C-832 |
| J034236.95+221230.2 | | 16-Oct-14 | Keck 1 | HIRES |
| J034444.80+404150.4 | | 26-Nov-12 | Lick 3-m | Hamilton |
| J035100.83+141339.2 | | 31-Aug-13 | Lick 3-m | Kast |
| J035134.51+072224.5 | | 16-Oct-14 | Keck 1 | HIRES |
| J035223.52-282619.6 | | 25-Sep-13 | MPG 2.2m | FEROS |
| J035223.52-282619.6 | | 16-Nov-13 | Keck 1 | HIRES |
| J035345.92-425018.0 | | 23-Sep-13 | MPG 2.2m | FEROS |
| J035716.56-271245.5 | | 17-Dec-13 | MPG 2.2m | FEROS |
| J035733.95+244510.2 | | 30-Oct-12 | Lick 3-m | Kast |
| J035829.67-432517.2 | | 14-Nov-14 | du Pont | B&C-832 |
| J040539.68-401410.5 | | 25-Sep-13 | MPG 2.2m | FEROS |
| J040649.38-450936.3 | | 24-Sep-13 | MPG 2.2m | FEROS |
| J040711.50-291834.3 | | 16-Nov-13 | Keck 1 | HIRES |
| J040743.83-682511.0 | | 12-Sep-14 | du Pont | echelle |
| J040809.80-611904.3 | | 23-Sep-13 | MPG 2.2m | FEROS |
| J040827.01-784446.7 | | 20-Dec-13 | MPG 2.2m | FEROS |
| J041050.04-023954.4 | | 26-Sep-13 | MPG 2.2m | FEROS |
| J041255.78-141859.2 | | 25-Nov-12 | Lick 3-m | Hamilton |
| J041255.78-141859.2 | | 23-Sep-13 | MPG 2.2m | FEROS |
| J041336.14-441332.4 | | 12-Sep-14 | du Pont | echelle |
| J041525.58-212214.5 | | 16-Oct-14 | Keck 1 | HIRES |
| J041749.66+001145.4 | | 20-Dec-13 | MPG 2.2m | FEROS |
| J041807.76+030826.0 | | 17-Dec-13 | MPG 2.2m | FEROS |
| J042139.19-723355.7 | | 18-Dec-13 | MPG 2.2m | FEROS |
| J042500.91-634309.8 | | 14-Nov-14 | du Pont | B&C-832 |
| J042736.03-231658.8 | | 16-Nov-13 | Keck 1 | HIRES |
| J042739.33+171844.2 | | 21-Oct-13 | Keck 1 | HIRES |
| J043213.46-285754.8 | | 14-Nov-14 | du Pont | B&C-832 |
| J043257.29+740659.3 | | 1-Sep-13 | Lick 3-m | Kast |
| J043657.44-161306.7 | | 19-Dec-13 | MPG 2.2m | FEROS |
| J043726.87+185126.2 | | 25-Nov-12 | Lick 3-m | Hamilton |
| J043923.21+333149.0 | | 16-Oct-14 | Keck 1 | HIRES |
| J043939.24-050150.9 | AB | 1-Sep-13 | Lick 3-m | Kast |
| J043939.24-050150.9 | AB | 17-Oct-13 | Keck 1 | HIRES |
| J044036.23-380140.8 | | 15-Dec-13 | du Pont | echelle |
| J044036.23-380140.8 | | 19-Dec-13 | MPG 2.2m | FEROS |
| J044120.81-194735.6 | | 1-Sep-13 | Lick 3-m | Kast |
| J044120.81-194735.6 | | 23-Sep-13 | MPG 2.2m | FEROS |
| J044120.81-194735.6 | | 17-Oct-13 | Keck 1 | HIRES |
| J044154.44+091953.1 | | 14-Nov-14 | du Pont | B&C-832 |
| J044336.19-003401.8 | | 21-Oct-13 | Keck 1 | HIRES |
| J044349.19+742501.6 | | 31-Oct-12 | Lick 3-m | Kast |
| J044356.87+372302.7 | | 17-Oct-13 | Keck 1 | HIRES |
| J044455.71+193605.3 | | 25-Nov-12 | Lick 3-m | Hamilton |
| J044530.77-285034.8 | | 15-Nov-14 | du Pont | B&C-832 |
| J044700.46-513440.4 | | 20-Dec-13 | MPG 2.2m | FEROS |
| J044721.05+280852.5 | | 30-Oct-12 | Lick 3-m | Kast |
| J044800.86+143957.7 | AB | 17-Oct-13 | Keck 1 | HIRES |
| J044802.59+143951.1 | AB | 21-Oct-13 | Keck 1 | HIRES |
| J045114.41-601830.5 | | 20-Dec-13 | MPG 2.2m | FEROS |
| J045420.20-400009.9 | | 15-Nov-14 | du Pont | B&C-832 |
| J045651.47-311542.7 | | 26-Sep-13 | MPG 2.2m | FEROS |
| J050333.31-382135.6 | | 15-Nov-14 | du Pont | B&C-832 |
| J050610.44-582828.5 | | 24-Sep-13 | MPG 2.2m | FEROS |
| J050827.31-210144.3 | | 16-Nov-13 | Keck 1 | HIRES |
| J051026.38-325307.4 | | 16-Nov-13 | Keck 1 | HIRES |
| J051255.82-212438.7 | | 15-Nov-14 | du Pont | B&C-832 |
| J051310.57-303147.7 | | 18-Dec-13 | MPG 2.2m | FEROS |
| J051403.20-251703.8 | | 24-Sep-13 | MPG 2.2m | FEROS |
| J051403.20-251703.8 | | 16-Nov-13 | Keck 1 | HIRES |
| J051650.66+022713.0 | | 15-Nov-14 | du Pont | B&C-832 |
| J051803.00-375721.2 | | 15-Dec-13 | du Pont | echelle |
| J052419.14-160115.5 | | 21-Oct-13 | Keck 1 | HIRES |
| J052535.85-250230.2 | | 15-Nov-14 | du Pont | B&C-832 |
| J052944.69-323914.1 | | 21-Oct-13 | Keck 1 | HIRES |
| J053100.27+231218.3 | | 21-Oct-13 | Keck 1 | HIRES |
| J053311.32-291419.9 | | 16-Nov-13 | Keck 1 | HIRES |
| J053328.01-425720.1 | AB | "2426-Sep-2013" | MPG 2.2m | FEROS |
| J053747.56-424030.8 | | 16-Oct-14 | Keck 1 | HIRES |
| J053925.08-424521.0 | | 20-Dec-13 | MPG 2.2m | FEROS |
| J054223.86-275803.3 | | 15-Nov-14 | du Pont | B&C-832 |
| J054433.76-200515.5 | | 18-Dec-13 | MPG 2.2m | FEROS |
| J054448.20-265047.4 | | 15-Dec-13 | du Pont | echelle |
| J054709.88-525626.1 | | 24-Sep-13 | MPG 2.2m | FEROS |
| J054719.52-335611.2 | | 15-Nov-14 | du Pont | B&C-832 |
| J055008.59+051153.2 | | 20-Mar-14 | Lick 3-m | Kast |
| J055041.58+430451.8 | | 20-Mar-14 | Lick 3-m | Kast |
| J055041.58+430451.8 | | 14-Apr-16 | Lick 3-m | Hamilton |
| J055208.04+613436.6 | | 20-Mar-14 | Lick 3-m | Kast |
| J055941.10-231909.4 | | 15-Nov-14 | du Pont | B&C-832 |
| J060156.10-164859.9 | AB | 15-Dec-13 | du Pont | echelle |
| J060156.10-164859.9 | AB | 19-Dec-13 | MPG 2.2m | FEROS |
| J060156.10-164859.9 | AB | 20-Dec-13 | MPG 2.2m | FEROS |
| J060156.10-164859.9 | AB | 14-Apr-16 | Lick 3-m | Hamilton |
| J060224.56-163450.0 | | 18-Dec-13 | MPG 2.2m | FEROS |
| J060329.60-260804.7 | | 28-Mar-13 | du Pont | B&C-600 |
| J060329.60-260804.7 | | 24-Mar-14 | MPG 2.2m | FEROS |
| J061313.30-274205.6 | | 18-Feb-13 | MPG 2.2m | FEROS |
| J061740.43-475957.2 | | 28-Mar-13 | du Pont | B&C-600 |
| J061851.01-383154.9 | | 19-Mar-14 | du Pont | echelle |
| J062047.17-361948.2 | | 18-Feb-13 | MPG 2.2m | FEROS |
| J062130.52-410559.1 | | 28-Mar-13 | du Pont | B&C-600 |
| J062407.62+310034.4 | | 25-Nov-12 | Lick 3-m | Hamilton |
| J063001.84-192336.6 | | 19-Mar-14 | du Pont | echelle |
| J065846.87+284258.9 | | 4-May-13 | Lick 3-m | Kast |
| J070657.72-535345.9 | | 28-Mar-13 | du Pont | B&C-600 |
| J070657.72-535345.9 | | 23-Dec-13 | MPG 2.2m | FEROS |
| J071036.50+171322.6 | | 20-Mar-14 | Lick 3-m | Kast |
| J072641.52+185034.0 | | 30-Oct-12 | Lick 3-m | Kast |
| J072821.16+334511.6 | | 25-Nov-12 | Lick 3-m | Hamilton |
| J072911.26-821214.3 | | 18-Dec-13 | MPG 2.2m | FEROS |
| J073138.47+455716.5 | | 5-May-13 | Lick 3-m | Kast |
| J075233.22-643630.5 | | 28-Mar-13 | du Pont | B&C-600 |
| J075808.25-043647.5 | | 24-Mar-14 | MPG 2.2m | FEROS |
| J075830.92+153013.4 | AB | 21-Oct-13 | Keck 1 | HIRES |
| J080352.54+074346.7 | | 5-May-13 | Lick 3-m | Kast |
| J080636.05-744424.6 | | 17-Feb-13 | MPG 2.2m | FEROS |
| J081443.62+465035.8 | | 30-Oct-12 | Lick 3-m | Kast |
| J081738.97-824328.8 | | 15-Dec-13 | du Pont | echelle |
| J082105.04-090853.8 | AB | 24-Mar-14 | MPG 2.2m | FEROS |
| J082558.91+034019.5 | | 26-Nov-12 | Lick 3-m | Hamilton |
| J083528.87+181219.9 | | 5-May-13 | Lick 3-m | Kast |
| J090227.87+584813.4 | | 30-Oct-12 | Lick 3-m | Kast |
| J092216.12+043423.3 | | 4-May-13 | Lick 3-m | Kast |
| J093212.63+335827.3 | | 5-May-13 | Lick 3-m | Kast |
| J094317.05-245458.3 | | 17-Dec-13 | MPG 2.2m | FEROS |
| J094508.15+714450.1 | | 20-Mar-14 | Lick 3-m | Kast |
| J094508.15+714450.1 | | 17-Jun-14 | Keck 1 | HIRES |
| J100146.28+681204.1 | | 30-Oct-12 | Lick 3-m | Kast |
| J100230.94-281428.2 | | 18-Feb-13 | MPG 2.2m | FEROS |
| J100230.94-281428.2 | | 24-Mar-14 | MPG 2.2m | FEROS |
| J101543.44+660442.3 | | 4-May-13 | Lick 3-m | Kast |
| J101905.68-304920.3 | | 19-Dec-13 | MPG 2.2m | FEROS |
| J101917.57-443736.0 | | 18-Feb-13 | MPG 2.2m | FEROS |
| J101917.57-443736.0 | | 19-Dec-13 | MPG 2.2m | FEROS |
| J102602.07-410553.8 | AB | 16-Feb-13 | MPG 2.2m | FEROS |
| J102602.07-410553.8 | AB | 17-Dec-13 | MPG 2.2m | FEROS |
| J102602.07-410553.8 | AB | 24-Mar-14 | MPG 2.2m | FEROS |
| J102636.95+273838.4 | | 22-May-13 | Lick 3-m | Kast |
| J103016.11-354626.3 | | 15-Dec-13 | du Pont | echelle |
| J103137.59-374915.9 | | 19-Mar-14 | du Pont | echelle |
| J103557.17+285330.8 | | 3-May-13 | Lick 3-m | Kast |
| J103952.70-353402.5 | | 20-Jun-14 | du Pont | B&C-832 |
| J104008.36-384352.1 | | 28-Mar-13 | du Pont | B&C-600 |
| J104044.98-255909.2 | | 15-Dec-13 | du Pont | echelle |
| J104044.98-255909.2 | | 14-Apr-16 | Lick 3-m | Hamilton |
| J104551.72-112615.4 | | 17-Feb-13 | MPG 2.2m | FEROS |
| J105515.87-033538.2 | | 3-May-13 | Lick 3-m | Kast |
| J105518.12-475933.2 | | 17-Dec-13 | MPG 2.2m | FEROS |
| J105524.25-472611.7 | AB | 19-Mar-14 | du Pont | echelle |
| J105711.36+054454.2 | | 3-May-13 | Lick 3-m | Kast |
| J105850.47-234620.8 | | 16-Feb-13 | MPG 2.2m | FEROS |
| J105850.47-234620.8 | | 28-Mar-13 | du Pont | B&C-600 |
| J110119.22+525222.9 | | 3-May-13 | Lick 3-m | Kast |
| J110335.71-302449.5 | | 18-Feb-13 | MPG 2.2m | FEROS |
| J110551.56-780520.7 | | 21-Jun-14 | du Pont | B&C-832 |
| J111052.06-725513.0 | | 21-Jun-14 | du Pont | B&C-832 |
| J111103.54-313459.0 | | 19-Mar-14 | du Pont | echelle |
| J111103.54-313459.0 | | 17-Jun-14 | Keck 1 | HIRES |
| J111128.13-265502.9 | | 18-Feb-13 | MPG 2.2m | FEROS |
| J111128.13-265502.9 | | 28-Mar-13 | du Pont | B&C-600 |
| J111229.74-461610.1 | | 15-Dec-13 | du Pont | echelle |
| J111309.15+300338.4 | | 4-May-13 | Lick 3-m | Kast |
| J111707.56-390951.3 | | 15-Dec-13 | du Pont | echelle |
| J112047.03-273805.8 | | 17-Feb-13 | MPG 2.2m | FEROS |
| J112105.43-384516.6 | | 17-Feb-13 | MPG 2.2m | FEROS |
| J112105.43-384516.6 | | 28-Mar-13 | du Pont | B&C-600 |
| J112512.28-002438.2 | | 22-May-13 | Lick 3-m | Kast |
| J112547.46-441027.4 | | 18-Feb-13 | MPG 2.2m | FEROS |
| J112651.28-382455.5 | | 18-Feb-13 | MPG 2.2m | FEROS |
| J112816.27-261429.6 | | 17-Feb-13 | MPG 2.2m | FEROS |
| J112816.27-261429.6 | | 28-Mar-13 | du Pont | B&C-600 |
| J112816.27-261429.6 | | 24-Mar-14 | MPG 2.2m | FEROS |
| J112955.84+520213.2 | | 4-May-13 | Lick 3-m | Kast |
| J113105.57+542913.5 | | 22-May-13 | Lick 3-m | Kast |
| J113114.81-482628.0 | | 19-Mar-14 | du Pont | echelle |
| J113120.31+132140.0 | | 5-May-13 | Lick 3-m | Kast |
| J114623.01-523851.8 | | 20-Jun-14 | du Pont | B&C-832 |
| J114728.37+664402.7 | | 20-Mar-14 | Lick 3-m | Kast |
| J115156.73+073125.7 | | 4-May-13 | Lick 3-m | Kast |
| J115438.73-503826.4 | | 20-Jun-14 | du Pont | B&C-832 |
| J115927.82-451019.3 | | 18-Feb-13 | MPG 2.2m | FEROS |
| J115949.51-424426.0 | | 17-Feb-13 | MPG 2.2m | FEROS |
| J115949.51-424426.0 | | 28-Mar-13 | du Pont | B&C-600 |
| J115949.51-424426.0 | | 18-Dec-13 | MPG 2.2m | FEROS |
| J115949.51-424426.0 | | 19-Mar-14 | du Pont | echelle |
| J115957.68-262234.1 | | 23-Dec-13 | MPG 2.2m | FEROS |
| J120001.54-173131.1 | | 18-Feb-13 | MPG 2.2m | FEROS |
| J120001.54-173131.1 | | 28-Mar-13 | du Pont | B&C-600 |
| J120001.54-173131.1 | | 19-Mar-14 | du Pont | echelle |
| J120001.54-173131.1 | | 17-Jun-14 | Keck 1 | HIRES |
| J120237.94-332840.4 | | 17-Feb-13 | MPG 2.2m | FEROS |
| J120237.94-332840.4 | | 28-Mar-13 | du Pont | B&C-600 |
| J120647.40-192053.1 | | 16-Nov-13 | Keck 1 | HIRES |
| J120929.80-750540.2 | | 21-Jun-14 | du Pont | B&C-832 |
| J121153.04+124912.9 | | 3-May-13 | Lick 3-m | Kast |
| J121341.59+323127.7 | AB | 3-May-13 | Lick 3-m | Kast |
| J121341.59+323127.7 | AB | 3-May-13 | Lick 3-m | Kast |
| J121429.15-425814.8 | | 19-Mar-14 | du Pont | echelle |
| J121511.25-025457.1 | | 4-May-13 | Lick 3-m | Kast |
| J121558.37-753715.7 | | 21-Jun-14 | du Pont | B&C-832 |
| 2M 12182363-3515098[b] | | 17-Feb-13 | MPG 2.2m | FEROS |
| J122643.99-122918.3 | | 3-May-13 | Lick 3-m | Kast |
| J122643.99-122918.3 | | 19-Mar-14 | du Pont | echelle |
| J122725.27-454006.6 | | 24-Mar-14 | MPG 2.2m | FEROS |
| J122813.57-431638.9 | | 28-Mar-13 | du Pont | B&C-600 |
| J123005.17-440236.1 | | 17-Feb-13 | MPG 2.2m | FEROS |
| J123234.07-414257.5 | | 20-Jun-14 | du Pont | B&C-832 |
| J123425.84-174544.4 | | 17-Feb-13 | MPG 2.2m | FEROS |
| J123704.99-441919.5 | | 28-Mar-13 | du Pont | B&C-600 |
| J124054.09-451625.4 | | 24-Mar-14 | MPG 2.2m | FEROS |
| J124612.32-384013.5 | | 20-Jun-14 | du Pont | B&C-832 |
| J124955.67-460737.3 | | 24-Mar-14 | MPG 2.2m | FEROS |
| J125049.12-423123.6 | | 17-Feb-13 | MPG 2.2m | FEROS |
| J125049.12-423123.6 | | 28-Mar-13 | du Pont | B&C-600 |
| J125326.99-350415.3 | | 17-Feb-13 | MPG 2.2m | FEROS |
| J125902.99-314517.9 | | 18-Feb-13 | MPG 2.2m | FEROS |
| J125902.99-314517.9 | | 28-Mar-13 | du Pont | B&C-600 |
| J130501.18-331348.7 | | 21-Jun-14 | du Pont | B&C-832 |
| J130522.37-405701.2 | | 24-Mar-14 | MPG 2.2m | FEROS |
| J130530.31-405626.0 | | 24-Mar-14 | MPG 2.2m | FEROS |
| J130618.16-342857.0 | | 20-Jun-14 | du Pont | B&C-832 |
| J130650.27-460956.1 | | 21-Jun-14 | du Pont | B&C-832 |
| J130731.03-173259.9 | | 21-Jun-14 | du Pont | B&C-832 |
| J131129.00-425241.9 | | 20-Jun-14 | du Pont | B&C-832 |
| J132112.77-285405.1 | | 19-Mar-14 | du Pont | echelle |
| J133238.94+305905.8 | | 20-Mar-14 | Lick 3-m | Kast |
| J133509.40+503917.5 | | 3-May-13 | Lick 3-m | Kast |
| J133901.87-214128.0 | | 24-Jun-14 | Lick 3-m | Kast |
| J134146.41+581519.2 | | 25-Jun-04 | Lick 3-m | Kast |
| J134907.28+082335.8 | | 3-May-13 | Lick 3-m | Kast |
| J135145.65-374200.7 | | 17-Feb-13 | MPG 2.2m | FEROS |
| J135511.38+665207.0 | | 20-Mar-14 | Lick 3-m | Kast |
| J135913.33-292634.2 | | 3-May-13 | Lick 3-m | Kast |
| J135913.33-292634.2 | | 19-Mar-14 | du Pont | echelle |
| J140337.56-501047.9 | | 28-Mar-13 | du Pont | B&C-600 |
| J141045.24+364149.8 | | 5-May-13 | Lick 3-m | Kast |
| J141045.24+364149.8 | | 25-Jun-04 | Lick 3-m | Kast |
| J141332.23-145421.1 | | 24-Jun-14 | Lick 3-m | Kast |
| J141510.77-252012.0 | | 22-May-13 | Lick 3-m | Kast |
| J141842.36+475514.9 | | 3-May-13 | Lick 3-m | Kast |
| J141903.13+645146.4 | | 23-May-13 | Lick 3-m | Kast |
| J143517.80-342250.4 | | 20-Jun-14 | du Pont | B&C-832 |
| J143648.16+090856.5 | | 20-Mar-14 | Lick 3-m | Kast |
| J143753.36-343917.8 | | 20-Jun-14 | du Pont | B&C-832 |
| J145014.12-305100.6 | | 23-May-13 | Lick 3-m | Kast |
| J145731.11-305325.0 | | 24-Jun-14 | Lick 3-m | Kast |
| J145949.90+244521.9 | | 5-May-13 | Lick 3-m | Kast |
| J150119.48-200002.1 | | 24-Jun-14 | Lick 3-m | Kast |
| J150230.94-224615.4 | | 24-Jun-14 | Lick 3-m | Kast |
| J150355.37-214643.1 | | 23-May-13 | Lick 3-m | Kast |
| J150355.37-214643.1 | | 17-Jun-14 | Keck 1 | HIRES |
| J150355.37-214643.1 | | 14-Apr-16 | Lick 3-m | Hamilton |
| J150601.66-240915.0 | | 30-Jun-13 | du Pont | B&C-832 |
| J150723.91+433353.6 | | 5-May-13 | Lick 3-m | Kast |
| J150820.15-282916.6 | | 20-Jun-14 | du Pont | B&C-832 |
| J150836.69-294222.9 | | 25-Jun-04 | Lick 3-m | Kast |
| J150939.16-133212.4 | | 22-May-13 | Lick 3-m | Kast |
| J150939.16-133212.4 | | 17-Jun-14 | Keck 1 | HIRES |
| J150939.16-133212.4 | | 14-Apr-16 | Lick 3-m | Hamilton |
| J151212.18-255708.3 | | 3-May-13 | Lick 3-m | Kast |
| J151242.69-295148.0 | | 30-Jun-13 | du Pont | B&C-832 |
| J151411.31-253244.1 | | 30-Jun-13 | du Pont | B&C-832 |
| J152150.76-251412.1 | | 4-May-13 | Lick 3-m | Kast |
| J153248.80-230812.4 | | 30-Jun-13 | du Pont | B&C-832 |
| J153549.35-065727.8 | | 24-Mar-14 | MPG 2.2m | FEROS |
| J154220.24+593653.0 | | 20-Mar-14 | Lick 3-m | Kast |
| J154220.24+593653.0 | | 17-Jun-14 | Keck 1 | HIRES |
| J154227.07-042717.1 | | 30-Jun-13 | du Pont | B&C-832 |
| J154349.42-364838.7 | | 30-Jun-13 | du Pont | B&C-832 |
| J154435.17+042307.5 | | 3-May-13 | Lick 3-m | Kast |
| J154435.17+042307.5 | | 21-Oct-13 | Keck 1 | HIRES |
| J154435.17+042307.5 | | 14-Apr-16 | Lick 3-m | Hamilton |
| J154656.43+013650.8 | | 3-May-13 | Lick 3-m | Kast |
| J155046.47+305406.9 | | 3-May-13 | Lick 3-m | Kast |
| J155515.35+081327.9 | | 24-Jun-14 | Lick 3-m | Kast |
| J155759.01-025905.8 | | 24-Jun-14 | Lick 3-m | Kast |
| J155947.24+440359.6 | | 4-May-13 | Lick 3-m | Kast |
| J160116.86-345502.7 | | 21-Jun-14 | du Pont | B&C-832 |
| J160549.19-311521.6 | | 4-May-13 | Lick 3-m | Kast |
| J160549.19-311521.6 | | 17-Jun-14 | Keck 1 | HIRES |
| J160828.45-060734.6 | | 4-May-13 | Lick 3-m | Kast |
| J160828.45-060734.6 | | 19-Mar-14 | du Pont | echelle |
| J160828.45-060734.6 | | 17-Jun-14 | Keck 1 | HIRES |
| J160954.85-305858.4 | | 30-Jun-13 | du Pont | B&C-832 |
| J161410.76-025328.8 | AB | 24-Mar-14 | MPG 2.2m | FEROS |
| J161743.18+261815.2 | | 25-Aug-12 | Lick 3-m | Hamilton |
| J162422.68+195922.0 | | 3-May-13 | Lick 3-m | Kast |
| J162548.69-135912.0 | | 28-Aug-12 | Lick 3-m | Hamilton |
| J162548.69-135912.0 | | 4-May-13 | Lick 3-m | Kast |
| J162602.80-155954.5 | | 30-Jun-13 | du Pont | B&C-832 |
| J162602.80-155954.5 | | 24-Mar-14 | MPG 2.2m | FEROS |
| J163051.34+472643.8 | | 20-Mar-14 | Lick 3-m | Kast |
| J163632.90+635344.9 | | 4-May-13 | Lick 3-m | Kast |
| J164539.37+702400.1 | | 31-Oct-12 | Lick 3-m | Kast |
| J170415.15-175552.5 | | 28-Aug-12 | Lick 3-m | Hamilton |
| J170415.15-175552.5 | | 30-Jun-13 | du Pont | B&C-832 |
| J171038.44-210813.0 | | 27-Aug-12 | Lick 3-m | Hamilton |
| J171117.68+124540.4 | | 23-May-13 | Lick 3-m | Kast |
| J171426.13-214845.0 | | 20-Jun-14 | du Pont | B&C-832 |
| J171441.70-220948.8 | | 1-Sep-13 | Lick 3-m | Kast |
| J171441.70-220948.8 | | 21-Oct-13 | Keck 1 | HIRES |
| J172130.71-150617.8 | | 20-Jun-14 | du Pont | B&C-832 |
| J172131.73-084212.3 | | 20-Jun-14 | du Pont | B&C-832 |
| J172309.67-095126.2 | | 31-Aug-13 | Lick 3-m | Kast |
| J172454.26+502633.0 | | 20-Mar-14 | Lick 3-m | Kast |
| J172454.26+502633.0 | | 14-Apr-16 | Lick 3-m | Hamilton |
| J172615.23-031131.9 | | 21-Jun-14 | du Pont | B&C-832 |
| J172615.23-031131.9 | | 24-Jun-14 | Lick 3-m | Kast |
| J172951.38+093336.9 | | 20-Mar-14 | Lick 3-m | Kast |
| J173353.07+165511.7 | | 28-Aug-12 | Lick 3-m | Hamilton |
| J173544.26-165209.9 | | 27-Aug-12 | Lick 3-m | Hamilton |
| J173544.26-165209.9 | | 3-May-13 | Lick 3-m | Kast |
| J173623.80+061853.0 | | 24-Jun-14 | Lick 3-m | Kast |
| J173826.94-055628.0 | | 21-Jun-14 | du Pont | B&C-832 |
| J174203.85-032340.4 | | 21-Jun-14 | du Pont | B&C-832 |
| J174426.59-074925.3 | | 24-Jun-14 | Lick 3-m | Kast |
| J174439.27+483147.1 | | 24-Jun-14 | Lick 3-m | Kast |
| J174536.31-063215.3 | | 21-Jun-14 | du Pont | B&C-832 |
| J174735.31-033644.4 | | 20-Jun-14 | du Pont | B&C-832 |
| J174811.33-030510.2 | | 24-Jun-14 | Lick 3-m | Kast |
| J174936.01-010808.7 | | 20-Jun-14 | du Pont | B&C-832 |
| J175022.27-094457.8 | | 20-Jun-14 | du Pont | B&C-832 |
| J175839.30+155208.6 | | 22-May-13 | Lick 3-m | Kast |
| J175839.30+155208.6 | | 17-Jun-14 | Keck 1 | HIRES |
| J175942.12+784942.1 | | 17-Jun-14 | Keck 1 | HIRES |
| J180508.62-015058.5 | | 21-Jun-14 | du Pont | B&C-832 |
| J180554.92-570431.3 | | 21-Jun-14 | du Pont | B&C-832 |
| J180658.07+161037.9 | | 3-May-13 | Lick 3-m | Kast |
| J180658.07+161037.9 | | 21-Oct-13 | Keck 1 | HIRES |
| J180733.00+613153.6 | | 30-Oct-12 | Lick 3-m | Kast |
| J180929.71-543054.2 | | 21-Jun-14 | du Pont | B&C-832 |
| J181059.88-012322.4 | | 20-Jun-14 | du Pont | B&C-832 |
| J181725.08+482202.8 | | 25-Aug-12 | Lick 3-m | Hamilton |
| J181725.08+482202.8 | | 17-Jun-14 | Keck 1 | HIRES |
| J182054.20+022101.5 | | 20-Jun-14 | du Pont | B&C-832 |
| J182905.79+002232.2 | | 20-Jun-14 | du Pont | B&C-832 |
| J184204.85-555413.3 | | 21-Jun-14 | du Pont | B&C-832 |
| J184206.97-555426.2 | | 30-May-14 | MPG 2.2m | FEROS |
| J184206.97-555426.2 | | 21-Jun-14 | du Pont | B&C-832 |
| J184536.02-205910.8 | | 24-Jun-14 | Lick 3-m | Kast |
| J190453.69-140406.0 | | 27-Aug-12 | Lick 3-m | Hamilton |
| J191019.82-160534.8 | | 30-Jun-13 | du Pont | B&C-832 |
| J191036.02-650825.5 | | 21-Jun-14 | du Pont | B&C-832 |
| J191235.95+630904.7 | | 1-Sep-13 | Lick 3-m | Kast |
| J191500.80-284759.1 | | 30-Jun-13 | du Pont | B&C-832 |
| J191534.83-083019.9 | | 27-Aug-12 | Lick 3-m | Hamilton |
| J191629.61-270707.2 | | 30-Jun-13 | du Pont | B&C-832 |
| J192240.05-061208.0 | | 1-Sep-13 | Lick 3-m | Kast |
| J192242.80-051553.8 | | 31-Aug-13 | Lick 3-m | Kast |
| J192250.70-631058.6 | | 26-Sep-13 | MPG 2.2m | FEROS |
| J192323.20+700738.3 | | 26-Aug-12 | Lick 3-m | Hamilton |
| J192323.20+700738.3 | | 3-May-13 | Lick 3-m | Kast |
| J192434.97-344240.0 | | 30-Jun-13 | du Pont | B&C-832 |
| J192434.97-344240.0 | | 30-May-14 | MPG 2.2m | FEROS |
| J192600.77-533127.6 | AB | 30-Jun-13 | du Pont | B&C-832 |
| J192600.77-533127.6 | AB | 30-May-14 | MPG 2.2m | FEROS |
| J192659.33-710923.8 | | 25-Sep-13 | MPG 2.2m | FEROS |
| J193052.51-545325.4 | | 24-Sep-13 | MPG 2.2m | FEROS |
| J193411.46-300925.3 | | 23-Sep-13 | MPG 2.2m | FEROS |
| J193711.26-040126.7 | | 24-Jun-14 | Lick 3-m | Kast |
| J194309.89-601657.8 | | 21-Jun-14 | du Pont | B&C-832 |
| J194309.89-601657.8 | | 12-Sep-14 | du Pont | echelle |
| J194444.21-435903.0 | | 21-Jun-14 | du Pont | B&C-832 |
| J194539.01+704445.9 | | 26-Aug-12 | Lick 3-m | Hamilton |
| J194539.01+704445.9 | | 3-May-13 | Lick 3-m | Kast |
| J194714.54+640237.9 | | 4-May-13 | Lick 3-m | Kast |
| J194816.54-272032.3 | | 3-May-13 | Lick 3-m | Kast |
| J194834.58-760546.9 | | 21-Jun-14 | du Pont | B&C-832 |
| J195227.23-773529.4 | AB | 24-Sep-13 | MPG 2.2m | FEROS |
| J195315.67+745948.9 | | 4-May-13 | Lick 3-m | Kast |
| J195331.72-070700.5 | | 20-Jun-14 | du Pont | B&C-832 |
| J195340.71+502458.2 | | 23-May-13 | Lick 3-m | Kast |
| J195602.95-320719.3 | | 20-Jun-14 | du Pont | B&C-832 |
| J200137.19-331314.5 | | 30-Jun-13 | du Pont | B&C-832 |
| J200137.19-331314.5 | | 24-Sep-13 | MPG 2.2m | FEROS |
| J200311.61-243959.2 | | 4-May-13 | Lick 3-m | Kast |
| J200311.61-243959.2 | | 23-Sep-13 | MPG 2.2m | FEROS |
| J200409.19-672511.7 | | 21-Jun-14 | du Pont | B&C-832 |
| J200423.80-270835.8 | | 31-Aug-13 | Lick 3-m | Kast |
| J200556.44-321659.7 | | 30-Jun-13 | du Pont | B&C-832 |
| J200837.87-254526.2 | | 30-Jun-13 | du Pont | B&C-832 |
| J200853.72-351949.3 | | 30-Jun-13 | du Pont | B&C-832 |
| J201000.06-280141.6 | | 30-Jun-13 | du Pont | B&C-832 |
| J201000.06-280141.6 | | 25-Sep-13 | MPG 2.2m | FEROS |
| J201931.84-081754.3 | | 4-May-13 | Lick 3-m | Kast |
| J202505.36+835954.2 | | 17-Jun-14 | Keck 1 | HIRES |
| J202716.80-254022.8 | | 30-Jun-13 | du Pont | B&C-832 |
| J203023.10+711419.8 | | 30-Oct-12 | Lick 3-m | Kast |
| J203023.10+711419.8 | | 21-Oct-13 | Keck 1 | HIRES |
| J203301.99-490312.6 | | 20-Jun-14 | du Pont | B&C-832 |
| J203337.63-255652.8 | | 30-Jun-13 | du Pont | B&C-832 |
| J203337.63-255652.8 | | 21-Oct-13 | Keck 1 | HIRES |
| J204406.36-153042.3 | | 24-Jun-14 | Lick 3-m | Kast |
| J204714.59+110442.2 | | 24-Jun-14 | Lick 3-m | Kast |
| J205131.01-154857.6 | | 30-Jun-13 | du Pont | B&C-832 |
| J205136.27+240542.9 | | 17-Jun-14 | Keck 1 | HIRES |
| J205832.99-482033.8 | | 20-Jun-14 | du Pont | B&C-832 |
| J210131.13-224640.9 | | 30-Jun-13 | du Pont | B&C-832 |
| J210338.46+075330.3 | | 25-Aug-12 | Lick 3-m | Hamilton |
| J210708.43-113506.0 | | 25-Sep-13 | MPG 2.2m | FEROS |
| J210722.53-705613.4 | | 25-Sep-13 | MPG 2.2m | FEROS |
| J210736.82-130458.9 | | 27-Aug-12 | Lick 3-m | Hamilton |
| J210957.48+032121.1 | | 17-Jun-14 | Keck 1 | HIRES |
| J211004.67-192031.2 | | 24-Sep-13 | MPG 2.2m | FEROS |
| J211005.41-191958.4 | | 23-Sep-13 | MPG 2.2m | FEROS |
| J211031.49-271058.1 | AB | 30-Jun-13 | du Pont | B&C-832 |
| J211031.49-271058.1 | AB | 21-Oct-13 | Keck 1 | HIRES |
| J211031.49-271058.1 | AB | 16-Oct-14 | Keck 1 | HIRES |
| J211635.34-600513.4 | | 26-Sep-13 | MPG 2.2m | FEROS |
| J212007.84-164548.2 | | 30-Jun-13 | du Pont | B&C-832 |
| J212007.84-164548.2 | | 1-Sep-13 | Lick 3-m | Kast |
| J212007.84-164548.2 | | 26-Sep-13 | MPG 2.2m | FEROS |
| J212007.84-164548.2 | | 26-Sep-13 | MPG 2.2m | FEROS |
| J212128.89-665507.1 | | 21-Jun-14 | du Pont | B&C-832 |
| J212230.56-333855.2 | | 30-Jun-13 | du Pont | B&C-832 |
| J212750.60-684103.9 | | 21-Jun-14 | du Pont | B&C-832 |
| J212750.60-684103.9 | | 12-Sep-14 | du Pont | echelle |
| J212750.60-684103.9 | | 12-Sep-14 | du Pont | echelle |
| J213507.39+260719.4 | | 1-Sep-13 | Lick 3-m | Kast |
| J213520.34-142917.9 | | 30-Jun-13 | du Pont | B&C-832 |
| J213644.54+670007.1 | | 30-Oct-12 | Lick 3-m | Kast |
| J213708.89-603606.4 | | 24-Sep-13 | MPG 2.2m | FEROS |
| J213740.24+013713.2 | | 16-Oct-14 | Keck 1 | HIRES |
| J213835.44-505111.0 | | 21-Jun-14 | du Pont | B&C-832 |
| J213847.58+050451.4 | | 30-Jun-13 | du Pont | B&C-832 |
| J214101.48+723026.7 | | 30-Oct-12 | Lick 3-m | Kast |
| J214126.66+204310.5 | | 31-Aug-13 | Lick 3-m | Kast |
| J214414.73+321822.3 | | 24-Jun-14 | Lick | Kast |
| J214905.04-641304.8 | | 21-Jun-14 | du Pont | B&C-832 |
| J215053.68-055318.9 | | 28-Aug-12 | Lick 3-m | Hamilton |
| J215128.95-023814.9 | | 30-Jun-13 | du Pont | B&C-832 |
| J215128.95-023814.9 | | 31-Aug-13 | Lick 3-m | Kast |
| J215717.71-341834.0 | | 20-Jun-14 | du Pont | B&C-832 |
| J220216.29-421034.0 | | 23-Sep-13 | MPG 2.2m | FEROS |
| J220254.57-644045.0 | | 23-Sep-13 | MPG 2.2m | FEROS |
| J220306.98-253826.6 | | 31-Oct-12 | Lick 3-m | Kast |
| J220730.16-691952.6 | | 24-Sep-13 | MPG 2.2m | FEROS |
| J220850.39+114412.7 | | 16-Nov-13 | Keck 1 | HIRES |
| J221217.17-681921.1 | | 21-Jun-14 | du Pont | B&C-832 |
| J221559.00-014733.0 | | 30-Jun-13 | du Pont | B&C-832 |
| J221833.85-170253.2 | | 30-Jun-13 | du Pont | B&C-832 |
| J221842.70+332113.5 | | 30-Oct-12 | Lick 3-m | Kast |
| J222024.21-072734.5 | | 30-Jun-13 | du Pont | B&C-832 |
| J224111.08-684141.8 | | 21-Jun-14 | du Pont | B&C-832 |
| J224221.02-410357.2 | | 20-Jun-14 | du Pont | B&C-832 |
| J224448.45-665003.9 | | 15-Nov-14 | du Pont | B&C-832 |
| J224500.20-331527.2 | | 14-Nov-14 | du Pont | B&C-832 |
| J224634.82-735351.0 | | 26-Sep-13 | MPG 2.2m | FEROS |
| J225914.87+373639.3 | | 16-Oct-14 | Keck 1 | HIRES |
| J225934.89-070447.1 | | 30-Jun-13 | du Pont | B&C-832 |
| J230209.10-121522.0 | | 31-Oct-12 | Lick 3-m | Kast |
| J230209.10-121522.0 | | 23-Sep-13 | MPG 2.2m | FEROS |
| J230209.10-121522.0 | | 23-Sep-13 | MPG 2.2m | FEROS |
| J230327.73-211146.2 | | 30-Jun-13 | du Pont | B&C-832 |
| J230740.98+080359.7 | | 1-Sep-13 | Lick 3-m | Kast |
| J231021.75+685943.6 | | 31-Aug-13 | Lick 3-m | Kast |
| J231211.37+150329.7 | | 31-Aug-13 | Lick 3-m | Kast |
| J231246.53-504924.8 | | 30-May-14 | MPG 2.2m | FEROS |
| J231246.53-504924.8 | | 20-Jun-14 | du Pont | B&C-832 |
| J231246.53-504924.8 | | 20-Jun-14 | du Pont | B&C-832 |
| J231457.86-633434.0 | AB | 26-Sep-13 | MPG 2.2m | FEROS |
| J231457.86-633434.0 | AB | 23-Dec-13 | MPG 2.2m | FEROS |
| J231457.86-633434.0 | AB | 26-Sep-13 | MPG 2.2m | FEROS |
| J231457.86-633434.0 | AB | 23-Dec-13 | MPG 2.2m | FEROS |
| J231543.66-140039.6 | | 27-Aug-12 | Lick 3-m | Hamilton |
| J231543.66-140039.6 | | 31-Oct-12 | Lick 3-m | Kast |
| J231543.66-140039.6 | | 31-Oct-12 | Lick 3-m | Kast |
| J231933.16-393924.3 | | 16-Nov-13 | Keck 1 | HIRES |
| J232008.15-634334.9 | | 14-Nov-14 | du Pont | B&C-832 |
| J232151.23+005037.3 | | 31-Aug-13 | Lick 3-m | Kast |
| J232656.43+485720.9 | | 25-Aug-12 | Lick 3-m | Hamilton |
| J232857.75-680234.5 | | 25-Sep-13 | MPG 2.2m | FEROS |
| J232904.42+032910.8 | | 30-Oct-12 | Lick 3-m | Kast |
| J232917.64-675000.6 | | 14-Nov-14 | du Pont | B&C-832 |
| J232959.47+022834.0 | | 30-Jun-13 | du Pont | B&C-832 |
| J233647.87+001740.1 | | 30-Jun-13 | du Pont | B&C-832 |
| J233647.87+001740.1 | | 1-Sep-13 | Lick 3-m | Kast |
| J234243.45-622457.1 | | 14-Nov-14 | du Pont | B&C-832 |
| J234326.88-344658.5 | | 30-Jun-13 | du Pont | B&C-832 |
| J234326.88-344658.5 | | 21-Dec-13 | MPG 2.2m | FEROS |
| J234333.91-192802.8 | | 31-Oct-12 | Lick 3-m | Kast |
| J234333.91-192802.8 | | 26-Nov-12 | Lick 3-m | Hamilton |
| J234333.91-192802.8 | | 26-Nov-12 | Lick 3-m | Hamilton |
| J234347.83-125252.1 | | 1-Sep-13 | Lick 3-m | Kast |
| J234857.35+100929.3 | | 30-Jun-13 | du Pont | B&C-832 |
| J234924.87+185926.7 | | 21-Oct-13 | Keck 1 | HIRES |
| J234926.23+185912.4 | | 30-Oct-12 | Lick 3-m | Kast |
| J235250.70-160109.7 | | 1-Sep-13 | Lick 3-m | Kast |

NOTES:
a) Binary systems indicated as “AB” or, for visual binaries with well-defined directional separations, “EW” or “NS”.
b) 2MASS (not WISE) designation.

TABLE 9 Spectroscopic Properties of All Observed Stars

| WISE Designation | Comp | Hα EW | err | Hα 10-% | err | Li EW | err |
|---|---|---|---|---|---|---|---|
| | | Å | Å | km/s | km/s | Å | Å |
| | | | | HIRES | | | |
| J003234.86+072926.4 | | -8.0 | 0.3 | 55.3 | 4.2 | <0.09 | |
| J003903.51+133016.0 | | -9.0 | 0.1 | 61.2 | 4.4 | <0.07 | |
| J010711.99-193536.4 | | -2.4 | 0.1 | 71.3 | 0.9 | 0.35 | 0.01 |
| J021935.52-455106.2 | W | -0.9 | 0.0 | 65.8 | 11.4 | <0.08 | |
| J021935.52-455106.2 | E | -1.8 | 0.1 | 85.0 | 4.8 | <0.05 | |
| J024852.67-340424.9 | | -9.1 | 0.1 | 72.1 | 5.7 | <0.03 | |
| J030251.62-191150.0 | | -7.3 | 0.6 | 67.7 | 0.7 | <0.05 | |
| J034236.95+221230.2 | | -12.9 | 0.3 | 51.9 | 6.3 | 0.68 | 0.01 |
| J035134.51+072224.5 | | -4.2 | 0.1 | 50.8 | 2.1 | <0.04 | |
| J040711.50-291834.3 | | -2.6 | 0.1 | 64.1 | 1.2 | 0.40 | 0.05 |
| J041525.58-212214.5 | | | | | | <0.04 | |
| J042736.03-231658.8 | | -8.9 | | 53.0 | | | 0.39 |
| J042739.33+171844.2 | | | | DBL | | <0.08 | |
| J043923.21+333149.0 | | -11.0 | 0.2 | 71.8 | 5.5 | <0.04 | |
| J044336.19-003401.8 | | -5.9 | 0.3 | 66.2 | 13.9 | 0.25 | 0.01 |
| J044800.86+143957.7 | AB | DBL | | DBL | | | 0.45 |
| J044802.59+143951.1 | A | -23.2 | | 71.0 | | | 0.63 |
| J050827.31-210144.3 | | -17.8 | | 107.0 | | | 0.49 |
| J052419.14-160115.5 | | -19.8 | 0.5 | 167.3 | 28.5 | 0.15 | 0.03 |
| J052944.69-323914.1 | | -6.1 | 0.2 | 53.9 | 3.3 | <0.09 | |
| J053100.27+231218.3 | | -9.3 | 0.2 | 108.0 | 5.3 | 0.13 | 0.01 |
| J053311.32-291419.9 | | -8.3 | | 71.0 | | <0.06 | |
| J053747.56-424030.8 | | -12.8 | 0.3 | 84.4 | 2.3 | 0.84 | 0.02 |
| J075830.92+153013.4 | B | -4.8 | | 56.0 | | <0.08 | |
| J075830.92+153013.4 | A | -4.8 | | 56.0 | | <0.08 | |
| J094508.15+714450.1 | | -12.4 | 0.6 | 73.1 | 10.2 | <0.10 | |
| J150355.37-214643.1 | | -3.9 | 0.1 | 73.4 | 9.2 | 0.53 | 0.01 |
| J150939.16-133212.4 | | -8.6 | 0.3 | 156.2 | 20.6 | 0.52 | 0.01 |
| J154220.24+593653.0 | | -32.3 | 0.7 | 129.6 | 29.2 | <0.06 | |
| J160549.19-311521.6 | | -2.3 | 0.1 | 62.9 | 7.0 | 0.54 | 0.03 |
| J175839.30+155208.6 | | -3.6 | 0.2 | 49.5 | 0.8 | <0.06 | |
| J175942.12+784942.1 | | lowSNR | | | | <0.08 | |
| J202505.36+835954.2 | | -5.7 | 0.3 | 58.2 | 3.1 | <0.08 | |
| J205136.27+240542.9 | | -5.2 | 0.3 | 60.5 | 5.6 | <0.07 | |
| J210957.48+032121.1 | | 0.3 | 0.0 | | | <0.06 | |
| J211031.49-271058.1 | B | -11.9 | 1.1 | 206.4 | 3.0 | 0.86 | 0.01 |
| J213740.24+013713.2 | | -12.1 | 0.9 | 106.2 | 3.7 | <0.08 | |
| J220850.39+114412.7 | | -6.8 | 0.0 | 64.5 | 11.8 | <0.05 | |
| J225914.87+373639.3 | | -7.9 | 0.2 | 79.9 | 7.7 | <0.06 | |
| J231933.16-393924.3 | | 0.2 | 0.0 | | | <0.03 | |
| J234924.87+185926.7 | | -0.8 | 0.1 | lowSNR | | <0.05 | |
| | | | | FEROS | | | |
| 2M 12182363-3515098[b] | | -0.5 | 0.1 | | | 0.31 | 0.01 |
| J001527.62-641455.2 | | -3.3 | 0.2 | | | <0.08 | |
| J004524.84-775207.5 | | | | | | <0.05 | |
| J004528.25-513734.4 | B | -0.9 | 0.1 | | | <0.03 | |
| J004528.25-513734.4 | A | 0.0 | 0.0 | | | <0.03 | |
| J010243.86-623534.8 | | -3.2 | 0.5 | 47.9 | 13.6 | <0.16 | |
| J010629.32-122518.4 | | -9.7 | 0.8 | 66.1 | 8.3 | <0.11 | |
| J012118.22-543425.1 | | 0.7 | 0.0 | | | <0.03 | |
| J012245.24-631845.0 | | -12.9 | 0.8 | 79.2 | 1.9 | <0.08 | |
| J015057.01-584403.4 | | -8.8 | 0.3 | | | <0.06 | |
| J015455.24-295746.0 | | 0.6 | 0.1 | | | 0.13 | 0.01 |
| J021330.24-465450.3 | | -7.7 | 0.5 | 58.4 | 3.1 | <0.09 | |
| J024746.49-580427.4 | | -3.1 | 0.2 | | | <0.05 | |
| J031650.45-350937.9 | | -8.6 | 0.2 | 88.4 | 0.0 | <0.12 | |
| J033640.91+032918.3 | | -12.2 | 0.9 | 122.9 | 0.6 | <0.08 | |
| J034115.60-225307.8 | | -2.8 | 0.2 | 65.4 | 5.3 | <0.08 | |
| J035223.52-282619.6 | | -5.1 | 0.8 | 117.4 | 1.6 | <0.04 | |
| J035345.92-425018.0 | | -2.4 | 0.1 | 50.8 | 3.4 | <0.10 | |
| J040539.68-401410.5 | | -8.6 | 0.3 | 59.4 | 0.8 | <0.08 | |
| J040649.38-450936.3 | | -5.7 | 0.3 | 55.1 | 0.9 | <0.10 | |
| J040809.80-611904.3 | | 0.5 | 0.0 | | | <0.04 | |
| J040827.01-784446.7 | | -2.2 | 0.1 | | | 0.08 | 0.01 |
| J041050.04-023954.4 | | -2.4 | 0.1 | 79.3 | 5.7 | <0.05 | |
| J041255.78-141859.2 | | -2.0 | 0.1 | | | <0.08 | |
| J041749.66+001145.4 | | -3.3 | 0.2 | | | 0.30 | 0.02 |
| J042139.19-723355.7 | | -4.4 | 0.2 | | | <0.07 | |
| J043657.44-161306.7 | | -8.2 | 0.3 | | | <0.10 | |
| J044120.81-194735.6 | | -1.4 | 0.0 | 67.3 | 6.5 | 0.32 | 0.01 |
| J044700.46-513440.4 | | -2.9 | 0.2 | | | <0.05 | |
| J045114.41-601830.5 | | -1.2 | 0.2 | | | <0.04 | |
| J045651.47-311542.7 | | 1.2 | 0.0 | | | 0.02 | 0.01 |
| J050610.44-582828.5 | | -4.6 | 0.5 | 59.3 | 6.5 | <0.07 | |
| J051310.57-303147.7 | | -2.0 | 0.1 | | | 0.17 | 0.01 |
| J051403.20-251703.8 | | -4.9 | 0.2 | 50.3 | 3.6 | <0.06 | |
| J053328.01-425720.1 | A | -4.6 | 0.2 | | | <0.25 | |
| J053925.08-424521.0 | | -2.3 | 0.4 | | | <0.05 | |
| J054433.76-200515.5 | | -2.2 | 0.1 | | | <0.04 | |
| J054709.88-525626.1 | | -1.9 | 0.2 | 155.9 | 9.7 | <0.20 | |
| J060224.56-163450.0 | | -2.8 | 0.1 | | | <0.05 | |
| J061313.30-274205.6 | | -4.9 | 0.4 | | | <0.07 | |
| J062047.17-361948.2 | | -1.7 | 0.1 | | | 0.35 | 0.03 |
| J072911.26-821214.3 | | -1.9 | 0.1 | | | <0.05 | |
| J075808.25-043647.5 | | | | | | | |
| J080636.05-744424.6 | | -2.0 | 0.3 | | | <0.06 | |
| J082105.04-090853.8 | A | | | | | <0.05 | |
| J082105.04-090853.8 | B | | | | | | |
| J082105.04-090853.8 | B | | | | | <0.05 | |
| J094317.05-245458.3 | | | | | | | |
| J100230.94-281428.2 | | -9.4 | 0.8 | 64.1 | 1.0 | <0.10 | |
| J101905.68-304920.3 | | -2.0 | 0.1 | | | <0.08 | |
| J101917.57-443736.0 | | -3.9 | 0.3 | 63.9 | 11.9 | 0.49 | 0.01 |
| J102602.07-410553.8 | B | -9.9 | 0.8 | 319.3 | 0.4 | | 0.23 |
| J102602.07-410553.8 | A | -9.9 | 0.8 | 319.3 | 0.4 | | 0.38 |
| J104551.72-112615.4 | | | | 124.7 | 8.2 | <0.26 | |
| J105518.12-475933.2 | | -1.4 | 0.2 | | | 0.45 | 0.02 |
| J105850.47-234620.8 | | | | | | | |
| J111128.13-265502.9 | | -17.2 | 0.8 | 56.1 | 4.2 | 0.92 | 0.05 |
| J112047.03-273805.8 | | -16.3 | 1.9 | 58.0 | 0.9 | 1.00 | 0.08 |
| J112105.43-384516.6 | | -3.5 | 0.1 | | | 0.54 | 0.02 |
| J112547.46-441027.4 | | -10.7 | 0.2 | 83.5 | 9.9 | <0.15 | |
| J112651.28-382455.5 | | -7.5 | 0.4 | | | 0.57 | 0.03 |
| J112816.27-261429.6 | | -3.0 | 0.5 | 108.3 | 19.9 | <0.08 | |
| J115927.82-451019.3 | | -10.6 | 0.2 | 66.4 | 5.2 | 0.72 | 0.03 |
| J115949.51-424426.0 | | 0.5 | 0.0 | 78.5 | 7.1 | <0.10 | |
| J115957.68-262234.1 | | -2.9 | 0.4 | | | 0.56 | 0.03 |
| J120001.54-173131.1 | | -7.8 | 0.2 | 101.2 | 2.5 | 0.77 | 0.02 |
| J120237.94-332840.4 | | -9.9 | 0.6 | | | 0.73 | 0.06 |
| J122725.27-454006.6 | | -1.1 | 0.1 | 99.9 | 0.2 | 0.42 | 0.04 |
| J123005.17-440236.1 | | -7.4 | 0.9 | | | 0.67 | 0.05 |
| J123425.84-174544.4 | | -1.5 | 0.1 | | | <0.08 | |
| J124054.09-451625.4 | | -3.8 | 0.3 | 58.3 | 2.6 | 0.11 | 0.01 |
| J124955.67-460737.3 | | -1.7 | 0.2 | 69.0 | 6.5 | 0.44 | 0.03 |
| J125049.12-423123.6 | | -16.0 | 1.2 | 206.5 | 0.6 | 0.23 | 0.05 |
| J125326.99-350415.3 | | -5.2 | 0.3 | 65.2 | 3.2 | <0.15 | |
| J125902.99-314517.9 | | -5.2 | 0.1 | 74.4 | 3.9 | 0.61 | 0.03 |
| J130522.37-405701.2 | | -4.6 | 0.3 | 60.9 | 6.4 | <0.08 | |
| J130530.31-405626.0 | | -1.5 | 0.1 | 67.8 | 2.0 | 0.49 | 0.03 |
| J153549.35-065727.8 | | -1.8 | 0.1 | 70.6 | 1.6 | 0.65 | 0.05 |
| J161410.76-025328.8 | AB | -3.7 | 0.2 | 181.1 | 2.5 | 0.11 | |
| J184206.97-555426.2 | | -7.0 | 1.1 | | | <0.14 | |
| J192250.70-631058.6 | | -6.0 | 0.4 | 65.7 | 10.6 | <0.10 | |
| J192659.33-710923.8 | | -3.4 | 0.2 | 95.8 | 0.3 | <0.15 | |
| J193052.51-545325.4 | | -2.6 | 0.2 | 87.8 | 4.5 | <0.08 | |
| J193411.46-300925.3 | | -13.1 | 1.7 | 75.5 | 0.9 | 0.65 | 0.14 |
| J195227.23-773529.4 | B | -3.0 | 1.4 | | | <0.30 | |
| J195227.23-773529.4 | A | -3.0 | 1.4 | | | <0.30 | |
| J200311.61-243959.2 | | -2.0 | 0.4 | 92.9 | 3.4 | <0.14 | |
| J210708.43-113506.0 | | -1.0 | 0.0 | | | 0.20 | 0.01 |
| J210722.53-705613.4 | | -3.9 | 0.2 | | | 0.14 | 0.03 |
| J211004.67-192031.2 | | -8.7 | 0.2 | 135.6 | 6.6 | <0.20 | |
| J211005.41-191958.4 | | -4.4 | 0.4 | 50.3 | 0.9 | <0.07 | |
| J211635.34-600513.4 | | -6.4 | 0.2 | 66.4 | 4.3 | <0.22 | |
| J213708.89-603606.4 | | -7.5 | 0.4 | 115.3 | 4.2 | <0.11 | |
| J220216.29-421034.0 | | -2.0 | 0.1 | 56.9 | 7.7 | <0.06 | |
| J220254.57-644045.0 | | -3.3 | 0.3 | 72.5 | 0.9 | <0.06 | |
| J220730.16-691952.6 | | -1.8 | 0.5 | 50.9 | 2.5 | <0.30 | |
| J224634.82-735351.0 | | -5.0 | 0.3 | 55.7 | 2.0 | <0.15 | |
| J230209.10-121522.0 | | -7.0 | 0.8 | 167.8 | 24.1 | <0.15 | |
| J231246.53-504924.8 | | -7.4 | 0.5 | | | <0.08 | |
| J231457.86-633434.0 | B | -1.6 | 0.2 | 59.9 | 2.0 | <0.06 | |
| J231457.86-633434.0 | A | -1.5 | 0.1 | 43.0 | 8.0 | <0.06 | |
| J232857.75-680234.5 | | -6.1 | 0.3 | 71.5 | 2.2 | <0.11 | |
| Hamilton | | | | | | | |
| J001709.96+185711.8 | S | 0.9 | 0.0 | | | <0.02 | |
| J001709.96+185711.8 | N | 1.1 | 0.1 | | | <0.02 | |
| J020012.84-084052.4 | | -4.0 | 0.1 | 55.4 | 2.3 | <0.08 | |
| J025154.17+222728.9 | | -5.8 | 0.1 | 63.7 | 10.1 | <0.07 | |
| J025913.40+203452.6 | | 4.0 | 0.2 | | | <0.02 | |
| J030002.98+550652.4 | | -1.1 | 0.2 | | | <0.05 | |
| J034444.80+404150.4 | | -1.1 | 0.2 | | | <0.13 | |
| J043726.87+185126.2 | | -1.0 | 0.0 | | | 0.49 | 0.01 |
| J044455.71+193605.3 | | -0.8 | 0.1 | | | <0.03 | |
| J062407.62+310034.4 | | -4.0 | 0.4 | | | <0.09 | |
| J072821.16+334511.6 | | -7.0 | 0.5 | | | <0.08 | |
| J082558.91+034019.5 | | -5.1 | 0.3 | | | <0.10 | |
| J161743.18+261815.2 | | | | | | <0.05 | |
| J171038.44-210813.0 | | -2.2 | 0.1 | 72.0 | 5.1 | 0.53 | 0.02 |
| J173353.07+165511.7 | | -10.3 | 0.9 | 73.3 | 3.5 | 0.17 | 0.03 |
| J190453.69-140406.0 | | | | | | <0.05 | |
| J191534.83-083019.9 | | -1.3 | 0.1 | 71.8 | 1.4 | 0.12 | 0.01 |
| J210338.46+075330.3 | | -1.1 | 0.1 | 73.0 | 1.0 | 0.24 | 0.01 |
| J210736.82-130458.9 | | -4.7 | 0.2 | 82.0 | 8.6 | <0.08 | |
| J215053.68-055318.9 | | -2.2 | 0.1 | 56.3 | 11.8 | <0.05 | |
| J232656.43+485720.9 | | -0.7 | 0.0 | 43.7 | 0.9 | <0.02 | |
| J044356.87+372302.7 | | -6.9 | 0.3 | 77.7 | 13.8 | 0.21 | 0.01 |
| J181725.08+482202.8 | | -2.3 | 0.4 | 51.8 | 7.0 | <0.07 | |
| J010126.59+463832.6 | | | | | | <0.10 | |
| J162548.69-135912.0 | | 1.5 | 0.1 | | | 0.13 | 0.01 |
| J173544.26-165209.9 | | -1.0 | 0.1 | | | <0.05 | |
| J192323.20+700738.3 | | | | | | <0.05 | |
| J194539.01+704445.9 | | -2.2 | 0.1 | 110.6 | 0.7 | 0.18 | 0.02 |
| J231543.66-140039.6 | | 0.6 | 0.0 | | | <0.03 | |
| J234333.91-192802.8 | | -1.2 | 0.0 | | | <0.05 | |
| duPont Echelle | | | | | | | |
| J001536.79-294601.2 | | -4.5 | 0.5 | 64.2 | 1.3 | <0.04 | |
| J001723.69-664512.4 | | -5.3 | 0.3 | 56.2 | 6.5 | <0.07 | |
| J003057.97-655006.4 | | -3.1 | 0.1 | 68.4 | 0.3 | <0.03 | |
| J004210.98-425254.8 | | -2.5 | 0.1 | 49.4 | 7.9 | <0.07 | |
| J013110.69-760947.7 | | -1.3 | 0.1 | 53.9 | 2.4 | <0.03 | |
| J020805.55-474633.7 | | 0.5 | 0.0 | ABS | | <0.06 | |
| J021258.28-585118.3 | | -3.2 | 0.2 | 59.7 | 11.9 | <0.04 | |
| J022424.69-703321.2 | | -3.1 | 0.1 | 52.6 | 6.8 | <0.03 | |
| J032047.66-504133.0 | | -0.8 | 0.2 | | | <0.03 | |
| J040743.83-682511.0 | | -1.8 | 0.1 | 68.0 | 5.2 | <0.02 | |
| J041336.14-441332.4 | | -2.4 | 0.2 | 65.0 | 3.3 | <0.03 | |
| J044036.23-380140.8 | | | | | | | |
| J051803.00-375721.2 | | -3.9 | 0.1 | 56.6 | 5.7 | <0.05 | |
| J054448.20-265047.4 | | | | giant | | <0.06 | |
| J060156.10-164859.9 | A | -1.5 | 0.0 | 54.3 | 7.8 | <0.05 | |
| J060156.10-164859.9 | B | -1.5 | 0.1 | 68.2 | 3.2 | <0.05 | |
| J061851.01-383154.9 | | -3.4 | 0.5 | 218.7 | 4.1 | <1.00 | |
| J063001.84-192336.6 | | -2.4 | 0.2 | 44.6 | 5.4 | <1.00 | |
| J081738.97-824328.8 | | -8.0 | 0.2 | 72.9 | 3.6 | <0.03 | |
| J103016.11-354626.3 | | -2.2 | 0.1 | 74.3 | 14.8 | <0.04 | |
| J103137.59-374915.9 | | -5.8 | 0.5 | 62.7 | 10.1 | 0.64 | 0.08 |
| J104044.98-255909.2 | | 0.4 | 0.0 | ABS | | <0.04 | |
| J105524.25-472611.7 | A | | | lowSNR | | <0.14 | |
| J111103.54-313459.0 | | -4.9 | 0.5 | 77.7 | 6.7 | <0.24 | |
| J111229.74-461610.1 | | | | giant | | <0.05 | |
| J111707.56-390951.3 | | -1.2 | 0.1 | 83.8 | 5.4 | 0.42 | 0.02 |
| J113114.81-482628.0 | | -8.7 | 0.3 | 165.4 | 2.3 | <0.50 | |
| J121429.15-425814.8 | | -3.6 | 0.1 | 121.7 | 4.4 | 0.43 | 0.08 |
| J122643.99-122918.3 | | -6.2 | 0.5 | 51.3 | 14.5 | <0.40 | |
| J132112.77-285405.1 | | -1.6 | 0.2 | giant? | | <0.14 | |
| J135913.33-292634.2 | | 0.3 | 0.0 | ABS | | <0.07 | |
| J160828.45-060734.6 | | -4.6 | 0.2 | 86.6 | 20.7 | 0.20 | 0.01 |
| Kast | | | | | | | |
| J000453.05-103220.0 | | -8.3 | 0.4 | 197.3 | 7.4 | 0.69 | 0.01 |
| J001552.28-280749.4 | | -1.6 | 0.4 | | | <0.21 | |
| J002101.27-134230.7 | | 0.5 | 0.2 | | | <0.06 | |
| J004826.70-184720.7 | | -8.4 | 0.3 | 56.1 | 3.6 | 0.59 | 0.02 |
| J005633.96-225545.4 | | -2.1 | 0.4 | | | <0.18 | |
| J010047.97+025029.0 | | -5.6 | | 168.0 | | <0.05 | |
| J010251.05+185653.7 | | -7.5 | 0.3 | | | <0.18 | |
| J011440.20+205712.9 | | -1.3 | 0.3 | | | <0.09 | |
| J011846.91+125831.4 | | -0.8 | 0.3 | | | <0.06 | |
| J014156.94-123821.6 | | -1.2 | 0.3 | | | <0.18 | |
| J015257.41+083326.3 | | -11.9 | 1.0 | 125.5 | 5.1 | 0.45 | 0.01 |
| J015350.81-145950.6 | | -7.7 | 0.4 | | | <0.21 | |
| J020302.74+221606.8 | | -1.1 | 0.3 | | | <0.21 | |
| J022240.88+305515.4 | | -5.0 | 0.3 | | | <0.24 | |
| J023005.14+284500.0 | | 0.1 | 0.3 | | | <0.21 | |
| J023139.36+445638.1 | | -7.8 | 0.3 | | | <0.18 | |
| J024552.65+052923.8 | | -3.0 | 0.3 | | | <0.21 | |
| J030444.10+220320.8 | | -9.5 | 0.2 | 74.9 | 8.1 | 0.71 | 0.03 |
| J030824.14+234554.2 | | -2.0 | 0.3 | | | <0.21 | |
| J033235.82+284354.6 | | -7.1 | 0.3 | | | <0.24 | |
| J035100.83+141339.2 | | -10.0 | 0.5 | | | <0.36 | |
| J035733.95+244510.2 | | -1.7 | 0.2 | | | <0.03 | |
| J043257.29+740659.3 | | -3.1 | 0.4 | | | <0.24 | |
| J044349.19+742501.6 | | -3.5 | 0.3 | | | <0.15 | |
| J044721.05+280852.5 | | -9.6 | 0.3 | | | <0.12 | |
| J055008.59+051153.2 | | -1.7 | 0.4 | | | <0.39 | |
| J055041.58+430451.8 | | -2.4 | 0.3 | | | <0.06 | |
| J055208.04+613436.6 | | -0.7 | 0.3 | | | <0.03 | |
| J065846.87+284258.9 | | 0.0 | 0.3 | | | <0.18 | |
| J071036.50+171322.6 | | -1.6 | 0.4 | | | <0.06 | |
| J072641.52+185034.0 | | -5.0 | 0.3 | | | <0.15 | |
| J073138.47+455716.5 | | -7.3 | 0.4 | | | <0.27 | |
| J080352.54+074346.7 | | -0.1 | 0.3 | | | <0.15 | |
| J081443.62+465035.8 | | -9.1 | 0.3 | | | <0.15 | |
| J083528.87+181219.9 | | -3.8 | 0.4 | | | <0.27 | |
| J090227.87+584813.4 | | -4.0 | 0.3 | | | <0.09 | |
| J092216.12+043423.3 | | -2.7 | 0.4 | | | <0.09 | |
| J093212.63+335827.3 | | -4.7 | 0.5 | | | <0.33 | |
| J100146.28+681204.1 | | -0.5 | 0.2 | | | <0.06 | |
| J101543.44+660442.3 | | -3.9 | 0.4 | | | <0.15 | |
| J102636.95+273838.4 | | -2.2 | 0.4 | | | <0.18 | |
| J103557.17+285330.8 | | -2.3 | 0.4 | | | <0.12 | |
| J105515.87-033538.2 | | -1.6 | 0.3 | | | <0.03 | |
| J105524.25-472611.7 | B | | | lowSNR | | <0.14 | |
| J105711.36+054454.2 | | -2.4 | 0.3 | | | <0.06 | |
| J110119.22+525222.9 | | -0.4 | 0.3 | | | <0.06 | |
| J110335.71-302449.5 | | -5.5 | 0.2 | | | <0.09 | |
| J111309.15+300338.4 | | -0.6 | 0.3 | | | <0.06 | |
| J112512.28-002438.2 | | -2.1 | 0.3 | | | <0.15 | |
| J112955.84+520213.2 | | -4.2 | 0.3 | | | <0.09 | |
| J113105.57+542913.5 | | -3.2 | 0.3 | | | <0.12 | |
| J113120.31+132140.0 | | -3.7 | 0.5 | | | <0.30 | |
| J114728.37+664402.7 | | -8.4 | 0.4 | | | <0.12 | |
| J115156.73+073125.7 | | -0.4 | 0.3 | | | <0.15 | |
| J121153.04+124912.9 | | -1.8 | 0.3 | | | <0.09 | |
| J121341.59+323127.7 | B | -0.6 | | | | 0.05 | |
| J121341.59+323127.7 | A | -0.9 | 0.3 | | | <0.15 | |
| J121511.25-025457.1 | | 0.0 | | | | <0.02 | |
| J133238.94+305905.8 | | -5.8 | 0.4 | | | <0.21 | |
| J133509.40+503917.5 | | -4.5 | 0.4 | | | <0.18 | |
| J133901.87-214128.0 | | -4.5 | 0.3 | | | <0.15 | |
| J134146.41+581519.2 | | -2.3 | 0.3 | | | <0.21 | |
| J134907.28+082335.8 | | -0.3 | 0.3 | | | <0.06 | |
| J135511.38+665207.0 | | 1.2 | 0.2 | | | <0.09 | |
| J140337.56-501047.9 | | -3.7 | 0.2 | | | 0.24 | 0.03 |
| J141045.24+364149.8 | | -3.7 | 0.3 | | | <0.24 | |
| J141332.23-145421.1 | | -8.0 | 0.4 | | | <0.15 | |
| J141510.77-252012.0 | | 0.2 | 0.3 | | | <0.09 | |
| J141842.36+475514.9 | | -0.6 | 0.3 | | | <0.09 | |
| J141903.13+645146.4 | | -6.8 | 0.4 | | | <0.18 | |
| J143648.16+090856.5 | | -5.9 | 0.3 | | | <0.21 | |
| J143713.21-340921.1 | | -7.8 | 0.4 | | | <0.15 | |
| J145014.12-305100.6 | | -1.0 | 0.2 | | | <0.03 | |
| J145731.11-305325.0 | | -5.6 | 0.5 | | | <0.12 | |
| J145949.90+244521.9 | | -4.9 | 0.3 | | | <0.33 | |
| J150119.48-200002.1 | | -7.7 | 0.4 | | | <0.27 | |
| J150230.94-224615.4 | | -4.8 | 0.5 | | | <0.21 | |
| J150723.91+433353.6 | | -1.5 | 0.4 | | | <0.33 | |
| J150836.69-294222.9 | | -1.2 | 0.6 | | | <0.33 | |
| J151212.18-255708.3 | | -3.9 | 0.3 | | | 0.47 | 0.01 |
| J152150.76-251412.1 | | -7.1 | 0.4 | | | <0.18 | |
| J154435.17+042307.5 | | -4.7 | 0.1 | 154.8 | 28.5 | 0.60 | 0.01 |
| J154656.43+013650.8 | | -4.3 | 0.4 | | | <0.12 | |
| J155046.47+305406.9 | | -5.9 | 0.4 | | | <0.27 | |
| J155515.35+081327.9 | | -0.7 | 1.0 | | | 0.07 | 0.01 |
| J155759.01-025905.8 | | -0.1 | 0.3 | | | <0.06 | |
| J155947.24+440359.6 | | -2.9 | 0.3 | | | <0.18 | |
| J162422.68+195922.0 | | -1.1 | 0.3 | | | <0.06 | |
| J163051.34+472643.8 | | -5.5 | 0.4 | | | <0.33 | |
| J163632.90+635344.9 | | 0.0 | 0.2 | | | <0.21 | |
| J164539.37+702400.1 | | -2.3 | 0.3 | | | <0.21 | |
| J171117.68+124540.4 | | -4.3 | 0.3 | | | <0.18 | |
| J171441.70-220948.8 | | -5.7 | 0.2 | 64.9 | 1.7 | 0.55 | 0.02 |
| J172309.67-095126.2 | | -1.6 | 0.3 | | | <0.09 | |
| J172454.26+502633.0 | | 0.4 | 0.2 | | | <0.09 | |
| J172951.38+093336.9 | | -4.9 | 0.3 | | | <0.24 | |
| J173623.80+061853.0 | | 0.6 | 0.5 | | | <0.09 | |
| J174426.59-074925.3 | | 1.6 | 0.2 | | | <0.06 | |
| J174439.27+483147.1 | | 0.0 | 0.6 | | | <0.06 | |
| J174811.33-030510.2 | | -4.0 | 0.3 | | | <0.24 | |
| J180658.07+161037.9 | | -5.5 | 0.2 | 201.4 | 22.0 | 0.58 | 0.01 |

| | | | | | |
|---|---|---|---|---|---|
| J180733.00+613153.6 | -3.3 | 0.3 | | <0.03 | |
| J184536.02-205910.8 | -0.7 | 0.2 | | <0.09 | |
| J191235.95+630904.7 | -3.0 | 0.3 | | <0.15 | |
| J192240.05-061208.0 | -4.6 | 0.4 | | <0.18 | |
| J192242.80-051553.8 | -6.2 | 0.4 | | <0.27 | |
| J193711.26-040126.7 | 0.7 | 0.2 | | <0.06 | |
| J194714.54+640237.9 | -1.8 | 0.3 | | <0.21 | |
| J194816.54-272032.3 | -4.3 | 0.3 | | <0.09 | |
| J195315.67+745948.9 | -0.3 | 0.3 | | <0.09 | |
| J195340.71+502458.2 | -6.9 | 0.4 | | <0.27 | |
| J200423.80-270835.8 | -1.6 | 0.4 | | <0.15 | |
| J201931.84-081754.3 | -1.7 | 0.3 | | <0.12 | |
| J203023.10+711419.8 | -1.5 | | 208.0 | <0.08 | |
| J204406.36-153042.3 | 0.0 | 0.4 | | <0.24 | |
| J204714.59+110442.2 | -3.1 | 0.2 | | <0.12 | |
| J213507.39+260719.4 | 0.2 | 0.3 | | <0.09 | |
| J213644.54+670007.1 | 1.3 | 0.2 | | <0.03 | |
| J214101.48+723026.7 | -2.8 | 0.3 | | <0.06 | |
| J214126.66+204310.5 | -4.8 | 0.4 | | <0.30 | |
| J214414.73+321822.3 | 0.3 | | | <0.02 | |
| J220306.98-253826.6 | -7.3 | 0.4 | | <0.12 | |
| J221842.70+332113.5 | -2.4 | 0.2 | | <0.12 | |
| J230740.98+080359.7 | -0.5 | 0.3 | | <0.09 | |
| J231021.75+685943.6 | 0.0 | 0.2 | | <0.24 | |
| J231211.37+150329.7 | -2.9 | 0.3 | | <0.27 | |
| J232151.23+005037.3 | 0.4 | 0.3 | | <0.21 | |
| J232904.42+032910.8 | -7.6 | 0.4 | | <0.15 | |
| J234347.83-125252.1 | -2.2 | 0.4 | | <0.15 | |
| J234926.23+185912.4 | -1.1 | 0.3 | | <0.06 | |
| J235250.70-160109.7 | -6.3 | 0.5 | | <0.24 | |
| duPont B&C 832 | | | | | |
| J001555.65-613752.2 | -5.8 | 0.3 | | <0.33 | |
| J012332.89-411311.4 | -5.9 | 0.5 | | <0.27 | |
| J012532.11-664602.6 | -7.8 | 0.4 | | <0.21 | |
| J014431.99-460432.1 | -23.4 | 4.1 | | <1.11 | |
| J020305.46-590146.6 | -4.1 | 1.0 | | <1.32 | |
| J033431.66-350103.3 | -11.5 | 0.5 | | <0.27 | |
| J034116.16-225244.0 | -2.8 | 0.3 | | <0.09 | |
| J035829.67-432517.2 | -8.5 | 0.3 | | <0.12 | |
| J042500.91-634309.8 | -6.0 | 0.3 | | <0.21 | |
| J043213.46-285754.8 | -10.6 | 1.0 | | <0.63 | |
| J044154.44+091953.1 | -9.2 | 0.4 | | <0.50 | |
| J044530.77-285034.8 | -5.9 | 0.3 | | <0.27 | |
| J045420.20-400009.9 | -4.3 | 0.2 | | <0.15 | |
| J050333.31-382135.6 | -8.4 | 0.3 | | <0.27 | |
| J051255.82-212438.7 | -9.8 | 0.5 | | <0.36 | |
| J051650.66+022713.0 | -11.2 | 0.5 | | <0.24 | |
| J052535.85-250230.2 | -2.6 | 0.3 | | <0.15 | |
| J054223.86-275803.3 | -8.0 | 0.3 | | <0.36 | |
| J054719.52-335611.2 | -3.2 | 0.2 | | <0.36 | |
| J055941.10-231909.4 | -10.5 | 0.3 | | <0.27 | |
| J103952.70-353402.5 | -2.3 | 0.2 | | <0.12 | |
| J110551.56-780520.7 | -0.2 | 0.2 | | <1.00 | |
| J111052.06-725513.0 | -9.9 | 0.9 | | <1.00 | |
| J114623.01-523851.8 | -8.4 | 0.5 | | <0.42 | |
| J115438.73-503826.4 | -7.7 | 0.3 | | <0.18 | |
| J120929.80-750540.2 | -6.9 | 0.3 | | <0.18 | |
| J121558.37-753715.7 | -9.4 | 0.6 | | <0.39 | |
| J123234.07-414257.5 | -10.7 | 0.6 | | <0.54 | |
| J124612.32-384013.5 | -6.2 | 0.4 | | <0.18 | |
| J130501.18-331348.7 | -0.7 | 0.2 | | <0.12 | |
| J130618.16-342857.0 | -8.2 | 0.4 | | 0.42 | 0.07 |
| J130650.27-460956.1 | -1.4 | 0.2 | | 0.08 | 0.02 |
| J130731.03-173259.9 | -7.9 | 0.5 | | <0.24 | |
| J131129.00-425241.9 | -3.4 | 0.2 | | 0.27 | 0.05 |
| J143517.80-342250.4 | -2.2 | 0.2 | | <0.09 | |
| J143753.36-343917.8 | -3.8 | 0.3 | | <0.21 | |
| J150601.66-240915.0 | -10.8 | 2.3 | | <1.11 | |
| J150820.15-282916.6 | -6.6 | 0.4 | | <0.33 | |
| J151242.69-295148.0 | -3.7 | 0.3 | | 0.29 | 0.07 |
| J151411.31-253244.1 | -0.6 | 0.2 | | 0.21 | 0.04 |
| J153248.80-230812.4 | -2.2 | 0.2 | | 0.26 | 0.03 |
| J154227.07-042717.1 | -1.3 | 0.2 | | 0.26 | 0.08 |
| J154349.42-364838.7 | -8.6 | 0.4 | | 0.31 | 0.06 |
| J160116.86-345502.7 | -2.0 | 0.5 | | <0.51 | |
| J160954.85-305858.4 | -0.9 | 0.2 | | 0.19 | 0.05 |
| J162602.80-155954.5 | 3.5 | 0.2 | | <0.03 | |
| J170415.15-175552.5 | 1.4 | 0.0 | | 0.12 | 0.01 |
| J171426.13-214845.0 | 1.9 | 0.1 | | <0.15 | |
| J172130.71-150617.8 | -5.9 | 0.5 | | <0.36 | |

| | | | | | | | |
|---|---|---|---|---|---|---|---|
| J172131.73-084212.3 | | -5.0 | 0.3 | | | <0.27 | |
| J172615.23-031131.9 | | -13.4 | 0.7 | | | <0.36 | |
| J173826.94-055628.0 | | 2.4 | 0.2 | | | <0.15 | |
| J174203.85-032340.4 | | -7.6 | 0.5 | | | <0.30 | |
| J174536.31-063215.3 | | 3.1 | 0.2 | | | <0.15 | |
| J174735.31-033644.4 | | 2.2 | 0.2 | | | <0.15 | |
| J174936.01-010808.7 | | -5.6 | 0.5 | | | <0.27 | |
| J175022.27-094457.8 | | 1.4 | 0.1 | | | <0.12 | |
| J180508.62-015058.5 | | 3.0 | 0.2 | | | <0.12 | |
| J180554.92-570431.3 | | -6.7 | 0.3 | | | <0.24 | |
| J180929.71-543054.2 | | -8.9 | 0.5 | | | 0.24 | 0.08 |
| J181059.88-012322.4 | | | | | | <0.33 | |
| J182054.20+022101.5 | | 2.3 | 0.2 | | | <0.06 | |
| J182905.79+002232.2 | | 3.2 | 0.3 | | | <0.06 | |
| J184204.85-555413.3 | | -5.9 | 0.3 | | | <0.24 | |
| J191019.82-160534.8 | | -1.0 | 0.1 | | | <0.09 | |
| J191036.02-650825.5 | | -4.4 | 0.3 | | | <0.27 | |
| J191500.80-284759.1 | | -11.9 | 0.9 | | | 0.98 | 0.21 |
| J191629.61-270707.2 | | -6.1 | 0.4 | | | <0.18 | |
| J192434.97-344240.0 | | -12.2 | 0.6 | | | <0.14 | |
| J194309.89-601657.8 | | -9.2 | 0.4 | | | <0.21 | |
| J194444.21-435903.0 | | -6.8 | 0.5 | | | <0.54 | |
| J194834.58-760546.9 | | -1.2 | 0.3 | | | <0.15 | |
| J195331.72-070700.5 | | -4.6 | 0.4 | | | <0.33 | |
| J195602.95-320719.3 | | -6.1 | 0.3 | | | <0.33 | |
| J200137.19-331314.5 | | -3.8 | 0.4 | 100.6 | 0.6 | 0.13 | 0.02 |
| J200409.19-672511.7 | | -5.2 | 0.2 | | | <0.12 | |
| J200556.44-321659.7 | | -2.5 | 0.2 | | | <0.09 | |
| J200837.87-254526.2 | | -11.4 | 0.4 | | | 0.35 | 0.09 |
| J200853.72-351949.3 | | -6.9 | 0.3 | | | <0.36 | |
| J201000.06-280141.6 | | -13.7 | 0.6 | | | <0.08 | |
| J202716.80-254022.8 | | -1.1 | 0.2 | | | <0.09 | |
| J203301.99-490312.6 | | -7.6 | 0.4 | | | <0.39 | |
| J203337.63-255652.8 | | -12.3 | 0.3 | 74.1 | 2.0 | 0.52 | 0.02 |
| J205131.01-154857.6 | | -42.1 | 3.8 | | | <0.72 | |
| J205832.99-482033.8 | | -10.5 | 1.2 | | | <0.60 | |
| J210131.13-224640.9 | | 0.1 | 0.2 | | | <0.12 | |
| J211031.49-271058.1 | A | -30.5 | 2.9 | 78.9 | 3.7 | 0.48 | 0.01 |
| J212007.84-164548.2 | | -6.6 | 0.5 | 60.5 | 3.7 | <0.20 | |
| J212128.89-665507.1 | | | | | | <0.06 | |
| J212230.56-333855.2 | | -18.3 | 0.5 | | | <0.24 | |
| J212750.60-684103.9 | | -9.4 | 0.8 | | | <0.10 | |
| J213520.34-142917.9 | | -6.2 | 0.3 | | | <0.39 | |
| J213835.44-505111.0 | | -7.1 | 0.4 | | | <0.15 | |
| J213847.58+050451.4 | | -6.3 | 0.6 | | | <0.48 | |
| J214905.04-641304.8 | | -7.7 | 0.5 | | | <0.24 | |
| J215128.95-023814.9 | | -3.3 | 0.3 | | | <0.24 | |
| J215717.71-341834.0 | | -1.6 | 0.3 | | | <0.27 | |
| J221217.17-681921.1 | | | | | | <0.39 | |
| J221559.00-014733.0 | | 0.2 | 0.1 | | | <0.12 | |
| J221833.85-170253.2 | | 0.2 | 0.2 | | | <0.18 | |
| J222024.21-072734.5 | | -3.9 | 0.2 | | | <0.27 | |
| J224111.08-684141.8 | | -9.9 | 0.9 | | | <1.17 | |
| J224221.02-410357.2 | | -5.3 | 0.4 | | | <0.45 | |
| J224448.45-665003.9 | | -10.2 | 0.4 | | | <0.48 | |
| J224500.20-331527.2 | | -8.1 | 0.2 | | | <0.15 | |
| J225934.89-070447.1 | | -5.4 | 0.9 | | | <1.05 | |
| J230327.73-211146.2 | | -5.6 | 0.4 | | | <0.30 | |
| J232008.15-634334.9 | | -5.7 | 0.7 | | | <0.66 | |
| J232917.64-675000.6 | | -7.4 | 0.4 | | | <0.24 | |
| J232959.47+022834.0 | | -5.6 | 0.5 | | | <0.36 | |
| J233647.87+001740.1 | | -0.1 | 0.2 | | | <0.12 | |
| J234243.45-622457.1 | | -11.2 | 0.7 | | | <0.39 | |
| J234326.88-344658.5 | | -2.3 | 0.2 | | | <0.12 | |
| J234857.35+100929.3 | | -0.1 | 0.2 | | | <0.18 | |
| duPont B&C 600 | | | | | | | |
| J060329.60-260804.7 | | -1.7 | 0.2 | 71.2 | 0.7 | <0.07 | |
| J061740.43-475957.2 | | -10.9 | 0.4 | | | <0.15 | |
| J062130.52-410559.1 | | -10.9 | 0.4 | | | <0.15 | |
| J070657.72-535345.9 | | -2.4 | 0.5 | | | <0.27 | |
| J075233.22-643630.5 | | -2.0 | 0.2 | | | 0.26 | 0.03 |
| J104008.36-384352.1 | | -4.1 | 0.3 | | | <0.06 | |
| J122813.57-431638.9 | | -6.0 | 0.4 | | | <0.27 | |
| J123704.99-441919.5 | | -10.0 | 1.1 | | | <0.39 | |

a) Some binary systems are well resolved in radial velocity (RV), enabling separate measurements of the two components. Cases in which the components are blended in RV are indicated “AB”.
b) 2MASS (not WISE) designation.

TABLE 10 Photometry of all Stars

| WISE Designation | | B [mag] | V [mag] | J [mag] | H [mag] | K [mag] | dist [pc] | $M_V$ [mag] | $M_K$ [mag] | NUV-W1 [mag] | J-W2 [mag] | W1-W3 [mag] | W1-W4 [mag] |
|---|---|---|---|---|---|---|---|---|---|---|---|---|---|
| J000453.05-103220.0 | | 17.08 | 15.47 | 10.53 | 9.96 | 9.68 | 64.67 | 12.63 | 6.84 | 11.39 | 1.25 | 0.38 | 0 |
| J001527.62-641455.2 | | 14.37 | 12.9 | 9.33 | 8.69 | 8.44 | 47.69 | 9.49 | 5.03 | 11.46 | 1.09 | 0.21 | 0.2 |
| J001536.79-294601.2 | | 15.84 | 14.27 | 9.78 | 9.25 | 8.9 | 36.26 | 11.87 | 6.5 | 11.58 | 1.22 | 0.3 | 0.46 |
| J001552.28-280749.4 | | 15.24 | 13.67 | 10.09 | 9.44 | 9.2 | 45.12 | 10.25 | 5.78 | 11.66 | 1.15 | 0.26 | 0.62 |
| J001555.65-613752.2 | | 16.2 | 14.71 | 10.59 | 10.0 | 9.77 | 54.92 | 11.09 | 6.15 | 11.36 | 1.17 | 0.28 | 0.62 |
| J001709.96+185711.8 | N | 12.25 | 11.09 | 9.1 | 8.5 | 8.42 | 96.34 | 7.9 | 5.23 | 10.59 | 1.05 | -0.18 | -0.29 |
| J001709.96+185711.8 | S | 12.25 | 11.09 | 9.1 | 8.5 | 8.42 | 96.34 | 7.9 | 5.23 | 10.59 | 1.05 | -0.18 | -0.29 |
| J001723.69-664512.4 | | 14.01 | 12.49 | 8.56 | 7.93 | 7.7 | 36.82 | 9.27 | 4.48 | 11.49 | 1.08 | 0.2 | 0.27 |
| J002101.27-134230.7 | | | 11.37 | 9.58 | 8.96 | 8.68 | 69.77 | 9.57 | 6.88 | 11.69 | 1.26 | -0.05 | 0.06 |
| J003057.97-655006.4 | | 15.88 | 14.27 | 9.82 | 9.24 | 8.95 | 45.37 | 11.72 | 6.4 | 11.46 | 1.22 | 0.32 | |
| J003234.86+072926.4 | | 14.19 | 12.82 | 8.4 | 7.79 | 7.51 | 35.42 | 9.58 | 4.27 | 11.42 | 1.18 | 0.22 | 0.44 |
| J003903.51+133016.0 | | 17.44 | 15.71 | 10.94 | 10.37 | 10.06 | 54.0 | 13.08 | 7.43 | 11.4 | 1.3 | 0.28 | |
| J004210.98-425254.8 | | 14.84 | 13.36 | 9.62 | 8.98 | 8.76 | 52.85 | 9.9 | 5.3 | 11.27 | 1.08 | 0.2 | -0.3 |
| J004524.84-775207.5 | | 14.24 | 12.79 | 9.53 | 8.88 | 8.66 | 62.74 | 10.84 | 6.71 | 11.6 | 1.15 | 0.2 | 0.69 |
| J004528.25-513734.4 | B | 13.45 | 11.97 | 8.48 | 7.87 | 7.62 | 41.16 | 8.78 | 4.43 | 12.44 | 1.05 | 0.18 | 0.19 |
| J004528.25-513734.4 | A | 13.45 | 11.97 | 8.48 | 7.87 | 7.62 | 41.16 | 8.78 | 4.43 | 12.44 | 1.05 | 0.18 | 0.19 |
| J004826.70-184720.7 | | | | 10.75 | 10.14 | 9.86 | 51.48 | | 5.99 | 12.1 | 1.25 | 0.3 | |
| J005633.96-225545.4 | | 15.55 | 14.1 | 10.77 | 10.13 | 9.91 | 133.86 | 9.5 | 5.31 | 11.69 | 1.08 | 0.22 | 1.22 |
| J010047.97+025029.0 | | 16.25 | 14.61 | 10.32 | 9.67 | 9.39 | | 10.55 | 5.33 | 10.86 | 1.19 | 0.27 | 1.32 |
| J010126.59+463832.6 | | 13.04 | 11.99 | 9.85 | 9.22 | 9.06 | 413.02 | 7.63 | 4.7 | 9.85 | 0.88 | 0.11 | 0.29 |
| J010243.86-623534.8 | | 15.6 | 14.04 | 9.64 | 9.04 | 8.8 | | 10.73 | 5.49 | 12.11 | 1.21 | 0.32 | 0.42 |
| J010251.05+185653.7 | | 15.58 | 14.08 | 9.51 | 8.92 | 8.67 | 38.18 | 10.28 | 4.87 | 11.6 | 1.19 | 0.3 | 0.51 |
| J010629.32-122518.4 | | 16.04 | 14.51 | 10.5 | 9.84 | 9.64 | 66.78 | 10.46 | 5.59 | 11.3 | 1.13 | 0.29 | |
| J010711.99-193536.4 | | 12.92 | 11.46 | 8.15 | 7.47 | 7.25 | 56.68 | 7.31 | 3.1 | 11.26 | 1.03 | 0.1 | 0.15 |
| J011440.20+205712.9 | | 13.91 | 12.56 | 9.86 | 9.25 | 9.06 | 116.49 | 8.63 | 5.13 | 11.12 | 1.02 | 0.08 | -0.05 |
| J011846.91+125831.4 | | 13.12 | 11.96 | 9.52 | 8.94 | 8.73 | 91.52 | 7.96 | 4.73 | 10.78 | 0.9 | 0.1 | 0.07 |
| J012118.22-543425.1 | | 12.49 | 11.2 | 8.62 | 7.97 | 7.81 | 47.46 | 7.73 | 4.34 | 12.25 | 0.96 | 0.05 | 0.2 |
| J012245.24-631845.0 | | 15.61 | 14.07 | 9.83 | 9.21 | 8.98 | 45.35 | 10.85 | 5.76 | 11.23 | 1.16 | 0.3 | 0.52 |
| J012332.89-411311.4 | | 17.15 | 15.55 | 10.8 | 10.19 | 9.92 | 39.98 | 12.51 | 6.88 | 12.25 | 1.28 | 0.34 | 1.17 |
| J012532.11-664602.6 | | 17.38 | 15.6 | 10.95 | 10.42 | 10.11 | 46.25 | 12.24 | 6.75 | 11.33 | 1.24 | 0.32 | |
| J013110.69-760947.7 | | 15.85 | 14.28 | 10.47 | 9.85 | 9.62 | 60.74 | 10.41 | 5.75 | 11.58 | 1.1 | 0.23 | 0.14 |
| J014156.94-123821.6 | | 14.96 | 13.37 | 9.85 | 9.23 | 8.99 | 49.74 | 9.74 | 5.36 | 12.03 | 1.1 | 0.22 | -0.32 |
| J014431.99-460432.1 | | 19.3 | | 11.88 | 11.3 | 10.98 | 38.54 | | 8.0 | 12.38 | 1.4 | 0.37 | |
| J015057.01-584403.4 | | 15.08 | 13.54 | 9.54 | 8.87 | 8.64 | 45.25 | 10.25 | 5.35 | 11.16 | 1.12 | 0.26 | 0.16 |
| J015257.41+083326.3 | | 15.64 | 14.06 | 9.24 | 8.64 | 8.36 | | 10.61 | 4.91 | 11.7 | 1.23 | 0.33 | 0.51 |
| J015350.81-145950.6 | | 13.49 | 12.01 | 7.94 | 7.3 | 7.07 | 33.84 | 8.97 | 4.03 | 11.28 | 1.21 | 0.14 | 0.21 |
| J015455.24-295746.0 | | 16.05 | 14.62 | 11.77 | 11.09 | 10.81 | 170.78 | 11.31 | 7.5 | 11.2 | 1.32 | 0.14 | |
| J020012.84-084052.4 | | 13.87 | 12.44 | 8.77 | 8.14 | 7.87 | 36.79 | 8.71 | 4.14 | 11.23 | 1.09 | 0.19 | 0.24 |
| J020302.74+221606.8 | | 15.47 | 13.96 | 10.45 | 9.82 | 9.55 | 68.47 | 10.92 | 6.51 | 12.36 | 1.23 | 0.21 | 0.38 |
| J020305.46-590146.6 | | | 16.47 | 13.0 | 12.43 | 12.13 | 196.12 | 13.57 | 9.23 | 10.82 | 1.31 | 0.11 | |
| J020805.55-474633.7 | | 14.75 | 13.31 | 10.41 | 9.79 | 9.57 | 63.07 | 8.83 | 5.09 | 12.49 | 1.01 | 0.15 | |
| J021258.28-585118.3 | | 14.41 | 12.93 | 9.33 | 8.65 | 8.44 | 47.91 | 10.07 | 5.58 | 10.92 | 1.13 | 0.2 | 0.32 |
| J021330.24-465450.3 | | 15.29 | 13.85 | 9.49 | 8.86 | 8.6 | | 11.47 | 6.22 | 10.76 | 1.2 | 0.25 | 0.29 |
| J021935.52-455106.2 | W | | | | | | 117.71 | | | | | | |
| J021935.52-455106.2 | E | 14.79 | 13.43 | 10.84 | 10.29 | 10.1 | 117.66 | 11.48 | 8.15 | 10.74 | 1.52 | -0.04 | 0.41 |
| J022240.88+305515.4 | | 15.69 | 14.15 | 9.92 | 9.33 | 9.06 | 44.92 | 10.96 | 5.87 | 11.79 | 1.18 | 0.27 | 0.68 |
| J022424.69-703321.2 | | 16.66 | 15.0 | 10.37 | 9.75 | 9.49 | 44.05 | 12.23 | 6.72 | 11.47 | 1.24 | 0.36 | 0.63 |
| J023005.14+284500.0 | | 15.11 | 13.65 | 10.57 | 9.99 | 9.81 | 76.32 | 9.49 | 5.65 | 12.45 | 1.13 | 0.12 | |
| J023139.36+445638.1 | | | 14.27 | 9.97 | 9.4 | 9.13 | 49.97 | 11.2 | 6.06 | 11.47 | 1.18 | 0.3 | 0.54 |
| J024552.65+052923.8 | | 15.11 | 13.6 | 10.08 | 9.38 | 9.17 | 59.68 | 9.75 | 5.32 | 12.19 | 1.15 | 0.15 | 0.92 |
| J024746.49-580427.4 | | 14.43 | 13.0 | 9.36 | 8.67 | 8.45 | 44.62 | 9.81 | 5.26 | 11.28 | 1.13 | 0.21 | 0.24 |
| J024852.67-340424.9 | | 15.18 | 13.63 | 9.31 | 8.63 | 8.4 | 41.69 | 10.39 | 5.16 | 10.84 | 1.26 | 0.28 | 0.48 |
| J025154.17+222728.9 | | 14.84 | 13.28 | 8.92 | 8.32 | 8.07 | 27.02 | 11.04 | 5.83 | 11.18 | 1.15 | 0.3 | 0.71 |
| J025913.40+203452.6 | | 13.81 | 12.36 | 9.74 | 9.34 | 9.14 | 351.99 | 7.74 | 4.52 | 11.42 | 0.82 | 0.0 | 0.38 |
| J030002.98+550652.4 | | 14.62 | 13.17 | 9.95 | 9.32 | 9.08 | 65.62 | 9.08 | 4.99 | 11.37 | 0.97 | 0.22 | |
| J030251.62-191150.0 | | | | 10.54 | 9.98 | 9.68 | 41.94 | | 6.54 | 11.62 | 1.27 | 0.36 | 0.7 |
| J030444.10+220320.8 | | 17.16 | 15.56 | 10.49 | 9.93 | 9.66 | | 11.69 | 5.79 | 11.84 | 1.26 | 0.33 | |
| J030824.14+234554.2 | | 14.62 | 13.12 | 9.71 | 9.06 | 8.85 | 72.28 | 9.59 | 5.32 | 11.37 | 1.08 | 0.19 | |
| J031650.45-350937.9 | | | 13.27 | 9.17 | 8.54 | 8.32 | 46.7 | 11.37 | 6.42 | 10.73 | 1.19 | 0.25 | 0.43 |
| J032047.66-504133.0 | | 14.42 | 12.96 | 9.41 | 8.79 | 8.56 | 43.85 | 9.71 | 5.31 | 12.2 | 1.08 | 0.18 | 0.58 |
| J033235.82+284354.6 | | 15.42 | 13.88 | 9.36 | 8.76 | 8.47 | | 10.74 | 5.33 | 12.45 | 1.22 | 0.28 | 0.1 |
| J033431.66-350103.3 | | 16.78 | 15.21 | 10.73 | 10.1 | 9.87 | 105.55 | 10.43 | 5.09 | 10.69 | 1.16 | 0.32 | 1.08 |
| J033640.91+032918.3 | | 15.52 | 13.94 | 9.3 | 8.68 | 8.44 | 26.78 | 12.43 | 6.93 | 11.65 | 1.25 | 0.3 | 0.29 |
| J034115.60-225307.8 | | 14.36 | 12.94 | 9.91 | 9.27 | 9.02 | 71.41 | 8.83 | 4.91 | 10.97 | 1.02 | 0.18 | -0.23 |
| J034116.16-225244.0 | | 14.6 | 13.18 | 9.98 | 9.3 | 9.11 | 71.61 | 8.99 | 4.92 | 10.7 | 1.04 | 0.18 | 0.06 |
| J034236.95+221230.2 | | 17.2 | 17.2 | 11.23 | 10.66 | 10.32 | 39.46 | 14.39 | 7.51 | 12.33 | 1.39 | 0.53 | |
| J034444.80+404150.4 | | 14.95 | 13.44 | 10.03 | 9.36 | 9.13 | 68.81 | 9.71 | 5.4 | 10.63 | 1.1 | 0.17 | 0.03 |
| J035100.83+141339.2 | | 15.51 | 13.96 | 9.44 | 8.75 | 8.58 | 40.03 | 10.67 | 5.29 | 11.79 | 1.26 | 0.3 | |
| J035134.51+072224.5 | | 16.55 | 14.86 | 10.75 | 10.12 | 9.85 | 36.18 | 13.24 | 8.23 | 10.86 | 1.55 | 0.19 | 0.4 |
| J035223.52-282619.6 | | 14.63 | 13.22 | 9.85 | 9.22 | 8.98 | 49.7 | 9.75 | 5.51 | 11.05 | 1.06 | 0.17 | 0.17 |
| J035345.92-425018.0 | | 15.57 | 14.1 | 10.55 | 9.96 | 9.75 | 91.11 | 9.29 | 4.94 | 11.79 | 1.08 | 0.2 | 0.37 |
| J035716.56-271245.5 | | 11.56 | 10.29 | 7.65 | 7.01 | 6.79 | 35.65 | 8.38 | 4.88 | 11.67 | 0.94 | 0.08 | 0.18 |
| J035733.95+244510.2 | | 14.12 | 12.62 | 9.59 | 8.91 | 8.73 | 68.74 | 8.38 | 4.49 | 11.01 | 0.96 | 0.1 | 0.19 |
| J035829.67-432517.2 | | 15.94 | 14.42 | 10.26 | 9.6 | 9.41 | 93.55 | 9.96 | 4.95 | 11.03 | 1.17 | 0.27 | 0.24 |
| J040539.68-401410.5 | | 16.0 | 14.31 | 9.82 | 9.26 | 8.98 | 50.19 | 10.93 | 5.6 | 11.71 | 1.23 | 0.32 | 0.66 |
| J040649.38-450936.3 | | 15.45 | 13.87 | 9.95 | 9.32 | 9.08 | 75.39 | 9.74 | 4.95 | 11.42 | 1.15 | 0.25 | 0.39 |
| J040711.50-291834.3 | | 13.41 | 12.06 | 9.06 | 8.35 | 8.19 | 75.86 | 7.79 | 3.92 | 10.9 | 1.0 | 0.11 | 0.12 |
| J040743.83-682511.0 | | 16.11 | 14.6 | 10.41 | 9.78 | 9.52 | 61.02 | 10.91 | 5.83 | 10.15 | 1.16 | 0.29 | 0.37 |
| J040809.80-611904.3 | | 13.92 | 12.51 | 9.81 | 9.15 | 8.93 | 75.24 | 8.69 | 5.11 | 12.47 | 0.99 | 0.11 | 0.2 |
| J040827.01-784446.7 | | 13.58 | 12.2 | 9.28 | 8.59 | 8.4 | 57.7 | 8.52 | 4.72 | 10.64 | 1.02 | 0.13 | 0.2 |
| J041050.04-023954.4 | | 14.58 | 13.12 | 10.15 | 9.51 | 9.33 | | 9.42 | 5.63 | 10.73 | 0.93 | 0.17 | 0.7 |
| J041255.78-141859.2 | | 14.22 | 12.76 | 9.51 | 8.83 | 8.62 | 57.55 | 9.1 | 4.96 | 11.23 | 1.09 | 0.15 | 0.07 |
| J041336.14-441332.4 | | 16.79 | 15.18 | 10.77 | 10.19 | 9.91 | 52.03 | 11.53 | 6.26 | 11.34 | 1.21 | 0.35 | 0.54 |
| J041525.58-212214.5 | | 16.79 | 15.25 | 11.13 | 10.5 | 10.32 | 88.27 | 12.59 | 7.66 | 12.05 | 1.33 | 0.23 | |
| J041749.66+001145.4 | | 13.26 | 12.17 | 9.6 | 9.01 | 8.82 | 102.06 | 8.05 | 4.7 | 10.13 | 0.88 | 0.13 | -0.21 |
| J041807.76+030826.0 | | 14.27 | 12.97 | 10.29 | 9.63 | 9.44 | 86.55 | 9.85 | 6.32 | 10.78 | 1.02 | 0.15 | |
| J042139.19-723355.7 | | 15.2 | 13.7 | 9.87 | 9.25 | 8.99 | 53.6 | 10.04 | 5.33 | 11.89 | 1.09 | 0.23 | 0.01 |
| J042500.91-634309.8 | | 16.54 | 14.85 | 10.78 | 10.21 | 9.9 | 64.43 | 12.04 | 7.09 | 11.14 | 1.3 | 0.28 | 0.34 |
| J042736.03-231658.8 | | 16.98 | 15.37 | 10.47 | 9.86 | 9.58 | 40.26 | 12.18 | 6.39 | 11.64 | 1.28 | 0.46 | 1.31 |
| J042739.33+171844.2 | | 15.58 | 14.11 | 10.65 | 9.97 | 9.74 | 104.6 | 9.41 | 5.04 | 11.84 | 1.22 | 0.1 | |
| J043213.46-285754.8 | | | 15.79 | 11.91 | 11.32 | 11.03 | 96.56 | 13.07 | 8.31 | 10.99 | 1.22 | 0.34 | |
| J043257.29+740659.3 | | 13.37 | 11.85 | 8.25 | 7.61 | 7.39 | 33.99 | 10.59 | 6.13 | 11.55 | 1.19 | 0.14 | 0.27 |
| J043657.44-161306.7 | | 14.55 | 13.13 | 9.12 | 8.47 | 8.26 | 39.41 | 9.45 | 4.58 | 10.61 | 1.15 | 0.23 | 0.41 |
| J043726.87+185126.2 | | 12.89 | 11.66 | 9.42 | 8.56 | 8.67 | 93.51 | 7.16 | 4.17 | 10.02 | 0.87 | 0.06 | |
| J043923.21+333149.0 | | 15.81 | 14.34 | 9.92 | 9.28 | 9.05 | | 9.58 | 4.29 | 10.76 | 1.16 | 0.31 | 0.92 |
| J043939.24-050150.9 | B | 12.57 | 11.43 | 8.85 | 8.29 | 8.04 | 203.44 | 8.0 | 4.61 | 9.9 | 0.85 | 0.16 | 0.08 |
| J043939.24-050150.9 | A | 12.57 | 11.43 | 8.85 | 8.29 | 8.04 | 203.44 | 8.0 | 4.61 | 9.9 | 0.85 | 0.16 | 0.08 |
| J044036.23-380140.8 | | 12.03 | 10.41 | 7.69 | 6.95 | 6.73 | 3844.65 | 7.63 | 3.95 | 12.21 | 1.03 | 0.16 | 0.27 |
| J044120.81-194735.6 | | 13.43 | 12.19 | 9.64 | 9.02 | 8.85 | 121.11 | 7.53 | 4.19 | 10.69 | 0.97 | 0.13 | 0.12 |
| J044154.44+091953.1 | | | 15.76 | 11.45 | 10.84 | 10.58 | 88.24 | 11.71 | 6.53 | 11.21 | 1.21 | 0.29 | 1.86 |
| J044336.19-003401.8 | | 16.69 | 15.12 | 10.61 | 10.01 | 9.77 | 102.04 | 11.09 | 5.74 | 12.39 | 1.22 | 0.26 | 0.88 |
| J044349.19+742501.6 | | 15.89 | 14.42 | 10.78 | 10.1 | 9.89 | 88.73 | 10.67 | 6.14 | 11.57 | 1.19 | 0.13 | |
| J044356.87+372302.7 | | 14.79 | 13.32 | 9.71 | 9.03 | 8.8 | 71.38 | 9.14 | 4.62 | 11.29 | 1.16 | 0.31 | 1.1 |
| J044455.71+193605.3 | | 14.19 | 12.9 | 10.24 | 9.6 | 9.42 | 126.15 | 8.48 | 5.0 | 10.89 | 0.98 | 0.13 | 0.52 |
| J044530.77-285034.8 | | 16.71 | 15.25 | 11.04 | 10.4 | 10.17 | 87.0 | 12.44 | 7.36 | 10.64 | 1.16 | 0.33 | |
| J044700.46-513440.4 | | 15.27 | 13.84 | 10.06 | 9.43 | 9.21 | 58.44 | 10.26 | 5.63 | 11.9 | 1.09 | 0.24 | -0.02 |
| J044721.05+280852.5 | | 15.51 | 13.97 | 9.73 | 9.12 | 8.87 | 61.98 | 9.96 | 4.86 | 11.25 | 1.16 | 0.28 | 0.64 |
| J044800.86+143957.7 | AB | 17.65 | 16.65 | 11.68 | 11.06 | 10.73 | 84.12 | 12.41 | 6.49 | 10.39 | 1.91 | 2.24 | 4.04 |
| J044802.59+143951.1 | A | | | 11.68 | 11.07 | 10.68 | 83.91 | | | | | | |
| J045114.41-601830.5 | | 14.69 | 13.25 | 10.38 | 9.73 | 9.52 | 100.33 | 8.32 | 4.59 | 11.13 | 0.98 | 0.13 | 0.22 |
| J045420.20-400009.9 | | 17.28 | 15.79 | 11.42 | 10.86 | 10.59 | 151.0 | 13.37 | 8.17 | 11.33 | 1.19 | 0.23 | |
| J045651.47-311542.7 | | 10.53 | 9.3 | 6.98 | 6.34 | 6.16 | 2282.95 | 8.0 | 4.86 | 12.09 | 0.94 | 0.07 | 0.14 |
| J050333.31-382135.6 | | 17.0 | 15.39 | 10.86 | 10.29 | 10.01 | 109.09 | 13.15 | 7.77 | 11.54 | 1.19 | 0.32 | |
| J050610.44-582828.5 | | 14.6 | 13.12 | 9.65 | 9.02 | 8.77 | 53.44 | 9.56 | 5.21 | 11.68 | 1.12 | 0.19 | 0.45 |
| J050827.31-210144.3 | | 15.94 | 14.67 | 9.72 | 9.11 | 8.83 | 48.3 | 13.6 | 7.76 | 11.33 | 1.4 | 0.36 | 0.2 |
| J051026.38-325307.4 | | 15.76 | 14.18 | 10.71 | 10.01 | 9.7 | 81.3 | 12.15 | 7.67 | 11.37 | 1.54 | 0.23 | 0.27 |
| J051255.82-212438.7 | | | 14.84 | 11.24 | 10.63 | 10.36 | 111.46 | 12.18 | 7.7 | 11.42 | 1.16 | 0.26 | |
| J051310.57-303147.7 | | 12.3 | 11.2 | 9.0 | 8.38 | 8.18 | 296.02 | 7.81 | 4.79 | 9.95 | 0.92 | 0.13 | 0.02 |
| J051403.20-251703.8 | | 15.65 | 14.13 | 10.24 | 9.59 | 9.37 | 63.96 | 9.97 | 5.21 | 10.39 | 1.13 | 0.26 | 0.63 |
| J051650.66+022713.0 | | 17.17 | 15.55 | 10.74 | 10.15 | 9.87 | 57.76 | 12.17 | 6.49 | 11.71 | 1.21 | 0.26 | |
| J051803.00-375721.2 | | 16.02 | 14.47 | 10.75 | 10.12 | 9.87 | 98.58 | 9.9 | 5.3 | 10.4 | 1.08 | 0.23 | 0.86 |
| J052419.14-160115.5 | | 15.17 | 13.57 | 8.67 | 8.13 | 7.81 | 31.19 | 10.98 | 5.22 | 11.19 | 1.25 | 0.35 | 0.44 |
| J052535.85-250230.2 | | 15.52 | 14.09 | 10.93 | 10.3 | 10.11 | 124.92 | 11.7 | 7.72 | 10.94 | 1.19 | 0.08 | |
| J052944.69-323914.1 | | 15.37 | 13.74 | 9.22 | 8.61 | 8.32 | 29.76 | 11.5 | 6.08 | 11.78 | 1.28 | 0.33 | 0.24 |
| J053100.27+231218.3 | | 16.8 | 15.15 | 10.58 | 9.94 | 9.69 | 105.07 | 10.12 | 4.66 | 11.39 | 1.24 | 0.26 | |
| J053311.32-291419.9 | | 16.19 | 14.56 | 10.22 | 9.61 | 9.32 | 63.45 | 10.63 | 5.39 | 10.59 | 1.22 | 0.31 | 0.28 |
| J053328.01-425720.1 | A | 14.19 | 12.56 | 8.0 | 7.4 | 7.12 | 10.42 | 11.91 | 6.47 | 11.92 | 1.22 | 0.3 | 0.48 |
| J053328.01-425720.1 | B | 14.19 | 12.56 | 8.0 | 7.4 | 7.12 | 10.42 | 11.91 | 6.46 | 11.92 | 1.22 | 0.3 | 0.48 |
| J053747.56-424030.8 | | 17.38 | 15.53 | 10.24 | 9.67 | 9.35 | 30.08 | 13.41 | 7.23 | 11.63 | 1.3 | 0.43 | 0.65 |
| J053925.08-424521.0 | | 14.53 | 13.02 | 9.45 | 8.8 | 8.6 | 49.34 | 8.52 | 4.1 | 10.78 | 1.08 | 0.18 | 0.33 |
| J054223.86-275803.3 | | 17.22 | 15.83 | 11.41 | 10.8 | 10.55 | 65.83 | 13.34 | 8.06 | 10.96 | 1.17 | 0.32 | |
| J054433.76-200515.5 | | 14.95 | 13.54 | 10.2 | 9.51 | 9.31 | 113.15 | 11.12 | 6.89 | 10.86 | 1.1 | 0.21 | |
| J054448.20-265047.4 | | 13.95 | 12.69 | 10.13 | 9.55 | 9.35 | 112.85 | 10.23 | 6.89 | 11.4 | 0.88 | 0.07 | |
| J054709.88-525626.1 | | 14.34 | 13.09 | 10.39 | 9.74 | 9.57 | 85.75 | 10.22 | 6.7 | 10.44 | 0.95 | 0.1 | 0.42 |
| J054719.52-335611.2 | | 17.21 | 15.62 | 11.31 | 10.72 | 10.48 | 73.09 | 13.91 | 8.77 | 11.38 | 1.16 | 0.31 | |
| J055008.59+051153.2 | | 14.06 | 12.61 | 9.37 | 8.7 | 8.47 | 64.06 | 8.65 | 4.51 | 11.29 | 1.05 | 0.12 | -0.57 |
| J055041.58+430451.8 | | 13.67 | 12.31 | 9.41 | 8.77 | 8.58 | 66.37 | 8.74 | 5.01 | 10.61 | 0.98 | 0.13 | 0.27 |
| J055208.04+613436.6 | | 13.1 | 11.98 | 9.33 | 8.73 | 8.56 | 60.06 | 8.15 | 4.73 | 11.47 | 0.9 | 0.13 | 0.17 |
| J055941.10-231909.4 | | 16.37 | 14.76 | 10.41 | 9.78 | 9.53 | 71.79 | 12.45 | 7.22 | 11.03 | 1.18 | 0.25 | 0.29 |
| J060156.10-164859.9 | B | 16.16 | 14.7 | 10.66 | 10.06 | 9.81 | 79.87 | 10.22 | 5.33 | 10.68 | 1.09 | 0.37 | 1.54 |
| J060156.10-164859.9 | A | 16.16 | 14.7 | 10.66 | 10.06 | 9.81 | 79.87 | 10.22 | 5.33 | 10.68 | 1.09 | 0.37 | 1.54 |
| J060224.56-163450.0 | | 13.51 | 12.1 | 8.99 | 8.38 | 8.17 | 40.22 | 8.82 | 4.89 | 11.18 | 0.95 | 0.16 | 0.25 |
| J060329.60-260804.7 | | 16.05 | 14.51 | 10.54 | 10.02 | 9.72 | 47.69 | 10.72 | 5.93 | 12.43 | 1.17 | 0.31 | 1.06 |
| J061313.30-274205.6 | | 13.77 | 12.26 | 8.0 | 7.43 | 7.15 | 33.77 | 9.8 | 4.69 | 11.62 | 1.15 | 0.28 | 0.37 |
| J061740.43-475957.2 | | | 14.12 | 10.44 | 9.85 | 9.61 | 87.31 | 9.65 | 5.14 | 10.3 | 1.15 | 0.3 | 0.36 |
| J061851.01-383154.9 | | 15.37 | 13.97 | 11.11 | 10.38 | 10.38 | 175.17 | 12.02 | 8.43 | 10.57 | 1.23 | 0.08 | 0.21 |
| J062047.17-361948.2 | | 13.46 | 12.29 | 10.0 | 9.37 | 9.2 | 140.35 | 7.91 | 4.82 | 9.6 | 0.92 | 0.1 | 0.54 |
| J062130.52-410559.1 | | 16.7 | 15.08 | 11.12 | 10.53 | 10.27 | 55.49 | 13.09 | 8.28 | 11.47 | 1.44 | 0.26 | |
| J062407.62+310034.4 | | | | 9.33 | 8.67 | 8.45 | 27.36 | | 6.29 | 10.17 | 1.21 | 0.19 | 0.12 |
| J063001.84-192336.6 | | 16.43 | 14.71 | 10.09 | 9.53 | 9.25 | 27.99 | 12.1 | 6.64 | 12.15 | 1.24 | 0.31 | -0.1 |
| J065846.87+284258.9 | | 13.63 | 12.15 | 8.64 | 8.0 | 7.78 | | 9.56 | 5.19 | 10.81 | 1.06 | 0.17 | 0.09 |
| J070657.72-535345.9 | | 12.78 | 11.42 | 8.54 | 7.9 | 7.67 | 46.82 | 7.74 | 3.99 | 11.1 | 0.97 | 0.11 | 0.2 |
| J071036.50+171322.6 | | 13.73 | 12.45 | 9.72 | 9.08 | 8.9 | 79.93 | 8.44 | 4.89 | 10.86 | 0.95 | 0.09 | |
| J072641.52+185034.0 | | 15.35 | 13.83 | 9.98 | 9.4 | 9.13 | | 10.05 | 5.35 | 10.97 | 1.09 | 0.26 | 0.29 |
| J072821.16+334511.6 | | | | 9.28 | 8.66 | 8.38 | 45.57 | | 4.62 | 11.03 | 1.17 | 0.25 | 0.34 |
| J072911.26-821214.3 | | 14.35 | 12.89 | 9.75 | 9.13 | 8.89 | 51.52 | 8.95 | 4.95 | 11.72 | 1.03 | 0.17 | 0.37 |
| J073138.47+455716.5 | | 15.39 | 13.85 | 9.78 | 9.2 | 8.92 | 55.93 | 10.67 | 5.74 | 11.03 | 1.15 | 0.28 | 0.57 |
| J075233.22-643630.5 | | 13.51 | 12.32 | 9.7 | 9.11 | 8.87 | 99.21 | 8.56 | 5.11 | 10.64 | 1.02 | 0.07 | -0.05 |
| J075808.25-043647.5 | | 15.2 | 13.66 | 10.17 | 9.56 | 9.31 | 57.17 | 9.68 | 5.33 | 12.24 | 1.09 | 0.29 | 0.92 |
| J075830.92+153013.4 | B | 15.9 | 14.42 | 9.97 | 9.38 | 9.1 | 31.17 | 12.89 | 7.57 | 11.08 | 1.34 | 0.2 | 0.69 |
| J075830.92+153013.4 | A | 15.9 | 14.42 | 9.97 | 9.38 | 9.1 | 31.17 | 12.89 | 7.57 | 11.08 | 1.34 | 0.2 | 0.69 |
| J080352.54+074346.7 | | 14.87 | 13.35 | 10.06 | 9.36 | 9.19 | 74.65 | 9.47 | 5.31 | 12.43 | 1.08 | 0.14 | 0.92 |
| J080636.05-744424.6 | | 15.08 | 13.52 | 10.03 | 9.36 | 9.13 | 68.59 | 9.36 | 4.97 | 11.99 | 1.08 | 0.2 | 0.41 |
| J081443.62+465035.8 | | 16.55 | 14.83 | 10.52 | 9.89 | 9.63 | | 12.13 | 6.93 | 10.88 | 1.26 | 0.27 | 0.57 |
| J081738.97-824328.8 | | 13.11 | 11.85 | 7.47 | 6.84 | 6.59 | 26.54 | 9.58 | 4.32 | 11.09 | 1.18 | 0.26 | 0.48 |
| J082105.04-090853.8 | B | 13.95 | 12.51 | 9.44 | 8.72 | 8.52 | 62.91 | 9.24 | 5.25 | 11.65 | 1.06 | 0.15 | -0.28 |
| J082105.04-090853.8 | A | 13.95 | 12.51 | 9.44 | 8.72 | 8.52 | 62.91 | 8.04 | 4.05 | 11.65 | 1.06 | 0.15 | -0.28 |
| J082558.91+034019.5 | | 15.74 | 14.1 | 10.01 | 9.43 | 9.12 | 35.14 | 12.48 | 7.5 | 11.41 | 1.32 | 0.22 | |
| J083528.87+181219.9 | | 15.48 | 13.98 | 10.32 | 9.7 | 9.48 | 49.23 | 9.81 | 5.31 | 10.4 | 1.08 | 0.33 | |
| J090227.87+584813.4 | | 14.59 | 13.3 | 9.85 | 9.23 | 8.95 | 68.01 | 9.88 | 5.53 | 10.8 | 1.12 | 0.21 | 0.5 |
| J092216.12+043423.3 | | 14.83 | 13.35 | 10.05 | 9.37 | 9.18 | 60.26 | 9.27 | 5.1 | 12.47 | 1.01 | 0.09 | 0.28 |
| J093212.63+335827.3 | | 15.41 | 13.85 | 9.9 | 9.26 | 9.02 | 49.88 | 10.95 | 6.12 | 11.18 | 1.19 | 0.24 | 0.27 |
| J094317.05-245458.3 | | 13.82 | 12.58 | 10.09 | 9.49 | 9.31 | 103.12 | 8.08 | 4.81 | 11.73 | 0.92 | 0.1 | 0.62 |
| J094508.15+714450.1 | | 16.6 | 15.63 | 10.68 | 10.07 | 9.77 | 29.44 | 13.77 | 7.91 | 12.09 | 1.33 | 0.41 | 0.25 |
| J100146.28+681204.1 | | 16.45 | 14.92 | 11.92 | 11.27 | 11.05 | 136.54 | 10.73 | 6.86 | 12.1 | 1.25 | 0.12 | |
| J100230.94-281428.2 | | 15.92 | 14.21 | 9.89 | 9.32 | 9.03 | 33.97 | 11.71 | 6.53 | 10.84 | 1.23 | 0.31 | 0.53 |
| J101543.44+660442.3 | | 13.14 | 11.66 | 8.71 | 8.04 | 7.87 | 37.87 | 8.6 | 4.81 | 10.55 | 1.02 | 0.14 | 0.04 |
| J101905.68-304920.3 | | 15.44 | 13.98 | 10.81 | 10.16 | 9.93 | 115.1 | 10.36 | 6.31 | 11.04 | 1.06 | 0.15 | |
| J101917.57-443736.0 | | 15.22 | 13.74 | 9.67 | 9.02 | 8.8 | 68.21 | 10.08 | 5.14 | 11.71 | 1.15 | 0.23 | 0.17 |
| J102602.07-410553.8 | B | 13.97 | 12.55 | 9.18 | 8.49 | 8.27 | 84.37 | 8.79 | 4.51 | 10.61 | 1.11 | 0.21 | 0.08 |
| J102602.07-410553.8 | A | 13.97 | 12.55 | 9.18 | 8.49 | 8.27 | 84.37 | 9.75 | 5.47 | 10.61 | 1.11 | 0.21 | 0.08 |
| J102636.95+273838.4 | | 15.35 | 13.94 | 10.88 | 10.25 | 10.01 | 103.93 | 9.84 | 5.91 | 10.89 | 1.17 | 0.13 | |
| J103016.11-354626.3 | | 15.11 | 13.7 | 10.56 | 9.96 | 9.72 | 94.52 | 9.13 | 5.15 | 10.83 | 1.03 | 0.13 | 0.71 |
| J103137.59-374915.9 | | 16.45 | 14.94 | 10.88 | 10.3 | 10.02 | 115.45 | 11.2 | 6.28 | 12.46 | 1.13 | 0.3 | 0.93 |
| J103557.17+285330.8 | | 14.75 | 13.17 | 9.25 | 8.62 | 8.37 | 29.06 | 10.53 | 5.73 | 12.25 | 1.14 | 0.25 | 0.25 |
| J103952.70-353402.5 | | 14.19 | 12.75 | 9.78 | 9.2 | 9.04 | 83.57 | 9.15 | 5.44 | 10.86 | 1.04 | 0.12 | |
| J104008.36-384352.1 | | 15.12 | 13.68 | 10.69 | 9.99 | 9.82 | 118.5 | 9.02 | 5.16 | 10.38 | 1.03 | 0.14 | |
| J104044.98-255909.2 | | 15.26 | 13.76 | 10.44 | 9.77 | 9.58 | 84.66 | 9.56 | 5.38 | 12.4 | 1.1 | 0.11 | 0.53 |
| J104551.72-112615.4 | | 15.58 | 14.08 | 9.6 | 9.0 | 8.74 | 3260.03 | 11.15 | 5.81 | 11.31 | 1.23 | 0.31 | 0.53 |
| J105515.87-033538.2 | | 14.25 | 12.77 | 9.56 | 8.96 | 8.68 | 53.56 | 11.25 | 7.16 | 11.47 | 1.28 | 0.18 | 0.18 |
| J105518.12-475933.2 | | 13.57 | 12.48 | 10.29 | 9.65 | 9.5 | 136.73 | 8.7 | 5.72 | 9.98 | 0.93 | 0.05 | |
| J105524.25-472611.7 | A | 15.4 | 13.87 | 10.64 | 10.04 | 9.81 | 114.16 | 9.35 | 5.29 | 11.39 | 1.07 | 0.15 | |
| J105524.25-472611.7 | B | | | | | | 114.16 | | | | | | |
| J105711.36+054454.2 | | 14.13 | 12.79 | 9.83 | 9.21 | 9.02 | 85.43 | 8.81 | 5.04 | 10.56 | 0.99 | 0.18 | 0.35 |
| J105850.47-234620.8 | | 17.2 | 15.58 | 10.3 | 9.71 | 9.43 | 43.8 | 12.41 | 6.26 | 12.12 | 1.33 | 0.43 | 0.77 |
| J110119.22+525222.9 | | 13.48 | 11.91 | 9.2 | 8.54 | 8.37 | 63.42 | 8.55 | 5.01 | 11.7 | 0.98 | 0.05 | -0.03 |
| J110335.71-302449.5 | | | | 9.81 | 9.19 | 8.97 | 524.08 | | | | | | |
| J110335.71-302449.5 | | 13.67 | 12.45 | 9.81 | 9.19 | 8.97 | 524.08 | 8.11 | 4.63 | 10.19 | 0.86 | 0.18 | 0.02 |
| J110551.56-780520.7 | | 15.42 | 14.01 | 11.16 | 10.53 | 10.26 | 226.91 | 9.49 | 5.74 | 12.11 | 1.27 | 0.0 | |
| J111052.06-725513.0 | | 16.89 | 15.36 | 10.95 | 10.37 | 10.07 | 101.23 | 11.52 | 6.23 | 11.46 | 1.29 | 0.37 | 0.36 |
| J111103.54-313459.0 | | 16.05 | 14.41 | 10.34 | 9.76 | 9.49 | 68.3 | 10.72 | 5.8 | 11.56 | 1.15 | 0.29 | 0.15 |
| J111128.13-265502.9 | | | | 10.33 | 9.81 | 9.45 | 49.34 | | 7.65 | 10.46 | 1.36 | 0.47 | 0.93 |
| J111229.74-461610.1 | | 14.51 | 13.13 | 10.4 | 9.72 | 9.54 | 94.08 | 8.45 | 4.86 | 11.05 | 0.94 | 0.12 | |
| J111309.15+300338.4 | | 15.15 | 13.65 | 10.15 | 9.52 | 9.27 | 70.53 | 9.76 | 5.38 | 11.32 | 1.1 | 0.21 | 0.15 |
| J111707.56-390951.3 | | 14.06 | 12.84 | 10.43 | 9.77 | 9.6 | 120.83 | 8.2 | 4.96 | 10.39 | 0.97 | 0.17 | 0.67 |
| J112047.03-273805.8 | | | | 11.82 | 11.22 | 10.93 | 87.49 | | 7.44 | 10.36 | 1.29 | 0.37 | |
| J112105.43-384516.6 | | 14.38 | 12.85 | 9.0 | 8.33 | 8.05 | 65.4 | 8.94 | 4.14 | | | | |
| J112512.28-002438.2 | | 15.81 | 14.25 | 10.29 | 9.76 | 9.5 | | 10.51 | 5.76 | 10.97 | 1.15 | 0.25 | 0.73 |
| J112547.46-441027.4 | | 16.25 | 14.63 | 10.34 | 9.75 | 9.48 | 46.34 | 11.11 | 5.96 | 11.18 | 1.2 | 0.34 | 0.52 |
| J112651.28-382455.5 | | 15.79 | 14.39 | 10.02 | 9.36 | 9.12 | 68.4 | 10.75 | 5.48 | 11.48 | 1.18 | 0.31 | 0.22 |
| J112816.27-261429.6 | | 17.5 | 15.7 | 12.54 | 11.82 | 11.56 | 248.3 | 12.12 | 7.98 | 10.72 | 1.47 | -0.2 | |
| J112955.84+520213.2 | | 14.82 | 13.3 | 10.21 | 9.55 | 9.34 | 76.41 | 9.12 | 5.16 | 12.18 | 1.03 | 0.16 | |
| J113105.57+542913.5 | | 15.85 | 14.25 | 10.62 | 9.94 | 9.74 | 94.46 | 10.09 | 5.58 | 11.71 | 1.13 | 0.22 | 0.53 |
| J113114.81-482628.0 | | 15.74 | 14.22 | 10.48 | 9.86 | 9.61 | 128.41 | 10.19 | 5.58 | 10.02 | 1.12 | 0.28 | 0.49 |
| J113120.31+132140.0 | | 15.63 | 14.12 | 10.47 | 9.85 | 9.61 | 78.36 | 9.76 | 5.25 | 10.89 | 1.06 | 0.23 | |
| J114623.01-523851.8 | | | 15.3 | 11.08 | 10.47 | 10.19 | 57.72 | 11.27 | 6.16 | 10.95 | 1.25 | 0.34 | |
| J114728.37+664402.7 | | 16.2 | 14.59 | 9.68 | 9.06 | 8.75 | 21.05 | 13.07 | 7.23 | 11.04 | 1.28 | 0.35 | 0.78 |
| J115156.73+073125.7 | | 13.82 | 12.42 | 8.81 | 8.14 | 7.89 | 50.75 | 9.82 | 5.29 | 10.69 | 1.07 | 0.12 | 0.29 |
| J115438.73-503826.4 | | 16.38 | 15.0 | 10.92 | 10.31 | 10.02 | 74.9 | 11.22 | 6.24 | 11.06 | 1.2 | 0.34 | 0.61 |
| J115927.82-451019.3 | | 16.1 | 14.52 | 9.93 | 9.35 | 9.07 | 72.25 | 10.76 | 5.31 | 11.18 | 1.23 | 0.35 | 0.56 |
| J115949.51-424426.0 | | 13.91 | 12.5 | 9.64 | 9.03 | 8.83 | 63.43 | 8.84 | 5.17 | 12.21 | 0.96 | 0.19 | 0.55 |
| J115957.68-262234.1 | | 15.1 | 13.63 | 10.19 | 9.48 | 9.33 | 115.23 | 10.14 | 5.84 | 10.16 | 1.05 | 0.21 | 0.15 |
| J120001.54-173131.1 | | 15.35 | 13.84 | 9.4 | 8.69 | 8.47 | 53.04 | 10.67 | 5.3 | 11.25 | 1.21 | 0.3 | 0.48 |
| J120237.94-332840.4 | | 16.57 | 15.77 | 10.69 | 10.12 | 9.85 | 62.58 | 12.23 | 6.31 | 12.01 | 1.24 | 0.32 | 0.46 |
| J120647.40-192053.1 | | 12.39 | 12.21 | 9.73 | 9.1 | 8.96 | 364.79 | | | | | | |
| J120929.80-750540.2 | | | 13.6 | 9.91 | 9.24 | 9.01 | | 9.03 | 4.44 | 9.7 | 1.16 | 0.26 | 0.63 |
| J121153.04+124912.9 | | 14.02 | 12.55 | 9.46 | 8.83 | 8.66 | 60.73 | 9.13 | 5.24 | 11.52 | 1.06 | 0.09 | -0.07 |
| J121341.59+323127.7 | B | 13.61 | 12.07 | 9.49 | 8.88 | 8.67 | 58.72 | 8.53 | 5.13 | 12.0 | 1.02 | 0.15 | 0.15 |
| J121341.59+323127.7 | A | 13.61 | 12.07 | 9.49 | 8.88 | 8.67 | 57.42 | 8.53 | 5.13 | 12.0 | 1.02 | 0.15 | 0.15 |
| J121429.15-425814.8 | | 13.81 | 12.65 | 10.0 | 9.33 | 9.16 | 91.4 | 8.37 | 4.88 | 9.66 | 0.94 | 0.08 | 0.44 |
| J121511.25-025457.1 | | 15.57 | 14.13 | 10.87 | 10.23 | 9.96 | 101.82 | 10.5 | 6.33 | 11.53 | 1.21 | 0.22 | |
| J121558.37-753715.7 | | | | 11.41 | 10.84 | 10.56 | 41.91 | | 7.72 | 11.02 | 1.26 | 0.39 | |
| J122643.99-122918.3 | | 14.22 | 12.66 | 8.87 | 8.12 | 7.87 | 30.26 | 12.26 | 7.47 | 11.74 | 1.31 | 0.21 | 0.38 |
| J122725.27-454006.6 | | 13.25 | 12.17 | 9.86 | 9.24 | 9.1 | 134.02 | 8.35 | 5.28 | 9.81 | 0.94 | 0.12 | -0.28 |
| J122813.57-431638.9 | | | 15.97 | 11.67 | 11.09 | 10.81 | 58.4 | 11.34 | 6.18 | 10.6 | 1.3 | 0.4 | |
| J123005.17-440236.1 | | 16.49 | 14.99 | 10.45 | 9.84 | 9.57 | 78.18 | 11.25 | 5.83 | 11.55 | 1.19 | 0.3 | 0.38 |
| J123234.07-414257.5 | | 17.26 | 15.72 | 10.96 | 10.35 | 10.12 | 97.49 | 12.02 | 6.42 | 10.63 | 1.21 | 0.42 | 1.08 |
| J123425.84-174544.4 | | 12.35 | 11.23 | 8.87 | 8.28 | 8.1 | 176.09 | 7.71 | 4.58 | 9.65 | 0.84 | 0.12 | 0.39 |
| J123704.99-441919.5 | | | 17.73 | 12.17 | 11.58 | 11.28 | 89.79 | 13.95 | 7.5 | 10.98 | 1.35 | 0.45 | |
| J124054.09-451625.4 | | 15.64 | 14.15 | 10.62 | 9.92 | 9.71 | 112.7 | 10.33 | 5.89 | 11.91 | 1.07 | 0.22 | 0.1 |
| J124612.32-384013.5 | | 16.81 | 15.29 | 10.98 | 10.38 | 10.1 | 100.64 | 11.55 | 6.36 | 11.46 | 1.15 | 0.17 | |
| J124955.67-460737.3 | | 15.08 | 13.62 | 10.26 | 9.59 | 9.38 | 107.74 | 9.8 | 5.56 | 10.97 | 1.08 | 0.21 | 0.09 |
| J125049.12-423123.6 | | | | 10.78 | 10.17 | 9.91 | 97.42 | | 6.19 | 11.54 | 1.3 | 0.26 | 0.44 |
| J125326.99-350415.3 | | 15.56 | 14.11 | 10.26 | 9.64 | 9.39 | 84.19 | 10.47 | 5.75 | 11.41 | 1.13 | 0.27 | 0.13 |
| J125902.99-314517.9 | | 16.37 | 14.78 | 10.32 | 9.73 | 9.49 | 73.78 | 11.24 | 5.95 | 11.96 | 1.16 | 0.31 | 0.91 |
| J130501.18-331348.7 | | 15.66 | 14.17 | 10.7 | 10.0 | 9.83 | 66.09 | 10.55 | 6.21 | 12.48 | 1.1 | 0.15 | |
| J130522.37-405701.2 | | 16.03 | 14.38 | 10.53 | 9.92 | 9.7 | 112.32 | 10.62 | 5.94 | 10.64 | 1.13 | 0.28 | |
| J130530.31-405626.0 | | 13.92 | 12.66 | 10.07 | 9.44 | 9.25 | 107.1 | 8.92 | 5.51 | 10.7 | 0.9 | 0.11 | 0.47 |
| J130618.16-342857.0 | | 16.89 | 15.32 | 10.5 | 9.9 | 9.63 | 84.61 | 11.7 | 6.01 | 12.05 | 1.27 | 0.38 | 0.75 |
| J130650.27-460956.1 | | 13.03 | 11.98 | 9.67 | 9.06 | 8.83 | 98.08 | 8.18 | 5.03 | 9.81 | 1.0 | 0.14 | 0.37 |
| J130731.03-173259.9 | | 16.18 | 14.55 | 10.24 | 9.68 | 9.41 | 33.72 | 11.46 | 6.32 | 10.97 | 1.16 | 0.3 | 0.78 |
| J131129.00-425241.9 | | 15.07 | 13.6 | 10.14 | 9.42 | 9.24 | 99.57 | 9.82 | 5.46 | 11.5 | 1.07 | 0.13 | -0.03 |
| J132112.77-285405.1 | | 13.98 | 12.8 | 10.32 | 9.68 | 9.53 | 115.03 | 8.17 | 4.9 | 10.24 | 0.95 | 0.08 | 0.75 |
| J133238.94+305905.8 | | | 14.28 | 9.62 | 9.08 | 8.76 | 22.43 | 12.6 | 7.08 | 11.6 | 1.27 | 0.32 | 0.62 |
| J133509.40+503917.5 | | 14.1 | 12.72 | 9.31 | 8.58 | 8.37 | 43.49 | 9.6 | 5.25 | 10.97 | 1.52 | 0.12 | 0.32 |
| J133901.87-214128.0 | | 15.99 | 14.49 | 10.4 | 9.73 | 9.51 | 84.67 | 10.83 | 5.85 | 11.35 | 1.16 | 0.36 | |
| J134146.41+581519.2 | | 14.36 | 12.55 | 8.73 | 8.17 | 7.88 | 21.79 | 10.72 | 6.05 | 12.45 | 1.18 | 0.24 | 0.43 |
| J134907.28+082335.8 | | 13.54 | 12.18 | 9.34 | 8.76 | 8.55 | 57.32 | 8.54 | 4.91 | 12.18 | 0.95 | 0.1 | 0.11 |
| J135145.65-374200.7 | | | | 10.08 | 9.35 | 9.16 | 94.79 | | | | | | |
| J135511.38+665207.0 | | | 16.57 | 11.7 | 11.06 | 10.81 | 66.35 | 12.64 | 6.88 | 11.91 | 1.26 | 0.33 | |
| J135913.33-292634.2 | | 14.02 | 12.57 | 9.59 | 8.96 | 8.7 | 74.99 | 9.06 | 5.19 | 12.38 | 1.04 | 0.11 | 0 |
| J140337.56-501047.9 | | | 14.43 | 10.66 | 9.99 | 9.8 | 97.18 | 10.25 | 5.62 | 11.53 | 1.13 | 0.25 | 0.83 |
| J141045.24+364149.8 | | 15.51 | 13.93 | 10.05 | 9.47 | 9.17 | 76.98 | 10.24 | 5.48 | 11.83 | 1.11 | 0.2 | 0.03 |
| J141332.23-145421.1 | | 16.18 | 14.56 | 10.29 | 9.67 | 9.42 | 32.96 | 11.27 | 6.13 | 11.36 | 1.19 | 0.31 | 1.13 |
| J141510.77-252012.0 | | 15.59 | 14.04 | 10.2 | 9.57 | 9.34 | 82.52 | 10.0 | 5.3 | 11.3 | 1.08 | 0.2 | 0.08 |
| J141842.36+475514.9 | | 13.77 | 12.54 | 9.63 | 8.96 | 8.79 | 77.79 | 8.75 | 5.0 | 10.81 | 0.98 | 0.11 | 0.19 |
| J141903.13+645146.4 | | 15.91 | 14.15 | 10.39 | 9.77 | 9.56 | 44.98 | 11.43 | 6.84 | 10.58 | 1.12 | 0.25 | 0.39 |
| J143517.80-342250.4 | | | 13.49 | 10.08 | 9.43 | 9.21 | 84.23 | 9.64 | 5.36 | 11.54 | 1.08 | 0.14 | -0.08 |
| J143648.16+090856.5 | | 16.04 | 14.44 | 10.29 | 9.69 | 9.43 | 36.57 | 11.15 | 6.14 | 11.4 | 1.19 | 0.28 | 0.55 |
| J143713.21-340921.1 | | 16.44 | 14.91 | 10.69 | 10.03 | 9.79 | 98.4 | 11.3 | 6.18 | 10.85 | 1.2 | 0.3 | 0.32 |
| J143753.36-343917.8 | | | 12.92 | 8.67 | 8.0 | 7.76 | 30.62 | 10.97 | 5.81 | 11.85 | 1.19 | 0.23 | 0.39 |
| J145014.12-305100.6 | | 15.2 | 13.67 | 10.38 | 9.71 | 9.48 | | 9.32 | 5.13 | 11.81 | 1.02 | 0.16 | |
| J145731.11-305325.0 | | 16.79 | 15.18 | 11.03 | 10.35 | 10.08 | 106.76 | 11.89 | 6.79 | 11.55 | 1.25 | 0.3 | 0.89 |
| J145949.90+244521.9 | | 15.79 | 14.19 | 10.3 | 9.7 | 9.49 | 65.5 | 10.08 | 5.38 | 11.54 | 1.1 | 0.26 | 0.32 |
| J150119.48-200002.1 | | 16.52 | 14.89 | 10.38 | 9.79 | 9.5 | 81.44 | 11.36 | 5.97 | 11.03 | 1.17 | 0.3 | |
| J150230.94-224615.4 | | 16.81 | 15.47 | 11.06 | 10.42 | 10.14 | | 12.13 | 6.8 | 10.64 | 1.25 | 0.27 | |
| J150355.37-214643.1 | | 15.61 | 14.13 | 10.44 | 9.75 | 9.53 | 114.33 | 9.98 | 5.38 | 11.17 | 1.1 | 0.24 | 0.96 |
| J150601.66-240915.0 | | | 15.86 | 11.76 | 11.15 | 10.91 | 152.11 | 11.29 | 6.34 | 10.25 | 1.21 | 0.28 | |
| J150723.91+433353.6 | | 15.41 | 13.73 | 9.6 | 9.05 | 8.78 | 30.27 | 10.88 | 5.93 | 12.26 | 1.17 | 0.32 | 0.64 |
| J150820.15-282916.6 | | 16.74 | 15.35 | 10.8 | 10.2 | 9.94 | 46.39 | 11.81 | 6.4 | 11.95 | 1.27 | 0.28 | 1.26 |
| J150836.69-294222.9 | | 16.64 | 15.13 | 10.91 | 10.28 | 10.05 | 126.33 | 11.36 | 6.28 | 11.39 | 1.21 | 0.35 | 1.14 |
| J150939.16-133212.4 | | 15.72 | 14.22 | 9.79 | 9.2 | 8.9 | 52.72 | 11.89 | 6.57 | 11.21 | 1.23 | 0.31 | -0.06 |
| J151212.18-255708.3 | | 14.24 | 12.81 | 9.66 | 9.01 | 8.76 | 46.95 | 9.21 | 5.16 | 11.87 | 1.03 | 0.13 | 0.39 |
| J151242.69-295148.0 | | 15.77 | 14.28 | 10.8 | 10.11 | 9.88 | 128.67 | 9.76 | 5.36 | 11.24 | 1.1 | 0.09 | |
| J151411.31-253244.1 | | 13.1 | 12.05 | 9.88 | 9.28 | 9.1 | 154.24 | 7.72 | 4.77 | 10.2 | 0.91 | 0.07 | 0.93 |
| J152150.76-251412.1 | | 14.94 | 13.36 | 9.51 | 8.9 | 8.65 | | 10.21 | 5.5 | 11.27 | 1.12 | 0.24 | 0.33 |
| J153248.80-230812.4 | | 15.16 | 13.67 | 10.64 | 9.94 | 9.73 | 159.12 | 9.19 | 5.25 | 11.13 | 1.06 | 0.19 | 0.75 |
| J153549.35-065727.8 | | 14.14 | 12.67 | 9.79 | 9.07 | 8.9 | 117.79 | 9.12 | 5.35 | 11.01 | 1.09 | 0.17 | 0.13 |
| J154220.24+593653.0 | | 16.58 | 14.86 | 10.6 | 9.98 | 9.72 | 43.91 | 11.54 | 6.4 | 10.86 | 1.22 | 0.27 | -0.13 |
| J154227.07-042717.1 | | 14.24 | 12.85 | 9.94 | 9.26 | 9.03 | 126.34 | 9.14 | 5.32 | 11.46 | 1.08 | 0.06 | 0.41 |
| J154349.42-364838.7 | | 16.77 | 15.15 | 10.92 | 10.32 | 10.07 | 148.19 | 10.87 | 5.79 | 10.64 | 1.15 | 0.22 | |
| J154435.17+042307.5 | | 14.43 | 13.01 | 9.69 | 9.0 | 8.81 | 91.9 | 9.19 | 4.99 | 11.17 | 0.98 | 0.2 | 0.38 |
| J154656.43+013650.8 | | 14.17 | 12.78 | 9.75 | 9.1 | 8.89 | | 8.77 | 4.88 | 11.45 | 0.94 | 0.11 | -0.19 |
| J155046.47+305406.9 | | 14.32 | 12.92 | 9.6 | 8.94 | 8.74 | 76.55 | 9.41 | 5.23 | 10.91 | 1.05 | 0.22 | 0.39 |
| J155515.35+081327.9 | | 10.35 | 9.51 | 6.89 | 6.24 | 6.1 | 1743.2 | 7.95 | 4.54 | 11.97 | 0.83 | 0.05 | 0.14 |
| J155759.01-025905.8 | | 14.2 | 12.86 | 10.26 | 9.56 | 9.33 | 97.52 | 9.44 | 5.91 | 11.9 | 1.17 | 0.1 | 0.47 |
| J155947.24+440359.6 | | 13.4 | 11.83 | 8.51 | 7.84 | 7.62 | 42.08 | 9.2 | 4.99 | 11.08 | 1.07 | 0.16 | 0.22 |
| J160116.86-345502.7 | | | | 12.06 | 11.51 | 11.14 | 26.44 | | 7.87 | 10.01 | 1.36 | 0.23 | |
| J160549.19-311521.6 | | 14.57 | 13.16 | 9.97 | 9.26 | 9.08 | 133.24 | 9.38 | 5.3 | 11.1 | 1.08 | 0.13 | 0.61 |
| J160828.45-060734.6 | | 15.22 | 13.64 | 9.66 | 9.03 | 8.78 | 88.49 | 10.49 | 5.63 | 11.53 | 1.13 | 0.3 | 0.84 |
| J160954.85-305858.4 | | 14.64 | 13.18 | 10.24 | 9.52 | 9.34 | 118.29 | 8.93 | 5.09 | 11.46 | 1.01 | 0.13 | 0.74 |
| J161410.76-025328.8 | AB | 13.74 | 12.7 | 10.11 | 9.47 | 9.31 | 513.14 | 8.04 | 4.65 | 9.51 | 0.87 | 0.21 | 0.65 |
| J161743.18+261815.2 | | 13.86 | 12.36 | 8.98 | 8.39 | 8.14 | 31.48 | 9.34 | 5.12 | 12.16 | 1.02 | 0.15 | 0.23 |
| J162422.68+195922.0 | | 14.83 | 13.28 | 9.32 | 8.77 | 8.48 | | 10.55 | 5.75 | 11.25 | 1.15 | 0.25 | 0.27 |
| J162548.69-135912.0 | | 13.7 | 12.53 | 9.86 | 9.26 | 9.11 | 1969.34 | 8.15 | 4.73 | 10.84 | 0.89 | 0.03 | 0.32 |
| J162602.80-155954.5 | | 13.95 | 12.73 | 9.91 | 9.44 | 9.24 | 303.77 | 8.02 | 4.53 | 10.56 | 0.82 | 0.24 | 1.51 |
| J163051.34+472643.8 | | 17.19 | 15.73 | 10.94 | 10.35 | 10.08 | 38.12 | 13.34 | 7.69 | 11.77 | 1.25 | 0.34 | 0.67 |
| J163632.90+635344.9 | | 15.13 | 13.63 | 10.19 | 9.5 | 9.13 | 73.2 | 12.06 | 7.56 | 12.33 | 1.34 | 0.13 | 0.17 |
| J164539.37+702400.1 | | | 14.68 | 11.86 | 10.98 | 10.83 | 112.34 | 11.99 | 8.14 | 12.13 | 1.52 | 0.18 | |
| J170415.15-175552.5 | | 13.01 | 11.81 | 9.18 | 8.48 | 8.32 | 1762.75 | 8.51 | 5.02 | 11.4 | 0.99 | 0.02 | 0.25 |
| J171038.44-210813.0 | | 12.97 | 11.69 | 8.85 | 8.22 | 8.03 | 96.88 | 8.75 | 5.09 | 10.05 | 1.01 | 0.13 | 0.02 |
| J171117.68+124540.4 | | 15.44 | 13.99 | 10.29 | 9.67 | 9.47 | 54.6 | 9.72 | 5.2 | 11.15 | 1.04 | 0.13 | |
| J171426.13-214845.0 | | 16.76 | 15.25 | 11.39 | 10.64 | 10.28 | 225.88 | 11.36 | 6.39 | 12.29 | 1.36 | -0.15 | |
| J171441.70-220948.8 | | 15.31 | 13.88 | 10.02 | 9.33 | 9.1 | 103.67 | 10.11 | 5.33 | 11.34 | 1.08 | 0.06 | |
| J172130.71-150617.8 | | | 15.4 | 11.26 | 10.67 | 10.42 | 83.82 | 12.44 | 7.46 | 11.76 | 1.19 | 0.33 | |
| J172131.73-084212.3 | | | 15.01 | 11.29 | 10.62 | 10.37 | 102.46 | 12.66 | 8.02 | 11.37 | 1.26 | 0.23 | |
| J172309.67-095126.2 | | 15.39 | 13.79 | 10.35 | 9.64 | 9.42 | 103.56 | 10.32 | 5.95 | 12.39 | 1.17 | 0.16 | |
| J172454.26+502633.0 | | 14.08 | 12.67 | 9.76 | 9.15 | 8.95 | 47.11 | 10.31 | 6.59 | 12.29 | 1.23 | 0.09 | 0.45 |
| J172615.23-031131.9 | | | 14.91 | 10.38 | 9.79 | 9.5 | 18.59 | 11.65 | 6.24 | 11.97 | 1.3 | 0.35 | 0.81 |
| J172951.38+093336.9 | | 16.5 | 15.0 | 10.41 | 9.81 | 9.47 | 69.44 | 12.84 | 7.31 | 11.45 | 1.29 | 0.32 | |
| J173353.07+165511.7 | | 16.13 | 14.38 | 8.9 | 8.3 | 8.0 | 16.42 | 15.32 | 8.94 | 11.69 | 1.33 | 0.38 | 0.53 |
| J173544.26-165209.9 | | 13.67 | 12.18 | 8.91 | 8.21 | 7.98 | 36.47 | 9.14 | 4.94 | 12.01 | 1.12 | 0.15 | 0.55 |
| J173623.80+061853.0 | | 10.79 | 9.62 | 7.19 | 6.54 | 6.4 | 1647.13 | 7.79 | 4.57 | 12.03 | 0.84 | 0.08 | 0.15 |
| J173826.94-055628.0 | | 14.04 | 12.78 | 10.25 | 9.99 | 9.6 | 487.33 | 9.8 | 6.62 | 11.22 | 0.98 | -0.06 | |
| J174203.85-032340.4 | | | 16.06 | 11.62 | 11.05 | 10.74 | | 14.45 | 9.13 | 10.87 | 1.23 | 0.49 | |
| J174426.59-074925.3 | | 15.22 | 13.89 | 11.32 | 10.7 | 10.47 | 891.3 | 10.96 | 7.54 | 11.53 | 1.33 | 0.35 | |
| J174439.27+483147.1 | | 10.47 | 9.4 | 7.11 | 6.53 | 6.41 | 1906.55 | 8.83 | 5.84 | 11.1 | 0.8 | 0.1 | 0.21 |
| J174536.31-063215.3 | | 13.56 | 12.42 | 10.16 | 9.74 | 9.55 | 450.42 | 9.76 | 6.89 | 10.06 | 0.96 | -0.05 | 0.7 |
| J174735.31-033644.4 | | 15.8 | 14.53 | 12.24 | 11.84 | 11.68 | 938.49 | 11.9 | 9.05 | 10.4 | 1.53 | 1.27 | 2.86 |
| J174811.33-030510.2 | | 16.42 | 14.67 | 10.23 | 9.67 | 9.38 | 43.04 | 11.68 | 6.39 | 11.98 | 1.22 | 0.35 | 0.18 |
| J174936.01-010808.7 | | 17.31 | 16.28 | 11.7 | 11.05 | 10.77 | 113.68 | 15.63 | 10.12 | 11.61 | 1.28 | 0.15 | |
| J175022.27-094457.8 | | 15.71 | 14.23 | 10.85 | 10.11 | 9.9 | 2914.59 | 11.99 | 7.66 | 12.28 | 1.09 | 0.09 | |
| J175839.30+155208.6 | | 15.81 | 14.23 | 10.29 | 9.77 | 9.58 | 46.43 | 12.44 | 7.79 | 11.94 | 1.4 | 0.23 | 0.47 |
| J175942.12+784942.1 | | | | 11.9 | 11.18 | 10.84 | 89.07 | | 7.91 | 11.76 | 1.44 | 0.16 | |
| J180508.62-015058.5 | | 14.38 | 12.99 | 9.85 | 9.42 | 9.17 | 616.94 | 11.18 | 7.36 | 11.85 | 0.88 | -0.04 | |
| J180554.92-570431.3 | | 15.14 | 13.63 | 9.56 | 8.88 | 8.63 | 56.45 | 9.89 | 4.89 | 10.58 | 1.19 | 0.26 | 0.09 |
| J180658.07+161037.9 | | 13.47 | 12.25 | 9.28 | 8.67 | 8.44 | 76.41 | 10.44 | 6.63 | 10.19 | 1.05 | 0.19 | 0.05 |
| J180733.00+613153.6 | | 14.38 | 13.08 | 10.32 | 9.68 | 9.51 | 154.74 | 10.04 | 6.47 | 10.46 | 0.94 | 0.15 | |
| J180929.71-543054.2 | | 17.02 | 15.21 | 10.0 | 9.41 | 9.12 | 38.96 | 12.02 | 5.93 | 12.32 | 1.32 | 0.4 | -0.02 |
| J181059.88-012322.4 | | | 17.32 | 12.66 | 11.74 | 11.45 | 2577.39 | 15.37 | 9.5 | 11.29 | 1.4 | 0.26 | |
| J181725.08+482202.8 | | 12.81 | 11.37 | 7.77 | 7.17 | 6.95 | 19.84 | 9.56 | 5.14 | 11.86 | 1.03 | 0.16 | 0.28 |
| J182054.20+022101.5 | | 13.27 | 11.92 | 9.18 | 8.71 | 8.52 | 1208.5 | 10.31 | 6.91 | 11.82 | 0.8 | -0.01 | 0.1 |
| J182905.79+002232.2 | | 14.59 | 13.49 | 10.5 | 10.0 | 9.78 | 438.23 | 10.93 | 7.22 | 10.22 | 1.12 | 0.7 | 2.58 |
| J184204.85-555413.3 | | 16.83 | 15.41 | 10.68 | 10.08 | 9.85 | 51.65 | 10.89 | 5.33 | 11.51 | 1.23 | 0.29 | |
| J184206.97-555426.2 | | 15.16 | 13.6 | 9.49 | 8.83 | 8.58 | 51.45 | 9.9 | 4.88 | 11.16 | 1.13 | 0.25 | 0.38 |
| J184536.02-205910.8 | | 14.26 | 12.85 | 9.83 | 9.33 | 9.01 | 70.1 | 8.62 | 4.78 | 12.28 | 1.2 | 0.03 | 0.84 |
| J190453.69-140406.0 | | 13.85 | 12.38 | 9.24 | 8.59 | 8.37 | 37.71 | 9.99 | 5.98 | 11.8 | 1.07 | 0.16 | |
| J191019.82-160534.8 | | | 13.0 | 9.58 | 8.95 | 8.75 | 42.56 | 9.2 | 4.95 | 11.83 | 1.08 | 0.18 | 0.15 |
| J191036.02-650825.5 | | | 10.2 | 10.48 | 9.8 | 9.59 | 79.63 | 4.96 | 4.35 | 10.99 | 1.12 | 0.18 | |
| J191235.95+630904.7 | | 15.23 | 13.73 | 10.09 | 9.42 | 9.2 | 98.66 | 9.79 | 5.26 | 11.28 | 1.06 | 0.18 | 0.46 |
| J191500.80-284759.1 | | | 15.06 | 10.86 | 10.28 | 9.96 | 71.92 | 10.7 | 5.6 | 12.06 | 1.29 | 0.29 | 1.21 |
| J191534.83-083019.9 | | 12.7 | 11.73 | 9.18 | 8.58 | 8.45 | 67.06 | 7.82 | 4.54 | 10.02 | 0.83 | 0.06 | -0.4 |
| J191629.61-270707.2 | | 16.63 | 15.12 | 10.67 | 10.06 | 9.81 | 67.91 | 10.8 | 5.49 | 12.1 | 1.2 | 0.34 | |
| J192240.05-061208.0 | | 15.8 | 14.26 | 10.59 | 9.91 | 9.69 | | 10.26 | 5.69 | 11.22 | 1.14 | 0.26 | |
| J192242.80-051553.8 | | 15.6 | 14.05 | 9.92 | 9.31 | 9.07 | | 10.67 | 5.69 | 10.46 | 1.15 | 0.22 | 0.06 |
| J192250.70-631058.6 | | 14.81 | 13.29 | 9.46 | 8.82 | 8.58 | 61.81 | 9.26 | 4.55 | 11.4 | 1.16 | 0.23 | 0.3 |
| J192323.20+700738.3 | | 13.19 | 12.09 | 9.7 | 9.14 | 8.92 | 103.66 | 8.05 | 4.88 | 10.86 | 0.94 | 0.1 | 0.26 |
| J192434.97-344240.0 | | 15.88 | 14.3 | 9.67 | 9.06 | 8.79 | 51.55 | 10.56 | 5.05 | 11.36 | 1.25 | 0.3 | -0.38 |
| J192600.77-533127.6 | A | 15.65 | 13.98 | 9.6 | 8.98 | 8.68 | 580.68 | 12.34 | 7.04 | 12.37 | 1.27 | 0.27 | -0.09 |
| J192600.77-533127.6 | B | 15.65 | 13.98 | 9.6 | 8.98 | 8.68 | 47.76 | 12.34 | 7.04 | 12.37 | 1.27 | 0.27 | -0.09 |
| J192659.33-710923.8 | | 14.83 | 13.34 | 10.05 | 9.45 | 9.22 | | 9.04 | 4.92 | 10.66 | 0.96 | 0.08 | -0.08 |
| J193052.51-545325.4 | | 14.81 | 13.4 | 10.34 | 9.69 | 9.47 | 76.33 | 8.94 | 5.01 | 10.42 | 0.98 | 0.12 | |
| J193411.46-300925.3 | | | | 11.72 | 11.15 | 10.81 | 67.73 | | 6.68 | 11.53 | 1.36 | 0.53 | |
| J193711.26-040126.7 | | 12.03 | 10.77 | 8.58 | 8.04 | 7.87 | 171.38 | 8.75 | 5.85 | 11.14 | 1.16 | -0.11 | -0.03 |
| J194309.89-601657.8 | | 15.96 | 14.7 | 10.41 | 9.8 | 9.54 | 56.25 | 10.86 | 5.7 | 10.93 | 1.16 | 0.26 | 0.14 |
| J194444.21-435903.0 | | | | 10.66 | 10.14 | 9.71 | 30.78 | | 6.7 | 11.11 | 1.37 | 0.31 | |
| J194539.01+704445.9 | | 13.75 | 12.57 | 10.1 | 9.46 | 9.28 | 408.13 | 8.17 | 4.88 | 10.55 | 0.94 | 0.13 | 0.08 |
| J194714.54+640237.9 | | 13.75 | 12.43 | 9.25 | 8.62 | 8.44 | 67.21 | 10.4 | 6.41 | 11.69 | 0.95 | 0.1 | -0.01 |
| J194816.54-272032.3 | | 14.63 | 13.16 | 9.67 | 9.0 | 8.8 | 64.65 | 8.87 | 4.51 | 11.91 | 1.1 | 0.19 | |
| J194834.58-760546.9 | | 13.96 | 12.5 | 9.84 | 9.17 | 9.0 | 59.46 | 8.82 | 5.32 | 11.12 | 1.03 | 0.21 | 0.72 |
| J195227.23-773529.4 | B | 15.37 | 13.75 | 9.64 | 9.03 | 8.77 | 38.92 | 10.66 | 5.68 | 11.49 | 1.17 | 0.29 | 0.2 |
| J195227.23-773529.4 | A | 15.37 | 13.75 | 9.64 | 9.03 | 8.77 | 38.92 | 10.92 | 5.94 | 11.49 | 1.17 | 0.29 | 0.2 |
| J195315.67+745948.9 | | 14.13 | 12.66 | 9.62 | 8.87 | 8.63 | 64.76 | 9.68 | 5.65 | 11.55 | 1.06 | 0.15 | 0.4 |
| J195331.72-070700.5 | | | 15.01 | 10.82 | 10.2 | 9.93 | 32.53 | 11.29 | 6.21 | 11.25 | 1.25 | 0.21 | |
| J195340.71+502458.2 | | 15.45 | 14.0 | 9.61 | 9.04 | 8.8 | 30.74 | 11.06 | 5.86 | 11.35 | 1.16 | 0.28 | 0.39 |
| J195602.95-320719.3 | | 14.79 | 13.23 | 8.96 | 8.34 | 8.11 | 51.19 | 9.34 | 4.22 | 11.41 | 1.21 | 0.26 | 0.29 |
| J200137.19-331314.5 | | 13.85 | 12.37 | 9.16 | 8.46 | 8.24 | 59.96 | 8.36 | 4.23 | 11.13 | 1.06 | 0.11 | 0.27 |
| J200311.61-243959.2 | | 15.1 | 13.64 | 10.01 | 9.41 | 9.21 | 71.67 | 9.8 | 5.37 | 11.99 | 1.1 | 0.2 | |
| J200409.19-672511.7 | | | 13.39 | 9.36 | 8.72 | 8.48 | 52.8 | 9.59 | 4.68 | 10.88 | 1.19 | 0.22 | 0.47 |
| J200423.80-270835.8 | | 14.47 | 13.51 | 10.04 | 9.34 | 9.18 | | 9.34 | 5.01 | 11.1 | 0.98 | 0.1 | |
| J200556.44-321659.7 | | | 12.38 | 8.81 | 8.16 | 7.94 | 49.56 | 8.72 | 4.28 | 11.28 | 1.03 | 0.11 | -0.09 |
| J200837.87-254526.2 | | 17.23 | 15.84 | 10.9 | 10.36 | 10.07 | 56.02 | 11.97 | 6.2 | 11.01 | 1.24 | 0.47 | |
| J200853.72-351949.3 | | | 13.05 | 9.17 | 8.48 | 8.32 | 44.57 | 9.67 | 4.94 | 12.09 | 1.27 | 0.26 | 0.35 |
| J201000.06-280141.6 | | 14.49 | 12.99 | 8.65 | 8.01 | 7.73 | 46.41 | 9.27 | 4.01 | 10.78 | 1.21 | 0.24 | 0.38 |
| J201931.84-081754.3 | | 14.05 | 12.7 | 9.7 | 9.04 | 8.89 | 64.69 | 8.79 | 4.98 | 10.95 | 0.98 | 0.13 | 0.56 |
| J202505.36+835954.2 | | 16.38 | 14.88 | 11.13 | 10.35 | 10.11 | 80.46 | 12.58 | 7.81 | 11.32 | 1.41 | 0.26 | 1.01 |
| J202716.80-254022.8 | | | 12.65 | 9.75 | 9.08 | 8.86 | 79.03 | 9.12 | 5.33 | 12.14 | 1.09 | 0.13 | 0.47 |
| J203023.10+711419.8 | | 16.03 | 14.53 | 11.67 | 10.91 | 10.74 | 149.54 | 9.99 | 6.2 | 12.03 | 1.2 | 0.42 | 1.23 |
| J203301.99-490312.6 | | | 15.26 | 10.11 | 9.52 | 9.19 | | 12.99 | 6.92 | 12.42 | 1.35 | 0.41 | 0.52 |
| J203337.63-255652.8 | | | 14.87 | 9.71 | 9.15 | 8.88 | 43.66 | 11.6 | 5.61 | 11.68 | 1.26 | 0.31 | 0.6 |
| J204406.36-153042.3 | | 16.98 | 15.28 | 11.73 | 11.13 | 10.83 | 85.59 | 12.4 | 7.95 | 12.03 | 1.46 | 0.12 | |
| J204714.59+110442.2 | | 16.61 | 14.89 | 10.66 | 10.07 | 9.84 | 51.54 | 11.4 | 6.35 | 11.92 | 1.21 | 0.3 | |
| J205131.01-154857.6 | | | 16.07 | 11.22 | 10.63 | 10.34 | 62.29 | 13.79 | 8.06 | 10.12 | 1.49 | 2.2 | 4.05 |
| J205136.27+240542.9 | | | | 10.41 | 9.89 | 9.59 | 55.89 | | 7.56 | 11.54 | 1.33 | 0.26 | 0.49 |
| J205832.99-482033.8 | | | | 12.52 | 11.91 | 11.64 | 112.66 | | 7.54 | 10.62 | 1.31 | 0.07 | |
| J210131.13-224640.9 | | 12.59 | 11.16 | 8.19 | 7.51 | 7.31 | 29.93 | 7.8 | 3.95 | 12.49 | 1.03 | 0.05 | -0.03 |
| J210338.46+075330.3 | | 13.19 | 12.17 | 9.96 | 9.43 | 9.24 | 97.79 | 7.82 | 4.89 | 9.84 | 0.95 | 0.11 | 0.51 |
| J210708.43-113506.0 | | 12.92 | 11.36 | 9.24 | 8.56 | 8.42 | 66.81 | | | | | | |
| J210722.53-705613.4 | | 15.33 | 13.8 | 9.76 | 9.12 | 8.9 | 48.31 | 10.55 | 5.65 | 10.95 | 1.13 | 0.26 | 0.62 |
| J210736.82-130458.9 | | 14.23 | 12.64 | 8.73 | 8.1 | 7.84 | 20.77 | 10.78 | 5.98 | 11.12 | 1.18 | 0.21 | |
| J210957.48+032121.1 | | | | 9.45 | 8.62 | 8.57 | 45.4 | | 6.41 | 12.19 | 1.31 | 0.1 | 0.03 |
| J211004.67-192031.2 | | 14.67 | 13.14 | 8.43 | 7.88 | 7.55 | 33.6 | 10.58 | 4.99 | 12.01 | 1.26 | 0.27 | 0.48 |
| J211005.41-191958.4 | | 13.17 | 11.75 | 8.11 | 7.45 | 7.2 | 32.36 | 9.12 | 4.57 | 11.5 | 1.12 | 0.11 | 0.15 |
| J211031.49-271058.1 | A | 16.84 | 15.15 | 10.3 | 9.71 | 9.41 | 40.39 | 11.47 | 5.73 | 11.89 | 1.48 | 0.26 | 0.39 |
| J211031.49-271058.1 | B | 16.84 | 15.15 | 10.3 | 9.71 | 9.41 | 40.39 | 13.77 | 8.03 | 11.89 | 1.48 | 0.26 | 0.39 |
| J211635.34-600513.4 | | 16.03 | 14.49 | 10.19 | 9.56 | 9.32 | 46.32 | 11.04 | 5.87 | 10.48 | 1.21 | 0.33 | 0.76 |
| J212007.84-164548.2 | | 16.03 | 14.56 | 10.15 | 9.57 | 9.3 | 48.26 | 11.07 | 5.81 | 11.85 | 1.2 | 0.29 | 0.47 |
| J212128.89-665507.1 | | 11.96 | 10.6 | 7.88 | 7.26 | 7.01 | 31.54 | 8.18 | 4.59 | 12.08 | 0.95 | 0.09 | 0.2 |
| J212230.56-333855.2 | | | 15.25 | 10.46 | 9.87 | 9.57 | | 11.63 | 5.95 | 11.18 | 1.29 | 0.32 | |
| J212750.60-684103.9 | | 16.9 | 15.16 | 10.42 | 9.83 | 9.58 | 49.77 | 11.52 | 5.94 | 11.5 | 1.24 | 0.37 | |
| J213507.39+260719.4 | | 13.51 | 12.14 | 9.41 | 8.73 | 8.56 | 51.6 | 8.65 | 5.07 | 12.02 | 1.0 | 0.04 | 0.31 |
| J213520.34-142917.9 | | | 15.34 | 11.72 | 11.27 | 10.66 | 70.04 | 11.16 | 6.48 | 11.85 | 1.58 | 0.24 | |
| J213644.54+670007.1 | | | | | | | 1220.55 | | | | | | |
| J213708.89-603606.4 | | 15.2 | 13.7 | 9.64 | 9.01 | 8.76 | 44.95 | 10.21 | 5.27 | 10.82 | 1.19 | 0.27 | 0.44 |
| J213740.24+013713.2 | | 14.91 | 13.14 | 8.8 | 8.14 | 7.88 | 35.91 | 10.24 | 4.98 | 10.97 | 1.27 | 0.29 | 0.32 |
| J213835.44-505111.0 | | 17.14 | 15.47 | 10.73 | 10.13 | 9.82 | 45.27 | 13.22 | 7.57 | 12.19 | 1.34 | 0.34 | 0.49 |
| J213847.58+050451.4 | | 16.84 | 15.28 | 10.72 | 10.19 | 9.87 | 85.81 | 11.79 | 6.38 | 11.78 | 1.21 | 0.33 | |
| J214101.48+723026.7 | | 15.36 | 13.9 | 10.67 | 10.01 | 9.82 | 101.07 | 9.39 | 5.31 | 11.29 | 1.08 | 0.15 | 0.32 |
| J214126.66+204310.5 | | 14.95 | 13.43 | 9.43 | 8.82 | 8.61 | 37.83 | 9.1 | 4.28 | 11.68 | 1.12 | 0.23 | 0.27 |
| J214414.73+321822.3 | | 15.11 | 13.68 | 10.93 | 10.27 | 10.04 | 132.84 | 11.33 | 7.69 | 12.26 | 1.38 | -0.02 | 0.63 |
| J214905.04-641304.8 | | 16.83 | 15.22 | 10.35 | 9.8 | 9.47 | 43.89 | 11.91 | 6.16 | 11.75 | 1.3 | 0.36 | 0.73 |
| J215053.68-055318.9 | | 14.04 | 12.65 | 9.38 | 8.68 | 8.51 | 55.0 | 9.36 | 5.22 | 11.09 | 1.05 | 0.18 | 0.66 |
| J215128.95-023814.9 | | 14.94 | 13.5 | 9.8 | 9.22 | 8.86 | 46.34 | 10.99 | 6.35 | 11.58 | 1.21 | 0.17 | 0.47 |
| J215717.71-341834.0 | | | 14.91 | 11.29 | 10.74 | 10.46 | | 11.27 | 6.82 | 11.9 | 1.21 | 0.26 | |
| J220216.29-421034.0 | | 13.55 | 12.15 | 8.93 | 8.23 | 7.99 | 44.46 | 8.86 | 4.7 | 11.24 | 1.07 | 0.14 | 0.23 |
| J220254.57-644045.0 | | 14.13 | 12.68 | 9.06 | 8.41 | 8.16 | 43.67 | 9.37 | 4.85 | 11.09 | 1.09 | 0.18 | -0.03 |
| J220306.98-253826.6 | | 16.5 | 14.91 | 10.58 | 10.03 | 9.69 | 49.44 | 11.42 | 6.2 | 11.38 | 1.29 | 0.27 | 0.9 |
| J220730.16-691952.6 | | 16.16 | 14.68 | 10.65 | 10.03 | 9.81 | 64.19 | 11.29 | 6.42 | 11.92 | 1.22 | 0.26 | |
| J220850.39+114412.7 | | 16.18 | 14.54 | 9.9 | 9.34 | 9.04 | 37.39 | 11.7 | 6.2 | 11.54 | 1.24 | 0.37 | 0.56 |
| J221217.17-681921.1 | | 15.99 | 14.63 | 10.67 | 10.0 | 9.77 | 54.93 | 9.67 | 4.81 | 12.37 | 1.18 | 0.22 | 0.63 |
| J221559.00-014733.0 | | 15.19 | 14.06 | 11.11 | 10.41 | 10.17 | 82.36 | 10.24 | 6.35 | 11.91 | 1.21 | 0.13 | |
| J221833.85-170253.2 | | 14.28 | 12.83 | 10.01 | 9.37 | 9.18 | 62.51 | 8.64 | 4.99 | 12.12 | 0.98 | 0.06 | |
| J221842.70+332113.5 | | 14.47 | 12.95 | 9.33 | 8.69 | 8.44 | 53.99 | 10.07 | 5.56 | 11.82 | 1.12 | 0.22 | 0.42 |
| J222024.21-072734.5 | | 14.82 | 13.35 | 9.81 | 9.15 | 8.94 | 59.15 | 9.77 | 5.36 | 11.27 | 1.09 | 0.15 | 0.34 |
| J224111.08-684141.8 | | | 16.69 | 12.61 | 11.96 | 11.61 | 118.41 | 12.45 | 7.37 | 10.56 | 1.44 | 0.29 | 2.08 |
| J224221.02-410357.2 | | | | 11.43 | 10.85 | 10.57 | 58.85 | | 7.3 | 11.63 | 1.25 | 0.29 | 1.14 |
| J224448.45-665003.9 | | | | 11.03 | 10.41 | 10.14 | 43.81 | | 6.83 | 11.95 | 1.34 | 0.41 | 0.89 |
| J224500.20-331527.2 | | | 13.37 | 8.68 | 8.06 | 7.79 | 20.83 | 11.77 | 6.19 | 12.05 | 1.25 | 0.34 | 0.44 |
| J224634.82-735351.0 | | 14.9 | 13.42 | 9.66 | 9.05 | 8.81 | 50.09 | 9.88 | 5.27 | 10.83 | 1.12 | 0.26 | -0.07 |
| J225914.87+373639.3 | | 16.94 | 15.36 | 10.38 | 9.89 | 9.54 | 41.87 | 13.32 | 7.5 | 11.42 | 1.32 | 0.33 | 0.75 |
| J225934.89-070447.1 | | | 14.87 | 11.01 | 10.38 | 10.13 | 63.8 | 12.41 | 7.67 | 11.84 | 1.37 | 0.48 | |
| J230209.10-121522.0 | | 16.53 | 14.87 | 10.48 | 9.91 | 9.65 | 46.15 | 11.68 | 6.46 | 11.11 | 1.18 | 0.3 | |
| J230327.73-211146.2 | | 17.57 | 15.86 | 11.91 | 11.2 | 11.03 | 88.16 | 11.58 | 6.75 | 10.01 | 1.24 | 0.31 | |
| J230740.98+080359.7 | | 13.92 | 12.67 | 9.56 | 8.9 | 8.71 | 48.19 | 9.0 | 5.04 | 12.08 | 0.99 | 0.11 | 0.25 |
| J231021.75+685943.6 | | 14.74 | 13.22 | 9.92 | 9.39 | 9.14 | 57.28 | 9.29 | 5.21 | 12.29 | 1.05 | 0.11 | |
| J231211.37+150329.7 | | 15.22 | 13.83 | 10.18 | 9.54 | 9.31 | 69.55 | 9.81 | 5.29 | 11.38 | 1.07 | 0.17 | |
| J231246.53-504924.8 | | 15.09 | 13.5 | 9.12 | 8.53 | 8.3 | 38.22 | 11.66 | 6.46 | 11.51 | 1.22 | 0.32 | 0.22 |
| J231457.86-633434.0 | B | | | 10.18 | 9.52 | 9.28 | 79.04 | | 6.69 | 10.41 | 1.1 | 0.22 | -0.13 |
| J231457.86-633434.0 | A | | | 10.18 | 9.52 | 9.28 | 79.04 | | 5.36 | 10.41 | 1.1 | 0.22 | -0.13 |
| J231543.66-140039.6 | | 12.93 | 11.59 | 9.09 | 8.41 | 8.28 | 40.69 | 8.19 | 4.88 | 12.41 | 0.94 | 0.07 | -0.18 |
| J231933.16-393924.3 | | 13.44 | 12.1 | 9.07 | 8.41 | 8.25 | 40.26 | 9.64 | 5.79 | | | | |
| J232008.15-634334.9 | | | | 11.83 | 11.21 | 10.92 | 46.94 | | 8.53 | 11.91 | 1.29 | 0.3 | |
| J232151.23+005037.3 | | 13.33 | 11.98 | 9.33 | 8.55 | 8.51 | 54.04 | 9.48 | 6.01 | 12.48 | 1.18 | 0.0 | -0.01 |
| J232656.43+485720.9 | | 13.54 | 12.55 | 9.75 | 9.1 | 8.94 | 116.78 | 8.6 | 4.99 | 9.84 | 0.98 | 0.12 | 0.13 |
| J232857.75-680234.5 | | 14.51 | 13.05 | 9.26 | 8.64 | 8.38 | 45.84 | 9.58 | 4.91 | 10.82 | 1.11 | 0.23 | 0.21 |
| J232904.42+032910.8 | | | 15.75 | 11.11 | 10.55 | 10.19 | 54.04 | 11.87 | 6.31 | 12.21 | 1.32 | 0.41 | |
| J232917.64-675000.6 | | 17.02 | 15.47 | 10.79 | 10.19 | 9.89 | 45.82 | 12.06 | 6.48 | 11.46 | 1.24 | 0.35 | |
| J232959.47+022834.0 | | | | 11.36 | 10.81 | 10.53 | 53.1 | | 6.39 | 11.88 | 1.22 | 0.33 | |
| J233647.87+001740.1 | | 14.02 | 12.57 | 9.54 | 8.83 | 8.65 | 51.3 | 9.14 | 5.22 | 12.3 | 1.05 | 0.1 | |
| J234243.45-622457.1 | | | | 11.3 | 10.76 | 10.42 | 43.59 | | 7.13 | 11.16 | 1.31 | 0.45 | 1.33 |
| J234326.88-344658.5 | | 13.29 | 11.81 | 8.47 | 7.86 | 7.63 | 38.42 | 8.77 | 4.59 | 11.66 | 1.09 | 0.17 | 0.19 |
| J234333.91-192802.8 | | 13.06 | 11.8 | 9.22 | 8.55 | 8.36 | 52.79 | 8.34 | 4.9 | 10.4 | 0.95 | 0.14 | 0.27 |
| J234347.83-125252.1 | | 15.4 | 13.87 | 10.26 | 9.67 | 9.41 | 80.29 | 9.96 | 5.5 | 11.81 | 1.12 | 0.18 | |
| J234857.35+100929.3 | | 14.37 | 13.06 | 9.94 | 9.29 | 9.12 | | 8.81 | 4.87 | 12.26 | 0.94 | 0.07 | |
| J234924.87+185926.7 | | 15.71 | 13.96 | 10.2 | 9.57 | 9.34 | 44.39 | 10.14 | 5.52 | 11.36 | 1.12 | 0.22 | 0.42 |
| J234926.23+185912.4 | | 13.44 | 12.04 | 9.17 | 8.53 | 8.32 | 44.34 | 8.84 | 5.12 | 11.4 | 1.02 | 0.07 | 0.03 |
| J235250.70-160109.7 | | 15.91 | 14.31 | 10.1 | 9.55 | 9.23 | 52.58 | 10.88 | 5.8 | 10.23 | 1.15 | 0.3 | 0.65 |
| 2M 12182363-3515098 | | | | 9.41 | 8.78 | 8.69 | | | 4.72 | | | | |

TABLE 11 Literature and Gaia Radial Velocity Measurements

| Name | RV | RV error | Source (Vizier Catalog Code) |
|---|---|---|---|
| J001527.62-641455.2 | 6.7 | 0.3 | J/AJ/147/146/stars |
| | 6.08 | 1.36 | J/A+A/649/A6/table1c |
| | 5.97 | 24.0 | Gaia DR3 |
| J001536.79-294601.2 | 0.61 | 1.31 | J/AJ/157/234/table2 |
| | 0.05 | 4.7 | III/283/ravedr6 |
| | 0.51 | 1.28 | Gaia DR3 |
| J001552.28-280749.4 | -0.44 | 0.65 | J/A+A/649/A6/table1c |
| | -0.57 | 0.65 | Gaia DR3 |
| J001555.65-613752.2 | 6.97 | 2.04 | III/279/rave$_on$ |
| | 24.47 | 2.02 | Gaia DR3 |
| J001709.96+185711.8 | 33.13 | 0.73 | J/A+A/649/A6/table1c |
| | 33.34 | 0.37 | Gaia DR3 |
| J001723.69-664512.4 | 10.7 | 0.2 | J/AJ/154/69/table4 |
| | 9.42 | 0.53 | Gaia DR3 |
| J002101.27-134230.7 | 1.14 | 0.81 | J/A+A/649/A6/table1c |
| | -0.17 | 0.43 | Gaia DR3 |
| | 2.73 | 0.36 | Gaia DR3 |
| J003057.97-655006.4 | 13.6 | 0.9 | J/A+A/649/A6/table1c |
| | 11.72 | 2.85 | Gaia DR3 |
| J003903.51+133016.0 | -8.77 | 2.69 | Gaia DR3 |
| J004210.98-425254.8 | 8.7 | 0.1 | J/AJ/147/146/stars |
| | 5.86 | 1.52 | Gaia DR3 |
| J004524.84-775207.5 | 6.14 | 1.04 | III/283/xgaia2 |
| | 4.03 | 2.64 | III/279/rave dr5 |
| | 5.43 | 0.72 | Gaia DR3 |
| J004528.25-513734.4 | 8.2 | 1.56 | J/A+A/649/A6/table1c |
| | 8.52 | 1.3 | Gaia DR3 |
| J004826.70-184720.7 | 7.21 | 0.68 | J/A+A/649/A6/table1c |
| | 11.81 | 3.42 | Gaia DR3 |
| J005633.96-225545.4 | 16.89 | 5.13 | Gaia DR3 |
| J010711.99-193536.4 | 9.3 | 0.5 | J/AJ/147/146/stars |
| J011440.20+205712.9 | 76.32 | 11.61 | Gaia DR3 |
| | 22.43 | 10.03 | Gaia DR3 |
| J011846.91+125831.4 | -5.59 | 3.8 | I/345/gaia2 |
| | -8.31 | 2.24 | Gaia DR3 |
| J012118.22-543425.1 | 24.0 | 0.3 | J/AJ/147/146/stars |
| | 26.42 | 1.49 | J/A+A/649/A6/table1c |
| | 24.54 | 0.13 | Gaia DR3 |
| J012245.24-631845.0 | 7.8 | 1.4 | J/A+A/649/A6/table1c |
| | 3.5 | 3.31 | Gaia DR3 |
| J012332.89-411311.4 | 6.91 | 1.7 | Gaia DR3 |
| J012532.11-664602.6 | 21.14 | 4.59 | Gaia DR3 |
| J013110.69-760947.7 | 28.92 | 5.98 | III/279/rave dr5 |
| | 29.63 | 10.18 | Gaia DR3 |
| J014156.94-123821.6 | 13.36 | 1.12 | J/A+A/649/A6/table1c |
| | 13.29 | 0.57 | Gaia DR3 |
| J014431.99-460432.1 | | | |
| J015057.01-584403.4 | 11.1 | 0.5 | J/AJ/147/146/stars |
| | 7.0 | 3.67 | Gaia DR3 |
| J015350.81-145950.6 | 10.5 | 0.4 | J/AJ/154/69/table4 |
| | 1.98 | 4.39 | III/279/rave dr5 |
| | 10.83 | 1.35 | Gaia DR3 |
| J015455.24-295746.0 | 35.18 | 2.64 | Gaia DR3 |
| J020012.84-084052.4 | 4.5 | 0.4 | J/A+A/649/A6/table1c |
| | 1.43 | 1.85 | Gaia DR3 |
| J020302.74+221606.8 | -10.96 | 7.42 | Gaia DR3 |
| | -3.57 | 2.25 | Gaia DR3 |
| J020305.46-590146.6 | | | |
| J020805.55-474633.7 | 2.77 | 0.77 | J/A+A/649/A6/table1c |
| | 3.21 | 0.58 | Gaia DR3 |
| J021258.28-585118.3 | 9.1 | 0.8 | J/AJ/147/146/stars |
| | 11.07 | 2.28 | Gaia DR3 |
| J022240.88+305515.4 | 4.7 | 1.23 | Gaia DR3 |
| J022424.69-703321.2 | 11.8 | 0.3 | J/A+A/649/A6/table1c |
| | 13.67 | 2.73 | Gaia DR3 |
| J023005.14+284500.0 | -26.16 | 1.07 | J/A+A/649/A6/table1c |
| | -27.16 | 4.46 | Gaia DR3 |
| | -26.13 | 1.11 | Gaia DR3 |
| J023139.36+445638.1 | -4.7 | 3.8 | J/A+A/649/A6/table1c |
| | -2.79 | 6.98 | Gaia DR3 |
| J024552.65+052923.8 | 6.77 | 1.55 | J/A+A/649/A6/table1c |
| | 6.49 | 1.33 | Gaia DR3 |
| J024746.49-580427.4 | 13.1 | 0.5 | J/AJ/147/146/stars |
| | 12.13 | 1.98 | Gaia DR3 |
| J024852.67-340424.9 | 14.6 | 0.3 | J/AJ/154/129/catalog |
| | 19.44 | 1.8 | Gaia DR3 |
| J025154.17+222728.9 | 9.14 | 0.08 | J/A+A/656/A162/table1 |
| | 10.57 | 1.0 | Gaia DR3 |
| J030002.98+550652.4 | -6.88 | 1.73 | J/A+A/649/A6/table1c |
| | -6.15 | 1.83 | Gaia DR3 |
| J030251.62-191150.0 | | | |
| J030824.14+234554.2 | 16.87 | 1.97 | J/A+A/649/A6/table1c |
| | 16.51 | 1.45 | Gaia DR3 |
| J031650.45-350937.9 | 12.57 | 5.29 | Gaia DR3 |
| J032047.66-504133.0 | 17.4 | 0.3 | J/AJ/147/146/stars |
| | 18.51 | 0.31 | Gaia DR3 |
| J033431.66-350103.3 | | | |
| J033640.91+032918.3 | 32.96 | 6.65 | J/A+A/614/A76/tablea1 |
| | 19.3 | 4.14 | Gaia DR3 |
| J034115.60-225307.8 | 17.64 | 1.63 | J/A+A/649/A6/table1c |
| | 17.78 | 1.21 | Gaia DR3 |
| J034116.16-225244.0 | 17.08 | 4.31 | J/A+A/649/A6/table1c |
| | 16.84 | 3.07 | Gaia DR3 |
| J034236.95+221230.2 | | | |
| J034444.80+404150.4 | 7.99 | 6.78 | Gaia DR3 |
| J035100.83+141339.2 | 10.8 | 2.53 | Gaia DR3 |
| J035134.51+072224.5 | 39.81 | 4.29 | Gaia DR3 |
| | 34.16 | 9.96 | Gaia DR3 |
| J035223.52-282619.6 | 23.23 | 6.12 | J/A+A/649/A6/table1c |
| | 16.72 | 3.25 | Gaia DR3 |
| | 16.7 | 4.13 | J/A+A/649/A6/table1c |
| J035345.92-425018.0 | 16.27 | 2.93 | Gaia DR3 |
| J035716.56-271245.5 | -10.78 | 0.62 | I/345/gaia2 |
| | -11.32 | 0.48 | Gaia DR3 |
| J035733.95+244510.2 | 12.85 | 1.29 | I/345/gaia2 |
| | 11.85 | 0.66 | Gaia DR3 |
| J035829.67-432517.2 | 10.99 | 4.6 | J/A+A/649/A6/table1c |
| | 11.72 | 6.84 | Gaia DR3 |
| J040539.68-401410.5 | 16.22 | 3.34 | Gaia DR3 |
| J040649.38-450936.3 | 20.24 | 2.47 | J/A+A/649/A6/table1c |
| | 22.62 | 3.57 | Gaia DR3 |
| J040711.50-291834.3 | 11.42 | 3.8 | III/283/ravedr6 |
| | 22.82 | 7.88 | Gaia DR3 |
| J040743.83-682511.0 | | | |
| J040809.80-611904.3 | 10.31 | 0.54 | J/A+A/649/A6/table1c |
| | 10.71 | 0.43 | Gaia DR3 |
| J040827.01-784446.7 | 16.8 | 0.5 | J/AJ/147/146/stars |
| | 12.69 | 2.24 | Gaia DR3 |
| J041255.78-141859.2 | 19.11 | 2.61 | III/283/ravedr6 |
| | 13.87 | 0.97 | Gaia DR3 |
| J041336.14-441332.4 | 16.4 | 1.4 | J/AJ/147/146/stars |
| J041525.58-212214.5 | 23.27 | 1.91 | Gaia DR3 |
| J041749.66+001145.4 | 19.19 | 3.09 | J/A+A/649/A6/table1r |
| | 17.02 | 2.71 | Gaia DR3 |
| J041807.76+030826.0 | | | |
| J042139.19-723355.7 | 15.0 | 0.3 | J/AJ/154/129/catalog |
| | 12.85 | 1.01 | Gaia DR3 |
| J042500.91-634309.8 | 35.05 | 8.95 | Gaia DR3 |
| J042736.03-231658.8 | 19.31 | 23.42 | Gaia DR3 |
| J042739.33+171844.2 | 17.88 | 1.33 | J/A+A/649/A6/table1r |
| | 19.57 | 1.25 | Gaia DR3 |
| J043213.46-285754.8 | | | |
| J043257.29+740659.3 | | | |
| J043657.44-161306.7 | 15.6 | 0.5 | J/A+A/649/A6/table1c |
| | 13.67 | 3.73 | Gaia DR3 |
| J043657.44-161306.7 | 15.6 | 0.5 | J/A+A/649/A6/table1c |
| | 13.67 | 3.73 | Gaia DR3 |
| J043726.87+185126.2 | 11.26 | 12.88 | J/A+A/649/A6/table1c |
| | 14.31 | 1.29 | Gaia DR3 |
| J043726.87+185126.2 | 15.41 | 10.4 | Gaia DR3 |
| J043939.24-050150.9 | -1.58 | 10.55 | Gaia DR3 |
| J044120.81-194735.6 | 23.49 | 0.89 | Gaia DR3 |
| J044120.81-194735.6 | 23.49 | 0.89 | Gaia DR3 |
| J044154.44+091953.1 | | | |
| J044336.19-003401.8 | 21.45 | 5.54 | Gaia DR3 |
| J044349.19+742501.6 | -27.62 | 3.61 | Gaia DR3 |
| J044349.19+742501.6 | -12.83 | 15.56 | Gaia DR3 |
| J044356.87+372302.7 | 6.4 | 0.2 | J/A+A/649/A6/table1c |
| | 3.02 | 3.34 | Gaia DR3 |
| J044455.71+193605.3 | | | |
| J044530.77-285034.8 | 31.04 | 2.31 | Gaia DR3 |
| J044700.46-513440.4 | 19.9 | 0.3 | J/AJ/147/146/stars |
| | 18.27 | 6.94 | Gaia DR3 |
| J044721.05+280852.5 | 21.13 | 5.24 | Gaia DR3 |
| J044800.86+143957.7 | 12.44 | 3.62 | Gaia DR3 |
| J044802.59+143951.1 | 20.74 | 2.85 | Gaia DR3 |
| J045114.41-601830.5 | 20.52 | 1.38 | J/A+A/649/A6/table1r |
| | 20.19 | 1.27 | Gaia DR3 |
| J045420.20-400009.9 | -7.43 | 3.9 | Gaia DR3 |
| J050333.31-382135.6 | | | |
| J050610.44-582828.5 | 19.63 | 0.72 | J/A+A/649/A6/table1c |
| | 18.27 | 1.18 | Gaia DR3 |
| J050827.31-210144.3 | 22.21 | 0.23 | J/A+A/656/A162/table1 |
| J051026.38-325307.4 | 20.68 | 6.27 | Gaia DR3 |
| J051255.82-212438.7 | 7.71 | 4.44 | Gaia DR3 |
| J051403.20-251703.8 | 23.45 | 33.36 | Gaia DR3 |
| J051650.66+022713.0 | 13.11 | 4.58 | Gaia DR3 |
| J051803.00-375721.2 | | | Gaia DR3 |
| J052419.14-160115.5 | 17.5 | 0.6 | J/AJ/154/69/table4 |
| | 27.47 | 3.21 | Gaia DR3 |
| | 20.79 | 6.48 | III/279/rave$_on$ |
| J052535.85-250230.2 | -2.37 | 5.57 | Gaia DR3 |
| | -4.78 | 3.41 | Gaia DR3 |
| J052944.69-323914.1 | 23.39 | 18.39 | Gaia DR3 |
| J053100.27+231218.3 | | | |
| J053311.32-291419.9 | 25.48 | 1.94 | Gaia DR3 |
| J053328.01-425720.1 | -4.85 | 0.34 | J/AJ/157/234/table2 |
| | -3.19 | 0.49 | Gaia DR3 |
| J053747.56-424030.8 | | | |
| J053925.08-424521.0 | 21.7 | 0.2 | J/AJ/147/146/stars |
| | 21.62 | 1.08 | |
| J054223.86-275803.3 | | | |
| J054433.76-200515.5 | 22.24 | 2.56 | J/A+A/649/A6/table1r |
| | 20.91 | 4.0 | Gaia DR3 |
| J054448.20-265047.4 | 43.28 | 14.68 | J/A+A/649/A6/table1r |
| | 35.86 | 6.74 | Gaia DR3 |
| J054709.88-525626.1 | 34.79 | 5.32 | J/A+A/649/A6/table1c |
| | 35.1 | 3.99 | Gaia DR3 |
| J054719.52-335611.2 | 28.29 | 3.25 | Gaia DR3 |
| J055008.59+051153.2 | 18.08 | 3.74 | J/A+A/657/A7/tablea3 |
| | 7.15 | 3.29 | Gaia DR3 |
| J055041.58+430451.8 | -2.46 | 9.18 | J/A+A/657/A7/tablea3 |
| | -2.42 | 5.39 | Gaia DR3 |
| J055041.58+430451.8 | -2.83 | 9.18 | J/A+A/649/A6/table1c |
| | -2.42 | 5.39 | Gaia DR3 |
| J055208.04+613436.6 | -1.17 | 0.54 | J/A+A/649/A6/table1c |
| | -0.71 | 0.3 | Gaia DR3 |
| J055941.10-231909.4 | | | |
| J060156.10-164859.9 | -6.76 | 33.63 | Gaia DR3 |
| J060224.56-163450.0 | -8.2 | 0.2 | J/A+A/649/A6/table1c |
| | -10.48 | 0.8 | Gaia DR3 |
| J060329.60-260804.7 | -2.89 | 2.17 | III/283/ravedr6 |
| | -1.85 | 1.68 | Gaia DR3 |
| J061313.30-274205.6 | 22.5 | 0.2 | J/AJ/154/129/catalog |
| | 23.05 | 1.22 | Gaia DR3 |
| J061740.43-475957.2 | | | |
| J061851.01-383154.9 | 24.33 | 7.15 | Gaia DR3 |
| | 22.36 | 9.18 | Gaia DR3 |
| J062047.17-361948.2 | 5.77 | 4.63 | III/283/ravedr6 |
| | 8.22 | 3.6 | Gaia DR3 |
| J062130.52-410559.1 | 1.68 | 2.01 | Gaia DR3 |
| J062407.62+310034.4 | 38.55 | 1.05 | Gaia DR3 |
| J063001.84-192336.6 | 23.05 | 9.0 | Gaia DR3 |
| J070657.72-535345.9 | 22.57 | 0.9 | J/AJ/157/234/table2 |
| | 21.48 | 0.85 | Gaia DR3 |
| | 22.94 | 1.22 | J/A+A/649/A6/table1c |
| J071036.50+171322.6 | 69.79 | 18.45 | I/345/gaia2 |
| | 33.15 | 9.91 | Gaia DR3 |
| J072821.16+334511.6 | 30.49 | -999.0 | J/MNRAS/475/1960/tablea6 |
| | 17.79 | 5.34 | Gaia DR3 |
| J072911.26-821214.3 | 25.26 | 1.04 | J/A+A/649/A6/table1c |
| | 25.66 | 0.75 | Gaia DR3 |
| J073138.47+455716.5 | 9.72 | 3.79 | Gaia DR3 |
| J075233.22-643630.5 | 17.05 | 2.6 | J/A+A/657/A7/tablea3 |
| | 18.88 | 4.27 | Gaia DR3 |
| J075808.25-043647.5 | 28.71 | 1.53 | J/A+A/649/A6/table1c |
| | 16.59 | 1.08 | Gaia DR3 |
| J075830.92+153013.4 | 50.96 | 5.38 | J/A+A/649/A6/table1c |
| | 23.34 | 3.2 | Gaia DR3 |
| J080352.54+074346.7 | 34.1 | 8.52 | J/A+A/649/A6/table1c |
| | 37.43 | 19.77 | Gaia DR3 |
| J080636.05-744424.6 | 17.0 | 1.04 | J/A+A/649/A6/table1c |
| | 16.59 | 1.27 | Gaia DR3 |
| J081738.97-824328.8 | 15.6 | 1.5 | J/AJ/154/69/table4 |
| J082105.04-090853.8 | 33.26 | 3.72 | J/A+A/649/A6/table1c |
| | 32.61 | 2.77 | Gaia DR3 |
| J082558.91+034019.5 | 34.59 | 2.18 | Gaia DR3 |
| | 35.61 | 1.77 | Gaia DR3 |
| J083528.87+181219.9 | -13.56 | 2.79 | J/A+A/649/A6/table1c |
| | -16.45 | 3.9 | Gaia DR3 |
| J090227.87+584813.4 | -4.48 | 1.84 | J/A+A/649/A6/table1c |
| | -6.55 | 5.2 | Gaia DR3 |
| J092216.12+043423.3 | -32.28 | 1.12 | J/A+A/649/A6/table1c |
| | -32.72 | 0.52 | Gaia DR3 |
| J093212.63+335827.3 | -15.03 | 4.25 | Gaia DR3 |
| | -19.02 | 4.11 | Gaia DR3 |
| J094317.05-245458.3 | -10.14 | 1.53 | J/A+A/649/A6/table1r |
| J094508.15+714450.1 | 9.14 | 1.71 | Gaia DR3 |
| J100146.28+681204.1 | -1.27 | 2.24 | Gaia DR3 |
| J100230.94-281428.2 | | | |
| J101543.44+660442.3 | -18.51 | 1.41 | Gaia DR3 |
| J101905.68-304920.3 | 12.87 | 6.17 | J/A+A/649/A6/table1r |
| | 22.91 | 5.88 | Gaia DR3 |
| J101917.57-443736.0 | 15.84 | 1.06 | Gaia DR3 |
| J102602.07-410553.8 | | | |
| J102636.95+273838.4 | 79.01 | 7.38 | Gaia DR3 |
| | 32.15 | 11.39 | Gaia DR3 |
| J103016.11-354626.3 | 13.92 | 6.05 | J/A+A/649/A6/table1c |
| | 11.68 | 4.3 | Gaia DR3 |
| J103137.59-374915.9 | | | |
| J103557.17+285330.8 | 11.65 | 0.58 | Gaia DR3 |
| J103952.70-353402.5 | 14.75 | 0.73 | J/A+A/649/A6/table1c |
| | 14.16 | 3.26 | Gaia DR3 |
| J103952.70-353402.5 | 21.3 | 5.49 | Gaia DR3 |
| J104008.36-384352.1 | 17.22 | 2.06 | J/A+A/649/A6/table1r |
| | 15.06 | 4.47 | Gaia DR3 |
| J104044.98-255909.2 | 20.56 | 1.75 | J/A+A/657/A7/tablea3 |
| | 19.33 | 1.09 | Gaia DR3 |
| J105515.87-033538.2 | 39.4 | 1.08 | Gaia DR3 |
| J105518.12-475933.2 | 16.37 | 3.71 | III/283/ravedr6 |
| | 16.34 | 3.78 | III/279/rave$_{o}n$ |
| | 12.97 | 3.27 | Gaia DR3 |
| J105524.25-472611.7 | -12.51 | 5.72 | J/A+A/649/A6/table1r |
| | -20.13 | 8.29 | Gaia DR3 |
| J105711.36+054454.2 | 10.74 | 1.82 | J/AJ/157/234/table2 |
| | 19.97 | 10.77 | Gaia DR3 |
| J105850.47-234620.8 | 11.02 | 37.24 | Gaia DR3 |
| J110119.22+525222.9 | 7.2 | 0.3 | J/A+A/649/A6/table1c |
| | 7.22 | 0.27 | Gaia DR3 |
| | 7.64 | 2.07 | Gaia DR3 |
| J111052.06-725513.0 | 20.31 | 4.57 | Gaia DR3 |
| J111103.54-313459.0 | 29.64 | 5.83 | Gaia DR3 |
| J111128.13-265502.9 | 7.89 | 2.87 | Gaia DR3 |
| J111229.74-461610.1 | 11.02 | 1.02 | J/A+A/649/A6/table1c |
| | 10.51 | 0.58 | Gaia DR3 |
| J111309.15+300338.4 | 7.63 | 3.73 | J/A+A/649/A6/table1c |
| | 0.46 | 2.89 | Gaia DR3 |
| J111707.56-390951.3 | 12.61 | 3.15 | J/A+A/649/A6/table1r |
| | 5.85 | 5.14 | Gaia DR3 |
| J112047.03-273805.8 | 19.58 | 4.31 | Gaia DR3 |
| J112105.43-384516.6 | 10.9 | 1.0 | J/AJ/156/137/sample |
| | 9.47 | 2.17 | J/ApJ/840/87/targets |
| J112547.46-441027.4 | | | |
| J112651.28-382455.5 | 12.89 | 1.46 | Gaia DR3 |
| J112955.84+520213.2 | -5.8 | 0.69 | J/A+A/649/A6/table1c |
| | -5.05 | 0.45 | Gaia DR3 |
| J113105.57+542913.5 | -1.08 | 32.02 | Gaia DR3 |
| J113114.81-482628.0 | | | |
| J113120.31+132140.0 | -3.0 | -999.0 | J/other/RAA/15.1154/mdwarfs |
| | -10.61 | 3.51 | Gaia DR3 |
| J114623.01-523851.8 | | | |
| J114728.37+664402.7 | -9.54 | -999.0 | J/A+A/612/A49/tableb1 |
| J115156.73+073125.7 | 54.86 | -999.0 | J/MNRAS/475/1960/tablea6 |
| J115438.73-503826.4 | | | |
| J115927.82-451019.3 | 14.53 | 3.55 | Gaia DR3 |
| J115949.51-424426.0 | 6.54 | 0.71 | J/AJ/157/234/table2 |
| | 3.43 | 1.45 | Gaia DR3 |
| | 2.12 | 0.86 | J/A+A/649/A6/table1c |
| | 2.96 | 1.13 | III/279/rave dr5 |
| J115957.68-262234.1 | 11.78 | 5.13 | Gaia DR3 |
| J120001.54-173131.1 | 1.65 | -999.0 | J/MNRAS/494/2429/table1 |
| | 12.29 | 4.15 | Gaia DR3 |
| | 20.12 | 2.37 | J/ApJ/840/87/targets |
| | 1.63 | 4.25 | J/A+A/657/A7/tablea3 |
| | 3.97 | 6.93 | III/279/rave dr5 |
| J120237.94-332840.4 | 6.03 | 1.85 | Gaia DR3 |
| J121153.04+124912.9 | 6.8 | -999.0 | J/other/RAA/15.1154/mdwarfs |
| | -1.44 | 1.05 | Gaia DR3 |
| | -0.25 | 0.74 | J/AJ/157/234/table2 |
| | -2.06 | 1.5 | Gaia DR3 |
| J121341.59+323127.7 | -16.75 | 6.93 | Gaia DR3 |
| J121429.15-425814.8 | 0.45 | 7.74 | J/A+A/649/A6/table1c |
| | -9.69 | 5.25 | Gaia DR3 |
| J121511.25-025457.1 | 0.0 | 1.59 | Gaia DR3 |
| | 2.71 | 3.73 | Gaia DR3 |
| J121558.37-753715.7 | | | |
| J122643.99-122918.3 | 2.2 | 0.78 | J/AJ/157/234/table2 |
| | -6.34 | 2.71 | J/A+A/649/A6/table1c |
| | -0.23 | 2.86 | Gaia DR3 |
| J122725.27-454006.6 | 13.47 | 3.62 | III/279/rave dr5 |
| | 10.91 | 3.0 | Gaia DR3 |
| J122813.57-431638.9 | | | |
| J123005.17-440236.1 | 4.58 | 4.85 | Gaia DR3 |
| J123234.07-414257.5 | 22.78 | 3.57 | Gaia DR3 |
| J123425.84-174544.4 | 28.14 | 3.42 | Gaia DR3 |
| J123704.99-441919.5 | | | |
| J124054.09-451625.4 | 12.53 | 3.82 | Gaia DR3 |
| J124612.32-384013.5 | 4.2 | 4.81 | Gaia DR3 |
| J124955.67-460737.3 | 8.29 | 2.21 | J/ApJ/868/32/members |
| | 6.62 | 3.85 | Gaia DR3 |
| J125049.12-423123.6 | 14.67 | 2.95 | Gaia DR3 |
| J125326.99-350415.3 | 3.59 | 7.45 | Gaia DR3 |
| J125902.99-314517.9 | 3.79 | 3.82 | Gaia DR3 |
| J130501.18-331348.7 | -2.42 | 1.72 | J/A+A/649/A6/table1c |
| | -2.74 | 1.26 | Gaia DR3 |
| J130522.37-405701.2 | 3.84 | 4.95 | J/A+A/649/A6/table1r |
| | 3.91 | 6.05 | Gaia DR3 |
| J130530.31-405626.0 | 9.74 | 4.13 | III/279/rave dr5 |
| | 7.77 | 4.06 | Gaia DR3 |
| J130618.16-342857.0 | 0.26 | 8.09 | Gaia DR3 |
| J130650.27-460956.1 | 12.42 | 2.41 | J/MNRAS/491/215/table2 |
| | 2.16 | 6.17 | I/345/gaia2 |
| | 5.07 | 4.4 | Gaia DR3 |
| J130731.03-173259.9 | -4.07 | 4.8 | Gaia DR3 |
| J131129.00-425241.9 | 4.63 | 2.81 | J/A+A/649/A6/table1c |
| | 5.99 | 3.16 | Gaia DR3 |
| J132112.77-285405.1 | 13.12 | 22.73 | Gaia DR3 |
| J133238.94+305905.8 | 9.88 | 3.02 | J/A+A/649/A6/table1c |
| | -11.8 | 31.32 | Gaia DR3 |
| J133509.40+503917.5 | -12.2 | -999.0 | J/other/RAA/15.1154/mdwarfs |
| | -20.94 | 3.4 | Gaia DR3 |
| | -12.82 | 2.76 | J/A+A/649/A6/table1c |
| J133901.87-214128.0 | 2.8 | -999.0 | J/MNRAS/494/2429/table1 |
| | 2.14 | 3.61 | Gaia DR3 |
| J134146.41+581519.2 | -10.7 | 0.62 | J/A+A/614/A76/tablea1 |
| | -13.11 | 3.23 | Gaia DR3 |
| | -10.55 | 0.72 | Gaia DR3 |
| J134907.28+082335.8 | 1.05 | 1.07 | J/A+A/649/A6/table1c |
| | 1.36 | 0.33 | Gaia DR3 |
| J135145.65-374200.7 | 3.17 | 2.16 | III/279/rave dr5 |
| | 5.0 | 1.57 | Gaia DR3 |
| J135511.38+665207.0 | -19.68 | 14.18 | Gaia DR3 |
| J135913.33-292634.2 | -1.26 | 1.43 | III/279/rave dr5 |
| | 2.19 | 3.25 | Gaia DR3 |
| J140337.56-501047.9 | | | |
| J141045.24+364149.8 | -1.84 | 1.18 | Gaia DR3 |
| | -1.65 | 2.66 | Gaia DR3 |
| J141332.23-145421.1 | -13.45 | 4.39 | Gaia DR3 |
| J141510.77-252012.0 | -0.37 | 1.12 | Gaia DR3 |
| J141842.36+475514.9 | -18.8 | 3.11 | J/ApJS/249/22/epochs |
| | 3.38 | 5.01 | Gaia DR3 |
| J141903.13+645146.4 | -15.64 | 13.05 | Gaia DR3 |
| J143517.80-342250.4 | 2.69 | 1.11 | J/A+A/649/A6/table1c |
| | 2.11 | 0.95 | Gaia DR3 |
| J143648.16+090856.5 | -20.53 | 2.51 | Gaia DR3 |
| J143753.36-343917.8 | -12.86 | 2.46 | Gaia DR3 |
| | -1.99 | 1.54 | Gaia DR3 |
| J145731.11-305325.0 | 22.92 | 3.77 | Gaia DR3 |
| J145949.90+244521.9 | 13.82 | 13.46 | J/A+A/649/A6/table1c |
| | 13.83 | 10.71 | Gaia DR3 |
| J150119.48-200002.1 | | | |
| J150355.37-214643.1 | | | |
| J150601.66-240915.0 | 18.92 | 7.27 | Gaia DR3 |
| J150723.91+433353.6 | -10.33 | 2.42 | Gaia DR3 |
| J150820.15-282916.6 | | | |
| J150836.69-294222.9 | | | |
| J150939.16-133212.4 | -256.16 | -999.0 | J/MNRAS/494/2429/table1 |
| | -6.91 | 3.49 | Gaia DR3 |
| | -8.83 | 5.88 | J/A+A/649/A6/table1c |
| | -256.16 | 42.15 | III/279/rave dr5 |
| J151212.18-255708.3 | -9.8 | 4.72 | J/A+A/649/A6/table1c |
| | -1.8 | 0.82 | Gaia DR3 |
| J151242.69-295148.0 | -17.91 | 14.53 | Gaia DR3 |
| J151411.31-253244.1 | | | |
| J153248.80-230812.4 | -15.68 | 9.32 | I/345/gaia2 |
| | -16.72 | 13.62 | Gaia DR3 |
| J153549.35-065727.8 | -10.34 | 3.94 | III/279/rave dr5 |
| | 16.27 | 8.85 | Gaia DR3 |
| J154220.24+593653.0 | -25.67 | 26.29 | Gaia DR3 |
| J154227.07-042717.1 | -3.09 | 4.88 | III/279/rave$_on$ |
| | -13.33 | 5.21 | Gaia DR3 |
| J154349.42-364838.7 | -0.27 | 3.51 | Gaia DR3 |
| J154435.17+042307.5 | -22.14 | -999.0 | J/MNRAS/494/2429/table1 |
| | -22.14 | 5.48 | J/A+A/649/A6/table1c |
| | -21.0 | 34.93 | Gaia DR3 |
| J155046.47+305406.9 | -19.16 | 3.63 | J/A+A/657/A7/tablea3 |
| | -22.87 | 2.21 | Gaia DR3 |
| J155759.01-025905.8 | -26.23 | 1.12 | J/A+A/649/A6/table1c |
| | -25.68 | 0.99 | Gaia DR3 |
| | -25.91 | 1.12 | J/A+A/657/A7/tablea3 |
| | -22.55 | 3.68 | Gaia DR3 |
| J155947.24+440359.6 | -23.89 | 1.96 | Gaia DR3 |
| J160116.86-345502.7 | | | |
| J160549.19-311521.6 | -2.23 | 3.65 | I/345/gaia2 |
| | -3.66 | 3.16 | Gaia DR3 |
| J160828.45-060734.6 | -12.32 | -999.0 | J/MNRAS/494/2429/table1 |
| | -12.68 | 4.0 | III/283/ravedr6 |
| | -19.13 | 3.35 | Gaia DR3 |
| J160954.85-305858.4 | -1.63 | 0.6 | Gaia DR3 |
| J161743.18+261815.2 | -17.72 | 2.66 | J/A+A/616/A37/table23 |
| | -18.33 | 1.4 | Gaia DR3 |
| J163051.34+472643.8 | -21.73 | 2.46 | Gaia DR3 |
| J163632.90+635344.9 | -55.4 | 1.33 | J/A+A/649/A6/table1c |
| | -54.4 | 1.62 | Gaia DR3 |
| J164539.37+702400.1 | -170.45 | 4.97 | Gaia DR3 |
| J164539.37+702400.1 | 11.26 | 6.22 | Gaia DR3 |
| J171038.44-210813.0 | -14.25 | 1.98 | J/A+A/649/A6/table1r |
| | -21.96 | 2.6 | Gaia DR3 |
| J171038.44-210813.0 | -21.2 | 6.28 | Gaia DR3 |
| J171117.68+124540.4 | -21.63 | 4.82 | J/A+A/649/A6/table1c |
| | -32.24 | 6.13 | Gaia DR3 |
| J171441.70-220948.8 | | | |
| J172131.73-084212.3 | -9.03 | 5.08 | Gaia DR3 |
| J172309.67-095126.2 | 24.46 | 10.82 | J/A+A/649/A6/table1r |
| | 5.16 | 4.16 | Gaia DR3 |
| J172309.67-095126.2 | 22.24 | 6.5 | Gaia DR3 |
| J172454.26+502633.0 | -34.75 | 0.83 | J/A+A/649/A6/table1c |
| | -35.81 | 0.43 | Gaia DR3 |
| J172615.23-031131.9 | | | |
| J172951.38+093336.9 | -18.71 | 38.59 | Gaia DR3 |
| J173353.07+165511.7 | -23.57 | 3.23 | J/A+A/614/A76/tablea1 |
| J173544.26-165209.9 | -20.18 | 0.86 | Gaia DR3 |
| J174811.33-030510.2 | -46.33 | 2.85 | Gaia DR3 |
| J174936.01-010808.7 | | | |
| J175839.30+155208.6 | -11.15 | 2.42 | Gaia DR3 |
| | -9.62 | 1.51 | Gaia DR3 |
| J175942.12+784942.1 | 6.22 | 3.24 | Gaia DR3 |
| J180554.92-570431.3 | -0.61 | 0.36 | J/A+A/649/A6/table1c |
| | -1.17 | 1.59 | Gaia DR3 |
| J180658.07+161037.9 | -11.43 | 4.33 | J/A+A/649/A6/table1c |
| | -12.74 | 2.95 | Gaia DR3 |
| J180733.00+613153.6 | -49.5 | 8.78 | Gaia DR3 |
| J180929.71-543054.2 | -2.0 | 0.8 | J/AJ/154/69/table4 |
| | -2.29 | 2.28 | Gaia DR3 |
| J181725.08+482202.8 | -24.05 | -999.0 | J/A+A/612/A49/tableb1 |
| | -24.45 | 0.22 | Gaia DR3 |
| | -24.24 | 0.02 | J/A+A/656/A162/table1 |
| J184204.85-555413.3 | 3.12 | 26.81 | Gaia DR3 |
| J184206.97-555426.2 | 1.17 | 0.17 | J/A+A/649/A6/table1c |
| | 0.2 | 1.28 | Gaia DR3 |
| | 0.3 | 0.5 | J/AJ/154/129/catalog |
| J184536.02-205910.8 | -30.09 | 0.14 | J/A+A/649/A6/table1c |
| | -22.54 | 2.93 | Gaia DR3 |
| | -27.89 | 1.22 | Gaia DR3 |
| J190453.69-140406.0 | -26.95 | 0.72 | J/A+A/649/A6/table1c |
| | -29.7 | 0.93 | Gaia DR3 |
| J191019.82-160534.8 | -34.34 | 1.23 | J/A+A/616/A37/table23 |
| | -35.23 | 0.65 | Gaia DR3 |
| J191036.02-650825.5 | -16.57 | 2.96 | III/283/ravedr6 |
| | -23.08 | 1.87 | Gaia DR3 |
| J191235.95+630904.7 | -44.75 | 3.87 | I/345/gaia2 |
| | -46.08 | 2.79 | Gaia DR3 |
| J191500.80-284759.1 | | | |
| J191534.83-083019.9 | -27.78 | 3.28 | J/A+A/649/A6/table1c |
| | -33.81 | 3.98 | Gaia DR3 |
| J191629.61-270707.2 | | | |
| J192250.70-631058.6 | 0.84 | 1.0 | J/AJ/157/234/table2 |
| | -0.63 | 2.69 | Gaia DR3 |
| J192323.20+700738.3 | 7.5 | 6.48 | Gaia DR3 |
| | -16.21 | 4.85 | Gaia DR3 |
| J192434.97-344240.0 | -3.7 | 0.2 | J/AJ/154/69/table4 |
| | -1.42 | 2.21 | Gaia DR3 |
| | -4.49 | 0.32 | J/A+A/649/A6/table1c |
| J192600.77-533127.6 | -1.3 | 2.5 | J/AJ/154/69/table4 |
| | 0.61 | 6.32 | Gaia DR3 |
| J193052.51-545325.4 | -25.68 | 4.14 | III/283/ravedr6 |
| | -32.12 | 4.9 | Gaia DR3 |
| J193411.46-300925.3 | | | |
| J193711.26-040126.7 | -9.19 | 7.63 | Gaia DR3 |
| J194309.89-601657.8 | | | |
| J194444.21-435903.0 | -0.03 | 3.03 | Gaia DR3 |
| | -7.19 | 5.54 | Gaia DR3 |
| J194714.54+640237.9 | -77.12 | 3.3 | J/A+A/649/A6/table1c |
| | -77.88 | 1.42 | Gaia DR3 |
| J194816.54-272032.3 | -5.82 | 1.03 | J/A+A/649/A6/table1c |
| | -6.2 | 1.26 | Gaia DR3 |
| J194834.58-760546.9 | 19.93 | 0.71 | J/A+A/649/A6/table1c |
| | 20.09 | 0.56 | Gaia DR3 |
| J195227.23-773529.4 | 32.66 | 8.11 | III/279/rave dr5 |
| | 31.94 | 25.12 | Gaia DR3 |
| J195315.67+745948.9 | -18.1 | 0.44 | J/A+A/649/A6/table1c |
| | -23.53 | 6.61 | Gaia DR3 |
| J195315.67+745948.9 | -17.71 | 0.8 | Gaia DR3 |
| J195331.72-070700.5 | -23.29 | 1.91 | Gaia DR3 |
| J195340.71+502458.2 | -25.44 | 39.77 | Gaia DR3 |
| J195602.95-320719.3 | -3.7 | 2.2 | J/AJ/154/129/catalog |
| J200137.19-331314.5 | -3.7 | 0.2 | J/AJ/154/129/catalog |
| | -4.13 | 0.46 | Gaia DR3 |
| J200137.19-331314.5 | -4.42 | 0.57 | J/A+A/649/A6/table1c |
| | -4.13 | 0.46 | Gaia DR3 |
| J200311.61-243959.2 | -13.4 | 1.26 | III/283/xgaia2 |
| | -10.65 | 1.33 | Gaia DR3 |
| J200311.61-243959.2 | -12.76 | 1.26 | J/A+A/657/A7/tablea3 |
| | -10.65 | 1.33 | Gaia DR3 |
| J200409.19-672511.7 | 21.16 | 6.68 | III/279/rave dr5 |
| | 1.86 | 3.23 | Gaia DR3 |
| J200556.44-321659.7 | -8.09 | 1.69 | J/A+A/649/A6/table1c |
| | -6.55 | 0.61 | Gaia DR3 |
| J200837.87-254526.2 | -5.74 | 1.57 | J/A+A/649/A6/table1c |
| J200837.87-254526.2 | -5.74 | 1.57 | J/A+A/649/A6/table1c |
| J200853.72-351949.3 | -3.7 | 0.2 | J/AJ/154/69/table4 |
| | -31.56 | 4.96 | Gaia DR3 |
| J200853.72-351949.3 | -5.29 | 2.86 | J/A+A/649/A6/table1c |
| | -4.16 | 1.42 | Gaia DR3 |
| J201000.06-280141.6 | -8.56 | 0.44 | J/A+A/649/A6/table1c |
| | -20.76 | 2.97 | Gaia DR3 |
| J201000.06-280141.6 | -5.8 | 0.6 | J/AJ/154/129/catalog |
| | -20.76 | 2.97 | Gaia DR3 |
| J201931.84-081754.3 | -55.04 | 8.94 | J/A+A/649/A6/table1c |
| | -64.49 | 7.39 | Gaia DR3 |
| J202505.36+835954.2 | -28.16 | 6.62 | Gaia DR3 |
| J202716.80-254022.8 | -8.51 | 5.33 | J/A+A/649/A6/table1c |
| | -9.74 | 2.82 | Gaia DR3 |
| J203023.10+711419.8 | -16.83 | 5.17 | Gaia DR3 |
| J203337.63-255652.8 | -7.6 | 0.4 | J/AJ/154/129/catalog |
| J203337.63-255652.8 | -8.48 | 0.52 | J/A+A/649/A6/table1c |
| J204406.36-153042.3 | -25.57 | 4.89 | Gaia DR3 |
| | -21.02 | 3.85 | Gaia DR3 |
| J204714.59+110442.2 | | | |
| J205131.01-154857.6 | | | |
| J205136.27+240542.9 | 3.21 | 4.16 | Gaia DR3 |
| J205832.99-482033.8 | | | |
| J210131.13-224640.9 | -30.98 | 0.37 | J/A+A/649/A6/table1c |
| | -30.94 | 0.34 | Gaia DR3 |
| J210338.46+075330.3 | -21.09 | 3.49 | I/345/gaia2 |
| J210338.46+075330.3 | -20.85 | 3.49 | J/A+A/657/A7/tablea3 |
| | -33.65 | 5.75 | Gaia DR3 |
| J210708.43-113506.0 | -7.01 | 0.97 | J/A+A/649/A6/table1c |
| | -6.75 | 0.97 | J/A+A/657/A7/tablea3 |
| | -8.51 | 0.68 | Gaia DR3 |
| J210722.53-705613.4 | 3.29 | 4.61 | J/ApJ/840/87/targets |
| | 4.04 | 2.33 | Gaia DR3 |
| J210736.82-130458.9 | -3.62 | 3.02 | III/279/rave$_{o}n$ |
| | -0.44 | 2.25 | Gaia DR3 |
| J210957.48+032121.1 | -19.57 | 0.4 | J/A+A/649/A6/table1c |
| | -19.1 | 0.82 | Gaia DR3 |
| J211004.67-192031.2 | 0.6 | 3.0 | J/AJ/154/69/table4 |
| | -14.46 | 4.19 | Gaia DR3 |
| J211005.41-191958.4 | -6.5 | 0.66 | J/A+A/649/A6/table1c |
| | -6.43 | 0.88 | Gaia DR3 |
| J211031.49-271058.1 | -4.3 | 0.2 | J/AJ/154/69/table4 |
| J211635.34-600513.4 | -2.6 | 10.43 | Gaia DR3 |
| J212007.84-164548.2 | -5.1 | 0.62 | J/A+A/649/A6/table1c |
| | -6.07 | 2.08 | Gaia DR3 |
| J212007.84-164548.2 | -6.07 | 2.08 | Gaia DR3 |
| J212128.89-665507.1 | 3.3 | 0.8 | J/AJ/154/69/table4 |
| | 5.0 | 0.44 | Gaia DR3 |
| J212750.60-684103.9 | 7.0 | 3.4 | J/A+A/649/A6/table1c |
| J213507.39+260719.4 | -36.27 | 0.3 | J/A+A/649/A6/table1c |
| | -36.15 | 0.2 | Gaia DR3 |
| J213520.34-142917.9 | -169.1 | 5.42 | Gaia DR3 |
| J213708.89-603606.4 | 0.2 | 0.4 | J/A+A/649/A6/table1c |
| | -0.53 | 2.22 | Gaia DR3 |
| J213740.24+013713.2 | -2.74 | 2.01 | J/A+A/614/A76/tablea1 |
| | -11.0 | 6.7 | Gaia DR3 |
| J213835.44-505111.0 | -12.09 | 3.61 | Gaia DR3 |
| J213847.58+050451.4 | -16.14 | 1.11 | J/A+A/649/A6/table1c |
| J214101.48+723026.7 | -28.26 | 8.49 | Gaia DR3 |
| J214126.66+204310.5 | 10.11 | 1.57 | Gaia DR3 |
| J214414.73+321822.3 | -29.93 | 1.27 | Gaia DR3 |
| | -28.4 | 2.95 | Gaia DR3 |
| J214905.04-641304.8 | 3.9 | 3.2 | J/A+A/649/A6/table1c |
| | 4.23 | 35.19 | Gaia DR3 |
| J215053.68-055318.9 | -3.07 | 5.49 | J/A+A/649/A6/table1c |
| | -10.48 | 2.4 | Gaia DR3 |
| J215128.95-023814.9 | -7.56 | 3.27 | J/A+A/649/A6/table1c |
| | -6.67 | 1.45 | Gaia DR3 |
| | -17.12 | 7.62 | Gaia DR3 |
| J220216.29-421034.0 | -2.8 | 0.3 | J/AJ/147/146/stars |
| | -3.38 | 0.77 | Gaia DR3 |
| J220254.57-644045.0 | 0.4 | 2.99 | J/A+A/649/A6/table1c |
| | -0.64 | 4.69 | Gaia DR3 |
| J220306.98-253826.6 | -37.13 | 9.87 | Gaia DR3 |
| | -12.61 | 4.86 | Gaia DR3 |
| | -13.53 | 3.35 | Gaia DR3 |
| J220850.39+114412.7 | -9.3 | 1.4 | J/AJ/154/69/table4 |
| J221217.17-681921.1 | -1.74 | 1.87 | Gaia DR3 |
| J221559.00-014733.0 | -53.1 | -999.0 | J/other/RAA/15.1154/mdwarfs |
| | -48.46 | 2.17 | Gaia DR3 |
| J221833.85-170253.2 | 13.92 | 0.33 | J/A+A/649/A6/table1c |
| | 14.16 | 0.44 | Gaia DR3 |
| J221842.70+332113.5 | -16.14 | 4.19 | J/A+A/649/A6/table1c |
| | -20.93 | 3.15 | Gaia DR3 |
| J222024.21-072734.5 | -13.05 | 3.29 | J/A+A/649/A6/table1c |
| | -17.04 | 9.75 | Gaia DR3 |
| J224111.08-684141.8 | | | |
| J224221.02-410357.2 | -7.03 | 2.42 | Gaia DR3 |
| J224448.45-665003.9 | 0.7 | 1.7 | J/A+A/649/A6/table1c |
| J224500.20-331527.2 | 2.03 | 0.04 | J/ApJ/762/88/members |
| | -4.65 | 3.83 | Gaia DR3 |
| J224634.82-735351.0 | 9.1 | 0.6 | J/AJ/147/146/stars |
| | 5.79 | 2.14 | Gaia DR3 |
| J225914.87+373639.3 | 654.36 | 6.81 | Gaia DR3 |
| J225934.89-070447.1 | -9.15 | 4.11 | III/279/rave$_{o}n$ |
| | -16.73 | 8.8 | Gaia DR3 |
| J230209.10-121522.0 | -12.89 | 2.53 | III/279/rave dr5 |
| | -13.31 | 3.37 | Gaia DR3 |
| J230327.73-211146.2 | -9.54 | 5.21 | J/A+A/649/A6/table1c |
| | 5.16 | 5.81 | Gaia DR3 |
| J230740.98+080359.7 | 3.29 | 0.62 | J/A+A/649/A6/table1c |
| | 4.19 | 0.52 | Gaia DR3 |
| J231021.75+685943.6 | 3.22 | 1.03 | J/A+A/649/A6/table1c |
| | 13.6 | 9.73 | Gaia DR3 |
| J231021.75+685943.6 | 3.26 | 0.75 | Gaia DR3 |
| J231211.37+150329.7 | -14.21 | 2.08 | I/345/gaia2 |
| | -14.28 | 1.67 | Gaia DR3 |
| J231246.53-504924.8 | | | |
| J231457.86-633434.0 | -56.27 | 13.98 | III/283/ravedr6 |
| | -18.99 | 15.56 | Gaia DR3 |
| J231543.66-140039.6 | -53.64 | 0.42 | J/A+A/649/A6/table1c |
| | -54.39 | 0.38 | Gaia DR3 |
| | -54.03 | 0.81 | III/283/ravedr6 |
| J231933.16-393924.3 | 1.4 | 0.34 | J/A+A/649/A6/table1c |
| | 1.5 | 0.32 | Gaia DR3 |
| J232008.15-634334.9 | | | |
| J232151.23+005037.3 | -32.5 | -999.0 | J/other/RAA/15.1154/mdwarfs |
| | -32.34 | 1.02 | III/279/rave dr5 |
| | -32.34 | 1.83 | Gaia DR3 |
| | -34.11 | 0.28 | Gaia DR3 |
| J232656.43+485720.9 | 16.48 | 1.09 | J/A+A/649/A6/table1r |
| | 16.59 | 0.93 | Gaia DR3 |
| J232857.75-680234.5 | 8.0 | 1.5 | J/AJ/147/146/stars |
| | 3.36 | 2.26 | Gaia DR3 |
| J232904.42+032910.8 | -3.35 | 9.52 | Gaia DR3 |
| J232917.64-675000.6 | 7.87 | 1.33 | Gaia DR3 |
| J232959.47+022834.0 | | | |
| J233647.87+001740.1 | -15.51 | 0.69 | J/A+A/649/A6/table1c |
| | -16.1 | 0.59 | Gaia DR3 |
| J234243.45-622457.1 | 5.1 | 4.6 | J/A+A/649/A6/table1c |
| J234326.88-344658.5 | 8.53 | 3.11 | III/279/rave dr5 |
| J234326.88-344658.5 | | | |
| J234333.91-192802.8 | -5.43 | 2.29 | Gaia DR3 |
| J234347.83-125252.1 | 3.61 | 2.41 | J/A+A/649/A6/table1c |
| | 5.17 | 2.05 | Gaia DR3 |
| J234924.87+185926.7 | | | |
| J234926.23+185912.4 | 6.28 | 0.38 | J/A+A/649/A6/table1c |
| | 6.37 | 0.31 | Gaia DR3 |
| J235250.70-160109.7 | 2.4 | 5.4 | Gaia DR3 |

TABLE 12 Radial Velocities: This Work and Adopted Values

| Name | | RV, This Work | error | Adopted RV | error | NYMG |
|---|---|---|---|---|---|---|
| J001527.62-641455.2 | | 6.3 | 0.9 | 6.6 | 0.4 | THA |
| J001536.79-294601.2 | | 1.0 | 0.7 | 0.8 | 0.7 | THA |
| J001723.69-664512.4 | | 11.4 | 0.8 | 10.5 | 0.3 | BPIC |
| J003057.97-655006.4 | | 7.2 | 2.5 | 13.0 | 1.0 | THA |
| J003903.51+133016.0 | | -7.0 | 1.5 | -7.6 | 1.6 | ABDOR |
| J004210.98-425254.8 | | 6.9 | 0.5 | 8.4 | 0.4 | COL |
| J004528.25-513734.4 | B | -8.0 | 1.3 | -8.0 | 1.3 | THA |
| J004528.25-513734.4 | A | 17.2 | 0.7 | 17.2 | 0.7 | THA |
| J004826.70-184720.7 | | 5.6 | 1.4 | 6.9 | 0.8 | BPIC |
| J010629.32-122518.4 | | 2.7 | 1.5 | 3.8 | 0.5 | COL |
| J010711.99-193536.4 | | 7.1 | 1.1 | 8.8 | 0.6 | BPIC |
| J012118.22-543425.1 | | 25.8 | 1.3 | 24.4 | 0.2 | ABDOR |
| J012245.24-631845.0 | | 18.3 | 5.3 | 8.9 | 1.6 | THA |
| J012332.89-411311.4 | | | | 6.9 | 1.7 | THA |
| J012532.11-664602.6 | | | | 21.1 | 4.6 | THA |
| J015057.01-584403.4 | | 9.3 | 1.5 | 10.8 | 0.7 | THA |
| J015350.81-145950.6 | | | | 10.5 | 0.6 | BPIC |
| J020012.84-084052.4 | | 5.5 | 1.4 | 4.5 | 0.5 | THA |
| J021258.28-585118.3 | | 8.4 | 1.9 | 9.1 | 0.9 | THA |
| J022240.88+305515.4 | | | | 4.7 | 1.3 | BPIC |
| J022424.69-703321.2 | | 12.5 | 1.2 | 11.8 | 0.4 | THA |
| J024552.65+052923.8 | | | | 6.6 | 1.1 | THA |
| J024746.49-580427.4 | | 12.2 | 1.0 | 12.9 | 0.6 | THA |
| J024852.67-340424.9 | | 12.5 | 2.0 | 14.6 | 0.4 | COL |
| J025154.17+222728.9 | | 11.4 | 2.6 | 9.2 | 0.2 | BPIC |
| J030251.62-191150.0 | | 12.3 | 1.2 | 12.3 | 1.2 | THA |
| J030824.14+234554.2 | | | | 16.7 | 1.3 | ARG |
| J031650.45-350937.9 | | 17.0 | 2.7 | 16.1 | 1.7 | THA |
| J032047.66-504133.0 | | 15.4 | 1.2 | 17.9 | 0.4 | COL |
| J033640.91+032918.3 | | 13.9 | 4.8 | 17.8 | 1.7 | COL |
| J034115.60-225307.8 | | 19.2 | 1.7 | 17.9 | 1.2 | COL |
| J034116.16-225244.0 | | | | 16.9 | 2.5 | COL |
| J034236.95+221230.2 | | 10.6 | 1.4 | 10.6 | 1.4 | BPIC |
| J035100.83+141339.2 | | | | 8.0 | 7.0 | BPIC |
| J035345.92-425018.0 | | 20.1 | 1.3 | 20.3 | 0.2 | COL |
| J035733.95+244510.2 | | | | 12.1 | 0.7 | BPIC |
| J035829.67-432517.2 | | | | 11.3 | 4.0 | THA |
| J040539.68-401410.5 | | 16.6 | 1.5 | 17.2 | 1.3 | THA |
| J040649.38-450936.3 | | 20.6 | 0.6 | 22.7 | 0.8 | COL |
| J040711.50-291834.3 | | 19.4 | 1.2 | 19.4 | 1.2 | COL |
| J040827.01-784446.7 | | 16.1 | 1.2 | 16.7 | 0.5 | CAR |
| J041255.78-141859.2 | | 16.4 | 1.1 | 16.8 | 0.2 | THA |
| J041336.14-441332.4 | | 16.8 | 4.5 | 16.4 | 1.4 | THA |
| J041807.76+030826.0 | | 15.7 | 1.7 | 15.7 | 1.7 | ABDOR |
| J042139.19-723355.7 | | 14.2 | 0.8 | 14.9 | 0.3 | THA |
| J042736.03-231658.8 | | 20.0 | 1.4 | 20.0 | 1.4 | COL |
| J043657.44-161306.7 | | 16.4 | 2.3 | 15.6 | 0.5 | THA |
| J044356.87+372302.7 | | 7.1 | 1.1 | 6.5 | 0.3 | BPIC |
| J044700.46-513440.4 | | 18.3 | 0.7 | 19.6 | 0.4 | THA |
| J044721.05+280852.5 | | | | 21.1 | 5.2 | BPIC |
| J044800.86+143957.7 | SB | 13.7 | 0.6 | 13.7 | 0.6 | BPIC |
| J044802.59+143951.1 | A | 15.3 | 0.4 | 15.3 | 0.4 | BPIC |
| J045114.41-601830.5 | | 20.0 | 1.8 | 20.2 | 0.8 | COL |
| J050610.44-582828.5 | | 18.0 | 1.6 | 19.3 | 0.2 | THA |
| J050827.31-210144.3 | | 28.9 | 3.0 | 22.2 | 0.3 | BPIC |
| J051026.38-325307.4 | | 23.1 | 0.4 | 23.1 | 0.4 | COL |
| J051403.20-251703.8 | | 23.7 | 1.7 | 24.5 | 1.4 | COL |
| J051650.66+022713.0 | | | | 13.1 | 4.6 | COL |
| J052419.14-160115.5 | | 25.4 | 2.5 | 18.1 | 0.8 | BPIC |
| J052944.69-323914.1 | | 22.2 | 0.8 | 22.2 | 0.8 | BPIC |
| J053747.56-424030.8 | | 18.5 | 3.3 | 18.5 | 3.3 | BPIC |
| J053925.08-424521.0 | | 21.6 | 0.5 | 21.7 | 0.2 | THA |
| J054433.76-200515.5 | | 22.4 | 2.5 | 22.3 | 1.8 | COL |
| J054709.88-525626.1 | | 27.4 | 6.6 | 31.5 | 3.0 | ABDOR |
| J055008.59+051153.2 | | | | 12.0 | 4.0 | COL |
| J061313.30-274205.6 | | 22.7 | 0.6 | 22.5 | 0.2 | BPIC |
| J070657.72-535345.9 | | | | 22.3 | 0.7 | CAR |
| J072821.16+334511.6 | | 4.9 | 2.7 | 7.0 | 4.0 | COL |
| J072911.26-821214.3 | | 26.1 | 0.7 | 25.8 | 0.5 | ABDOR |
| J073138.47+455716.5 | | | | 9.7 | 3.8 | COL |
| J075233.22-643630.5 | | | | 17.5 | 4.0 | CAR |
| J080636.05-744424.6 | | 18.3 | 0.8 | 18.5 | 1.0 | CAR |
| J081738.97-824328.8 | | 15.8 | 1.4 | 15.5 | 1.3 | BPIC |
| J101543.44+660442.3 | | | | -18.5 | 1.4 | ABDOR |
| J103952.70-353402.5 | | | | 14.7 | 0.8 | ARG |
| J111128.13-265502.9 | | 1.2 | 10.3 | 11.0 | 2.5 | TWA |
| J112105.43-384516.6 | | 12.0 | 0.8 | 11.6 | 1.0 | TWA |
| J112547.46-441027.4 | | 20.4 | 2.6 | 21.5 | 2.0 | ABDOR |
| J112651.28-382455.5 | | 11.4 | 0.6 | 12.2 | 1.0 | TWA |
| J115927.82-451019.3 | | 11.4 | 1.2 | 11.6 | 2.0 | TWA |
| J120237.94-332840.4 | | 8.1 | 5.2 | 7.5 | 2.0 | TWA |
| J121511.25-025457.1 | | | | 0.0 | 1.6 | ABDOR |
| J121511.25-025457.1 | | | | 2.7 | 3.8 | ABDOR |
| J123005.17-440236.1 | | 8.5 | 1.8 | 10.5 | 2.0 | TWA |
| J125049.12-423123.6 | | 8.7 | 12.9 | 11.0 | 2.0 | TWA |
| J133509.40+503917.5 | | | | -16.0 | 3.5 | ABDOR |
| J135913.33-292634.2 | | -2.2 | 0.4 | -2.1 | 0.5 | ARG |
| J155947.24+440359.6 | | | | -23.9 | 2.0 | ABDOR |
| J180554.92-570431.3 | | | | -0.7 | 0.4 | BPIC |
| J180929.71-543054.2 | | | | -2.0 | 0.8 | BPIC |
| J184206.97-555426.2 | | 1.2 | 0.9 | 1.2 | 0.2 | BPIC |
| J191534.83-083019.9 | | -27.0 | 2.9 | -28.0 | 2.4 | ARG |
| J192250.70-631058.6 | | 4.4 | 1.7 | 2.0 | 1.5 | THA |
| J192434.97-344240.0 | | -5.3 | 1.3 | -4.0 | 0.4 | BPIC |
| J193411.46-300925.3 | | -5.4 | 8.5 | -4.7 | 1.5 | BPIC |
| J194309.89-601657.8 | | 1.4 | 1.0 | 1.4 | 1.0 | THA |
| J194816.54-272032.3 | | | | -6.0 | 0.8 | BPIC |
| J194834.58-760546.9 | | | | 20.1 | 0.4 | ABDOR |
| J195227.23-773529.4 | B | -5.4 | 2.2 | -5.4 | 2.2 | ARG |
| J195227.23-773529.4 | A | 69.6 | 1.5 | 70.7 | 0.1 | ARG |
| J195331.72-070700.5 | | | | -23.3 | 1.9 | ARG |
| J195602.95-320719.3 | | | | -3.7 | 0.2 | BPIC |
| J200137.19-331314.5 | | -4.1 | 1.5 | -3.7 | 0.2 | BPIC |
| J200409.19-672511.7 | | | | 6.0 | 4.0 | THA |
| J200556.44-321659.7 | | | | -6.7 | 0.7 | BPIC |
| J200837.87-254526.2 | | | | -5.7 | 1.6 | BPIC |
| J200853.72-351949.3 | | | | -3.7 | 0.2 | BPIC |
| J201000.06-280141.6 | | -5.7 | 3.2 | -7.2 | 0.8 | BPIC |
| J203337.63-255652.8 | | -5.7 | 0.7 | -7.6 | 0.5 | BPIC |
| J210338.46+075330.3 | | -18.0 | 1.0 | -18.3 | 1.0 | ABDOR |
| J210722.53-705613.4 | | -2.7 | 2.6 | 6.0 | 0.6 | THA |
| J211004.67-192031.2 | | -6.0 | 3.3 | -5.7 | 1.5 | BPIC |
| J211005.41-191958.4 | | -6.1 | 1.1 | -5.7 | 0.3 | BPIC |
| J211031.49-271058.1 | A | | | -4.3 | 0.2 | BPIC |
| J211031.49-271058.1 | B | -6.3 | 2.3 | -6.3 | 2.3 | BPIC |
| J211635.34-600513.4 | | 2.4 | 2.3 | 2.0 | 1.5 | THA |
| J212007.84-164548.2 | | -7.1 | 3.2 | -5.1 | 0.6 | BPIC |
| J212128.89-665507.1 | | | | 4.6 | 0.6 | BPIC |
| J213708.89-603606.4 | | 1.9 | 0.5 | 2.0 | 0.2 | THA |
| J213740.24+013713.2 | | -5.5 | 9.7 | -2.7 | 2.0 | BPIC |
| J213835.44-505111.0 | | | | -12.1 | 3.6 | ARG |
| J214905.04-641304.8 | | | | 3.9 | 3.2 | THA |
| J220216.29-421034.0 | | -2.8 | 1.0 | -2.4 | 0.3 | THA |
| J220254.57-644045.0 | | -0.2 | 4.1 | 2.0 | 1.5 | THA |
| J220850.39+114412.7 | | -9.4 | 3.5 | -9.3 | 1.3 | BPIC |
| J224448.45-665003.9 | | | | 0.7 | 1.7 | THA |

| | | | | | |
|---|---|---|---|---|---|
| J224500.20-331527.2 | | | 2.0 | 0.1 | BPIC |
| J224634.82-735351.0 | 7.5 | 1.7 | 8.9 | 0.7 | THA |
| J230209.10-121522.0 | -11.4 | 1.2 | -11.0 | 1.5 | THA |
| J230327.73-211146.2 | | | -3.0 | 6.0 | ABDOR |
| J232857.75-680234.5 | 8.5 | 2.4 | 6.8 | 2.5 | THA |
| J232917.64-675000.6 | | | 7.9 | 1.3 | THA |
| J234243.45-622457.1 | | | 5.1 | 4.6 | THA |
| J234326.88-344658.5 | | | 8.5 | 3.1 | THA |

TABLE 13 Candidate NYMG Members: Gaia DR3 Data and Gaia2MASS/WISE Colors

| WISE Desig | Parallax mas | PM, RA mas $yr^{-1}$ | PM, Dec mas $yr^{-1}$ | $B_p - R_p$ mag | $M_G$ mag | $K - W2$ | $M_G - M_K$ | NYMG |
|---|---|---|---|---|---|---|---|---|
| J000453.05-103220.0 | 15.46 | 64.62 | -18.99 | 3.14 | 9.68 | 0.4 | 4.05 | BPMG |
| J001555.65-613752.2 | 18.21 | 68.17 | -41.96 | 2.71 | 9.73 | 0.35 | 3.66 | ABDMG |
| J001723.69-664512.4 | 27.16 | 102.89 | -16.83 | 2.48 | 8.5 | 0.22 | 3.64 | BPMG |
| J003057.97-655006.4 | 22.04 | 85.66 | -50.22 | 2.94 | 9.61 | 0.35 | 3.94 | THA |
| J003234.86+072926.4 | 28.43 | 104.36 | -62.91 | 2.76 | 8.92 | 0.29 | 4.14 | BPMG |
| J003903.51+133016.0 | 18.52 | 87.02 | -80.77 | 2.97 | 10.64 | 0.42 | 4.25 | ABDMG |
| J004528.25-513734.4 | 24.29 | 99.26 | -58.5 | 2.27 | 7.93 | 0.19 | 3.38 | THA |
| J004826.70-184720.7 | 19.42 | 73.28 | -47.15 | 3.16 | 10.39 | 0.36 | 4.08 | BPMG |
| J010251.05+185653.7 | 26.19 | 97.34 | -59.77 | 2.95 | 9.67 | 0.35 | 3.91 | BPMG |
| J012245.24-631845.0 | 22.05 | 91.9 | -33.35 | 2.75 | 9.45 | 0.31 | 3.76 | THA |
| J012332.89-411311.4 | 25.01 | 105.54 | -51.2 | 3.12 | 10.91 | 0.4 | 4 | THA |
| J012532.11-664602.6 | 21.62 | 90.39 | -29.05 | 3.05 | 10.73 | 0.4 | 3.95 | THA |
| J014431.99-460432.1 | 25.95 | 109.43 | -42.51 | 3.99 | 12.71 | 0.5 | 4.66 | THA |
| J015057.01-584403.4 | 22.1 | 93.59 | -27.08 | 2.64 | 9.12 | 0.22 | 3.75 | THA |
| J015350.81-145950.6 | 29.56 | 106.03 | -45.38 | 2.65 | 8.87 | 0.34 | 4.45 | BPMG |
| J015350.81-145950.6 | 29.56 | 106.03 | -45.38 | 2.65 | 8.87 | 0.34 | 4.45 | BPMG |
| J020012.84-084052.4 | 27.18 | 115.61 | -62.81 | 2.4 | 8.61 | 0.19 | 3.57 | THA |
| J021258.28-585118.3 | 20.87 | 88.34 | -17.35 | 2.34 | 8.5 | 0.24 | 3.46 | THA |
| J022240.88+305515.4 | 22.26 | 75.29 | -65.14 | 2.74 | 9.56 | 0.32 | 3.76 | BPMG |
| J022424.69-703321.2 | 22.7 | 94.31 | -3.73 | 3.06 | 10.28 | 0.36 | 4.01 | THA |
| J023139.36+445638.1 | 20.01 | 91.94 | -58.28 | 2.97 | 9.57 | 0.34 | 3.94 | ARG |
| J024552.65+052923.8 | 16.76 | 72.33 | -38.05 | 2.34 | 8.7 | 0.24 | 3.41 | THA |
| J024746.49-580427.4 | 22.41 | 92.32 | -6.53 | 2.39 | 8.71 | 0.22 | 3.5 | THA |
| J024852.67-340424.9 | 23.98 | 97.33 | -24.27 | 2.84 | 9.12 | 0.35 | 3.82 | THA |
| J025154.17+222728.9 | 37.01 | 108.77 | -111.15 | 2.82 | 9.73 | 0.3 | 3.82 | BPMG |
| J030251.62-191150.0 | 23.84 | 95.09 | -42.28 | 3.2 | 10.66 | 0.41 | 4.1 | THA |
| J031650.45-350937.9 | 21.41 | 83.53 | -17.41 | 2.62 | 8.65 | 0.34 | 3.68 | THA |
| J032047.66-504133.0 | 22.81 | 83.29 | 7.15 | 2.33 | 8.77 | 0.23 | 3.42 | COL |
| J033640.91+032918.3 | 37.35 | 122.13 | -120.24 | 3.03 | 10.26 | 0.39 | 3.95 | COL |
| J034115.60-225307.8 | 14 | 51.53 | -15.29 | 1.98 | 7.87 | 0.13 | 3.12 | COL |
| J034116.16-225244.0 | 13.96 | 49.71 | -16.38 | 2.05 | 8.01 | 0.17 | 3.17 | COL |
| J034236.95+221230.2 | 25.34 | 60.01 | -81.58 | 3.96 | 12.01 | 0.48 | 4.67 | BPMG |
| J035345.92-425018.0 | 10.98 | 37.05 | 1.81 | 2.28 | 8.34 | 0.28 | 3.39 | COL |
| J035733.95+244510.2 | 14.55 | 34.34 | -46.52 | 1.98 | 7.67 | 0.1 | 3.13 | BPMG |
| J040539.68-401410.5 | 19.92 | 72.33 | 5.75 | 2.94 | 9.75 | 0.39 | 4.27 | THA |
| J040649.38-450936.3 | 13.26 | 41.79 | 9.77 | 2.56 | 8.34 | 0.28 | 3.65 | COL |
| "J040827.01-784446.7 " | 17.33 | 54.28 | 43.87 | 1.91 | 7.67 | 0.14 | 3.08 | CAR |
| J041255.78-141859.2 | 17.38 | 60.75 | -25.35 | 2.13 | 8.1 | 0.2 | 3.28 | THA |
| "J041749.66+001145.4 " | 9.8 | 33.3 | -26.8 | 1.62 | 6.61 | 0.1 | 2.84 | THA |
| J042139.19-723355.7 | 18.66 | 61.35 | 33.63 | 2.47 | 8.92 | 0.21 | 3.58 | THA |
| J042736.03-231658.8 | 24.84 | 64 | -21.6 | 3.19 | 10.66 | 0.39 | 4.11 | COL |
| J043257.29+740659.3 | 29.62 | 78.19 | -134 | 2.8 | 9.88 | 0.33 | 5.13 | COL |
| J043257.29+740659.3 | 29.62 | 78.19 | -134 | 2.8 | 9.88 | 0.33 | 5.13 | COL |
| J044356.87+372302.7 | 14.01 | 22.94 | -61.93 | 2.34 | 8.04 | 0.25 | 3.51 | BPMG |
| J044700.46-513440.4 | 17.11 | 52.14 | 17.33 | 2.43 | 8.89 | 0.24 | 3.52 | THA |
| J044721.05+280852.5 | 16.13 | 25.78 | -54.99 | 2.71 | 8.68 | 0.3 | 3.77 | BPMG |
| J044800.86+143957.7 | 11.89 | 19.47 | -43.19 | 2.87 | 10.3 | 0.96 | 4.2 | BPMG |
| J044802.59+143951.1 | 11.92 | 19.94 | -43.49 | 3.17 | 10.32 | -1 | 4.26 | BPMG |
| J050610.44-582828.5 | 18.71 | 50.05 | 29.86 | 2.32 | 8.53 | 0.24 | 3.4 | THA |
| J050827.31-210144.3 | 20.7 | 34.31 | -12.68 | 3.37 | 9.65 | 0.51 | 4.24 | COL |
| J051026.38-325307.4 | 12.3 | 27.66 | 2.8 | 2.83 | 9.86 | 0.53 | 4.71 | COL |
| J051026.38-325307.4 | 12.3 | 27.66 | 2.8 | 2.83 | 9.86 | 0.53 | 4.71 | COL |
| J051403.20-251703.8 | 15.63 | 31.24 | -7.6 | 2.54 | 8.97 | 0.26 | 3.63 | COL |
| J052419.14-160115.5 | 32.06 | 31.99 | -24 | 3.17 | 10.03 | 0.39 | 4.69 | BPMG |
| J052944.69-323914.1 | 33.6 | 15.26 | 10.77 | 3.03 | 9.9 | 0.38 | 3.95 | BPMG |
| J053747.56-424030.8 | 33.24 | 11.65 | 34.94 | 3.47 | 11.31 | 0.41 | 4.35 | BPMG |
| J054719.52-335611.2 | 13.68 | 9.15 | -18.71 | 2.82 | 9.96 | 0.33 | 3.8 | ABDMG |
| J055008.59+051153.2 | 15.61 | 16.99 | -42.1 | 2.14 | 7.78 | 0.15 | 3.35 | COL |
| J061313.30-274205.6 | 29.62 | -14.16 | -1.23 | 2.74 | 8.3 | 0.3 | 3.79 | BPMG |
| J063001.84-192336.6 | 35.73 | -16.14 | -28.33 | 3.03 | 11.65 | 0.4 | 4.63 | BPMG |
| "J070657.72-535345.9 " | 21.36 | -7.41 | 38.91 | 1.87 | 7.36 | 0.1 | 3.04 | CAR |
| J072911.26-821214.3 | 19.41 | -30.82 | 52.76 | 2.1 | 8.55 | 0.17 | 3.22 | ABDMG |
| J073138.47+455716.5 | 17.88 | -13.69 | -92.77 | 2.67 | 9.03 | 0.29 | 3.85 | COL |
| "J075233.22-643630.5 " | 10.08 | -7.33 | 27.28 | 1.67 | 6.87 | 0.19 | 2.98 | CAR |
| "J080636.05-744424.6 " | 14.58 | -16.21 | 49.46 | 2.29 | 8.39 | 0.18 | 3.44 | CAR |
| J081738.97-824328.8 | 37.68 | -86.55 | 109.47 | 2.67 | 8.66 | 0.3 | 4.19 | BPMG |
| J101917.57-443736.0 | 14.66 | -57.53 | -0.64 | 2.64 | 8.34 | 0.28 | 3.71 | TWA |
| J103952.70-353402.5 | 11.97 | -61.39 | 13.93 | 1.97 | 7.98 | 0.3 | 3.55 | ARG |
| J103952.70-353402.5 | 11.97 | -61.39 | 13.93 | 1.97 | 7.98 | 0.3 | 3.55 | ARG |
| J111128.13-265502.9 | 20.27 | -83.82 | -20.97 | 3.75 | 10.56 | 0.48 | 4.58 | TWA |
| J111229.74-461610.1 | 10.63 | -56.99 | 7.6 | 1.78 | 7.62 | 0.08 | 2.94 | ARG |
| J112105.43-384516.6 | 15.29 | -63.05 | -14.6 | 2.31 | 7.61 | -0.95 | 3.63 | TWA |
| J112547.46-441027.4 | 21.58 | -90.9 | -65.2 | 2.87 | 10.64 | 0.34 | 4.49 | ABDMG |
| J112651.28-382455.5 | 14.62 | -60.7 | -15.49 | 2.79 | 8.77 | 0.28 | 3.82 | TWA |
| J114623.01-523851.8 | 17.32 | -87.21 | 12.33 | 3.07 | 10.44 | 0.36 | 4.05 | ARG |
| J115927.82-451019.3 | 13.84 | -56.45 | -18.34 | 3.03 | 8.77 | 0.37 | 3.99 | TWA |
| J120001.54-173131.1 | 18.85 | -78.94 | -28.16 | 2.88 | 8.79 | 0.28 | 3.94 | TWA |
| J120237.94-332840.4 | 15.98 | -66.24 | -23.39 | 3.26 | 10.02 | 0.4 | 4.15 | TWA |
| J122813.57-431638.9 | 17.12 | -98.72 | -21.63 | 3.13 | 11.67 | 0.44 | 4.69 | ARG |
| J122813.57-431638.9 | 17.12 | -98.72 | -21.63 | 3.13 | 11.67 | 0.44 | 4.69 | ARG |
| J123005.17-440236.1 | 12.79 | -52.18 | -21.9 | 2.91 | 9.02 | 0.31 | 3.91 | TWA |
| J123704.99-441919.5 | 11.14 | -44.41 | -19.22 | 3.6 | 10.89 | 0.46 | 4.37 | TWA |
| J125049.12-423123.6 | 10.26 | -38.72 | -19.98 | 3.05 | 9.04 | 0.43 | 4.07 | TWA |
| J133509.40+503917.5 | 22.99 | -94.81 | -41.94 | 2.69 | 10.06 | 0.58 | 4.88 | ABDMG |
| J150939.16-133212.4 | 18.97 | -52.62 | -49.37 | 2.9 | 9.2 | 0.34 | 3.91 | TWA |
| J163051.34+472643.8 | 26.23 | -60.74 | 3.34 | 3.08 | 11.18 | 0.39 | 4 | ABDMG |
| J180554.92-570431.3 | 17.71 | 0.89 | -72.95 | 2.62 | 8.6 | 0.26 | 3.73 | BPMG |
| J180929.71-543054.2 | 25.67 | 4.71 | -108.17 | 3.39 | 10.38 | 0.44 | 4.22 | BPMG |
| J184204.85-555413.3 | 19.36 | 11.12 | -78.05 | 3.08 | 10.28 | 0.4 | 3.99 | BPMG |
| J184206.97-555426.2 | 19.44 | 12.01 | -79.07 | 2.64 | 8.77 | 0.22 | 3.75 | BPMG |
| J191500.80-284759.1 | 13.9 | 16.52 | -49.36 | 3.29 | 9.92 | 0.39 | 4.25 | BPMG |
| J191534.83-083019.9 | 14.91 | 10.11 | -9.9 | 1.52 | 6.99 | 0.1 | 2.67 | ARG |
| J191629.61-270707.2 | 14.72 | 16.97 | -50.87 | 2.84 | 9.47 | 0.34 | 3.82 | BPMG |
| J192250.70-631058.6 | 16.18 | -7.12 | -77.26 | 2.53 | 8.25 | 0.28 | 3.62 | THA |
| J192434.97-344240.0 | 19.4 | 26.37 | -75.6 | 2.99 | 9.21 | 0.37 | 3.98 | BPMG |
| J192600.77-533127.6 | 20.94 | 26.96 | -81.54 | 2.91 | 9.17 | 0.35 | 3.89 | BPMG |
| J193411.46-300925.3 | 14.77 | 20.84 | -52.41 | 3.46 | 11.02 | 0.45 | 4.37 | BPMG |
| J194309.89-601657.8 | 17.78 | -3.96 | -82.63 | 2.87 | 9.66 | 0.29 | 3.87 | THA |
| J194816.54-272032.3 | 15.47 | 25.15 | -53.38 | 2.29 | 8.15 | 0.23 | 3.41 | BPMG |
| J194834.58-760546.9 | 16.82 | 25.96 | -73.73 | 1.83 | 8.01 | 0.19 | 2.89 | ABDMG |
| J195227.23-773529.4 | 25.69 | 66.2 | -107.26 | 2.7 | 9.54 | 0.3 | 3.72 | ARG |
| J195602.95-320719.3 | 19.54 | 33.13 | -73 | 2.78 | 8.36 | 0.36 | 3.8 | BPMG |
| J200137.19-331314.5 | 16.68 | 29.23 | -61.39 | 2.07 | 7.57 | 0.14 | 3.22 | BPMG |
| J200556.44-321659.7 | 20.18 | 38.44 | -70.45 | 2.08 | 7.73 | 0.16 | 3.26 | BPMG |
| J200837.87-254526.2 | 17.85 | 35.14 | -59.42 | 3.18 | 10.4 | 0.41 | 4.08 | BPMG |
| J200853.72-351949.3 | 22.33 | 50.03 | -83.27 | 2.98 | 9.57 | 0.42 | 4.51 | BPMG |
| J201000.06-280141.6 | 21.55 | 47.58 | -74.51 | 2.79 | 8.87 | 0.29 | 4.48 | BPMG |
| J203337.63-255652.8 | 22.9 | 53.83 | -74.29 | 3.36 | 9.88 | 0.43 | 4.2 | BPMG |
| J205131.01-154857.6 | 16.05 | 18.2 | -60.27 | 3.15 | 11.13 | 0.61 | 4.76 | BPMG |
| J210722.53-705613.4 | 20.7 | 26.72 | -92.28 | 2.65 | 9.18 | 0.27 | 3.7 | THA |
| J211004.67-192031.2 | 29.77 | 88.44 | -95.08 | 3.05 | 8.99 | 0.38 | 4.07 | BPMG |
| J211005.41-191958.4 | 30.9 | 90.61 | -90.99 | 2.42 | 8.26 | 0.21 | 3.61 | BPMG |
| J211031.49-271058.1 | 24.76 | 68.01 | -75.68 | 3.28 | 10.55 | 0.59 | 4.17 | BPMG |
| J211635.34-600513.4 | 21.59 | 30.97 | -97.77 | 2.84 | 9.8 | 0.34 | 3.81 | THA |
| J212007.84-164548.2 | 20.72 | 59.81 | -58.13 | 2.88 | 9.72 | 0.35 | 3.84 | BPMG |
| J212128.89-665507.1 | 31.7 | 104.06 | -99.49 | 1.84 | 7.5 | 0.08 | 2.98 | BPMG |
| J212750.60-684103.9 | 20.09 | 31.52 | -88.29 | 3.11 | 10.12 | 0.4 | 4.03 | THA |
| J213708.89-603606.4 | 22.25 | 39.83 | -97.47 | 2.67 | 9.18 | 0.31 | 3.69 | THA |
| J213740.24+013713.2 | 27.85 | 85.32 | -50.93 | 2.97 | 9.39 | 0.35 | 4.28 | BPMG |
| J213835.44-505111.0 | 22.09 | 99.56 | -60.37 | 3.07 | 10.64 | 0.43 | 4.1 | ARG |
| J213835.44-505111.0 | 22.09 | 99.56 | -60.37 | 3.07 | 10.64 | 0.43 | 4.1 | ARG |
| J214905.04-641304.8 | 22.78 | 44.05 | -97.16 | 3.24 | 10.39 | 0.42 | 4.13 | THA |
| J220216.29-421034.0 | 22.49 | 53.26 | -92.66 | 2.08 | 7.99 | 0.13 | 3.24 | THA |
| J220254.57-644045.0 | 22.9 | 49.98 | -95.73 | 2.33 | 8.44 | 0.19 | 3.49 | THA |
| J220306.98-253826.6 | 20.33 | 76.52 | -111.56 | 2.98 | 10.73 | 0.4 | 4.5 | ABDMG |
| J220850.39+114412.7 | 26.75 | 90.1 | -51.58 | 3.06 | 10.15 | 0.38 | 3.98 | BPMG |
| J221217.17-681921.1 | 18.21 | 61.74 | -62.59 | 2.61 | 9.75 | 0.28 | 3.68 | ABDMG |
| J224448.45-665003.9 | 22.82 | 64.75 | -84.34 | 3.49 | 11.24 | 0.45 | 4.31 | THA |
| J224500.20-331527.2 | 48 | 176.82 | -120.88 | 3.07 | 10.23 | 0.36 | 4.03 | BPMG |
| J224634.82-735351.0 | 19.96 | 56.64 | -70.97 | 2.45 | 8.84 | 0.27 | 3.53 | THA |
| J230209.10-121522.0 | 21.67 | 66.66 | -62.53 | 2.92 | 10.19 | 0.35 | 3.86 | THA |
| J232857.75-680234.5 | 21.81 | 73.47 | -68.22 | 2.46 | 8.67 | 0.23 | 3.6 | THA |
| J232917.64-675000.6 | 21.82 | 73.64 | -68.04 | 3.03 | 10.54 | 0.34 | 3.95 | THA |
| J234243.45-622457.1 | 22.94 | 81.02 | -70.55 | 3.35 | 11.44 | 0.43 | 4.22 | THA |
| J234326.88-344658.5 | 26.03 | 91.78 | -76.92 | 2.2 | 8.09 | 0.25 | 3.38 | THA |

TABLE 14 Spectroscopic Indicators of Stellar Magnetic Activity

| WISE Desig | $\chi$ | $L_{H\alpha}/L_{bol}$ | $L_{NUV}/L_{bol}$ | $L_X/L_{bol}$ | EW(Ca II H) [Å] | EW(Ca II K) [Å] |
|---|---|---|---|---|---|---|
| J000453.05-103220.0 | 2.34E-04 | -2.71 | -3.36 | | -18.5 | -17.1 |
| J001527.62-641455.2 | 2.25E-04 | -3.13 | -3.48 | -3.07 | -11.5 | -7.9 |
| J001536.79-294601.2 | 2.36E-04 | -2.97 | -3.46 | | | |
| J001552.28-280749.4 | 2.24E-04 | -3.45 | -3.54 | | | |
| J001555.65-613752.2 | 2.35E-04 | -2.86 | -3.39 | | | |

| | | | | | | | |
|---|---|---|---|---|---|---|---|
| J001709.96+185711.8 | N | 9.55E-03 | -2.00 | -3.18 | | | |
| J001709.96+185711.8 | S | 9.55E-03 | -2.08 | -3.18 | | | |
| J001723.69-664512.4 | | 2.28E-04 | -2.92 | -3.48 | -3.05 | | |
| J002101.27-134230.7 | | 6.68E-03 | -2.46 | -3.65 | | | |
| J003057.97-655006.4 | | 2.32E-04 | -3.15 | -3.41 | -3.08 | | |
| J003234.86+072926.4 | | 1.22E-04 | -3.010 | -3.41 | -2.95 | | |
| J003903.51+133016.0 | | 1.89E-07 | -5.77 | -3.350 | | | |
| J004210.98-425254.8 | | 2.27E-04 | -3.25 | -3.40 | -3.23 | | |
| J004524.84-775207.5 | | 2.26E-04 | | -3.51 | | -1.9 | -1.3 |
| J004528.25-513734.4 | B | 2.30E-04 | -3.70 | -3.88 | | -4.2 | -5.5 |
| J004528.25-513734.4 | A | 2.30E-04 | | -3.88 | | -9.5 | -14.0 |
| J004826.70-184720.7 | | 2.28E-04 | -2.72 | | | -20.4 | -24.8 |
| J005633.96-225545.4 | | 2.27E-04 | -3.33 | -3.57 | | | |
| J010047.97+025029.0 | | 1.20E-10 | -9.17 | -3.19 | | -17.8 | -16.7 |
| J010126.59+463832.6 | | 4.95E-02 | | -2.97 | | | |
| J010243.86-623534.8 | | 2.32E-04 | -3.12 | -3.66 | | | |
| J010251.05+185653.7 | | 4.32E-05 | -3.49 | -3.47 | -3.26 | -20.3 | -22.4 |
| J010629.32-122518.4 | | 2.26E-04 | -2.66 | -3.38 | | -10.4 | -7.0 |
| J010711.99-193536.4 | | 2.21E-04 | -3.27 | -3.42 | -3.07 | | |
| J011440.20+205712.9 | | 2.39E-03 | -2.51 | -3.37 | | | |
| J011846.91+125831.4 | | 8.00E-04 | -3.21 | -3.31 | | | |
| J012118.22-543425.1 | | 2.30E-04 | -3.80 | -3.86 | | -4.2 | -2.2 |
| J012245.24-631845.0 | | 0.00E+00 | | -3.34 | | lowSNR | lowSNR |
| J012332.89-411311.4 | | 2.29E-04 | -2.87 | -3.70 | | | |
| J012532.11-664602.6 | | 2.39E-04 | -2.73 | -3.34 | | | |
| J013110.69-760947.7 | | 2.30E-04 | -3.538 | -3.51 | | | |
| J014156.94-123821.6 | | 2.29E-04 | -3.58 | -3.69 | | | |
| J014431.99-460432.1 | | 2.30E-04 | -2.27 | | | | |
| J015057.01-584403.4 | | 2.22E-04 | -2.71 | -3.34 | | -14.2 | -8.1 |
| J015257.41+083326.3 | | 6.13E-05 | -3.14 | -3.49 | | | |
| J015350.81-145950.6 | | 5.89E-06 | -4.34 | -3.32 | -3.06 | | |
| J015455.24-295746.0 | | 2.17E-04 | -3.89 | -3.34 | | lowSNR | lowSNR |
| J020012.84-084052.4 | | 1.75E-04 | -3.16 | -3.39 | -3.06 | | |
| J020302.74+221606.8 | | 2.00E-05 | -4.66 | -3.805 | | | 2.5 |
| J020305.46-590146.6 | | 2.33E-04 | -3.02 | -3.15 | | | |
| J020805.55-474633.7 | | 2.31E-04 | -3.96 | -3.94 | | | |
| J021258.28-585118.3 | | 2.22E-04 | -3.15 | -3.25 | | | |
| J021330.24-465450.3 | | 2.26E-04 | -2.76 | -3.13 | -3.04 | lowSNR | lowSNR |
| J021935.52-455106.2 | W | | | | | | |
| J021935.52-455106.2 | E | | | | | | |
| J022240.88+305515.4 | | 1.60E-05 | -4.10 | -3.56 | | -13.6 | -14.5 |
| J022424.69-703321.2 | | 2.28E-04 | -3.16 | -3.41 | | | |
| J023005.14+284500.0 | | 2.40E-04 | -4.67 | -3.83 | | | |
| J023139.36+445638.1 | | 3.02E-07 | -5.63 | -3.42 | | -23.2 | -24.3 |
| J024552.65+052923.8 | | 1.60E-04 | -3.32 | -3.76 | | | |
| J024746.49-580427.4 | | 2.20E-04 | -3.16 | -3.40 | | -12.0 | -14.1 |
| J024852.67-340424.9 | | 2.20E-04 | -2.70 | -3.17 | -2.93 | | |
| J025154.17+222728.9 | | 2.32E-04 | -2.87 | -3.32 | -2.25 | | |
| J025913.40+203452.6 | | 7.93E-04 | -2.50 | -3.50 | | | |
| J030002.98+550652.4 | | 2.28E-04 | -3.59 | -3.49 | | -13.4 | -11.5 |
| J030251.62-191150.0 | | 2.35E-04 | -2.77 | | | | |
| J030444.10+220320.8 | | 2.37E-04 | -2.65 | -3.50 | | -31.7 | -29.9 |
| J030824.14+234554.2 | | 2.26E-04 | -3.35 | -3.45 | | | |
| J031650.45-350937.9 | | 2.29E-04 | -2.70 | -3.12 | | | |
| J032047.66-504133.0 | | 2.30E-04 | -3.76 | -3.76 | | | |
| J033235.82+284354.6 | | 2.29E-04 | -2.79 | -3.81 | -3.36 | | |
| J033431.66-350103.3 | | 2.28E-04 | -2.58 | -3.12 | | | |
| J033640.91+032918.3 | | 2.29E-04 | -2.55 | -3.46 | -3.48 | -9.6 | -10.3 |
| J034115.60-225307.8 | | 2.25E-04 | -3.20 | -3.33 | | -10.0 | -6.1 |
| J034116.16-225244.0 | | 2.23E-04 | -3.21 | -3.19 | -2.50 | | |
| J034236.95+221230.2 | | 2.30E-04 | -2.53 | -3.66 | | | |
| J034444.80+404150.4 | | 5.37E-08 | -7.21 | -3.151 | -3.04 | lowSNR | lowSNR |
| J035100.83+141339.2 | | 2.23E-04 | -2.65 | -3.53 | -2.91 | | |
| J035134.51+072224.5 | | 2.25E-04 | -3.03 | -3.06 | | | |
| J035223.52-282619.6 | | 2.28E-04 | -2.94 | -3.33 | -3.06 | | |
| J035345.92-425018.0 | | 2.36E-04 | -3.24 | -3.58 | | lowSNR | lowSNR |
| J035716.56-271245.5 | | 2.27E-04 | | -3.65 | -3.73 | | |
| J035733.95+244510.2 | | 2.24E-04 | -3.41 | -3.36 | | -8.4 | -10.1 |
| J035829.67-432517.2 | | 2.26E-04 | -2.72 | -3.25 | | | |
| J040539.68-401410.5 | | 2.36E-04 | -2.69 | -3.49 | | lowSNR | lowSNR |
| J040649.38-450936.3 | | 2.28E-04 | -2.89 | -3.43 | | -6.9 | -4.6 |
| J040711.50-291834.3 | | 2.21E-04 | -3.25 | -3.30 | -3.12 | | |
| J040743.83-682511.0 | | 2.26E-04 | -3.39 | -2.93 | | | |
| J040809.80-611904.3 | | 2.24E-04 | -3.93 | -3.95 | | lowSNR | lowSNR |
| J040827.01-784446.7 | | 2.22E-04 | -3.31 | -3.19 | -3.19 | -10.8 | -6.7 |
| J041050.04-023954.4 | | 2.30E-04 | -3.26 | -3.24 | | -8.9 | -7.2 |
| J041255.78-141859.2 | | 2.22E-04 | -3.344 | -3.40 | -3.28 | lowSNR | -13.6 |
| J041336.14-441332.4 | | 2.33E-04 | -3.251 | -3.36 | | | |
| J041525.58-212214.5 | | 2.32E-04 | | -3.57 | | | |
| J041749.66+001145.4 | | 3.57E-04 | -2.93 | -3.04 | -3.03 | -2.6 | -2.3 |
| J041807.76+030826.0 | | 2.26E-04 | | -3.26 | | -9.1 | -1.7 |
| J042139.19-723355.7 | | 2.28E-04 | -3.00 | -3.65 | -2.92 | -1.8 | -8.1 |
| J042500.91-634309.8 | | 2.32E-04 | -2.86 | -3.28 | | | |
| J042736.03-231658.8 | | | | -3.46 | | | |
| J042739.33+171844.2 | | 2.83E-08 | | -3.60 | | | |
| J043213.46-285754.8 | | 2.30E-04 | -2.61 | -3.27 | | | |
| J043257.29+740659.3 | | 2.27E-04 | -3.16 | -3.45 | -3.03 | | -8.9 |
| J043657.44-161306.7 | | 2.26E-04 | -2.73 | -3.11 | -2.58 | -19.3 | -9.2 |
| J043726.87+185126.2 | | 0.00E+00 | | -3.02 | -2.79 | -11.4 | -8.6 |
| J043923.21+333149.0 | | 2.27E-04 | -2.60 | -3.15 | | | |
| J043939.24-050150.9 | B | 2.38E-04 | -3.02 | -2.97 | | | |
| J043939.24-050150.9 | A | 2.38E-04 | -3.02 | -2.97 | | | |
| J044036.23-380140.8 | | 2.12E-04 | | -3.87 | | | |
| J044120.81-194735.6 | | 2.34E-04 | -3.50 | -3.24 | -3.21 | -6.1 | -4.7 |
| J044154.44+091953.1 | | 9.71E-06 | -4.05 | -3.33 | | | |
| J044336.19-003401.8 | | 1.83E-05 | -3.97 | -3.77 | | | |
| J044349.19+742501.6 | | 2.22E-04 | -3.11 | -3.49 | | -14.6 | -14.5 |
| J044356.87+372302.7 | | 2.20E-04 | -2.82 | -3.40 | -3.16 | | |
| J044455.71+193605.3 | | 2.30E-04 | -3.72 | -3.31 | | -5.4 | -4.5 |
| J044530.77-285034.8 | | 2.27E-04 | -2.88 | -3.11 | | | |
| J044700.46-513440.4 | | 2.29E-04 | -3.17 | -3.64 | | -7.5 | -6.6 |
| J044721.05+280852.5 | | 2.30E-04 | -2.66 | -3.34 | -3.15 | -31.5 | -30.9 |
| J044800.86+143957.7 | AB | 2.23E-04 | | -2.82 | | | |
| J044802.59+143951.1 | A | | | | | | |
| J045114.41-601830.5 | | 2.26E-04 | -3.57 | -3.40 | | -6.0 | -13.3 |
| J045420.20-400009.9 | | 2.37E-04 | -2.99 | -3.36 | | | |
| J045651.47-311542.7 | | 2.30E-04 | -3.55 | -3.86 | | | |
| J050333.31-382135.6 | | 2.34E-04 | -2.71 | -3.45 | | | |
| J050610.44-582828.5 | | 2.27E-04 | -2.98 | -3.56 | -3.20 | -8.6 | -6.6 |
| J050827.31-210144.3 | | 2.28E-04 | -2.39 | -3.31 | -3.21 | | |
| J051026.38-325307.4 | | 2.12E-04 | | -3.34 | | | |
| J051255.82-212438.7 | | 2.29E-04 | -2.65 | -3.47 | | | |
| J051310.57-303147.7 | | 2.32E-04 | -3.34 | -3.00 | -3.11 | -11.9 | -6.6 |
| J051403.20-251703.8 | | 2.26E-04 | -2.95 | -3.03 | | -15.6 | -14.3 |
| J051650.66+022713.0 | | 2.31E-04 | -2.59 | -3.51 | | | |
| J051803.00-375721.2 | | 2.27E-04 | -3.05 | -3.05 | | | |
| J052419.14-160115.5 | | 2.36E-04 | -2.33 | -3.28 | -3.18 | | |
| J052535.85-250230.2 | | 2.31E-04 | -3.22 | -3.23 | | | |
| J052944.69-323914.1 | | 2.27E-04 | -2.86 | -3.52 | -3.14 | | |
| J053100.27+231218.3 | | 2.25E-04 | -2.68 | -3.38 | | | |
| J053311.32-291419.9 | | 2.27E-04 | -2.73 | -3.07 | -2.94 | | |
| J053328.01-425720.1 | A | | | -3.59 | | | |
| J053328.01-425720.1 | B | | | -3.59 | | | |
| J053747.56-424030.8 | | 2.31E-04 | -2.53 | -3.43 | | | |
| J053925.08-424521.0 | | 2.27E-04 | -3.28 | -3.19 | -2.84 | -12.4 | -7.1 |
| J054223.86-275803.3 | | 2.30E-04 | -2.74 | -3.21 | -2.32 | | |
| J054433.76-200515.5 | | 2.21E-04 | -3.32 | -3.25 | -2.91 | -9.1 | -9.6 |
| J054448.20-265047.4 | | 2.39E-04 | | -3.55 | | | |
| J054709.88-525626.1 | | 2.29E-04 | -3.36 | -3.14 | | -7.1 | -6.1 |
| J054719.52-335611.2 | | 2.34E-04 | -3.12 | -3.39 | | | |
| J055008.59+051153.2 | | 2.22E-04 | -3.42 | -3.44 | -3.27 | | |
| J055041.58+430451.8 | | 2.30E-04 | -3.25 | -3.19 | -3.00 | | |
| J055208.04+613436.6 | | 2.38E-04 | -3.81 | -3.55 | -3.36 | | |
| J055941.10-231909.4 | | 2.27E-04 | -2.62 | -3.24 | | | |
| J060156.10-164859.9 | B | | | -3.14 | | | |
| J060156.10-164859.9 | A | 2.32E-04 | -3.45 | -3.14 | | | |
| J060224.56-163450.0 | | | | -3.41 | -3.23 | -10.1 | -8.5 |
| J060329.60-260804.7 | | 2.41E-04 | -3.38 | -3.81 | | -2.0 | -3.9 |
| J061313.30-274205.6 | | 2.34E-04 | -2.94 | -3.51 | -2.99 | -13.1 | -4.9 |
| J061740.43-475957.2 | | 2.34E-04 | -2.59 | -2.99 | | | |
| J061851.01-383154.9 | | 2.29E-04 | -3.11 | -3.04 | | | |
| J062047.17-361948.2 | | 2.33E-04 | -3.41 | -2.84 | -3.10 | -11.8 | -10.7 |
| J062130.52-410559.1 | | 2.33E-04 | -2.59 | -3.34 | | | |
| J062407.62+310034.4 | | 2.24E-04 | -3.04 | | | -16.3 | -7.7 |
| J063001.84-192336.6 | | 6.36E-05 | -3.81 | -3.66 | | | |
| J065846.87+284258.9 | | 1.81E-04 | -5.44 | -3.22 | -3.08 | 6.7 | |
| J070657.72-535345.9 | | 2.27E-04 | -3.27 | -3.40 | -3.13 | | |
| J071036.50+171322.6 | | 2.30E-04 | -3.42 | -3.30 | | 4.9 | |
| J072641.52+185034.0 | | 2.33E-04 | -2.94 | -3.26 | | -16.6 | -16.3 |
| J072821.16+334511.6 | | 2.26E-04 | -2.80 | | | | |
| J072911.26-821214.3 | | 2.29E-04 | -3.36 | -3.61 | | -8.5 | -9.4 |
| J073138.47+455716.5 | | 1.73E-04 | -2.90 | -3.28 | -3.05 | 6.8 | |
| J075233.22-643630.5 | | 2.34E-04 | -3.34 | -3.21 | -3.39 | | |
| J075808.25-043647.5 | | 2.30E-04 | | -3.78 | | -3.9 | -2.8 |
| J075830.92+153013.4 | B | 1.73E-06 | -5.08 | -3.22 | | | |
| J075830.92+153013.4 | A | 1.73E-06 | -5.08 | -3.22 | | | |
| J080352.54+074346.7 | | 2.31E-02 | -2.94 | -3.87 | | 5.3 | |
| J080636.05-744424.6 | | | | -3.71 | -2.61 | -18.0 | -10.0 |
| J081443.62+465035.8 | | 2.26E-04 | -2.69 | -3.17 | | -28.5 | -27.8 |
| J081738.97-824328.8 | | 2.27E-04 | -2.74 | -3.28 | -2.94 | | |
| J082105.04-090853.8 | B | | | -3.59 | | -1.3 | -2.3 |
| J082105.04-090853.8 | A | 2.16E-04 | | -3.59 | | -4.0 | -8.0 |
| J082558.91+034019.5 | | 3.44E-05 | -3.76 | -3.37 | | -20.1 | -8.8 |
| J083528.87+181219.9 | | 9.48E-02 | -0.44 | -3.04 | | 9.5 | |
| J090227.87+584813.4 | | 8.17E-05 | -3.48 | -3.21 | | -15.0 | -14.9 |
| J092216.12+043423.3 | | 3.12E-03 | -2.07 | -3.89 | | | |
| J093212.63+335827.3 | | 7.10E-04 | -2.48 | -3.32 | -3.10 | | |
| J094317.05-245458.3 | | 2.37E-04 | | -3.670 | | -2.1 | -2.5 |
| J094508.15+714450.1 | | 2.26E-04 | -2.55 | -3.63 | | | |
| J100146.28+681204.1 | | 2.10E-06 | -5.95 | -3.69 | | -2.4 | -4.1 |
| J100230.94-281428.2 | | 2.34E-04 | -2.66 | -3.16 | -3.16 | -18.1 | -11.5 |
| J101543.44+660442.3 | | | | -3.16 | -3.08 | | |
| J101905.68-304920.3 | | 2.25E-04 | -3.34 | -3.33 | | -7.0 | -6.7 |
| J101917.57-443736.0 | | 2.26E-04 | -3.05 | -3.54 | | -14.1 | -11.9 |
| J102602.07-410553.8 | B | 2.20E-04 | -2.66 | -3.15 | -3.26 | -14.6 | -15.4 |
| J102602.07-410553.8 | A | 2.20E-04 | -2.66 | -3.15 | -3.26 | -22.0 | -17.9 |
| J102636.95+273838.4 | | 2.99E-05 | -4.18 | -3.24 | | 4.9 | 4.8 |
| J103016.11-354626.3 | | 2.32E-04 | -3.28 | -3.24 | | | |
| J103137.59-374915.9 | | 2.33E-04 | -2.87 | -3.85 | | | |
| J103557.17+285330.8 | | 7.37E-03 | -1.77 | -3.76 | -3.38 | | |
| J103952.70-353402.5 | | 2.42E-04 | -3.25 | -3.21 | | | |
| J104008.36-384352.1 | | 2.21E-04 | -3.048 | -3.09 | | | |
| J104044.98-255909.2 | | 2.25E-04 | -4.04 | -3.85 | | | |
| J104551.72-112615.4 | | 2.31E-04 | | -3.34 | | -20.4 | -21.3 |
| J105515.87-033538.2 | | 3.33E-04 | -3.28 | -3.45 | | | |
| J105518.12-475933.2 | | | | -3.00 | | -5.6 | -4.5 |
| J105524.25-472611.7 | A | | | -3.44 | | | |
| J105524.25-472611.7 | B | | | | | | |
| J105711.36+054454.2 | | 8.78E-03 | -1.68 | -3.16 | -2.93 | | |
| J105850.47-234620.8 | | 2.31E-04 | | -3.61 | | -9.9 | -17.1 |
| J110119.22+525222.9 | | 8.57E-04 | -3.51 | -3.63 | | | |
| J110335.71-302449.5 | | | | | | | |
| J110335.71-302449.5 | | | | -3.08 | | | |
| J110551.56-780520.7 | | 2.25E-04 | -4.27 | -3.70 | | | |
| J111052.06-725513.0 | | 2.31E-04 | -2.64 | -3.40 | | | |
| J111103.54-313459.0 | | 2.33E-04 | -2.94 | -3.48 | | | |
| J111128.13-265502.9 | | 2.37E-04 | -2.39 | | | -17.1 | -9.9 |
| J111229.74-461610.1 | | 2.24E-04 | | -3.392 | | | |
| J111309.15+300338.4 | | 4.32E-03 | -2.59 | -3.43 | | | |
| J111707.56-390951.3 | | 2.28E-04 | -3.56 | -3.14 | | | |
| J112047.03-273805.8 | | 2.29E-04 | -2.43 | | | 0.0 | 0.0 |
| J112105.43-384516.6 | | | | | | -18.2 | -12.7 |
| J112512.28-002438.2 | | 1.06E-04 | -3.65 | -3.22 | -2.62 | 4.8 | 5.0 |
| J112547.46-441027.4 | | 2.32E-04 | -2.61 | -3.31 | -2.28 | -18.1 | -7.0 |
| J112651.28-382455.5 | | 2.23E-04 | -2.78 | -3.44 | | -15.0 | -12.2 |
| J112816.27-261429.6 | | 2.12E-04 | -3.19 | -3.08 | | 0.0 | 0.0 |
| J112955.84+520213.2 | | 1.28E-02 | -1.27 | -3.80 | | 8.6 | |
| J113105.57+542913.5 | | 6.50E-05 | -3.68 | -3.57 | | 2.8 | |
| J113114.81-482628.0 | | 2.28E-04 | -2.70 | -2.89 | | | |
| J113120.31+132140.0 | | 5.95E-05 | -3.66 | -3.26 | | 9.6 | |
| J114623.01-523851.8 | | 2.28E-04 | -2.72 | -3.21 | | | |
| J114728.37+664402.7 | | 3.33E-08 | -6.56 | -3.21 | | | -17.2 |
| J115156.73+073125.7 | | | | -3.16 | -2.94 | | |
| J115438.73-503826.4 | | 2.27E-04 | -2.76 | -3.28 | | | |
| J115927.82-451019.3 | | 2.33E-04 | -2.61 | -3.29 | | -15.3 | -8.5 |
| J115949.51-424426.0 | | 2.34E-04 | -3.91 | -3.82 | | -4.3 | -7.4 |
| J115957.68-262234.1 | | 2.21E-04 | -3.19 | -2.97 | | -8.1 | -7.2 |
| J120001.54-173131.1 | | 2.16E-04 | -2.77 | -3.35 | | -23.0 | -13.9 |
| J120237.94-332840.4 | | 2.35E-04 | -2.63 | -3.60 | | -10.0 | -8.9 |
| J120647.40-192053.1 | | | | | | | |
| J120929.80-750540.2 | | 2.22E-04 | -2.81 | -2.75 | -3.18 | | |
| J121153.04+124912.9 | | 7.35E-03 | -1.88 | -3.49 | | | |
| J121341.59+323127.7 | B | | | -3.77 | | | |
| J121341.59+323127.7 | A | 2.15E-06 | -5.73 | -3.77 | | | |
| J121429.15-425814.8 | | 2.26E-04 | -3.09 | -2.84 | | | |
| J121511.25-025457.1 | | 3.68E-07 | | -3.48 | | | |
| J121558.37-753715.7 | | 2.34E-04 | -2.66 | | | | |
| J122643.99-122918.3 | | 2.08E-04 | -2.89 | -3.56 | | | |
| J122725.27-454006.6 | | 2.36E-04 | -3.60 | -2.90 | | -8.5 | -6.8 |
| J122813.57-431638.9 | | 2.33E-04 | -2.85 | -3.04 | | | |
| J123005.17-440236.1 | | 2.29E-04 | -2.77 | -3.45 | | -13.2 | -14.2 |
| J123234.07-414257.5 | | 2.31E-04 | -2.61 | -3.06 | | | |
| J123425.84-174544.4 | | 2.38E-04 | -3.44 | -2.88 | | -12.1 | -9.1 |
| J123704.99-441919.5 | | 2.30E-04 | -2.64 | -3.15 | | | |
| J124054.09-451625.4 | | 2.19E-04 | -3.09 | -3.68 | | -9.2 | -9.9 |
| J124612.32-384013.5 | | 2.29E-04 | -2.85 | -3.44 | | | |
| J124955.67-460737.3 | | 2.23E-04 | -3.41 | -3.29 | | -9.7 | -11.8 |
| J125049.12-423123.6 | | 2.29E-04 | -2.44 | | | -13.6 | -14.0 |
| J125326.99-350415.3 | | 2.28E-04 | -2.93 | -3.44 | | -21.3 | -12.5 |
| J125902.99-314517.9 | | 2.34E-04 | -2.91 | -3.61 | | -10.4 | -10.7 |
| J130501.18-331348.7 | | 2.21E-04 | -3.79 | -3.87 | | | |
| J130522.37-405701.2 | | 2.32E-04 | -2.97 | -3.12 | | -6.1 | -6.4 |
| J130530.31-405626.0 | | 2.31E-04 | -3.47 | -3.27 | -2.96 | -9.0 | -7.3 |
| J130618.16-342857.0 | | 2.30E-04 | -2.72 | -3.63 | | | |
| J130650.27-460956.1 | | 2.31E-04 | -3.49 | -2.92 | | | |
| J130731.03-173259.9 | | 2.37E-04 | -2.73 | -3.22 | | | |
| J131129.00-425241.9 | | 2.18E-04 | -3.13 | -3.51 | -2.94 | | |
| J132112.77-285405.1 | | 2.32E-04 | -3.42 | -3.07 | | | |
| J133238.94+305905.8 | | 8.37E-05 | -3.32 | -3.43 | | | -19.8 |
| J133509.40+503917.5 | | 4.88E-02 | -0.66 | -3.163 | -2.91 | | |
| J133901.87-214128.0 | | | | -3.40 | -2.57 | | |
| J134146.41+581519.2 | | 2.30E-04 | -3.27 | -3.83 | -3.51 | | |
| J134907.28+082335.8 | | 8.17E-02 | -1.69 | -3.81 | | | |
| J135145.65-374200.7 | | | | | | -13.0 | -9.1 |
| J135511.38+665207.0 | | 4.48E-05 | -4.26 | -3.55 | | 5.4 | 7.6 |
| J135913.33-292634.2 | | 2.26E-04 | -4.14 | -3.89 | | | |
| J140337.56-501047.9 | | | | -3.49 | | | |
| J141045.24+364149.8 | | 2.12E-08 | -7.11 | -3.61 | | | |
| J141332.23-145421.1 | | 7.54E-05 | -3.22 | -3.39 | -2.83 | | |
| J141510.77-252012.0 | | 2.28E-04 | -4.30 | -3.39 | | 4.6 | 5.4 |
| J141842.36+475514.9 | | 3.52E-04 | -3.70 | -3.27 | | | |
| J141903.13+645146.4 | | 4.80E-05 | -3.49 | -3.10 | -2.97 | | |
| J143517.80-342250.4 | | 2.26E-04 | -3.31 | -3.51 | | | |
| J143648.16+090856.5 | | 4.80E-05 | -3.55 | -3.40 | | | -11.5 |
| J143713.21-340921.1 | | | | -3.18 | | | |
| J143753.36-343917.8 | | 2.21E-04 | -3.07 | -3.58 | | | |
| J145014.12-305100.6 | | 2.22E-04 | -3.67 | -3.66 | | | 4.2 |
| J145731.11-305325.0 | | 2.18E-04 | -2.92 | -3.47 | | -7.3 | -8.5 |
| J145949.90+244521.9 | | 3.41E-05 | -3.78 | -3.48 | | | |
| J150119.48-200002.1 | | | | -3.25 | | | |
| J150230.94-224615.4 | | 2.23E-04 | -2.97 | -3.08 | | | |
| J150355.37-214643.1 | | 2.20E-04 | -3.07 | -3.35 | | | |
| J150601.66-240915.0 | | 2.31E-04 | -2.60 | -2.93 | | | |
| J150723.91+433353.6 | | 1.04E-04 | -3.81 | -3.74 | -2.96 | | |
| J150820.15-282916.6 | | 2.31E-04 | -2.82 | -3.59 | | | |
| J150836.69-294222.9 | | 2.28E-04 | -3.58 | -3.38 | | | |
| J150939.16-133212.4 | | 2.30E-04 | -2.70 | -3.31 | | | |
| J151212.18-255708.3 | | 2.24E-04 | -3.06 | -3.68 | | | |
| J151242.69-295148.0 | | 2.19E-04 | -3.09 | -3.41 | | | |
| J151411.31-253244.1 | | 2.37E-04 | -3.86 | -3.10 | | | |
| J152150.76-251412.1 | | 2.30E-04 | -2.78 | -3.38 | -2.98 | lowSNR | 14.0 |
| J153248.80-230812.4 | | 2.19E-04 | -3.32 | -3.39 | | | |
| J153549.35-065727.8 | | 2.18E-04 | -3.41 | -3.34 | | -9.6 | -8.4 |
| J154220.24+593653.0 | | 2.05E-05 | -3.18 | -3.18 | -2.74 | | |
| J154227.07-042717.1 | | 2.20E-04 | -3.54 | -3.51 | | | |
| J154349.42-364838.7 | | 2.32E-04 | -2.70 | -3.10 | | | |
| J154435.17+042307.5 | | 1.45E-03 | -2.17 | -3.41 | | | |
| J154656.43+013650.8 | | 4.18E-02 | -0.75 | -3.53 | | | |
| J155046.47+305406.9 | | 1.60E-06 | -5.02 | -3.27 | | | |
| J155515.35+081327.9 | | 1.00E-02 | -2.16 | -3.79 | | | |
| J155759.01-025905.8 | | 2.17E-04 | -4.96 | -3.69 | | | |
| J155947.24+440359.6 | | 2.23E-04 | -3.18 | -3.34 | -3.17 | 1.3 | 2.0 |
| J160116.86-345502.7 | | 2.31E-04 | -3.35 | | | | |
| J160549.19-311521.6 | | 2.19E-04 | -3.31 | -3.35 | | | |
| J160828.45-060734.6 | | 2.27E-04 | -2.99 | -3.48 | | | |
| J160954.85-305858.4 | | 2.18E-04 | -3.73 | -3.538 | | | |
| J161410.76-025328.8 | AB | | | -2.82 | | -3.4 | -3.4 |
| J161743.18+261815.2 | | 2.77E-04 | | -3.77 | | | |
| J162422.68+195922.0 | | 5.64E-09 | -8.21 | -3.35 | -2.88 | | |
| J162548.69-135912.0 | | | | -3.29 | | | |
| J162602.80-155954.5 | | 2.58E-04 | -3.04 | -3.17 | | | |
| J163051.34+472643.8 | | 5.65E-08 | -6.51 | -3.51 | | | -12.7 |
| J163632.90+635344.9 | | | | -3.79 | | 7.4 | 3.0 |
| J164539.37+702400.1 | | 1.96E-04 | -3.34 | -3.72 | | 12.0 | 12.5 |
| J170415.15-175552.5 | | 2.22E-04 | -3.50 | -3.52 | | | |
| J171038.44-210813.0 | | 2.31E-04 | -3.30 | -2.97 | -2.96 | | |
| J171117.68+124540.4 | | 2.32E-04 | -3.01 | -3.35 | -2.74 | | 5.8 |
| J171426.13-214845.0 | | 2.01E-04 | -3.43 | -3.764 | | | |
| J171441.70-220948.8 | | 2.19E-04 | -2.90 | -3.43 | -2.97 | | |
| J172130.71-150617.8 | | 2.33E-04 | -2.86 | -3.54 | | | |
| J172131.73-084212.3 | | 2.21E-04 | -2.96 | -3.39 | | | |
| J172309.67-095126.2 | | 2.16E-04 | -3.47 | -3.83 | | | |
| J172454.26+502633.0 | | 2.34E-04 | -4.02 | -3.79 | | | |
| J172615.23-031131.9 | | 2.30E-04 | -2.51 | -3.60 | -2.93 | | |
| J172951.38+093336.9 | | 2.25E-04 | -2.96 | -3.41 | | -15.9 | -17.1 |
| J173353.07+165511.7 | | 2.28E-04 | -2.63 | -3.43 | -3.28 | | |
| J173544.26-165209.9 | | 2.17E-04 | -3.66 | -3.71 | | | |
| J173623.80+061853.0 | | 6.38E-03 | -2.43 | -3.83 | | | |
| J173826.94-055628.0 | | 2.82E-04 | -3.17 | -3.38 | | | |
| J174203.85-032340.4 | | 2.32E-04 | -2.75 | -3.17 | | | |
| J174426.59-074925.3 | | 2.30E-04 | -3.44 | -3.46 | | | |
| J174439.27+483147.1 | | 1.21E-03 | -4.92 | -3.47 | | | |
| J174536.31-063215.3 | | 2.68E-04 | -3.09 | -2.94 | | | |
| J174735.31-033644.4 | | 2.74E-04 | -3.21 | -2.91 | | | |
| J174811.33-030510.2 | | 2.35E-04 | -3.03 | -3.61 | | -8.0 | -10.3 |
| J174936.01-010808.7 | | 5.02E-05 | -3.55 | -3.45 | | | |
| J175022.27-094457.8 | | 2.13E-04 | -3.54 | -3.81 | | | |
| J175839.30+155208.6 | | 2.50E-04 | -3.05 | -3.48 | | | |
| J175942.12+784942.1 | | 1.86E-07 | | | | | |
| J180508.62-015058.5 | | 2.61E-04 | -3.11 | -3.65 | | | |
| J180554.92-570431.3 | | 2.19E-04 | -2.84 | -3.09 | | | |
| J180658.07+161037.9 | | 2.31E-04 | -2.90 | -3.01 | | | |
| J180733.00+613153.6 | | 2.31E-04 | -3.12 | -3.15 | | -12.9 | -11.5 |
| J180929.71-543054.2 | | 2.30E-04 | -2.69 | -3.69 | | | |
| J181059.88-012322.4 | | 1.83E-04 | | -3.33 | | | |
| J181725.08+482202.8 | | 2.34E-04 | -3.28 | -3.64 | -3.64 | | |
| J182054.20+022101.5 | | 2.59E-04 | -3.23 | -3.68 | | | |
| J182905.79+002232.2 | | | | -2.93 | | | |
| J184204.85-555413.3 | | 2.33E-04 | -2.86 | -3.40 | -2.17 | | |
| J184206.97-555426.2 | | 2.22E-04 | -2.81 | -3.35 | -2.70 | -6.1 | -6.0 |
| J184536.02-205910.8 | | 2.43E-04 | -3.788 | -3.76 | | -2.2 | -4.1 |
| J190453.69-140406.0 | | 2.26E-04 | | -3.62 | | | |
| J191019.82-160534.8 | | 2.30E-04 | -3.66 | -3.61 | | | |
| J191036.02-650825.5 | | 2.22E-04 | -3.01 | -3.85 | | | |
| J191235.95+630904.7 | | 2.23E-04 | -3.18 | -3.40 | | -9.1 | -10.3 |
| J191500.80-284759.1 | | 2.30E-04 | -2.56 | -3.65 | | | |
| J191534.83-083019.9 | | 2.41E-04 | -3.50 | -2.99 | -2.99 | | |
| J191629.61-270707.2 | | 2.30E-04 | -2.86 | -3.65 | | | |
| J192240.05-061208.0 | | 2.99E-07 | -5.86 | -3.37 | -2.76 | | -2.6 |
| J192242.80-051553.8 | | 3.42E-05 | -3.68 | -3.03 | -2.81 | | |
| J192250.70-631058.6 | | 2.26E-04 | -2.87 | -3.42 | -2.82 | -5.9 | -4.7 |
| J192323.20+700738.3 | | 2.41E-04 | | -3.32 | | | |
| J192434.97-344240.0 | | | | -3.36 | | -2.6 | -2.6 |
| J192600.77-533127.6 | A | 2.25E-04 | -3.37 | -3.78 | | | |
| J192600.77-533127.6 | B | 2.25E-04 | -3.37 | -3.78 | | | |
| J192659.33-710923.8 | | 2.33E-04 | -3.11 | -3.18 | | -4.5 | -3.2 |
| J193052.51-545325.4 | | 2.26E-04 | -3.23 | -3.11 | | -11.9 | -11.7 |
| J193411.46-300925.3 | | 2.30E-04 | -2.52 | | | lowSNR | lowSNR |
| J193711.26-040126.7 | | 2.48E-04 | -3.77 | -3.37 | | -0.2 | -0.5 |
| J194309.89-601657.8 | | 2.29E-04 | -2.68 | -3.22 | | | |
| J194444.21-435903.0 | | 2.32E-04 | -2.80 | | | | |
| J194539.01+704445.9 | | 2.30E-04 | -3.31 | -3.20 | | | |
| J194714.54+640237.9 | | 2.32E-04 | -3.37 | -3.61 | -3.30 | | |
| J194816.54-272032.3 | | 2.24E-04 | -3.02 | -3.65 | | | |
| J194834.58-760546.9 | | 2.26E-04 | -3.57 | -3.39 | -2.86 | | |
| J195227.23-773529.4 | B | 2.29E-04 | -3.16 | -3.45 | | | |
| J195227.23-773529.4 | A | 2.29E-04 | -3.16 | -3.45 | | | |
| J195315.67+745948.9 | | 2.09E-04 | -4.28 | -3.59 | | | |
| J195331.72-070700.5 | | 2.27E-04 | -2.98 | -3.33 | | | |
| J195340.71+502458.2 | | 2.37E-04 | -2.79 | -3.36 | -2.68 | | -5.4 |
| J195602.95-320719.3 | | 2.30E-04 | -2.86 | -3.39 | -2.85 | | |
| J200137.19-331314.5 | | 2.18E-04 | -3.09 | -3.37 | -3.33 | -7.2 | -6.2 |
| J200311.61-243959.2 | | 2.35E-04 | -3.33 | -3.65 | | | |
| J200409.19-672511.7 | | 2.26E-04 | -2.93 | -3.19 | | | |
| J200423.80-270835.8 | | 2.22E-04 | -3.45 | -3.35 | | | |
| J200556.44-321659.7 | | | | -3.40 | | | |
| J200837.87-254526.2 | | 2.39E-04 | -2.57 | -3.20 | | | |
| J200853.72-351949.3 | | 2.24E-04 | -2.81 | -3.66 | | | |
| J201000.06-280141.6 | | 2.23E-04 | -2.52 | -3.16 | -3.08 | -25.8 | -13.5 |
| J201931.84-081754.3 | | 2.29E-04 | -3.41 | -3.31 | -3.10 | | 5.0 |
| J202505.36+835954.2 | | 2.05E-04 | -2.93 | -3.36 | | | |
| J202716.80-254022.8 | | 2.23E-04 | -3.63 | -3.78 | | | |
| J203023.10+711419.8 | | 2.12E-04 | -3.50 | -3.72 | | -5.9 | -7.4 |
| J203301.99-490312.6 | | 2.28E-04 | -2.76 | -3.74 | | | |
| J203337.63-255652.8 | | 2.37E-04 | -2.54 | -3.44 | | | |
| J204406.36-153042.3 | | 8.57E-05 | -5.46 | -3.59 | | | |
| J204714.59+110442.2 | | 2.35E-04 | -3.14 | -3.58 | | 8.7 | 8.8 |
| J205131.01-154857.6 | | 2.30E-04 | -2.01 | -2.82 | | | |
| J205136.27+240542.9 | | 8.89E-05 | -3.33 | | | | |
| J205832.99-482033.8 | | 2.29E-04 | -2.62 | | | | |
| J210131.13-224640.9 | | 2.22E-04 | -4.75 | -3.93 | | | |
| J210338.46+075330.3 | | 1.44E-02 | -1.78 | -2.91 | -3.19 | | |
| J210708.43-113506.0 | | | | | -3.57 | | |
| J210722.53-705613.4 | | 2.27E-04 | -3.06 | -3.25 | | -6.6 | -3.9 |
| J210736.82-130458.9 | | 2.26E-04 | -2.97 | -3.32 | -3.10 | | |
| J210957.48+032121.1 | | 1.08E-02 | -2.49 | | | | |
| J211004.67-192031.2 | | 2.34E-04 | -2.69 | -3.60 | -2.74 | -13.6 | -14.3 |
| J211005.41-191958.4 | | 2.22E-04 | -3.01 | -3.48 | -2.93 | -30.9 | -19.8 |
| J211031.49-271058.1 | A | 2.30E-04 | -2.15 | -3.49 | -3.02 | | |
| J211031.49-271058.1 | B | | | -3.49 | | | |
| J211635.34-600513.4 | | 2.28E-04 | -2.84 | -3.02 | | -1.7 | -0.6 |
| J212007.84-164548.2 | | 2.33E-04 | -2.81 | -3.56 | | | |
| J212128.89-665507.1 | | 2.28E-04 | | -3.81 | -3.04 | | |
| J212230.56-333855.2 | | 2.30E-04 | -2.38 | -3.27 | | | |
| J212750.60-684103.9 | | 2.33E-04 | -2.66 | -3.411 | | | |
| J213507.39+260719.4 | | 1.42E-04 | -4.57 | -3.75 | | | 3.6 |
| J213520.34-142917.9 | | 2.30E-04 | -2.84 | -3.54 | | | |
| J213644.54+670007.1 | | | | | | | |
| J213708.89-603606.4 | | 2.27E-04 | -2.77 | -3.17 | -2.73 | -7.4 | -7.3 |
| J213740.24+013713.2 | | 2.21E-04 | -2.57 | -3.22 | -3.04 | | |
| J213835.44-505111.0 | | 2.27E-04 | -2.80 | -3.65 | | | |
| J213847.58+050451.4 | | 2.41E-05 | -3.82 | -3.53 | | | |
| J214101.48+723026.7 | | 1.03E-04 | -3.54 | -3.42 | | -13.1 | -13.5 |
| J214126.66+204310.5 | | 2.05E-04 | -3.01 | -3.52 | -3.16 | | |
| J214414.73+321822.3 | | 2.23E-04 | -4.25 | -3.73 | | | |
| J214905.04-641304.8 | | 2.34E-04 | -2.74 | -3.49 | -2.62 | | |
| J215053.68-055318.9 | | 4.57E-04 | -3.00 | -3.35 | -2.99 | | |
| J215128.95-023814.9 | | 4.15E-07 | -5.86 | -3.50 | -3.34 | | |
| J215717.71-341834.0 | | 2.38E-04 | -3.41 | -3.63 | | | |
| J220216.29-421034.0 | | 2.17E-04 | -3.37 | -3.42 | -3.11 | -22.4 | -15.0 |
| J220254.57-644045.0 | | 2.24E-04 | -3.13 | -3.34 | | -10.6 | -8.7 |
| J220306.98-253826.6 | | 2.33E-04 | -2.77 | -3.37 | | -38.6 | -25.6 |
| J220730.16-691952.6 | | 2.31E-04 | -3.39 | -3.58 | | lowSNR | lowSNR |
| J220850.39+114412.7 | | 1.42E-03 | -2.02 | -3.43 | | | |
| J221217.17-681921.1 | | 2.22E-04 | | -3.80 | | | |
| J221559.00-014733.0 | | 1.29E-04 | -4.53 | -3.66 | | | |
| J221833.85-170253.2 | | 2.30E-04 | -4.36 | -3.80 | | | |
| J221842.70+332113.5 | | 2.25E-04 | -3.26 | -3.61 | -3.45 | -9.3 | -11.2 |
| J222024.21-072734.5 | | 1.65E-03 | -2.19 | -3.38 | -3.21 | | |
| J224111.08-684141.8 | | 2.17E-04 | -2.67 | -3.04 | | | |
| J224221.02-410357.2 | | 2.33E-04 | -2.91 | | | | |
| J224448.45-665003.9 | | 2.27E-04 | -2.63 | | | | |
| J224500.20-331527.2 | | 2.27E-04 | -2.73 | -3.65 | | | |
| J224634.82-735351.0 | | 2.31E-04 | -2.94 | -3.21 | | -4.8 | -4.9 |
| J225914.87+373639.3 | | 2.43E-04 | -2.72 | -3.32 | | | |
| J225934.89-070447.1 | | 5.30E-05 | -3.54 | -3.54 | | | |
| J230209.10-121522.0 | | 2.36E-04 | -2.78 | -3.27 | | -3.1 | -2.8 |
| J230327.73-211146.2 | | 2.20E-04 | -2.91 | -2.84 | | | |
| J230740.98+080359.7 | | 9.06E-02 | -1.39 | -3.75 | | | |
| J231021.75+685943.6 | | 2.43E-04 | | -3.79 | | | |
| J231211.37+150329.7 | | 1.10E-02 | -1.50 | -3.44 | | | |
| J231246.53-504924.8 | | 2.35E-04 | -2.76 | -3.42 | | lowSNR | lowSNR |
| J231457.86-633434.0 | B | 2.23E-04 | -3.44 | | | -10.0 | -2.4 |
| J231457.86-633434.0 | A | 2.23E-04 | -3.49 | | | -12.4 | -3.2 |
| J231543.66-140039.6 | | 2.27E-04 | -3.86 | -3.94 | | | |
| J231933.16-393924.3 | | | | | | | |
| J232008.15-634334.9 | | 2.26E-04 | -2.89 | | | | |
| J232151.23+005037.3 | | 1.23E-03 | -3.34 | -3.86 | | | |
| J232656.43+485720.9 | | 4.42E-04 | -3.49 | -2.89 | | | |
| J232857.75-680234.5 | | 2.28E-04 | -2.86 | -3.21 | -3.00 | | |
| J232904.42+032910.8 | | 1.47E-05 | -3.95 | -3.69 | | -43.2 | -29.3 |
| J232917.64-675000.6 | | 2.28E-04 | -2.77 | -3.41 | | | |
| J232959.47+022834.0 | | 2.64E-05 | -3.83 | | | | |
| J233647.87+001740.1 | | 2.73E-04 | -4.72 | -3.84 | | | |
| J234243.45-622457.1 | | 2.35E-04 | -2.58 | | | | |
| J234326.88-344658.5 | | 2.31E-04 | -3.28 | -3.56 | | | |
| J234333.91-192802.8 | | 2.25E-04 | -3.57 | -3.15 | -3.05 | -10.0 | -12.1 |
| J234347.83-125252.1 | | 2.33E-04 | -3.29 | -3.61 | | | |
| J234857.35+100929.3 | | 2.43E-04 | -4.66 | -3.83 | | | |
| J234924.87+185926.7 | | 1.86E-09 | -8.83 | -3.42 | | | |
| J234926.23+185912.4 | | 9.06E-02 | -1.01 | -3.49 | | -7.8 | -8.1 |
| J235250.70-160109.7 | | 2.35E-04 | -2.83 | -2.95 | | | -8.2 |

NOTES:
a) $\chi$ is a multiplicative factor used to convert from EW of H$\alpha$ to L$_{H\alpha}$ (see Section 4.4).